\documentclass[11pt]{article}
\usepackage{amssymb,amsfonts,latexsym,amsmath,theorem,mathrsfs,graphicx,xcolor,braket}
\usepackage[square,sort&compress,numbers]{natbib}
\usepackage{subcaption}
\usepackage{hyperref}
\usepackage[utf8]{inputenc}
\usepackage[english]{babel}

\usepackage{array}
\usepackage{booktabs}

\newcolumntype{M}[1]{>{\centering\arraybackslash}m{#1}}

\title{\textbf{Manipulation and trapping of magnetic skyrmions with domain walls in chiral magnetic thin films}}
\author{P.~Leask\footnote{palea@kth.se} \\ \\
\small \textit{Department of Physics, KTH Royal Institute of Technology, 10691 Stockholm, Sweden}}
\date{\today}

\begin{document}

\maketitle

\begin{abstract}
In this article we use chiral domain walls to manipulate and trap magnetic skyrmion quasi-particles in chiral magnetic materials such as ultrathin Co/Pt films.
The magnetic skyrmions can be orientated such that their interaction with domain walls is repulsive, allowing for them to be stored between domain walls.
In certain orientations, the skyrmion can be absorbed into the domain wall, forming a domain wall kink.
In other orientations, it can absorbed to form a kink-antikink domain wall with zero topological charge.
The magnetic skyrmion can even be orientated in such a way that it remains trapped but creates a topological defect-antidefect pair in one of the walls.
By altering the phases of one of the chiral domain walls, one can erase or store the magnetic skyrmion as an isolated soliton or as a domain wall skyrmion/antiskyrmion.
This adds a valuable asset to an ever growing toolbox of spintronic nano-devices.
\end{abstract}



\section{Introduction}
\label{sec: Introduction}

Topological magnetic textures such as magnetic skyrmions and chiral domain walls represent a promising direction for future spintronics \cite{Li_2021}.
Due to their nanoscale size \cite{Sampaio_2013} and ability to be controlled electrically \cite{Zhang_2023}, they are rapidly emerging as an attractive source for robust information carriers in spintronic storage devices \cite{Je_2021}.
In most cases, these topological spin textures are localized in multilayers and thin films \cite{Zhang_2020,Gruber_2022,Desautels_2019,Sharma_2022}.

The concept of a skyrmion was first proposed in the context of nuclear physics by Skyrme \cite{Skyrme_1961}.
Then it was later studied in easy-axis chiral magnetic materials with the Dzyaloshinskii--Moriya interaction (DMI) by Bogdanov and Hubert \cite{Bogdanov_1994}.
The magnetic spin textures in these systems can be regarded as localized quasi-particles due to their topological nature, and are characterized by an integer-valued topological winding number.
Alongside magnetic skyrmions, other exotic spin textures such as domain walls \cite{Hongo_2020,Cheng_2019,Ross_Nitta_2023,Kuchkin_2020} play a pivotal role in emerging magnetic nano-devices.
These domain walls have also been predicted to form networks and arise as magnetic skyrmion crystals \cite{Lee_2024}.
In particular, chiral magnetic domain walls offer novel functionalities in guiding and manipulating skyrmions by opposing the skyrmion Hall effect \cite{Song_2020}.
A similar affect happens when you place magnetic skyrmions in wedge-shaped nanostructures of cubic helimagnets \cite{Leonov_2023}.

In this article we focus on the application of chiral magnetic domain walls as a guide for manipulating and controlling magnetic skyrmion quasi-particles.
We are particularly interested in such interactions in ultrathin CoPt films where skyrmions can be manipulated and guided by parallel magnetic domain walls.
We show that under correct phase orientations of the domain walls and skyrmion, the skyrmion can remain stable between the domain walls or it can be stored in the domain wall.
The latter allows for better control as it restricts the movement of the skyrmion along the domain wall.



\section{The model}
\label{sec: The model}

The Hamiltonian for a Heisenberg magnet with easy-axis anisotropy is given by
\begin{equation}
    H = \sum_{\braket{i,j}} \left\{ J \vec{m}(\vec{r}_i) \cdot \vec{m}(\vec{r}_j) + D(\vec{r}_i, \vec{r}_j) \cdot \left[ \vec{m}(\vec{r}_i) \times \vec{m}(\vec{r}_j) \right] \right\} - \frac{\mu^2}{2} \sum_i m_z^2(\vec{r}_i),
\end{equation}
where $\vec{m} \in S^2$ is the unit magnetization vector.
For a ferromagnet we require $J<0$.
We consider the generalized form of the Dzyaloshinskii-Moriya interaction (DMI) vector \cite{Amari_Nitta_Ross_2024}
\begin{align}
    D(\vec{r}_i, \vec{r}_i \pm a \vec{e}_1) = \, & \mp \kappa \left( \cos\theta, -\sin\theta, 0 \right), \\
    D(\vec{r}_i, \vec{r}_i \pm a \vec{e}_2) = \, & \mp \kappa \left( \sin\theta, \cos\theta, 0 \right), 
\end{align}
where $a$ is the lattice constant and $\kappa,\theta$ are constants with $\theta$ differentiating between different
types of spin-orbit coupling in the underlying lattice spin system.
The Dresselhaus spin-orbit coupling gives rise to the DMI term with $\theta=0$, whereas the Rashba spin-orbit coupling results in a DMI term with $\theta=\pi/2$.

\begin{figure}[t]
    \centering
    \begin{subfigure}[b]{0.3\textwidth}
    \includegraphics[width=\textwidth]{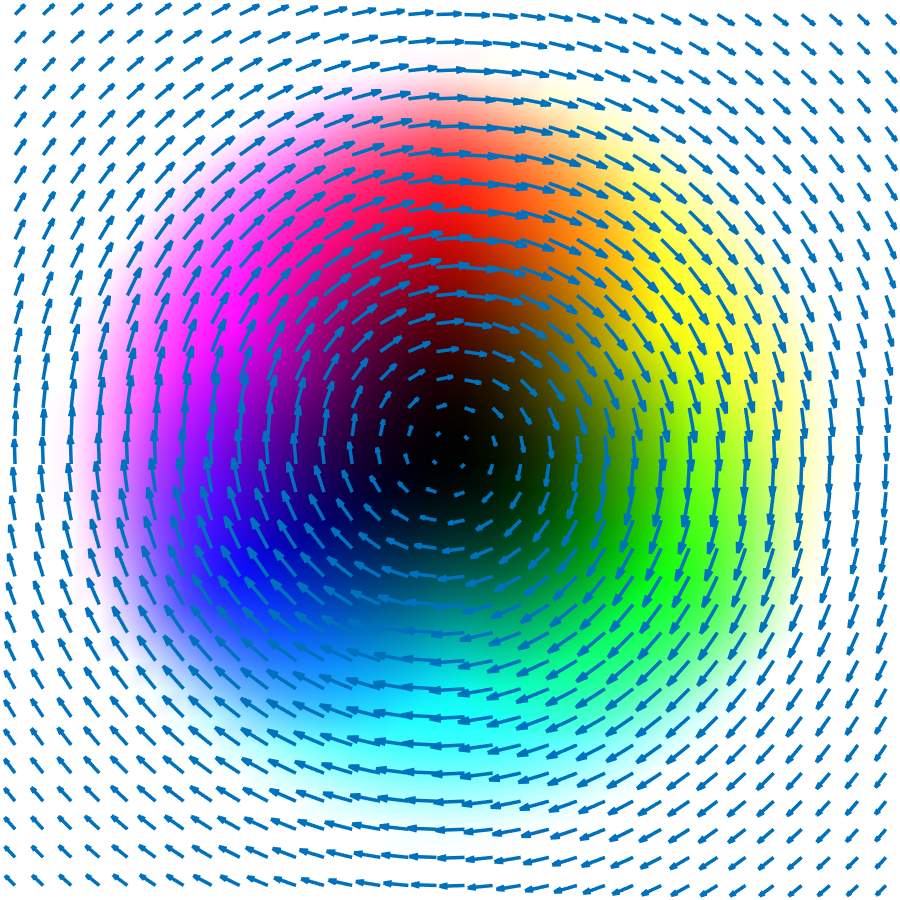}
    \caption{Bloch}
    \end{subfigure}
    ~
    \begin{subfigure}[b]{0.3\textwidth}
    \includegraphics[width=\textwidth]{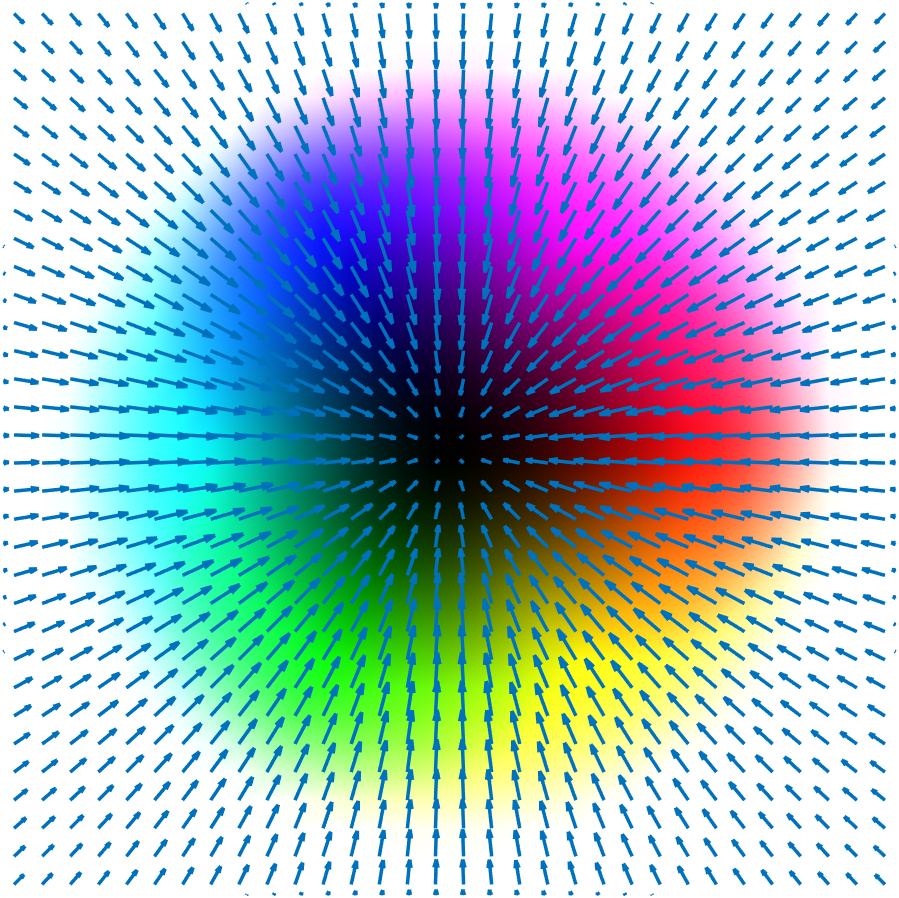}
    \caption{N\'eel}
    \end{subfigure}
    \caption{Magnetic skyrmions corresponding to the (a) Dresselhaus and (b) Rashba spin-orbit coupling DMI terms. The magnetization vector color scheme is described in detail in \cite{Leask_2022}.}
    \label{fig: Skyrmion types}
\end{figure}

In the continuum limit, the Hamiltonian becomes (up to a constant)
\begin{equation}
    \mathcal{H} = \frac{|J|}{2} \partial_i \vec{m} \cdot \partial_i \vec{m} + \frac{\mu^2}{2a}\left( 1 - m_z^2 \right) + \frac{\kappa}{a}
    \begin{cases}
        \vec{m} \cdot \left( \vec{\nabla} \times \vec{m} \right), & \textup{Dresselhaus} \, (\theta=0)\\
        \vec{m} \cdot \vec{\nabla}m_z - m_z \vec{\nabla} \cdot \vec{m}, & \textup{Rashba} \, (\theta=\pi/2)
    \end{cases}.
\end{equation}
It is known that the Dresselhaus DMI term gives rise to Bloch skyrmions, whereas the Rashba DMI term yields N\'eel skyrmions.
The resulting magnetic skyrmions for both of these DMI terms are shown in Fig.~\ref{fig: Skyrmion types}.

Now, consider the model as a gauged nonlinear $O(3)$ sigma model with an easy-axis potential and gauge covariant derivative $D_i \vec{m} = \partial_i \vec{m} + \vec{A}_i \times \vec{m}$ \cite{Schroers_2020},
\begin{align}
    \mathcal{H} = \, & \frac{1}{2}D_i \vec{m} \cdot D_i \vec{m} + \mu^2 \left( 1 - m_z^2 \right) \nonumber \\
    = \, & \frac{1}{2} \partial_i \vec{m} \cdot \partial_i \vec{m} + \partial_i \vec{m} \cdot \left( \vec{A}_i \times \vec{m} \right) + \frac{1}{2} \vec{A}_i \cdot \vec{A}_i - \frac{1}{2} \left( \vec{A}_i \cdot \vec{m} \right)^2 + \mu^2 \left( 1 - m_z^2 \right)
\end{align}
The type of spin-orbit coupling decides the choice of background gauge field, these are \cite{Amari_Nitta_2023}
\begin{equation}
    A_i^j =
    \begin{cases}
        -\kappa \delta_i^j, & \textup{Dresselhaus} \\
        -\kappa \epsilon_i^j, & \textup{Rashba}
    \end{cases}.
\end{equation}
This equivalently gives the same form of Hamiltonian as defined above.
Both spin-orbit coupling choices give the same non-abelian field strength $\vec{F}_{12} = \partial_1 \vec{A}_2 - \partial_2 \vec{A}_1 + \vec{A}_1 \times \vec{A}_2 = (0,0,\kappa^2)$,
and are related by a gauge transformation.
The Hamiltonian is invariant under this gauge transformation, so both descriptions are equivalent.

In this article we choose to employ the Dresselhaus description and express the ferromagnet energy as
\begin{equation}
    H = \int_{\mathbb{R}^2} \mathcal{H} \textup{d}^2x, \quad \mathcal{H} = \frac{|J|}{2} \left( \nabla \vec{m} \right)^2 + D \vec{m} \cdot \left( \vec{\nabla} \times \vec{m} \right) + K_m\left( 1 - m_z^2 \right),
\label{eq: Ferromagnetic Hamiltonian}
\end{equation}
where $D$ is the DMI coupling and $K_m$ the magnetic anisotropy parameter.
Our focus is on ultrathin CoPt films, with the magnetic parameters $|J|=30\,\textup{pJm}^{-1}$, $D=3\,\textup{mJm}^{-2}$ and $K_m=0.8\,\textup{MJm}^{-3}$ \cite{Sampaio_2013}.
Magnetic skyrmions are solutions of the field equations associated to this Hamiltonian,
\begin{equation}
    \frac{\delta \mathcal{H}}{\delta m_i} = - |J| \partial_j \partial_j m_i + 2D\epsilon_{ijk} \partial_j m_k - 2K_m m_z \delta_i^3 + \lambda m_i = 0,
\label{eq: Euler-Lagrange equation}
\end{equation}
where $\lambda$ is a Lagrange multiplier.
Throughout we impose the vacuum boundary condition $\vec{m}_\infty = (0,0,1)$, which compactifies the domain $\mathbb{R}^2 \cup \{\infty\} \cong S^2$.
So the map $\vec{m}: S^2\rightarrow S^2$ has an an associated topological charge given by
\begin{equation}
    Q = \frac{1}{4\pi} \int_{\mathbb{R}^2} \vec{m} \cdot \partial_1 \vec{m} \times \partial_2 \vec{m} \, \textup{d}^2x \in \pi_2(S^2) = \mathbb{Z}.
\end{equation}
This topological charge is useful for keeping track of when a skyrmion has been annihilated or an anti-skyrmion has formed.

As magnetic skyrmions are local energy minimizers, we must numerically relax the Hamiltonian \eqref{eq: Ferromagnetic Hamiltonian}.
This is achieved by discretising $H$ with a fourth order finite difference method, which gives a discrete approximation $H_{\textup{dis}}[\vec{m}]$ to the Hamiltonian $H[\vec{m}]$.
We employ a grid of $N^2$ lattice points with lattice spacing $h$.
Then we can regard the discretised Hamiltonian as a function $H_{\textup{dis}}:\mathcal{C} \rightarrow \mathbb{R}$, where the discretised configuration space is the manifold $\mathcal{C}=(S^2)^{N^2} \subset \mathbb{R}^{3N^2}$ \cite{Winyard_2020}.

To determine minimizers of the discretised Hamiltonian we use arrested Newton flow, which is an accelerated gradient descent based algorithm \cite{Gudnason_2020}.
This is carried out using highly parallelized custom CUDA C software.
The method is detailed as follows: starting from rest, we solve Newton's equations of motion for a particle on the discretised configuration space $\mathcal{C}$ with potential energy $H_{\textup{dis}}$.
That is, we are solving the system of second order ODEs
\begin{equation}
    \ddot{m}_i = -\frac{\delta H_{\textup{dis}}}{\delta m_i}[\vec{m}], \quad \vec{m}(0)=\vec{m},
\label{eq: Numerical minimization procedure - 2nd order ODEs}
\end{equation}
with the initial velocity $\vec{v}(0)=0$.
This problem can be reduced to a coupled system of first order ODEs, which we solve using a fourth order Runge--Kutta method.

The main advantage in implementing the arrested Newton flow algorithm is that the field will naturally relax to a local minimum.
After each time step $t \mapsto t + \delta t$, we check to see if the energy is increasing.
If $H_{\textup{dis}}(t + \delta t) > H_{\textup{dis}}(t)$, we take out all the kinetic energy in the system by setting $\vec{v}(t + \delta t)=0$ and restart the flow.
The flow then terminates when every component of the energy gradient $\delta H_{\textup{dis}}/\delta \vec{m}$ is zero to within a given tolerance (we have used $1e-5$).
All numerics carried out in this article were simulated on a $1024^2$ grid with $0.029$ lattice spacing.


\section{Magnetic skyrmions and domain walls}
\label{sec: Magnetic skyrmions and domain walls}

\begin{figure}[t]
    \centering
    \includegraphics[width=0.95\textwidth]{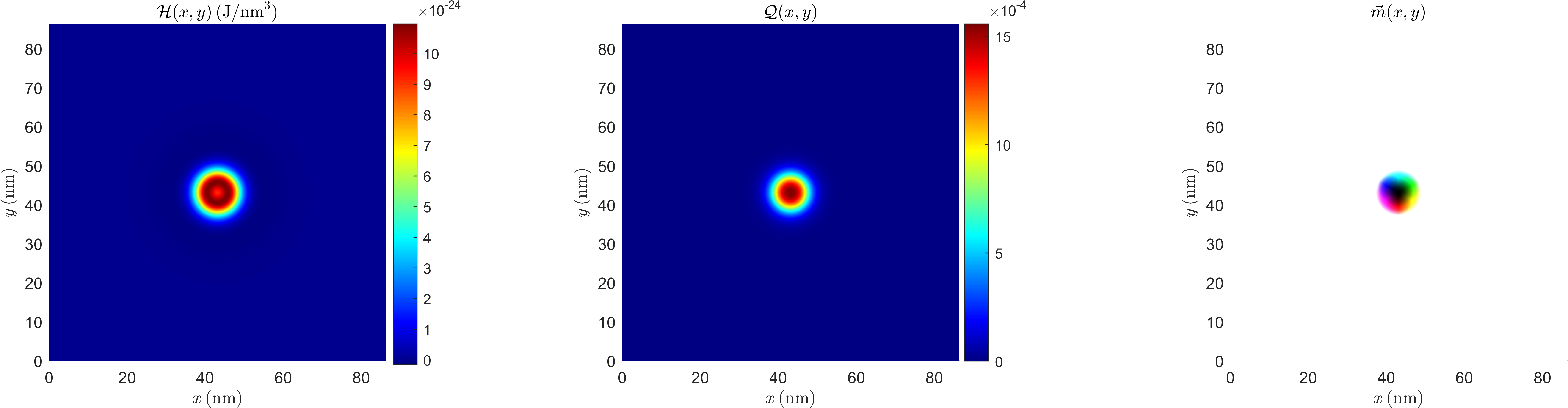}
    \caption{The axially symmetric $Q=1$ magnetic (Bloch) skyrmion in a CoPt thin-film showing the Hamiltonian density $\mathcal{H}$, topological charge density $\mathcal{Q}$ and the magnetization vector $\vec{m}$. The magnetic parameters are $|J|=30\,\textup{pJm}^{-1}$, $D=3\,\textup{mJm}^{-2}$ and $K_m=0.8\,\textup{MJm}^{-3}$.}
    \label{fig: Q=1 skyrmion}
\end{figure}

In order to describe magnetic skyrmions and domain walls, we parameterize the magnetization vector as
\begin{equation}
    \vec{m}= \left( \sin f \cos\theta, \sin f \sin\theta, \cos f \right),
\label{eq: hedgehog ansatz}
\end{equation}
where $f: \mathbb{R} \rightarrow \mathbb{R}$ is some suitably chosen profile function.
The charge $Q=1$ hedgehog skyrmion is axially symmetric with a monotonically decreasing radial profile function $f(r)$, satisfying $f(0)=\pi$ and $f(\infty)=0$.
Here, $r$ and $\theta$ are polar coordinates in the plane.
Upon substitution of the hedgehog ansatz \eqref{eq: hedgehog ansatz} into the Hamiltonian \eqref{eq: Ferromagnetic Hamiltonian}, the resulting energy functional can only be minimized numerically with some appropriate initial profile function (we use $f(r) = \pi \exp(-r)$).
Upon numerical relaxation, the minimal energy $Q=1$ magnetic skyrmion is plotted in Fig.~\ref{fig: Q=1 skyrmion}.

\begin{figure}[t]
    \centering
    \includegraphics[width=0.8\textwidth]{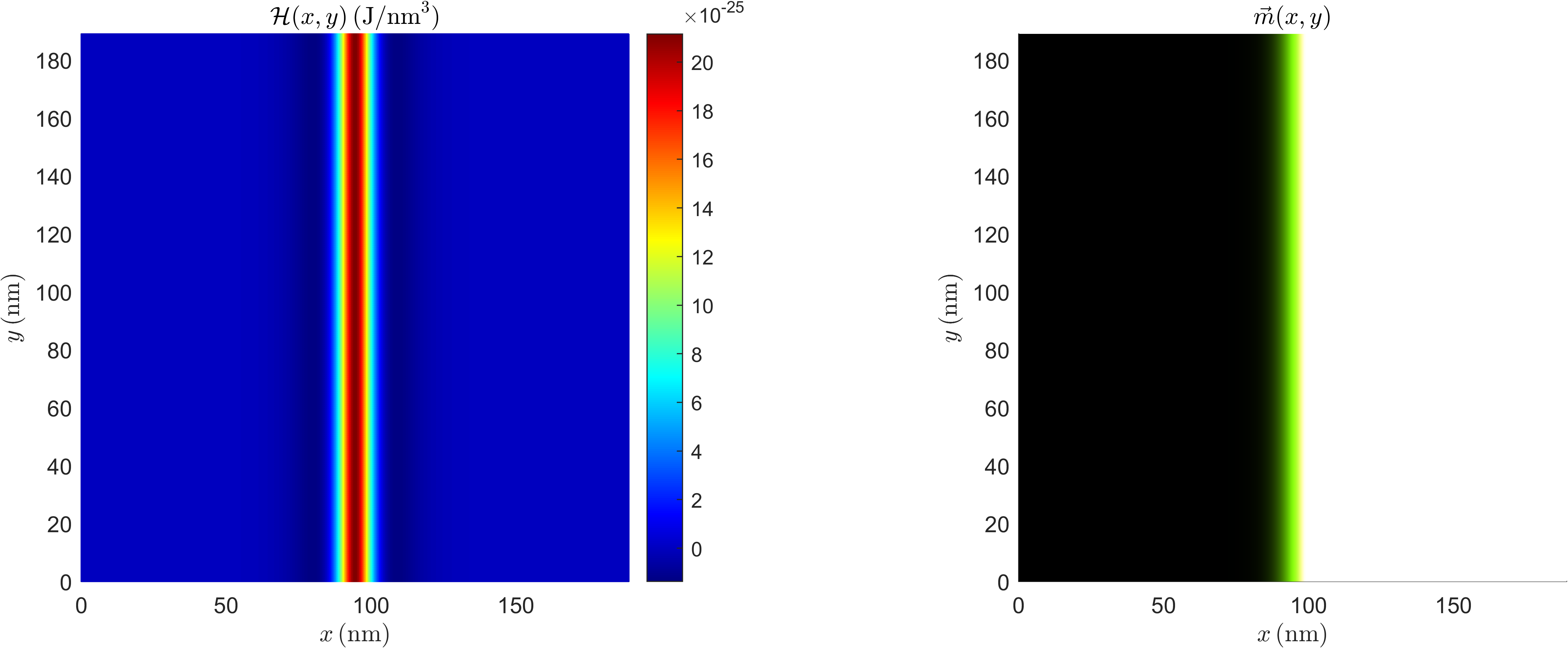}
    \caption{An example of a magnetic domain wall skyrmion in a CoFeB thinfilm showing the Hamiltonian density $\mathcal{H}$ and the magnetization vector $\vec{m}$. The magnetic parameters are $|J|=40\,\textup{pJm}^{-1}$, $D=4\,\textup{mJm}^{-2}$ and $K_m=0.8\,\textup{MJm}^{-3}$ \cite{Tomasello_2014}}
    \label{fig: Domain wall skyrmion}
\end{figure}

On the other hand, for a domain wall parallel to the $y$-axis, we set $\theta=\phi=\textup{constant}$ and let the profile function be $f=f(x)$ with boundary conditions $f(-\infty)=\pi$ and $f(\infty)=0$.
This gives us a domain wall with the boundary conditions $\vec{m}_{\textup{DW}} = (0,0,\pm1)$ as $x \rightarrow \pm\infty$\footnote{In \cite{Amari_Nitta_Ross_2024}, this particular domain wall orientation is referred to as an antidomain, but we will simply refer to it as a domain wall here.}.
Substituting this domain wall ansatz into the Hamiltonian \eqref{eq: Ferromagnetic Hamiltonian} yields the chiral sine-Gordon model \cite{Amari_Nitta_Ross_2024}
\begin{equation}
    \mathcal{H}_{\textup{DW}} = \frac{|J|}{2} (\partial_x f)^2 + D \sin\phi (\partial_x f) + K_m \sin^2f.
\label{eq: Chiral sine-Gordon energy}
\end{equation}
The Euler--Lagrange field equation associated to this reduced Hamiltonian is the sine-Gordon equation equation
\begin{equation}
    \partial_{xx}f = \frac{K_m}{|J|} \sin2f,
\end{equation}
which has the well-known sine-Gordon kink solution
\begin{equation}
    f(x) = 2\arctan\left[ \exp\left( kx + X \right) \right], \quad k = \sqrt{\frac{2K_m}{|J|}},
\end{equation}
where $X$ is a moduli parameter corresponding to translation of the domain wall \cite{Nitta_Ross_2023}.
An example domain wall solution is shown in Fig.~\ref{fig: Domain wall skyrmion}.

To be able to describe the interaction of magnetic skyrmions with domain walls, it will be convenient to work in the $\mathbb{C}P^1$ formulation of the model.
That is, we define the stereographic projection of the magnetization vector $\vec{m}$ by the map
\begin{equation}
    u: S^2 \rightarrow \mathbb{C}P^1, \quad u[\vec{m}] = \frac{m_1 + im_2}{1 + m_3}.
\label{eq: Riemann sphere coordinate}
\end{equation}
In this formalism, the $Q=1$ hedgehog skyrmion is given by
\begin{equation}
    u_{\textup{Sk}} = \tan \left( \frac{f(r)}{2} \right) e^{i(\theta+\phi)}
\label{eq: CP1 hedgehog}
\end{equation}
and the $Q=0$ domain wall by \cite{Nitta_Ross_2023}
\begin{equation}
    u_{\textup{DW}} = e^{\pm k(x-X) + i\chi},
\label{eq: CP1 domain wall}
\end{equation}
where $\phi, \chi \in [0, 2\pi)$ are moduli parameters controlling the phase of the skyrmion and domain wall.
To combine solutions we can approximate the solution as a linear superposition of static solutions, $u = \sum_i u_i$.
Then, in terms of the stereographic coordinate $u$, the magnetization vector is simply
\begin{equation}
    \vec{m} = \frac{1}{1 + |u|^2}\left( u + \bar{u}, i(\bar{u} - u), 1 - |u|^2 \right).
\label{eq: Initial configurations - Skyrme field}
\end{equation}


\section{Trapping and manipulating magnetic skyrmions with domain walls}
\label{sec: Trapping magnetic skyrmions with domain walls}

\begin{figure}[t]
    \centering
    \begin{subfigure}[b]{0.3\textwidth}
    \includegraphics[width=\textwidth]{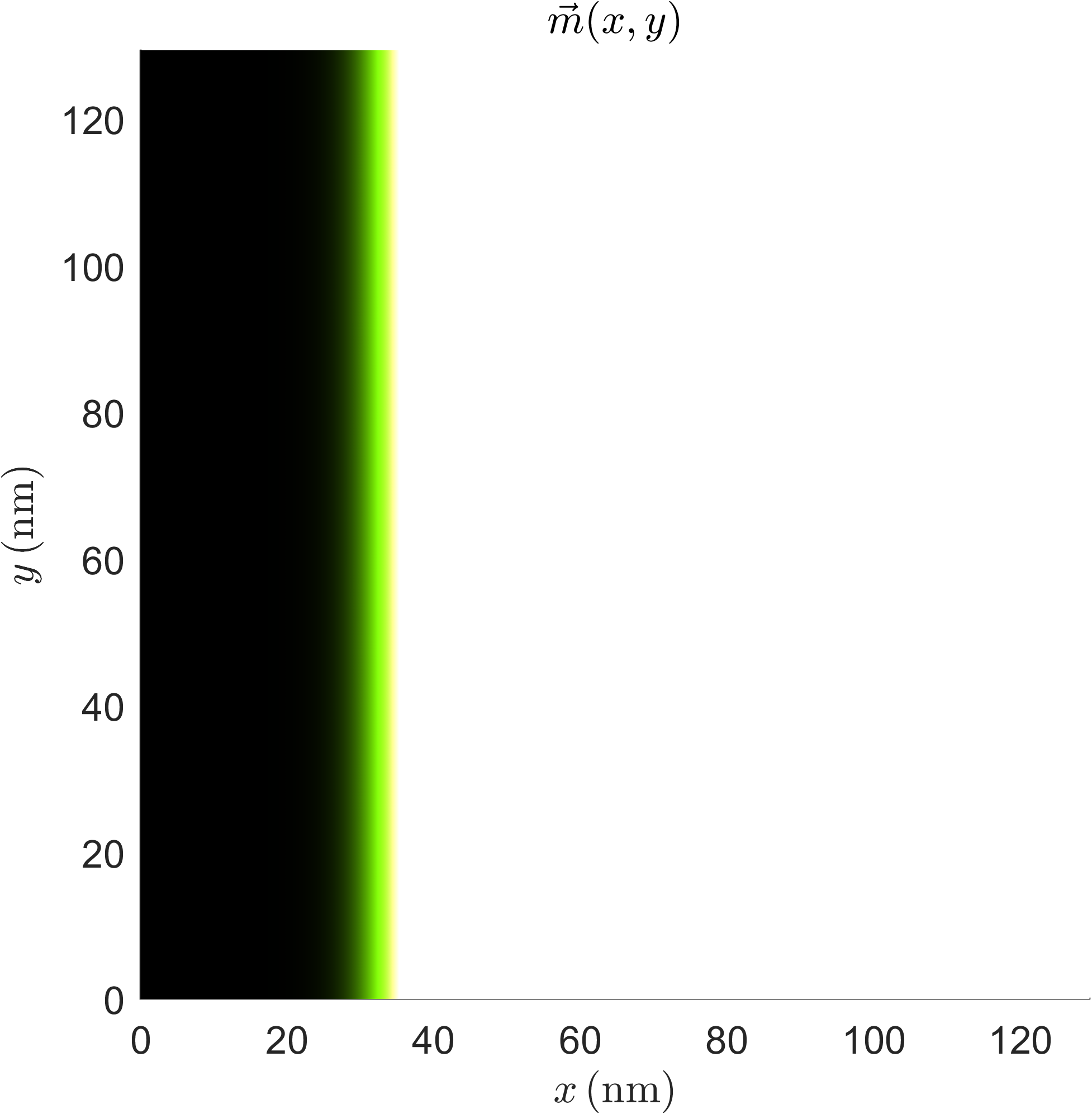}
    \caption{$Q=0$}
    \end{subfigure}
    ~
    \begin{subfigure}[b]{0.3\textwidth}
    \includegraphics[width=\textwidth]{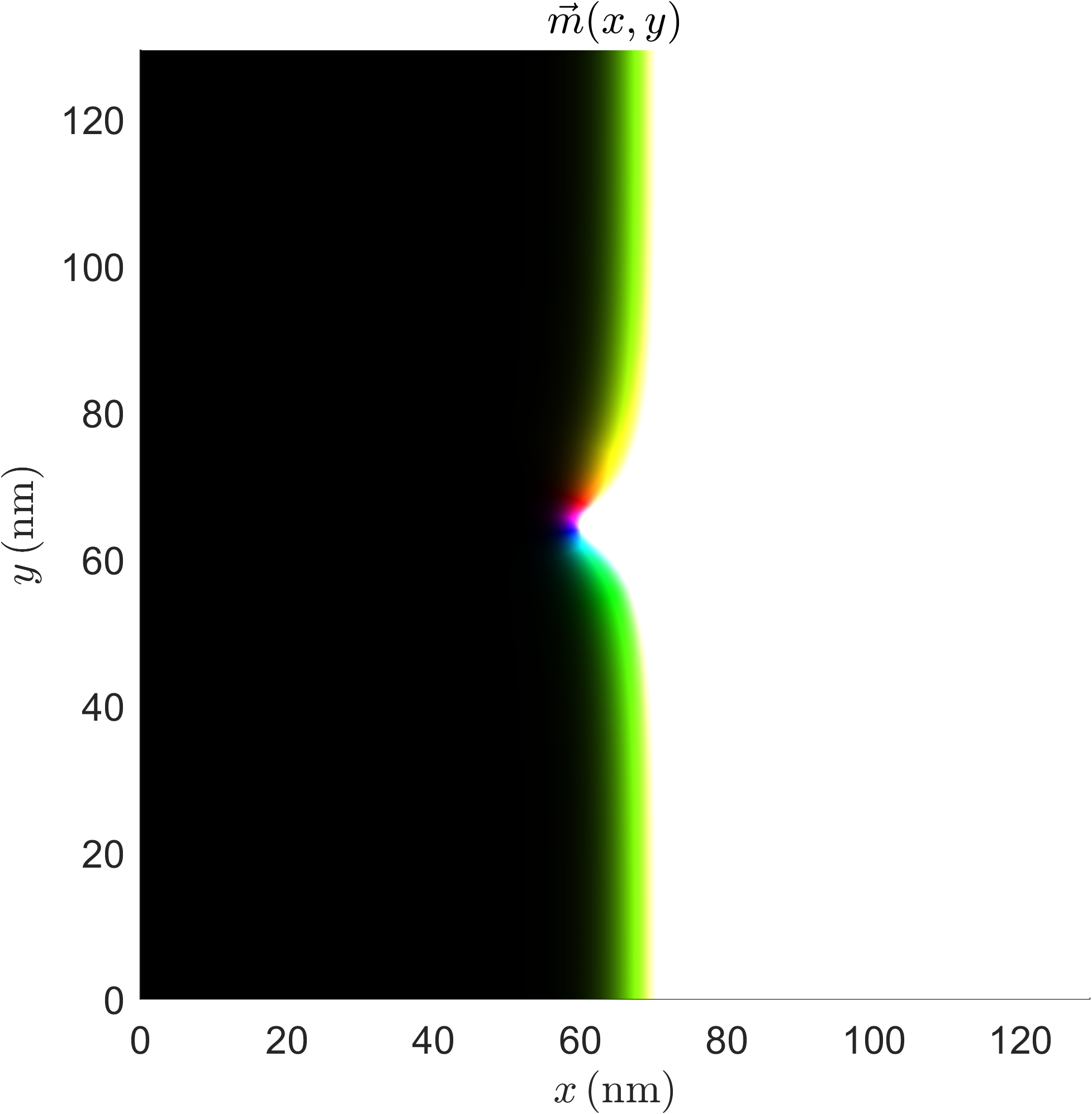}
    \caption{$Q=+1$}
    \end{subfigure}
        ~
    \begin{subfigure}[b]{0.3\textwidth}
    \includegraphics[width=\textwidth]{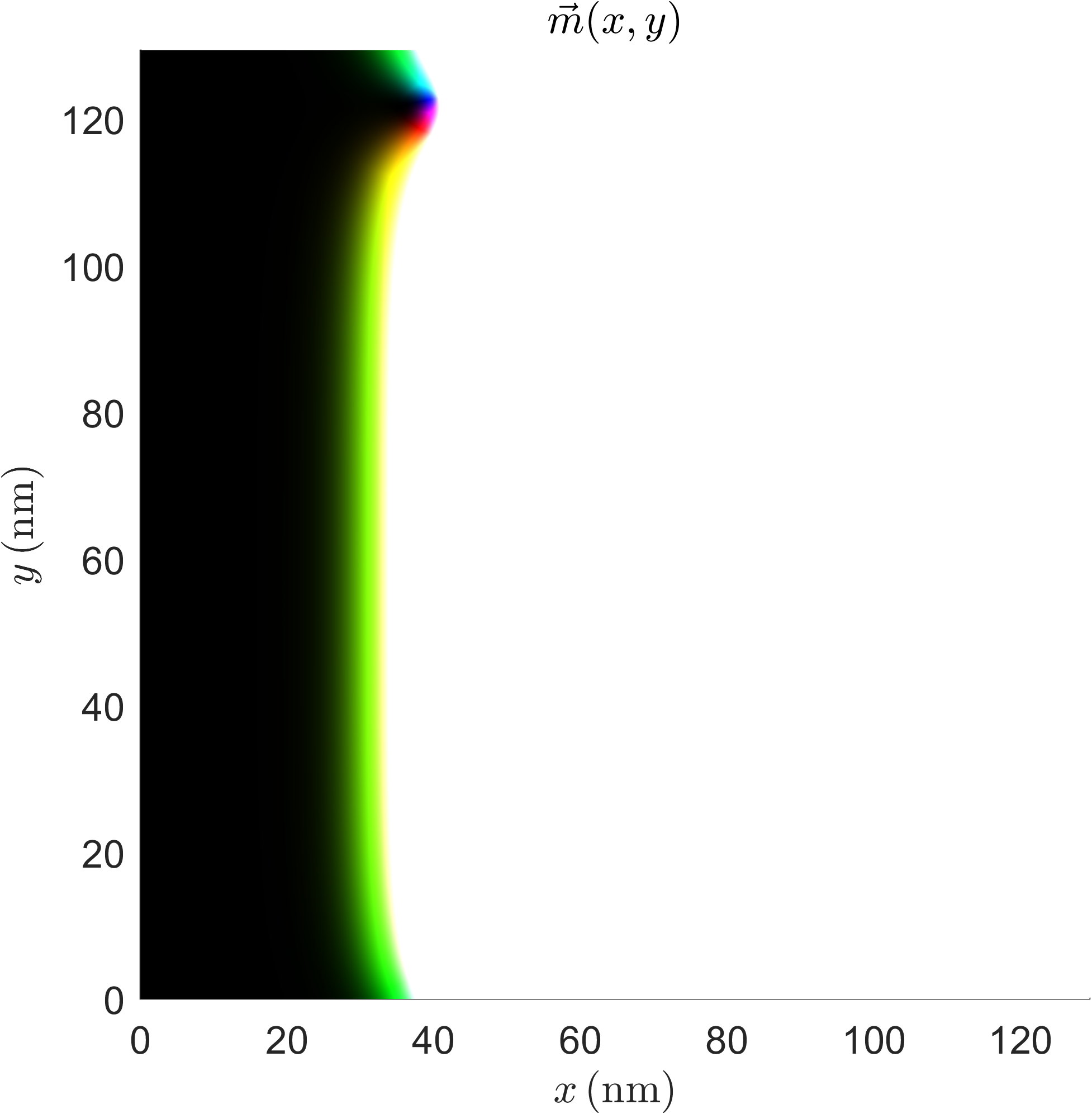}
    \caption{$Q=-1$}
    \end{subfigure}
    \\
    \begin{subfigure}[b]{0.3\textwidth}
    \includegraphics[width=\textwidth]{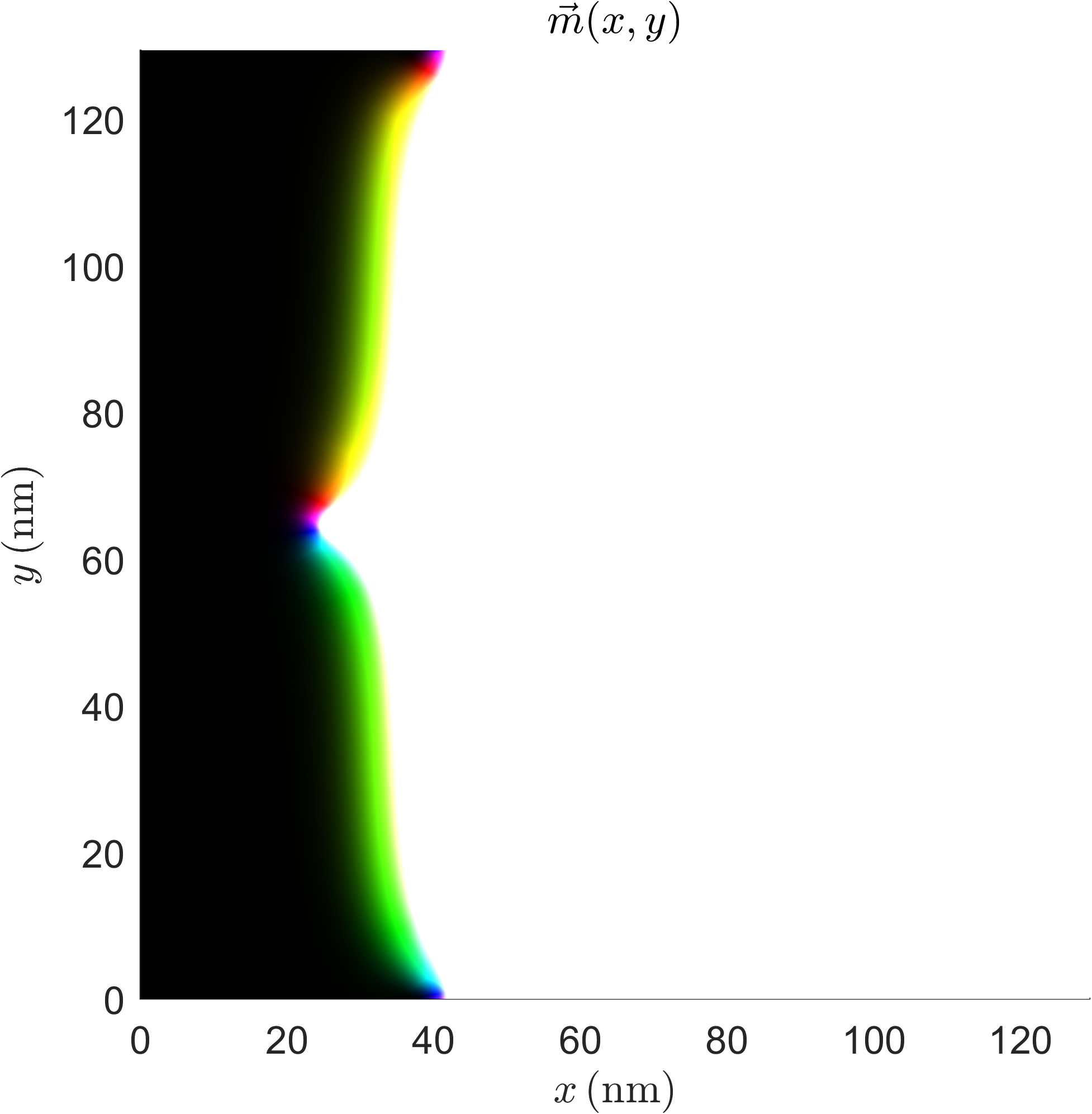}
    \caption{$Q=0$}
    \end{subfigure}
    ~
    \begin{subfigure}[b]{0.3\textwidth}
    \includegraphics[width=\textwidth]{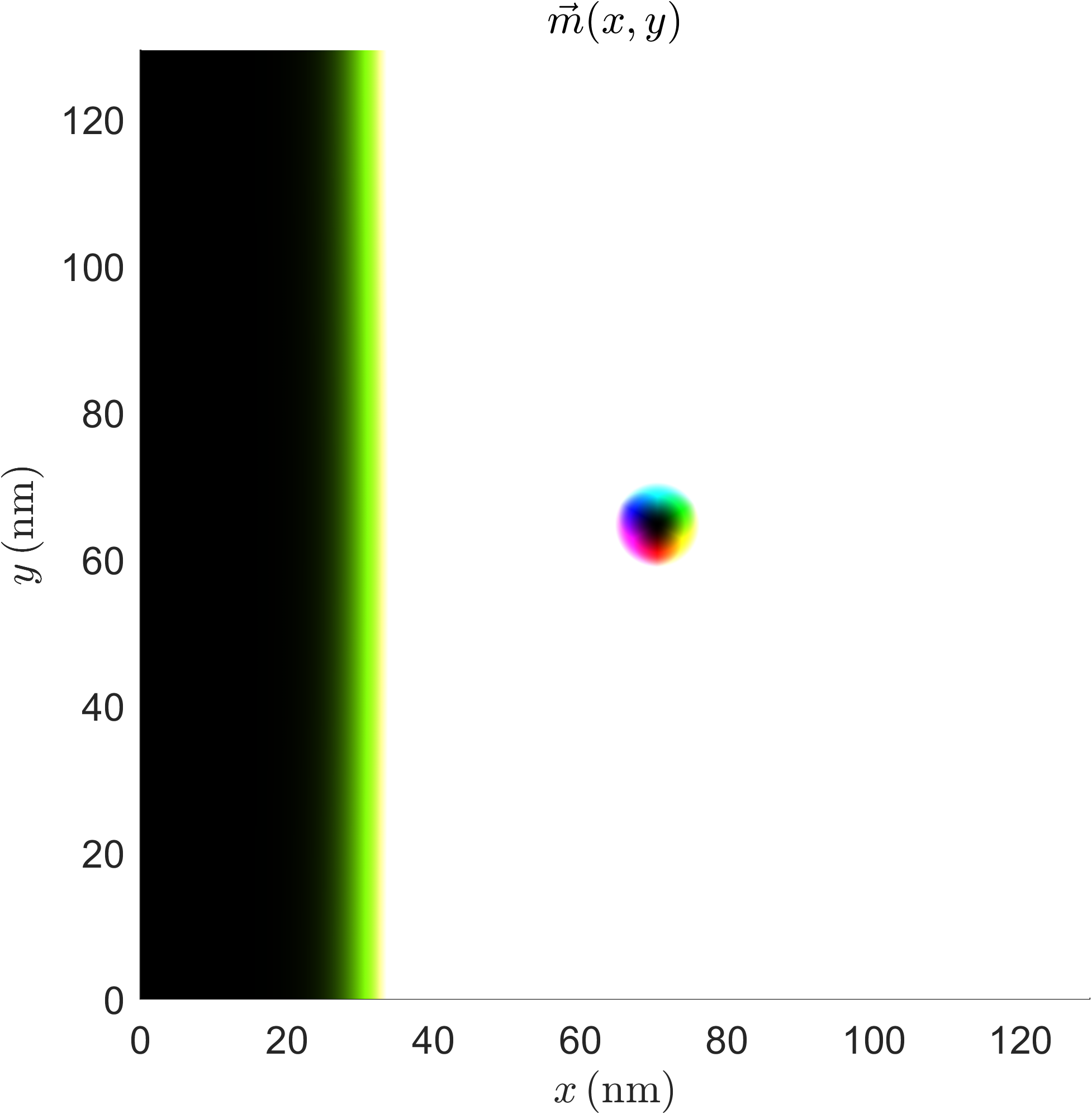}
    \caption{$Q=+1$}
    \end{subfigure}
        ~
    \begin{subfigure}[b]{0.3\textwidth}
    \includegraphics[width=\textwidth]{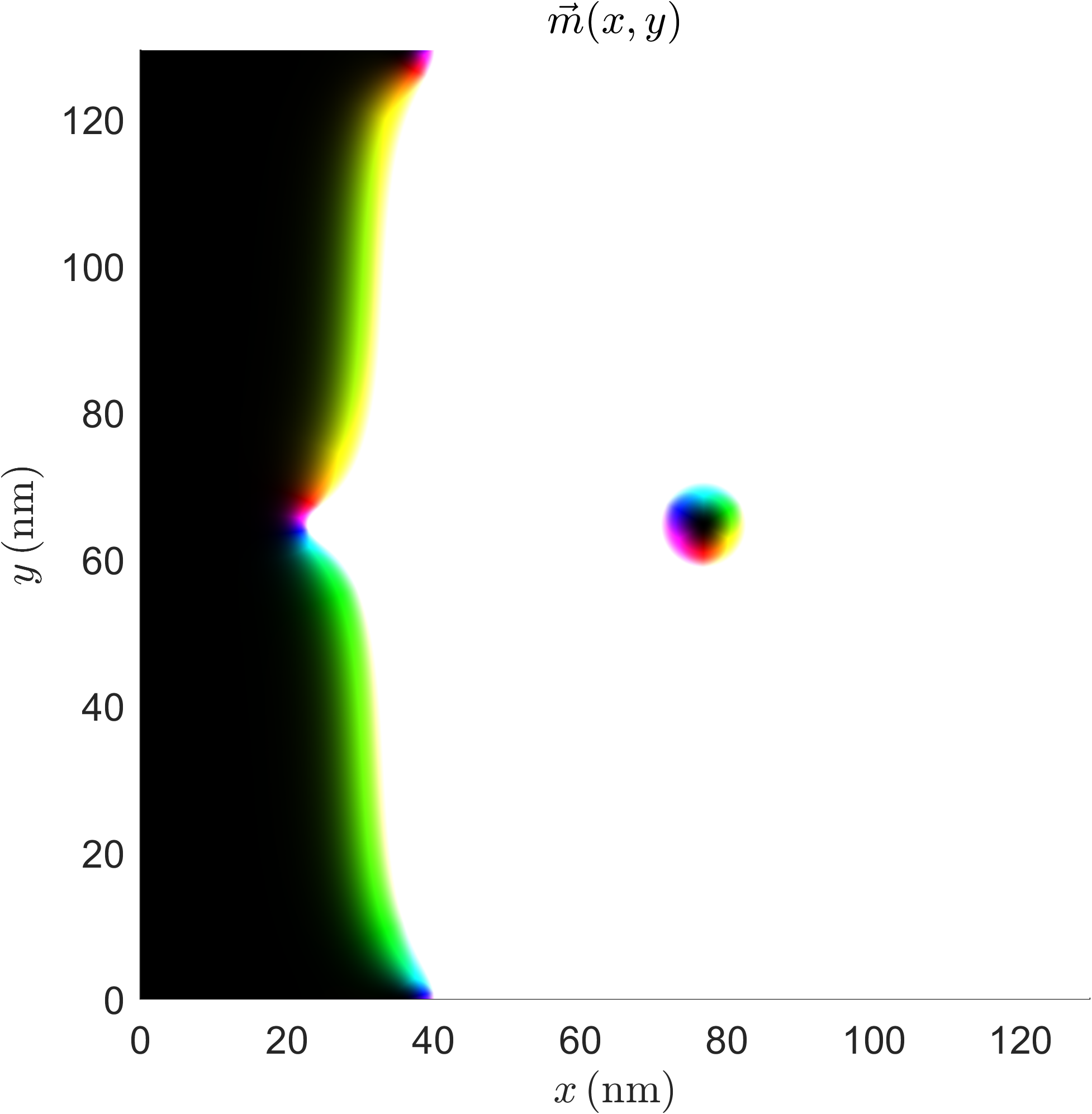}
    \caption{$Q=+1$}
    \end{subfigure}
    \caption{Different possible outcomes for the interaction of a magnetic skyrmion with a single domain wall. First is the (a) annihilation of the skyrmion into the domain wall, leaving only the domain wall. Second is the (b) absorption of the skyrmion into the domain wall, creating a domain wall kink. There is also the possibility of (c) absorption creating a domain wall antikink. Then there is (d) annihilation of the skyrmion but producing a domain wall kink-antikink. The last possibilities yield metastable skyrmions in the presence of (e) a domain wall and (f) a domain wall kink-antikink.}
    \label{fig: Single domain wall interaction}
\end{figure}

Our main interest lies in utilizing domain walls to manipulate and trap magnetic skyrmions in such a way that they could be easily guided by the control of chiral domain walls \cite{Song_2020}.
Before we consider how a magnetic skyrmion interacts with two domain walls, we first consider its interaction with a single domain wall.
There are a number of possibilities and the ones we have found are shown in Fig.~\ref{fig: Single domain wall interaction}.

The most obvious expected outcome is the total annihilation/erasure of the skyrmion into the domain wall, resulting in just the $Q=0$ domain wall.
Then there is the absorption of the skyrmion into the wall, forming a domain wall (anti-)kink with topological charge $Q=(-)1$.
A new unexpected state is the absorption of the skyrmion to form a domain wall kink-antikink with $Q=0$.
Finally, there are two metastable states with the magnetic skyrmion coexisting with the domain wall.
One of these was predicted by Ross and Nitta \cite{Nitta_Ross_2023}.
The other solution is an isolated skyrmion in the presence of a domain wall kink-antikink.
Both of these states are repulsive, so addition of another domain wall could allow for adequate control of the skyrmion.

Now we consider the addition of another domain wall with the same phase as the first domain wall.
The system we are interested in is a magnetic skyrmion in the presence of a domain wall - anti-domain wall, which is still described by the two moduli parameters $(\chi,\phi)$.
In terms of the stereographic coordinate, this is given by the initial configuration
\begin{equation}
    u = e^{k(x+X) + i\chi} + e^{-k(x-X) + i\chi} + \tan \left( \frac{f(r)}{2} \right) e^{i(\theta+\phi)}.
\label{eq: Double DW skyrmion}
\end{equation}
The boundary conditions are now $\vec{m} = (0,0,-1)$ as $x \rightarrow \pm\infty$.
We vary both phases $(\chi,\phi)$ between $0$ and $2\pi$ in increments of $\pi/3$ to see the effect the phase of each has on one another.
Results of this are detailed in Tables \ref{tbl: chi = 0}-\ref{tbl: chi = 5pi/3}.

\begin{figure}[t]
    \centering
    \begin{subfigure}[b]{0.3\textwidth}
    \includegraphics[width=\textwidth]{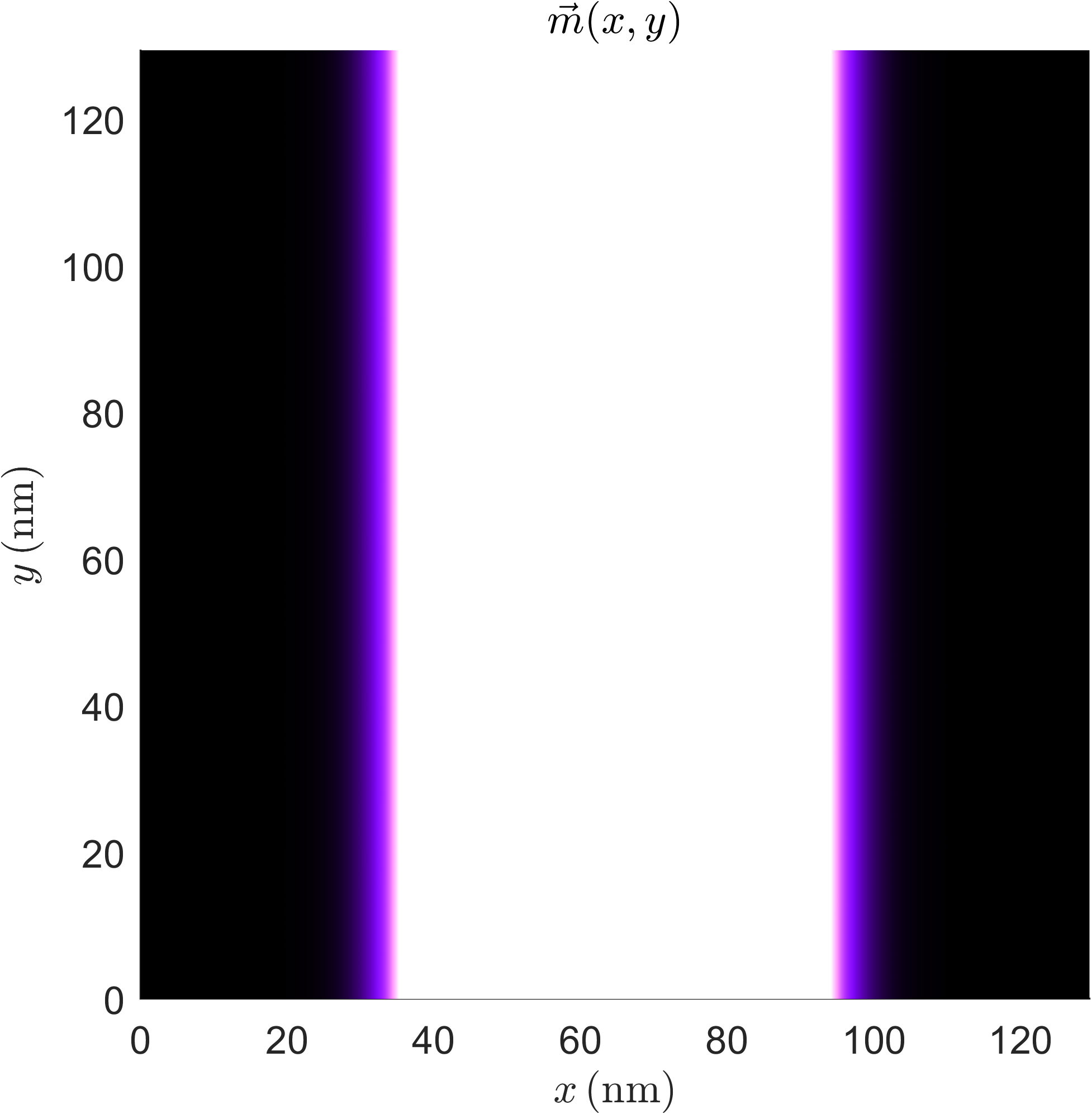}
    \caption{$Q=0$}
    \end{subfigure}
    ~
    \begin{subfigure}[b]{0.3\textwidth}
    \includegraphics[width=\textwidth]{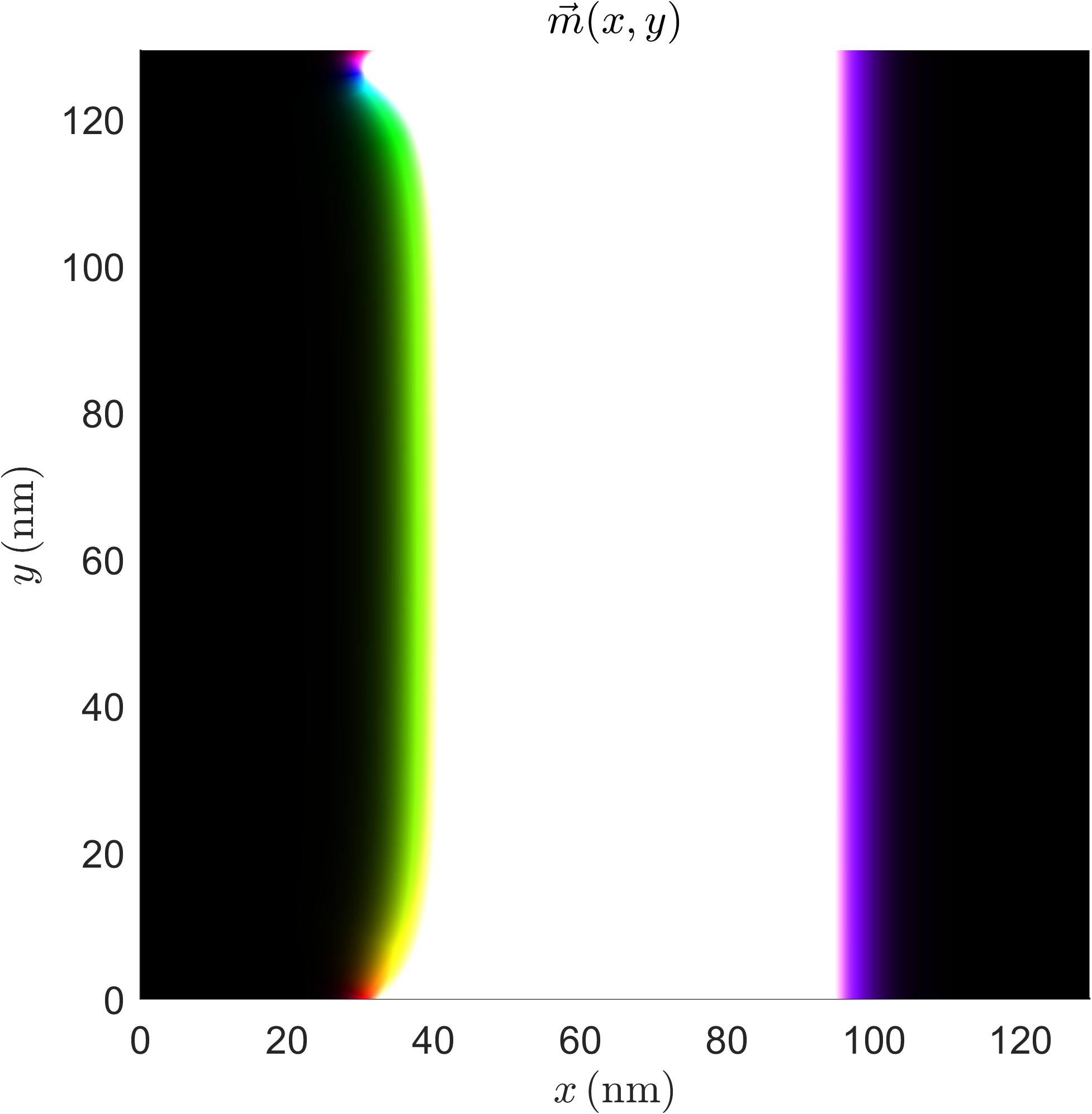}
    \caption{$Q=+1$}
    \end{subfigure}
        ~
    \begin{subfigure}[b]{0.3\textwidth}
    \includegraphics[width=\textwidth]{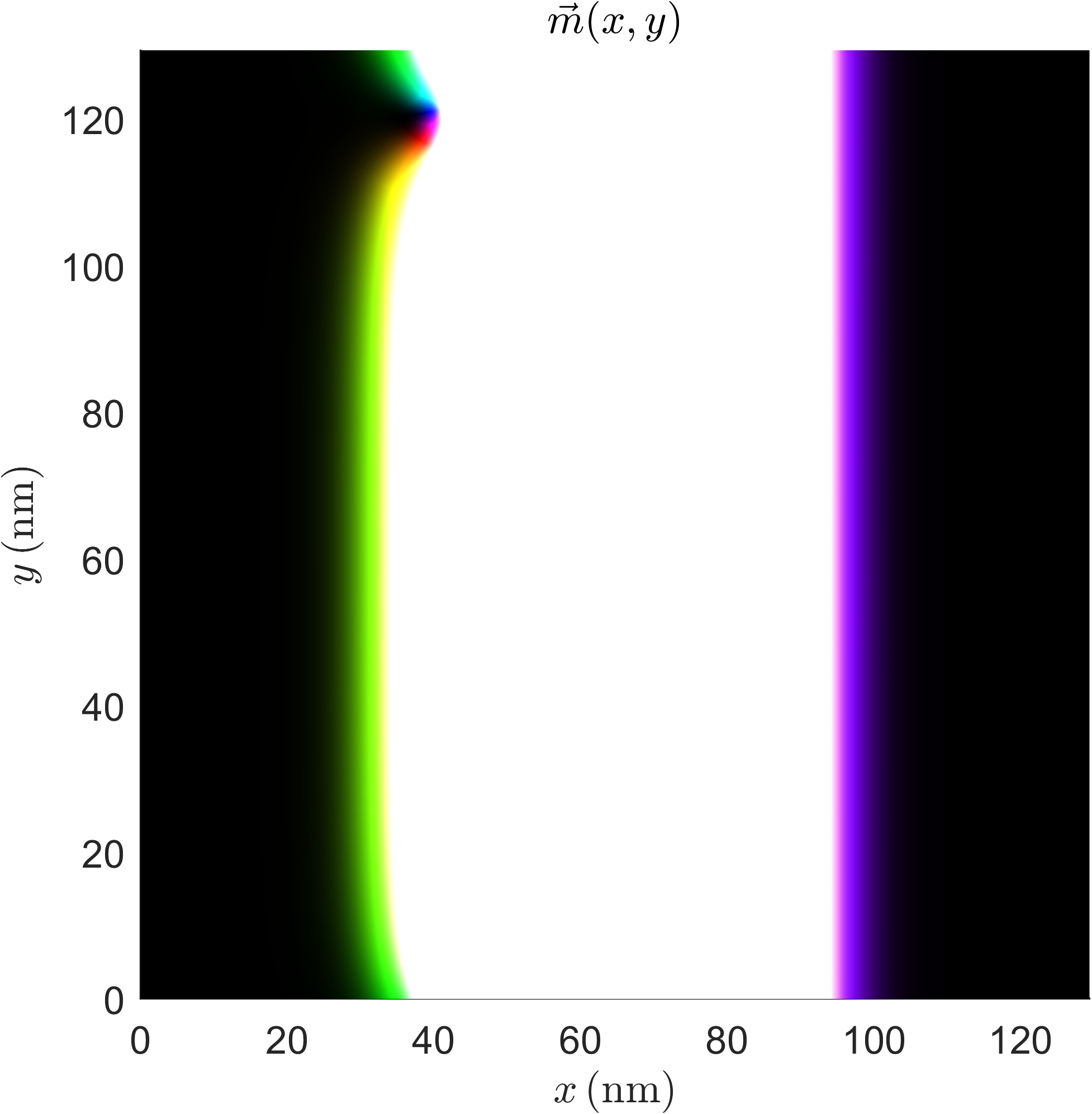}
    \caption{$Q=-1$}
    \end{subfigure}
    \\
    \begin{subfigure}[b]{0.3\textwidth}
    \includegraphics[width=\textwidth]{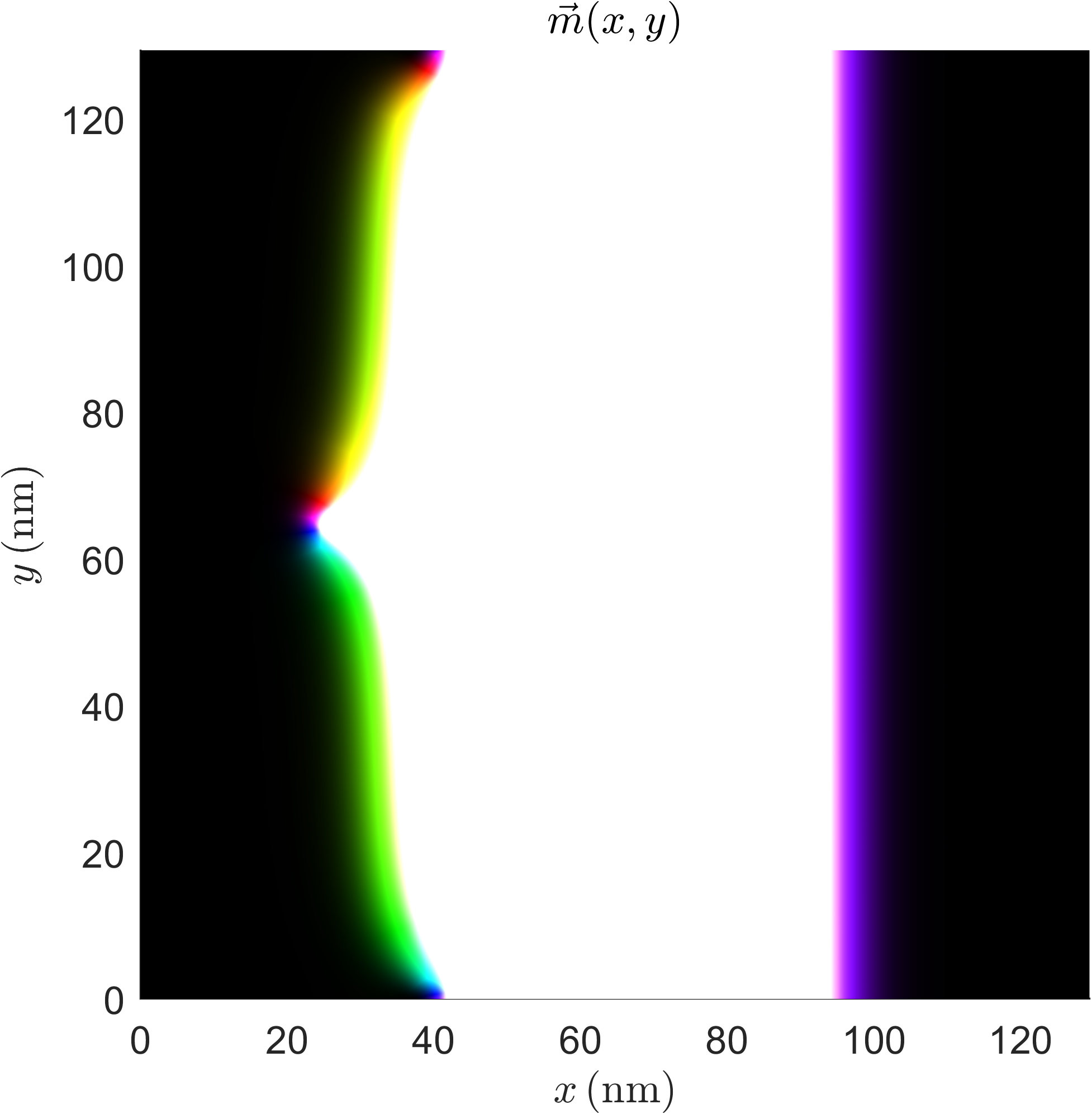}
    \caption{$Q=0$}
    \end{subfigure}
    ~
    \begin{subfigure}[b]{0.3\textwidth}
    \includegraphics[width=\textwidth]{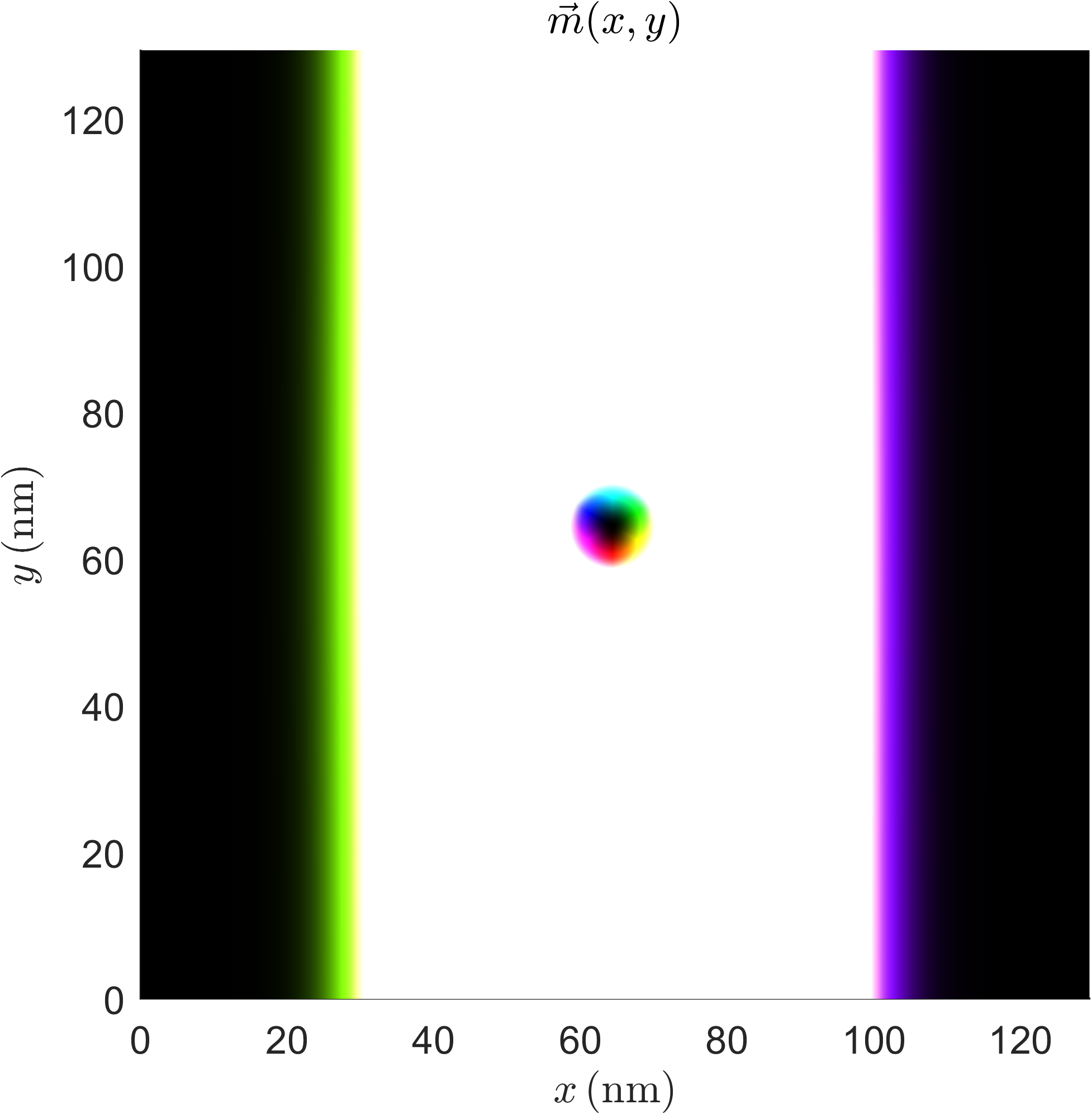}
    \caption{$Q=+1$}
    \end{subfigure}
        ~
    \begin{subfigure}[b]{0.3\textwidth}
    \includegraphics[width=\textwidth]{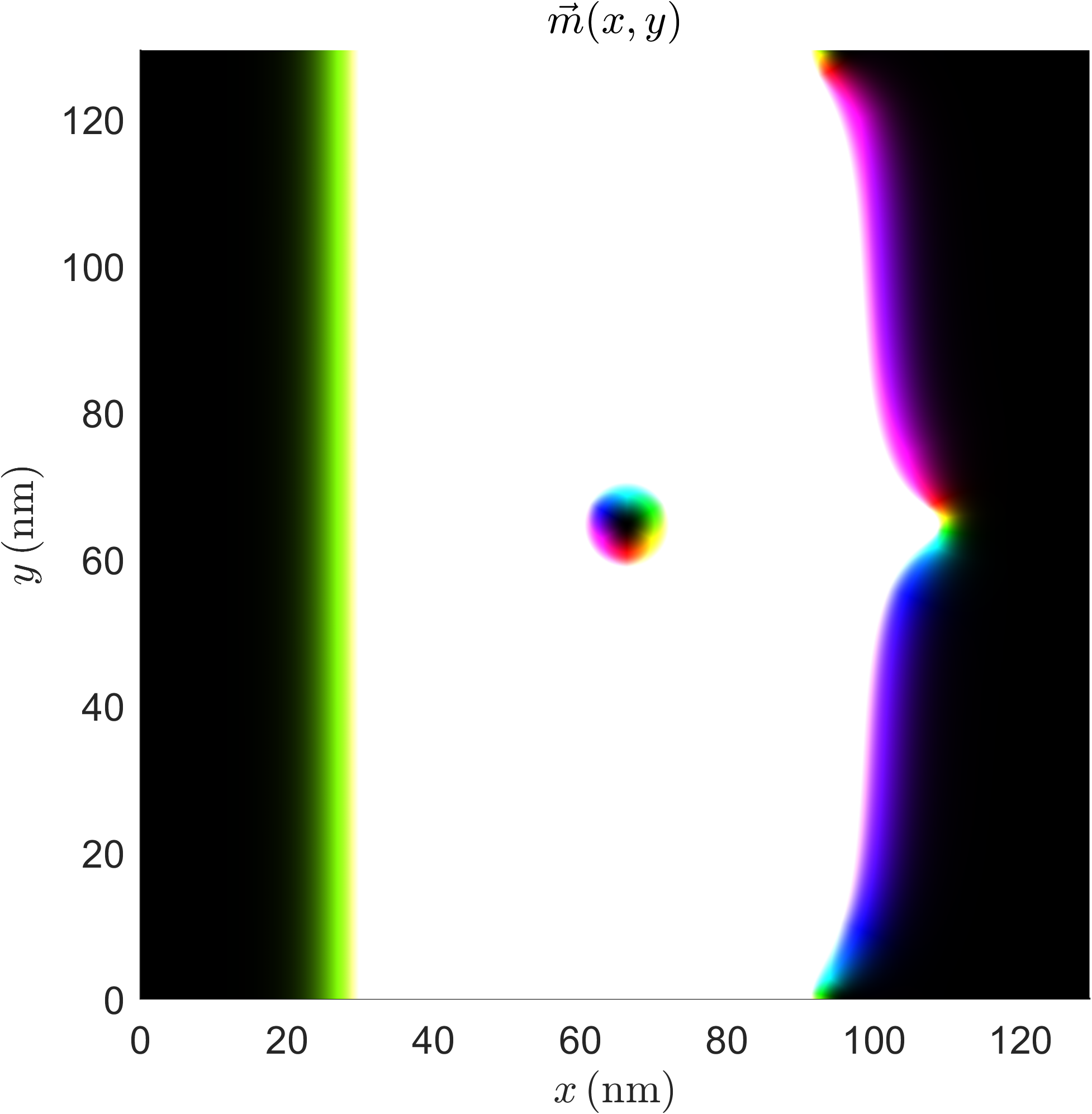}
    \caption{$Q=+1$}
    \end{subfigure}
    \caption{All of the possible outcomes for the interaction of a magnetic skyrmion with two chiral domain wall. First is the (a) erasure of the skyrmion into one of the domain walls, leaving only the domain wall - anti-domain wall system. Second is the absorption of the skyrmion into a domain wall, creating either a (b) a domain wall kink or (c) a domain wall antikink. Then there is (d) annihilation of the skyrmion but producing a domain wall kink-antikink in one of the walls. The last two possibilities are stable skyrmions between (e) two domain walls and (f) a domain wall and a domain wall kink-antikink.}
    \label{fig: Double domain wall interaction}
\end{figure}

It appears that rotations of $\pi/3$ in the both phases $(\chi,\phi)$ are essential for the creation of domain wall kinks/anti-kinks.
The six possible outcomes\footnote{There are another six possibilities where the same outcome happens for the opposite domain wall, but this is obviously equivalent.} are shown in Fig.~\ref{fig: Double domain wall interaction}.
Changing the phase of the domain wall $\chi$ controls which domain wall the skyrmion can be absorbed into.
Most importantly, the correct choice of phase can dictate whether the skyrmion is erased or stored in the domain wall.


\section{Conclusion}
\label{sec: Conclusion}

We have shown that a magnetic skyrmion quasi-particle can be controlled and manipulated via use of chiral domain walls.
Correct phase orientations allow for the skyrmion to be stored in a magnetic domain wall, creating either a domain wall kink ($Q=+1$) or antikink ($Q=-1$).
The choice of phase also determines which wall the skyrmion is stored in, offering more control.
Additionally, we have demonstrated that the skyrmion can be stably stored between domain walls, causing a phase change or the creation of a defect-antidefect pair in one of the walls.
Thus it should be possible to change the phases of one of the domain walls, thereby allowing for the erasure or storage of the magnetic skyrmion.
This is a welcomed addition to an ever growing toolbox of spintronic devices.
While we have demonstrated the domain wall manipulation technique in Co/Pt multilayered ultrathin films, the method can easily be applied to other multilayered thin films, making them ideal candidates for integration into spintronic devices.

Future directions for this project would be the reverse engineered version, whereby one could try separate the magnetic skyrmion from the domain wall.
This would be extremely advantageous for spintronic nano-devices as once the skyrmion is absorbed into the domain wall (forming, say, a domain wall kink), then its motion is restricted along the domain wall, allowing for easy guidance.
As we have shown that the skyrmion can be stably stored between domain walls, one should also study the dynamics of magnetic skyrmions trapped between chiral domain walls.


\section*{Acknowledgments}

The author acknowledges funding from the Roland Gustafssons Stiftelse för teoretisk fysik.
Whilst this manuscript was under final preparation, the author was informed of a recent independent preprint with similar results to this article on arXiv \cite{Amari_Gudnason_Nitta_2024}.
The results of their work cover the full range of interaction for a magnetic skyrmion with a single domain wall in Pt(Co/Ni)\textsubscript{M}/Ir\textsubscript{N} multilayers \cite{Maxwell_2021}, which is consistent with our findings for ultrathin Co/Pt films.

\begin{table}
    \centering
    \begin{tabular}{ccM{40mm}M{40mm}c}
        \toprule
        $(\chi,\phi)$ & $Q_{\textup{i}}$ & Initial Configuration & Final Configuration & $Q_{\textup{f}}$ \\
        \midrule
        $(0,0)$ & $+1$ & \includegraphics[width=25mm]{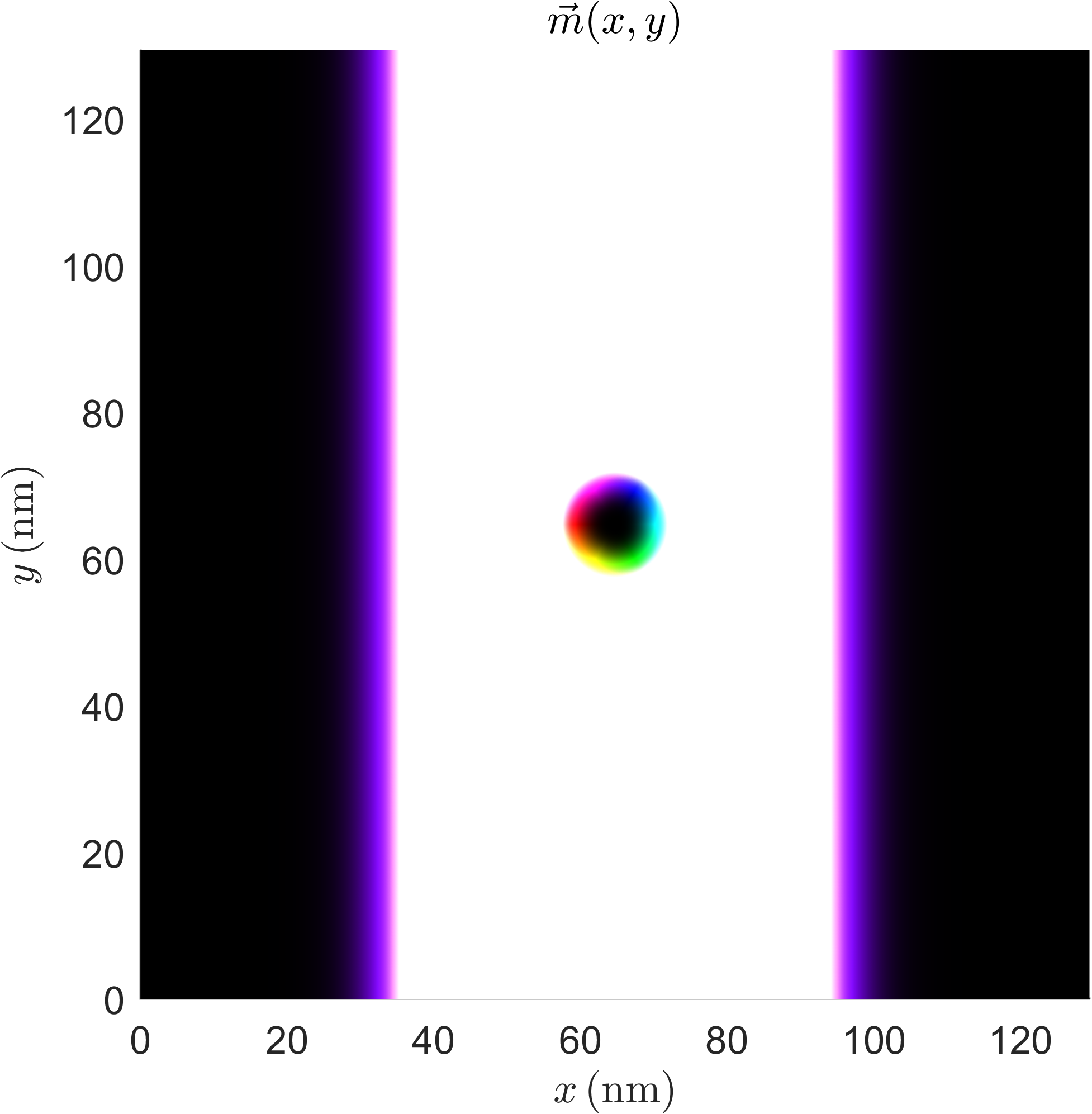} & \includegraphics[width=25mm]{Graphics/Alpha=0/alpha___0__beta___0__Final_.png} & $+1$ \\
        $(0,\pi/3)$ & $+1$ & \includegraphics[width=25mm]{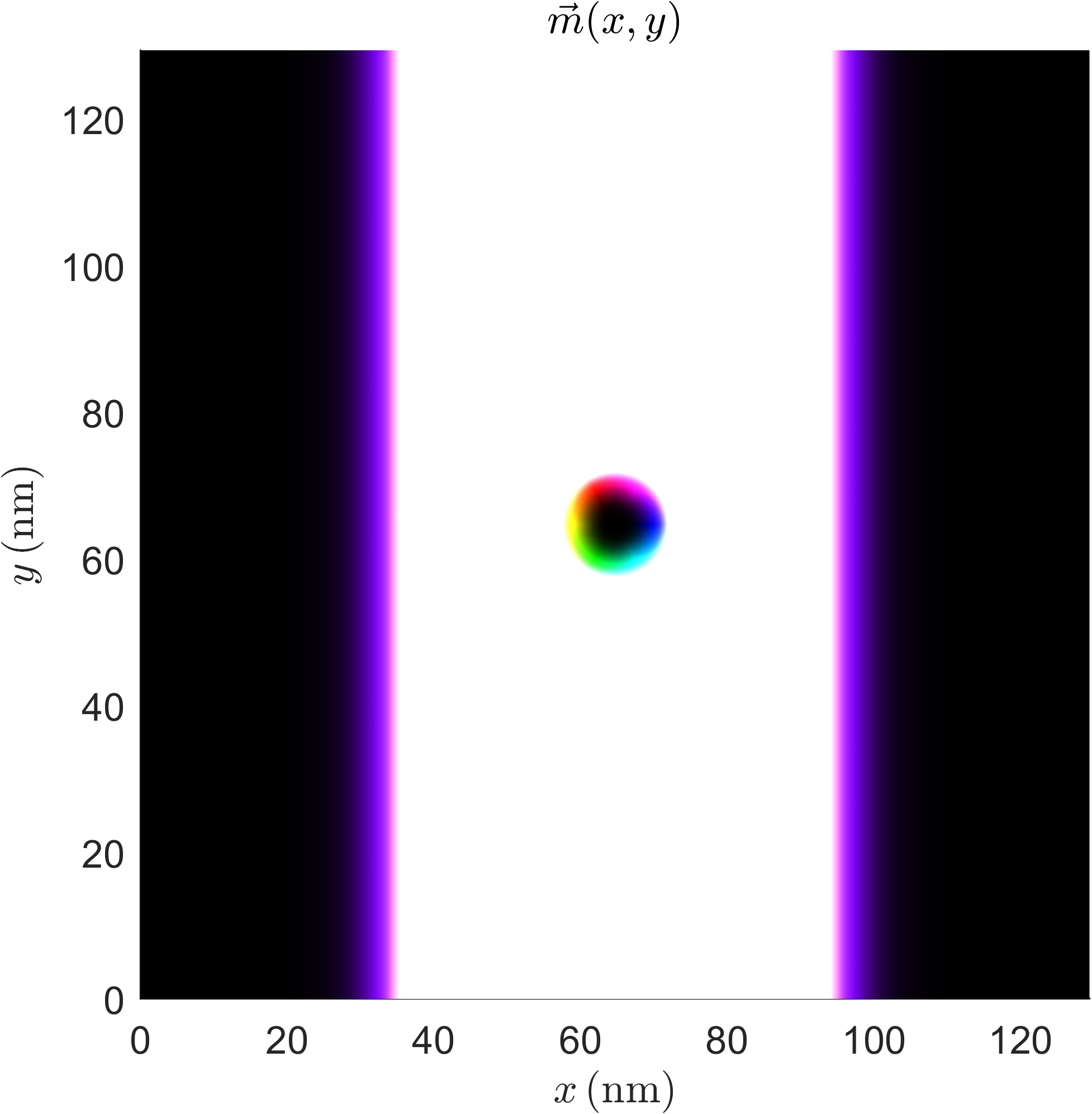} & \includegraphics[width=25mm]{Graphics/Alpha=0/alpha___0__beta___pi3__Final_.png} & $-1$ \\
        $(0,\pi/2)$ & $+1$ & \includegraphics[width=25mm]{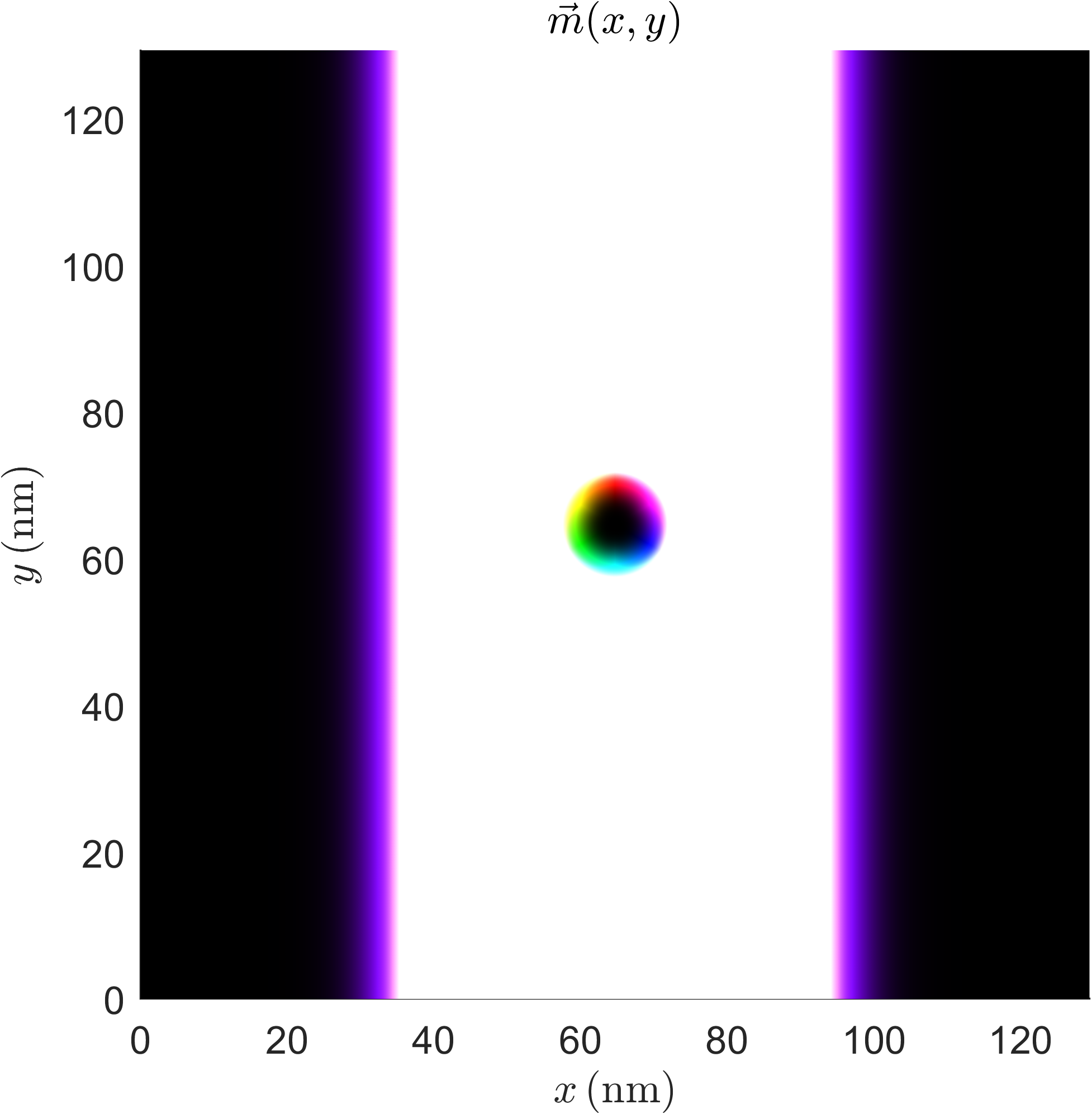} & \includegraphics[width=25mm]{Graphics/Alpha=0/alpha___0__beta___pi2__Final_.png} & $0$ \\
        $(0,2\pi/3)$ & $+1$ & \includegraphics[width=25mm]{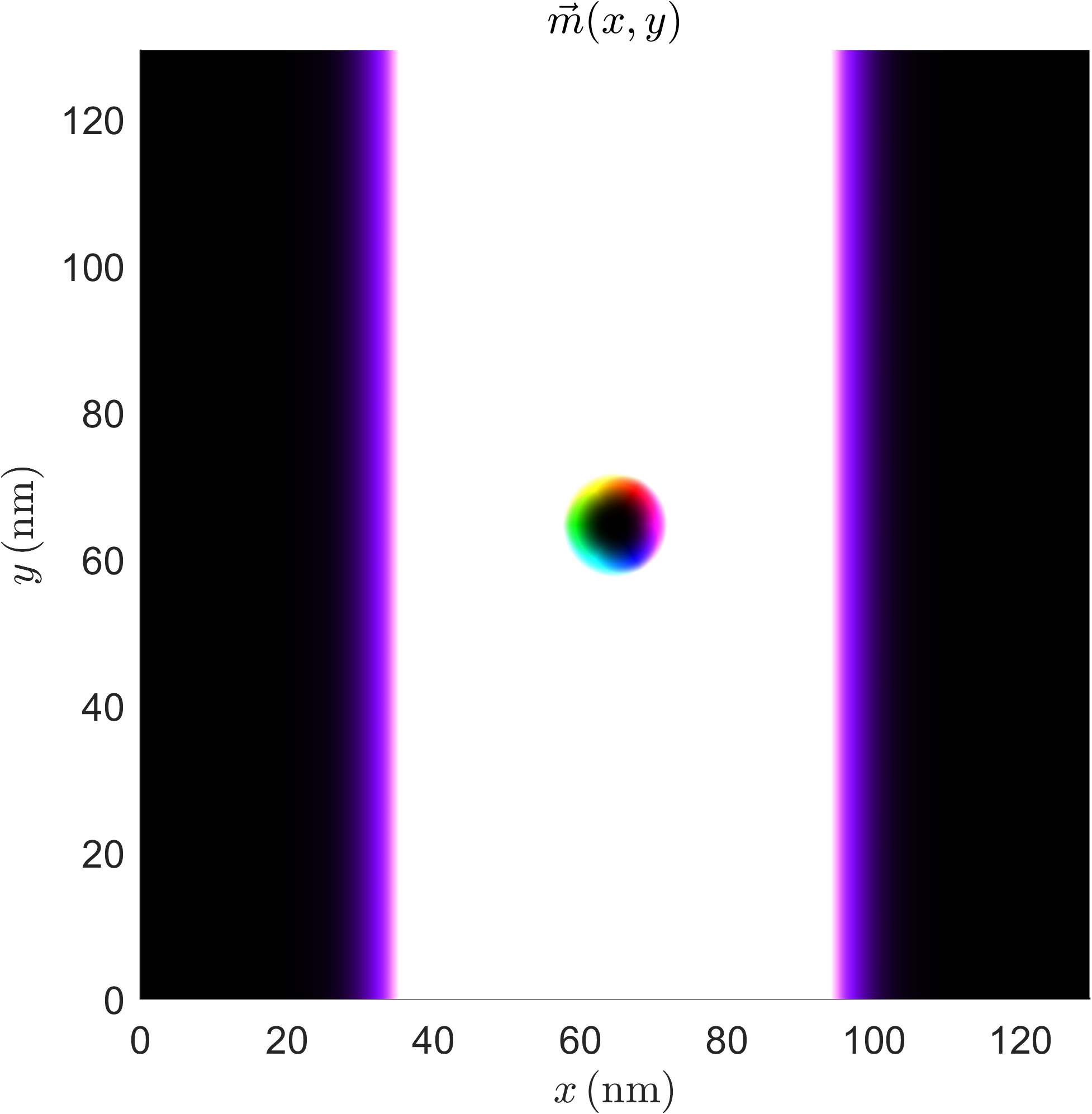} & \includegraphics[width=25mm]{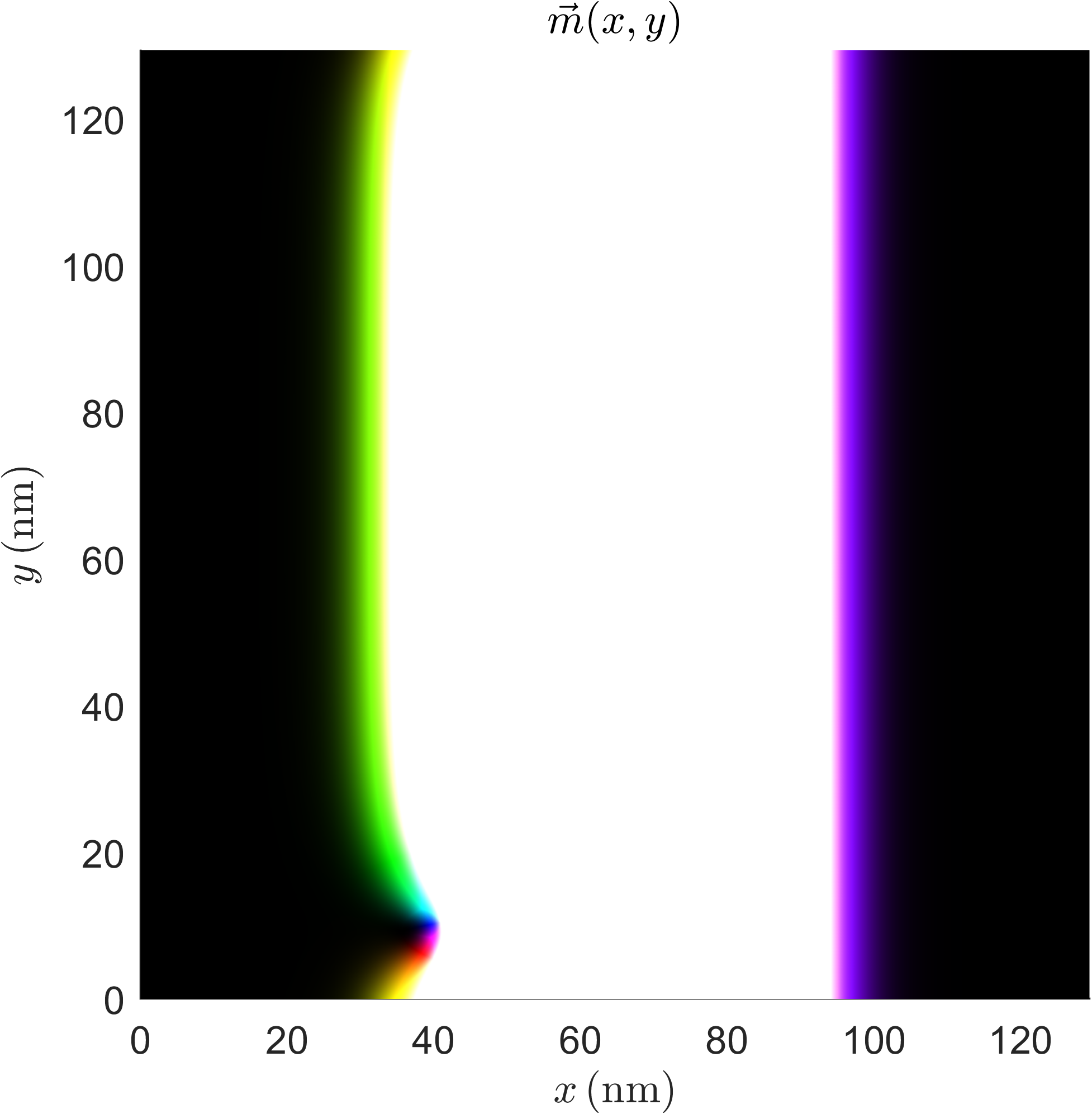} & $-1$ \\
        $(0,\pi)$ & $+1$ & \includegraphics[width=25mm]{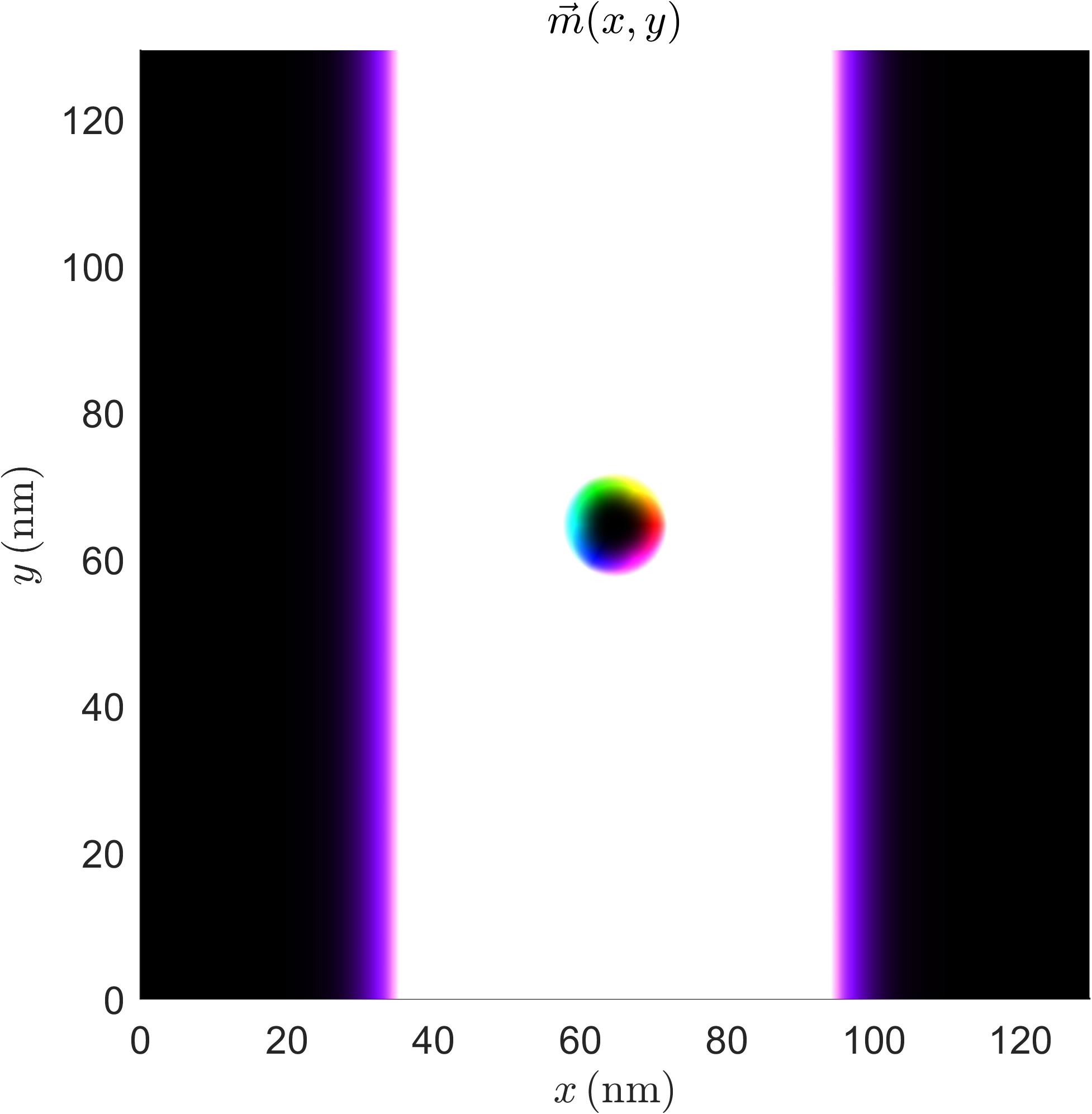} & \includegraphics[width=25mm]{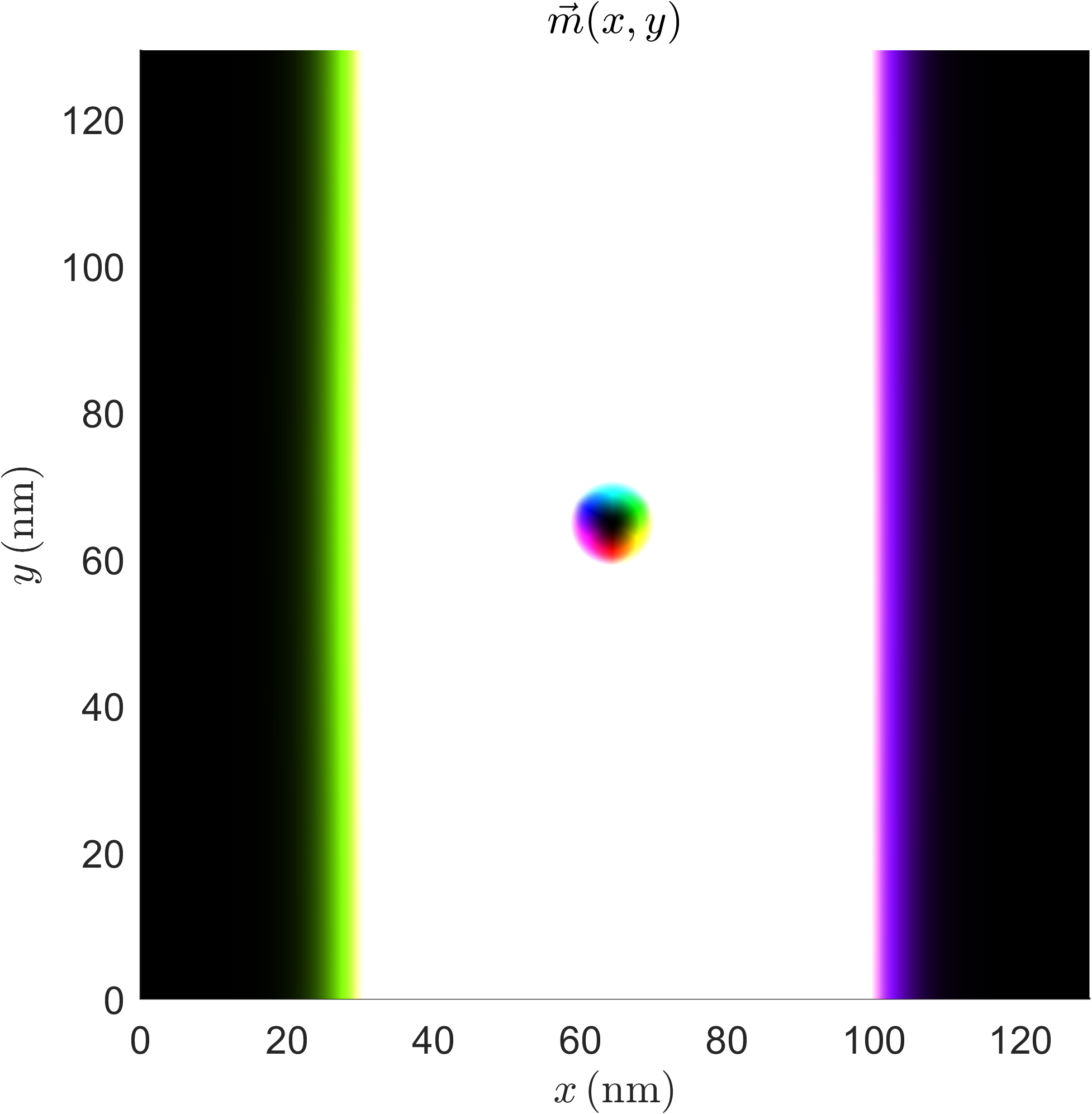} & $+1$ \\
        $(0,4\pi/3)$ & $+1$ & \includegraphics[width=25mm]{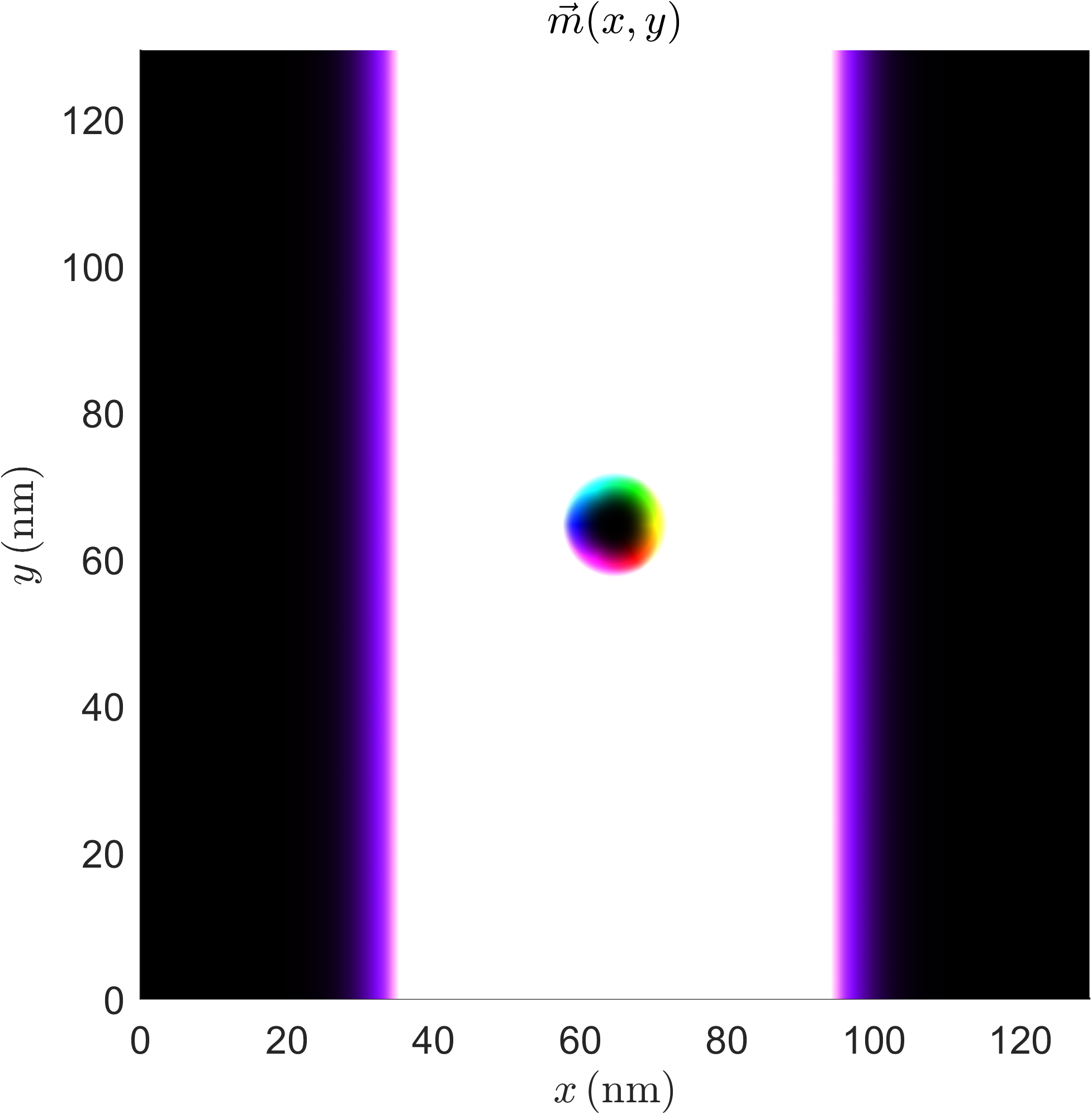} & \includegraphics[width=25mm]{Graphics/Alpha=0/alpha___0__beta___4pi3__Final_.png} & $+1$ \\
        $(0,3\pi/2)$ & $+1$ & \includegraphics[width=25mm]{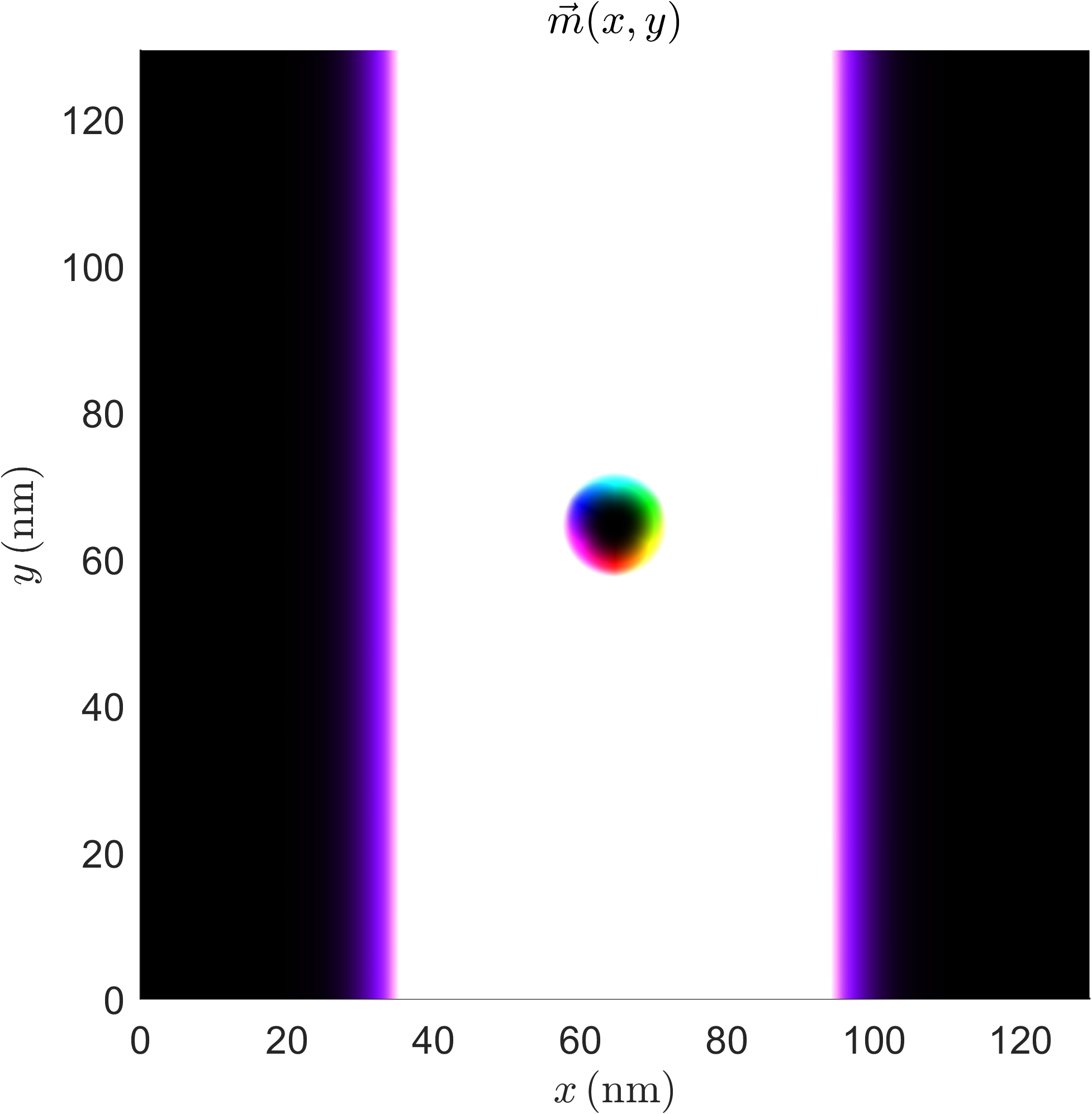} & \includegraphics[width=25mm]{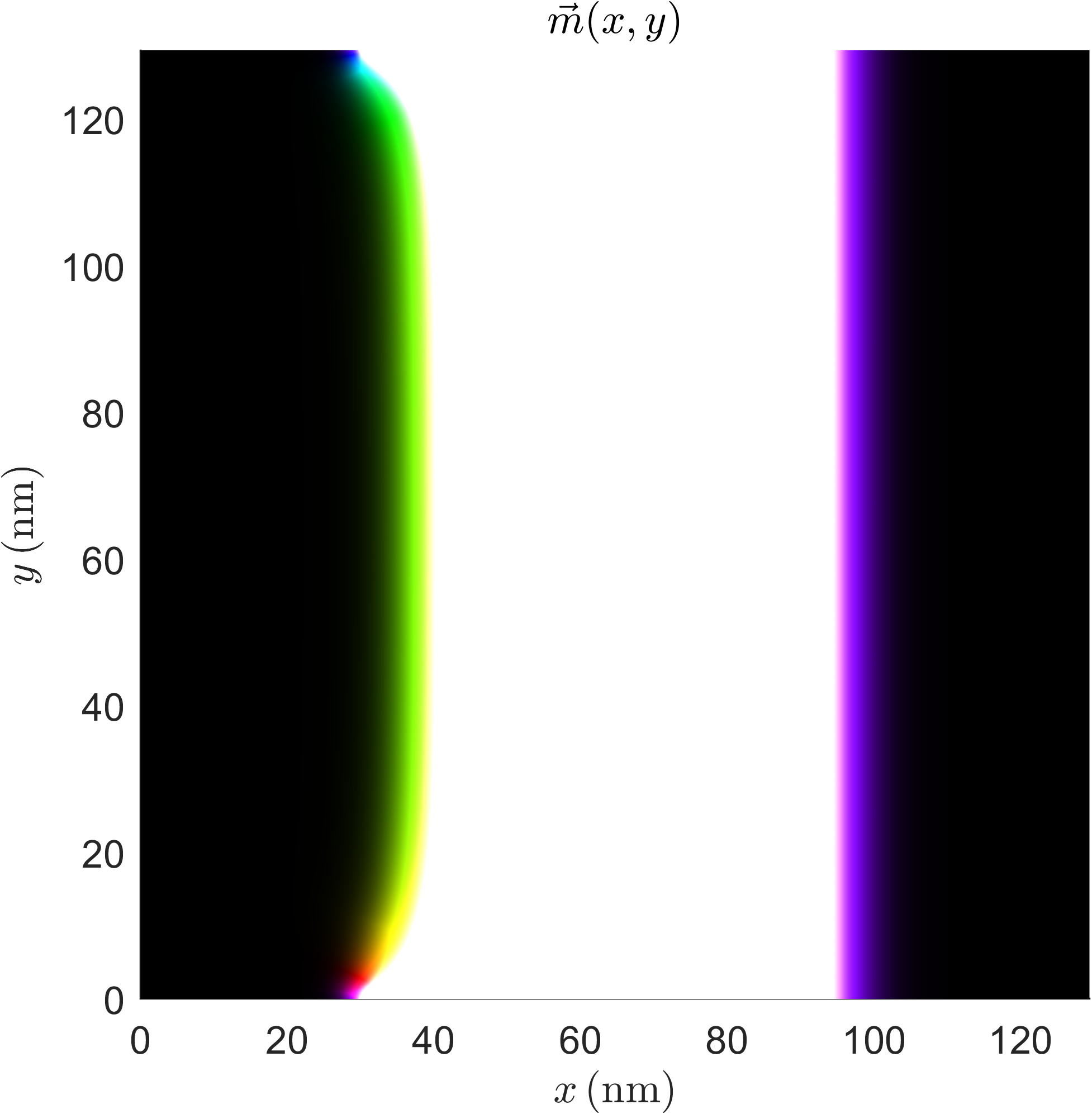} & $+1$ \\
        $(0,5\pi/3)$ & $+1$ & \includegraphics[width=25mm]{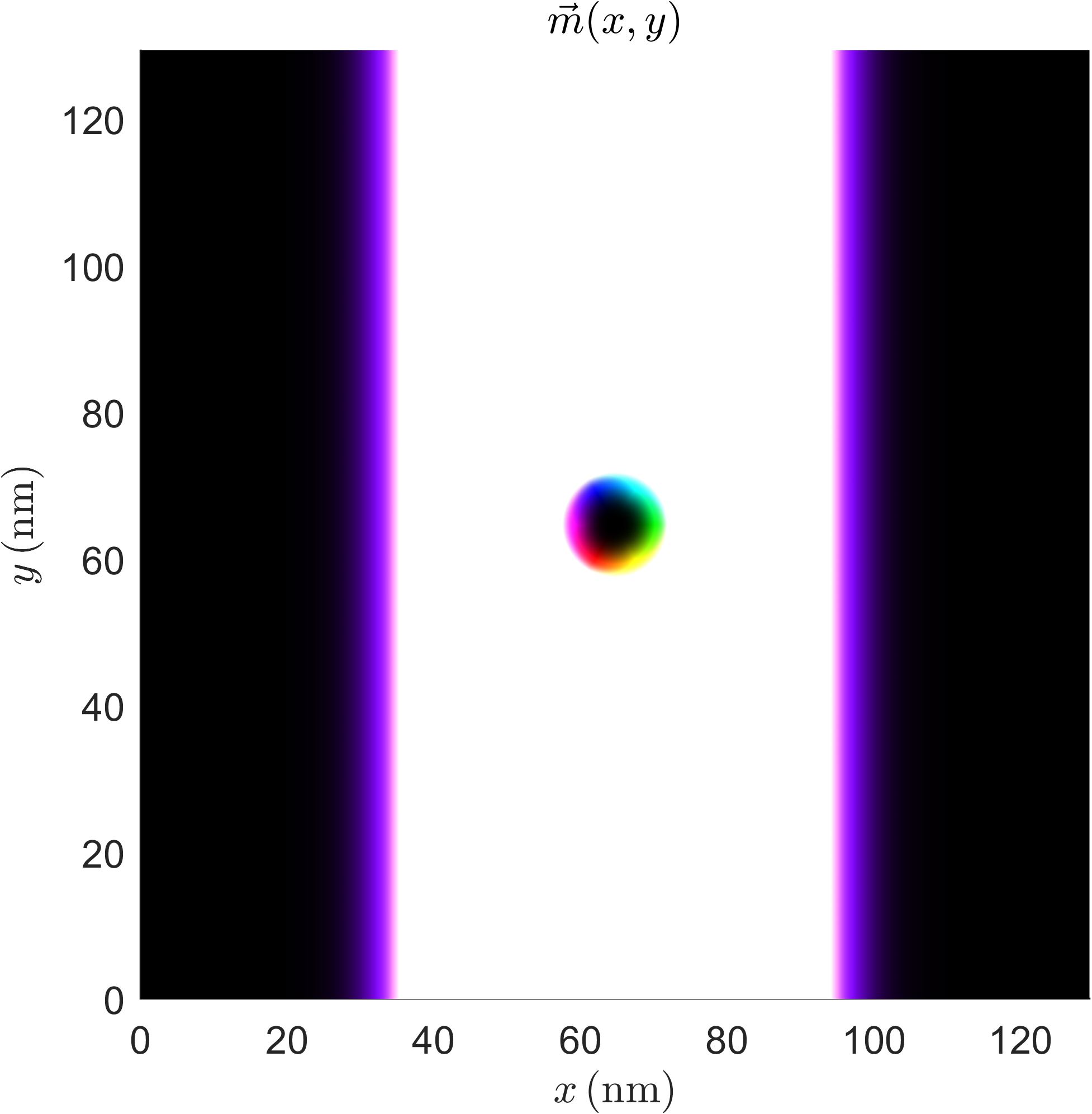} & \includegraphics[width=25mm]{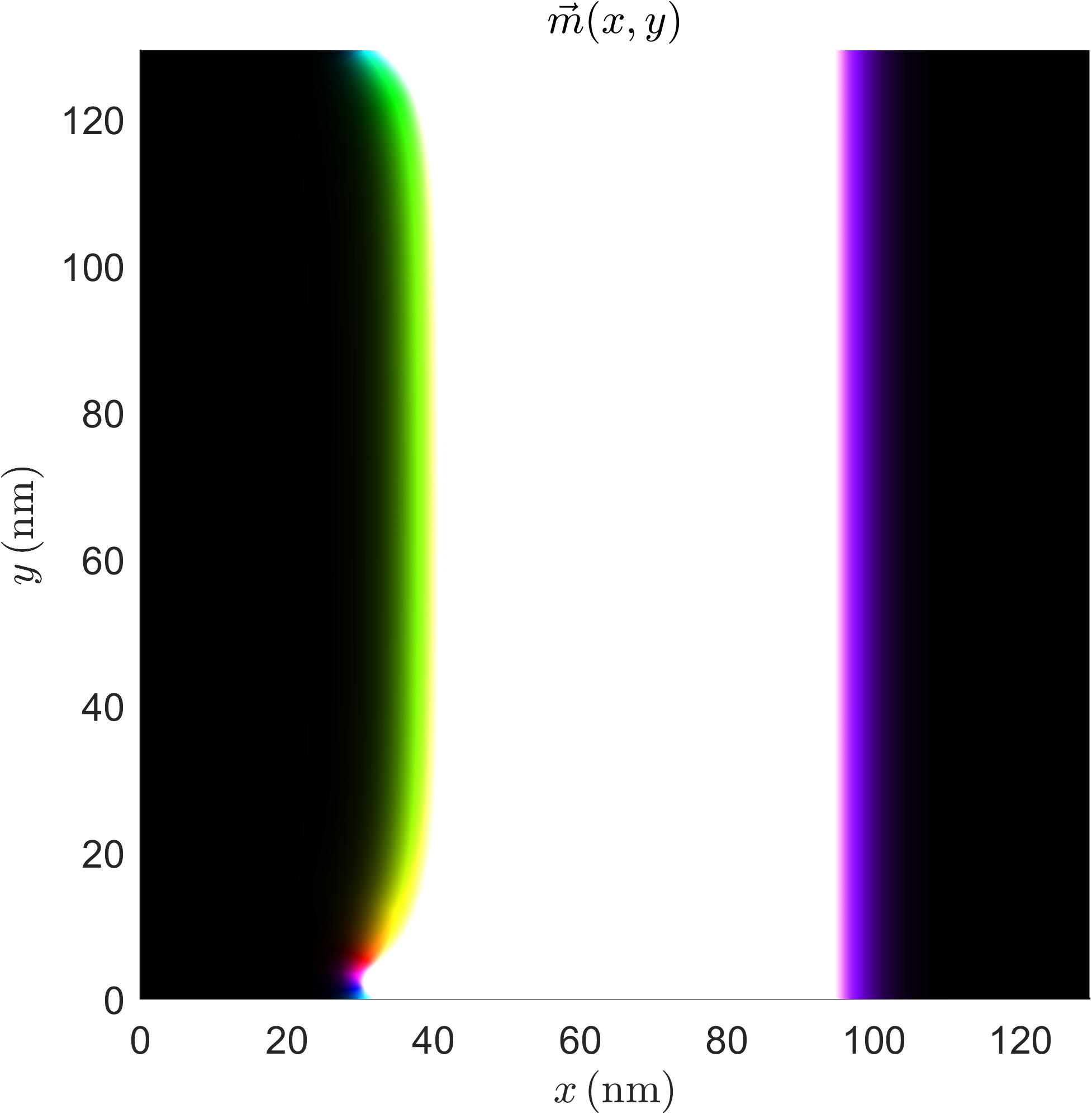} & $+1$ \\
        \bottomrule
    \end{tabular}
    \caption{Initial and final states for the domain wall phase $\chi=0$ as the skyrmion is rotated by $\pi/3$ from $\phi=0$ until $\phi=5\pi/3$.}
    \label{tbl: chi = 0}
\end{table}

\begin{table}
    \centering
    \begin{tabular}{ccM{40mm}M{40mm}c}
        \toprule
        $(\chi,\phi)$ & $Q_{\textup{i}}$ & Initial Configuration & Final Configuration & $Q_{\textup{f}}$ \\
        \midrule
        $(\pi/3,0)$ & $+1$ & \includegraphics[width=25mm]{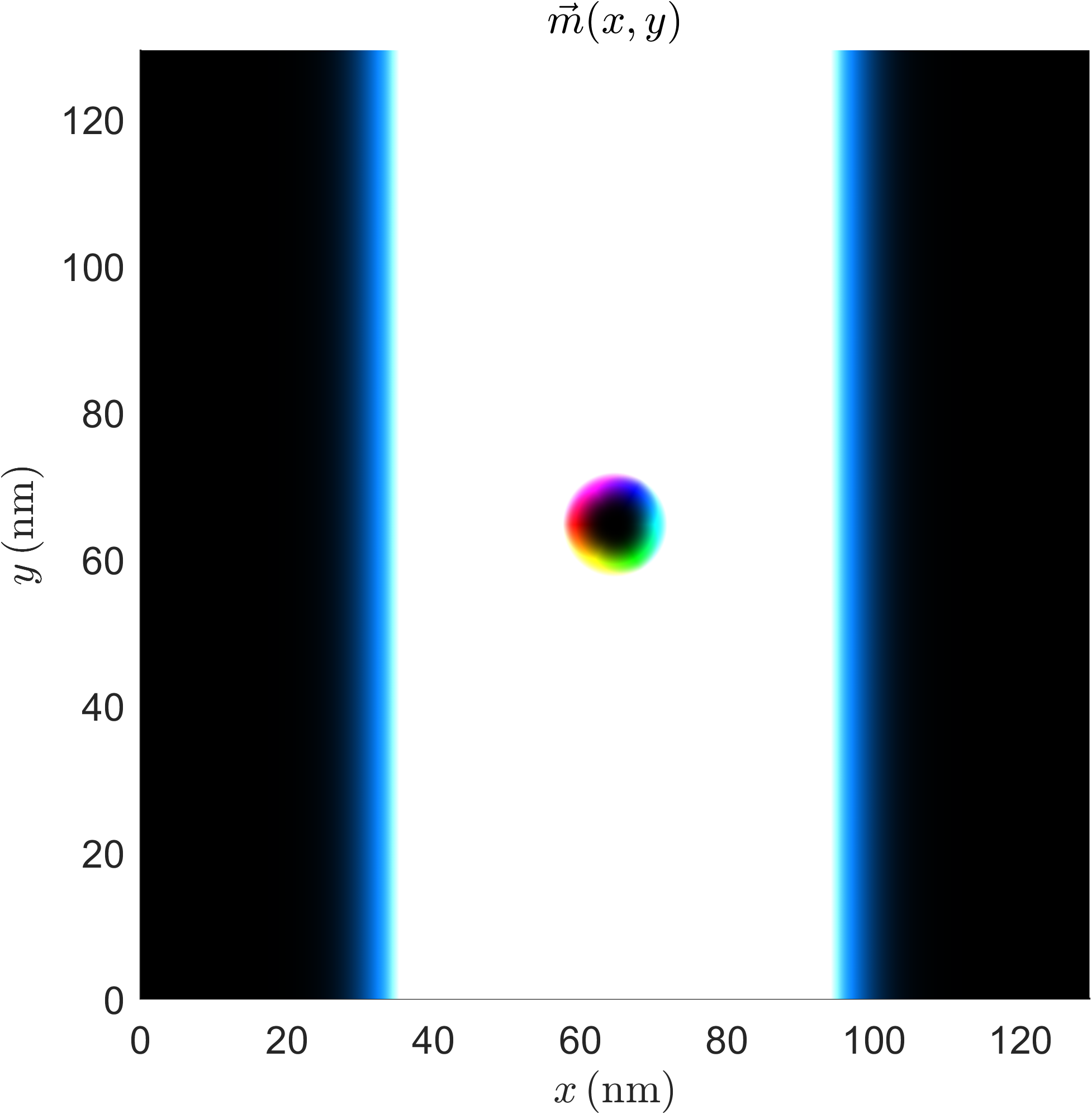} & \includegraphics[width=25mm]{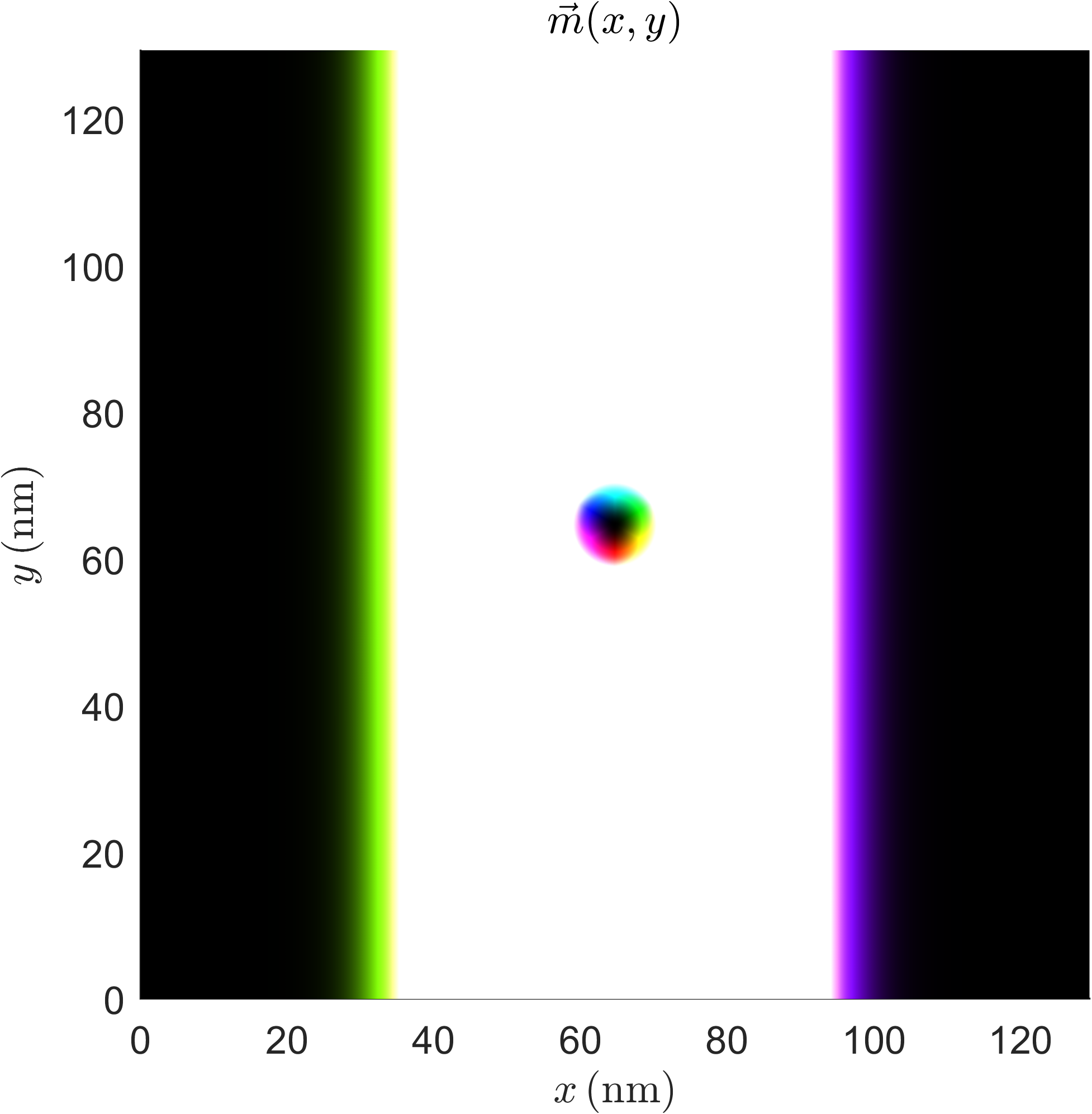} & $+1$ \\
        $(\pi/3,\pi/3)$ & $+1$ & \includegraphics[width=25mm]{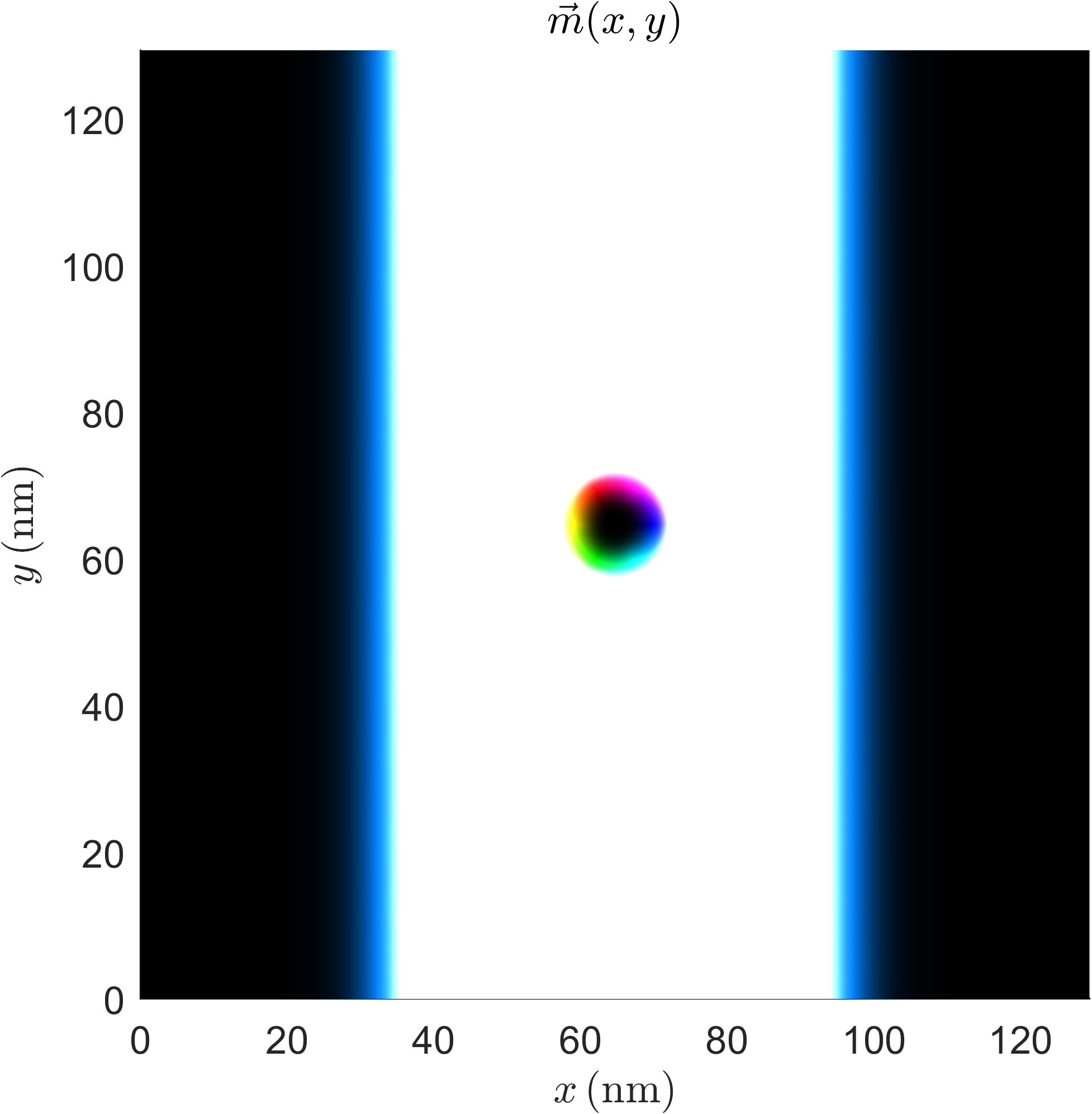} & \includegraphics[width=25mm]{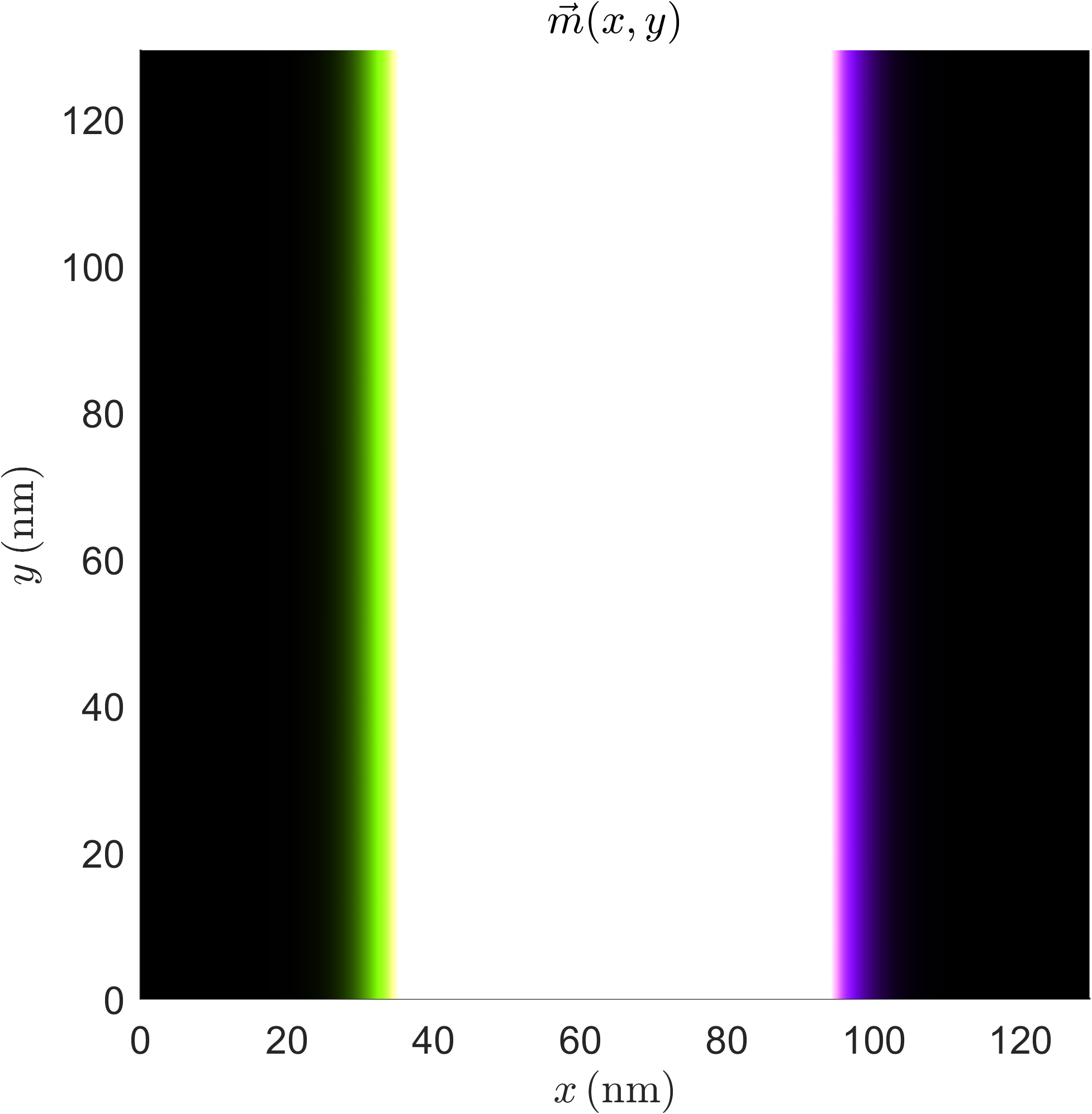} & $0$ \\
        $(\pi/3,\pi/2)$ & $+1$ & \includegraphics[width=25mm]{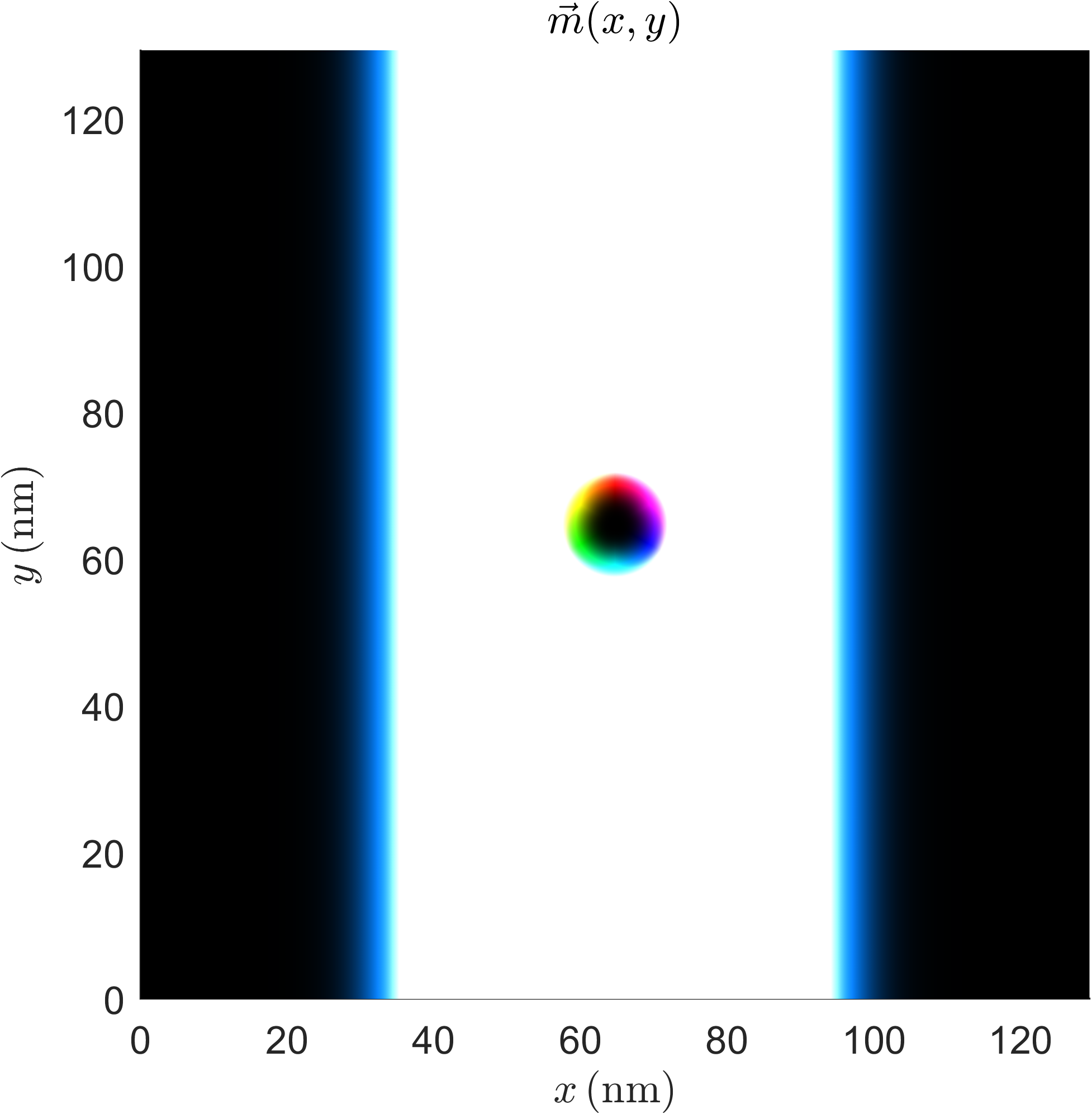} & \includegraphics[width=25mm]{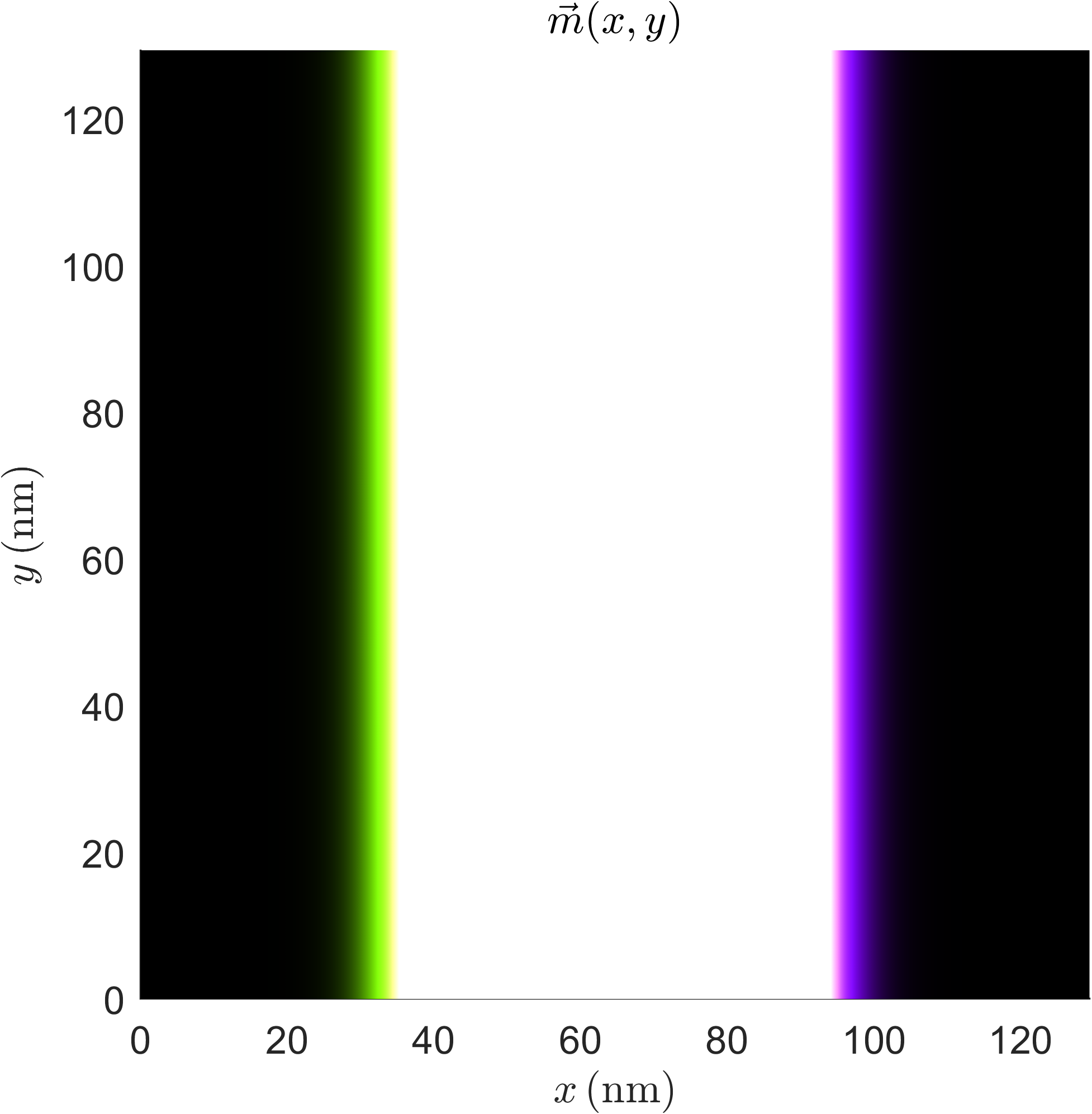} & $0$ \\
        $(\pi/3,2\pi/3)$ & $+1$ & \includegraphics[width=25mm]{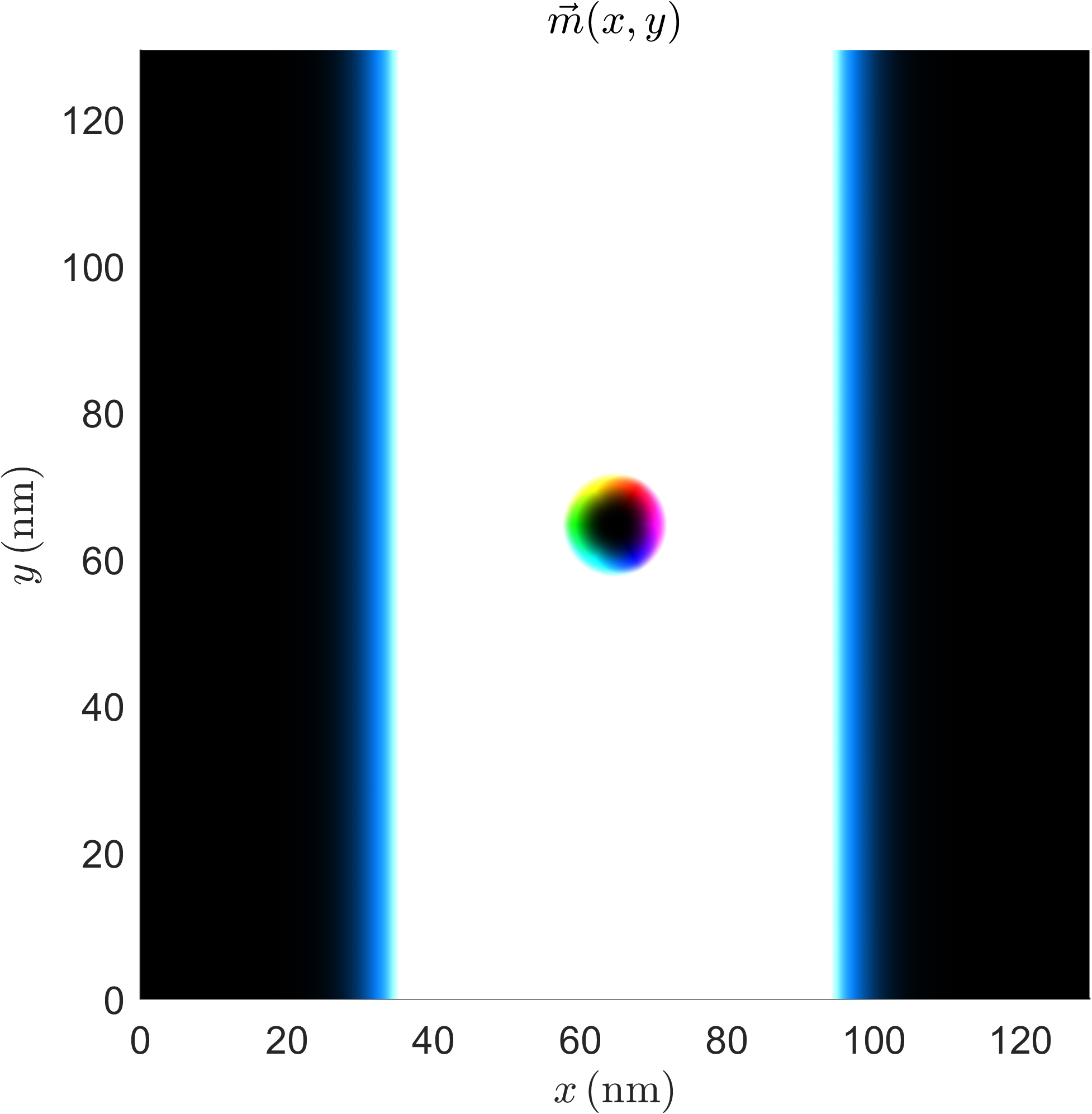} & \includegraphics[width=25mm]{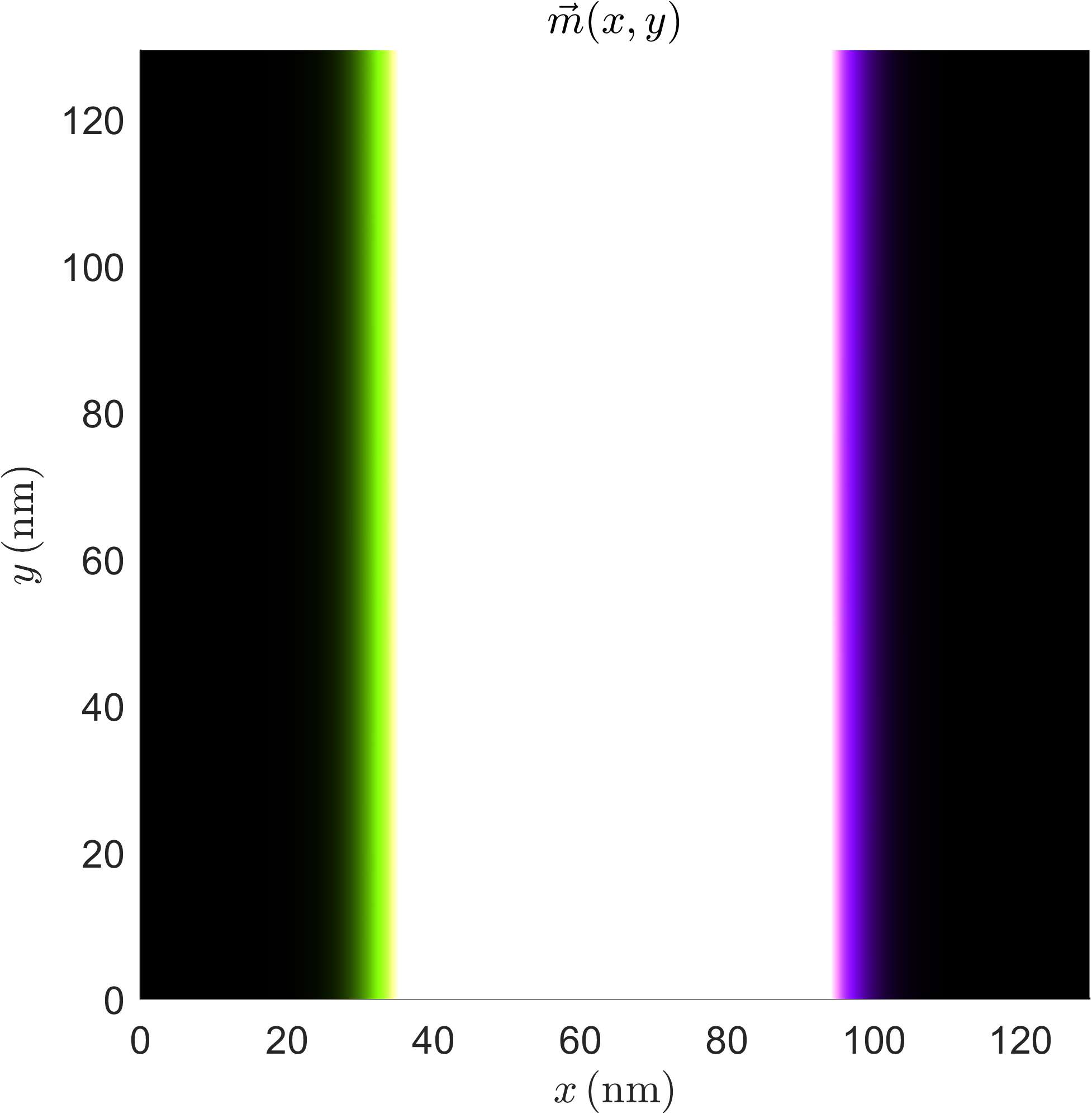} & $0$ \\
        $(\pi/3,\pi)$ & $+1$ & \includegraphics[width=25mm]{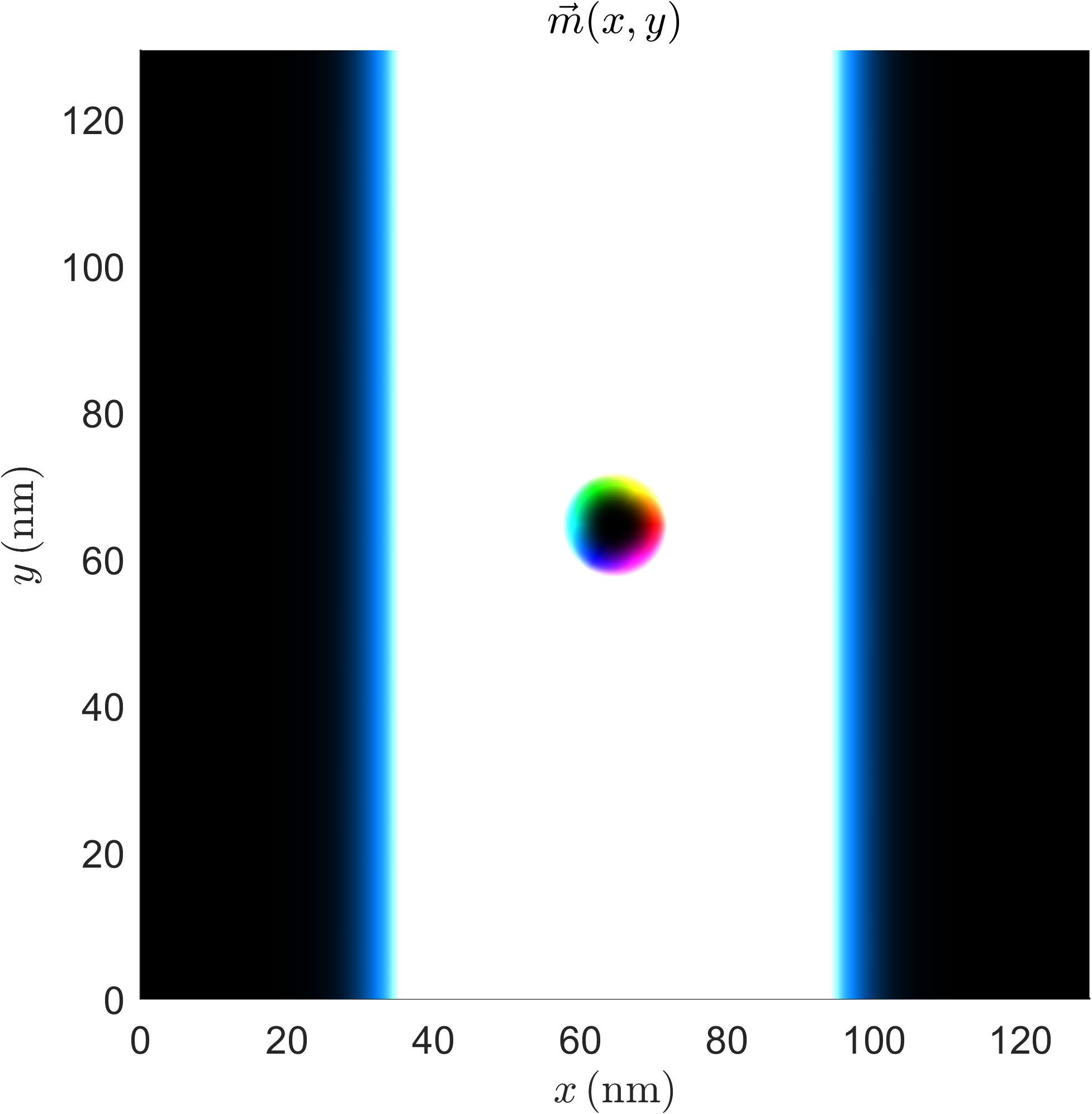} & \includegraphics[width=25mm]{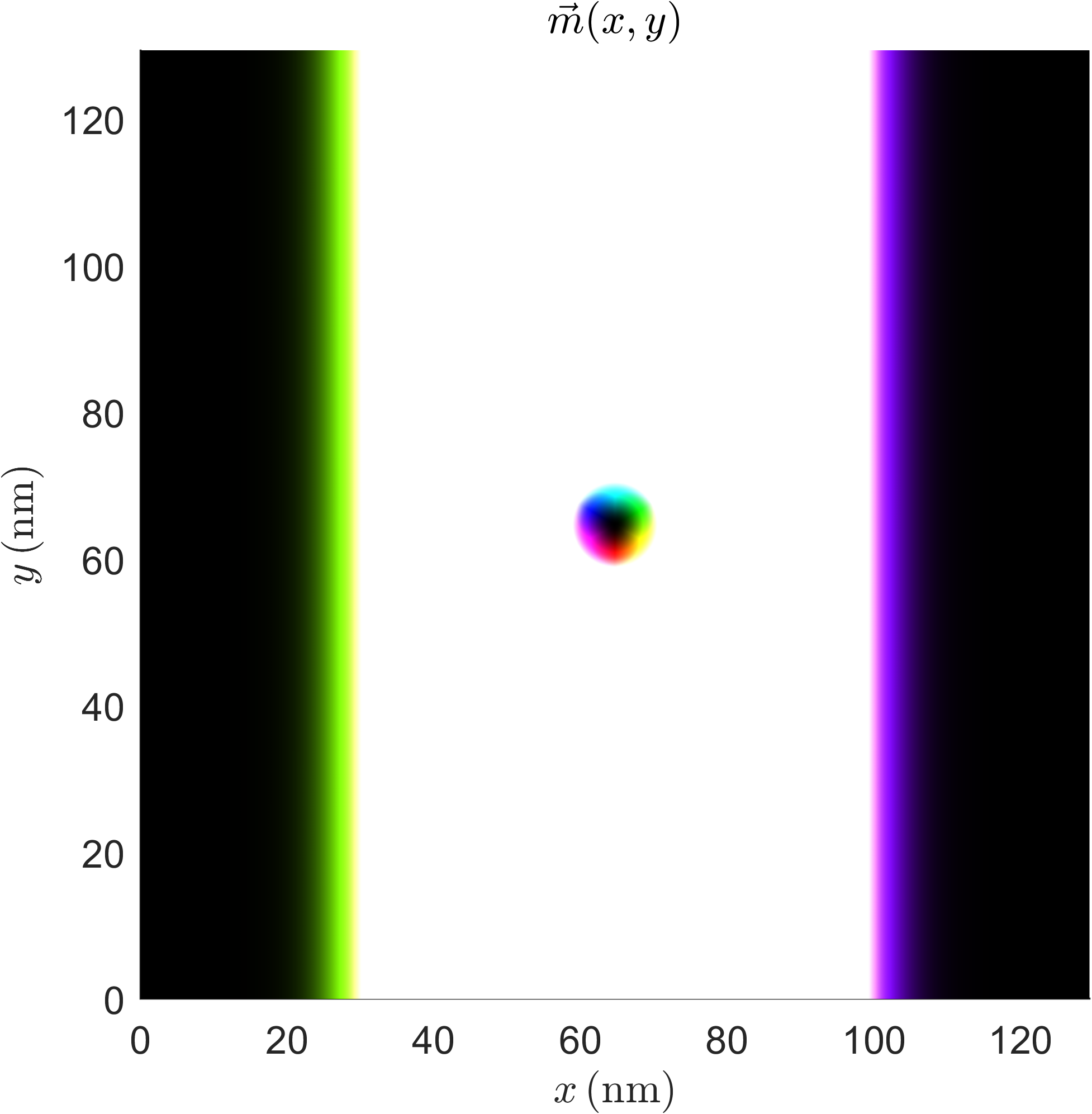} & $+1$ \\
        $(\pi/3,4\pi/3)$ & $+1$ & \includegraphics[width=25mm]{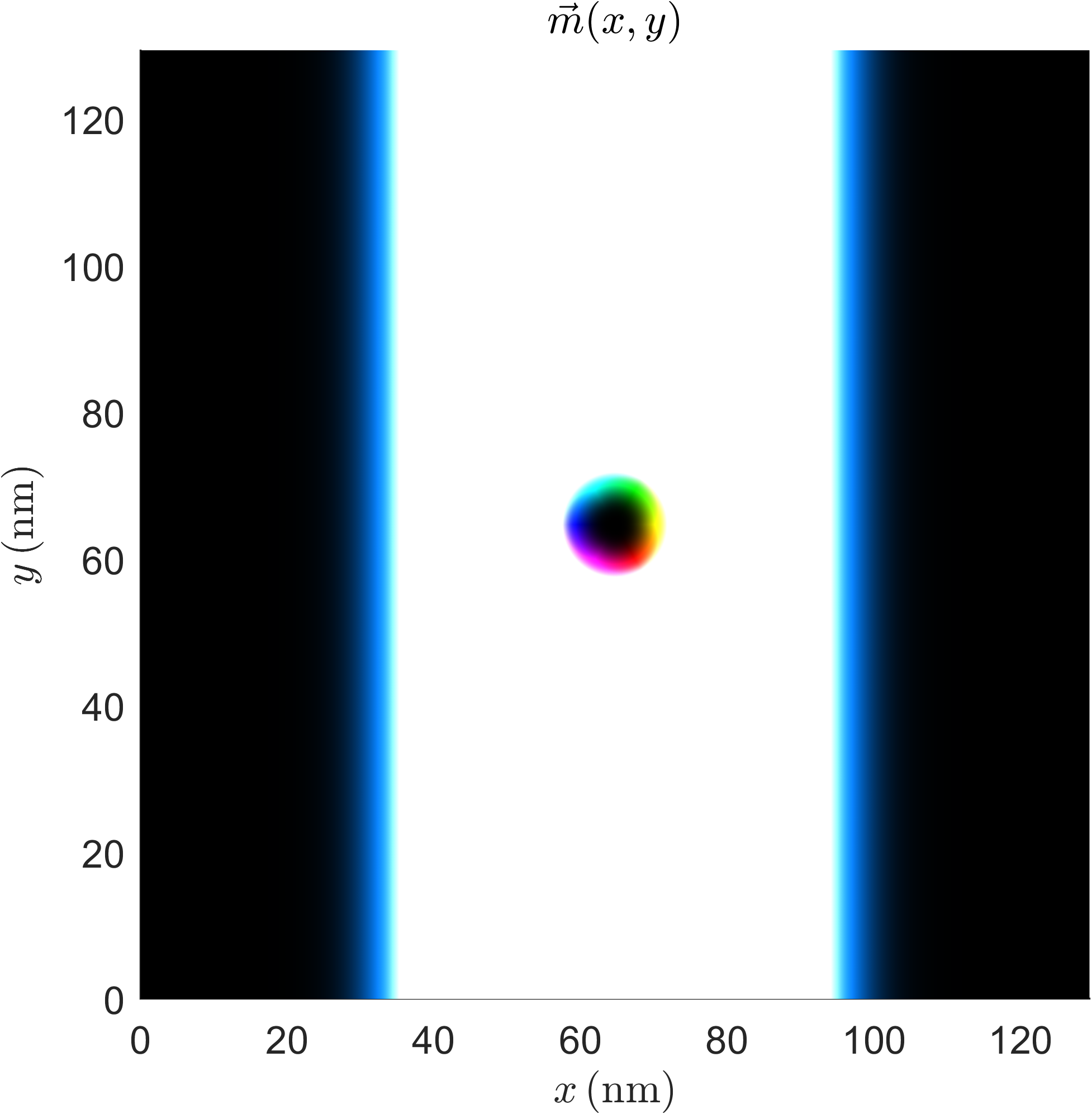} & \includegraphics[width=25mm]{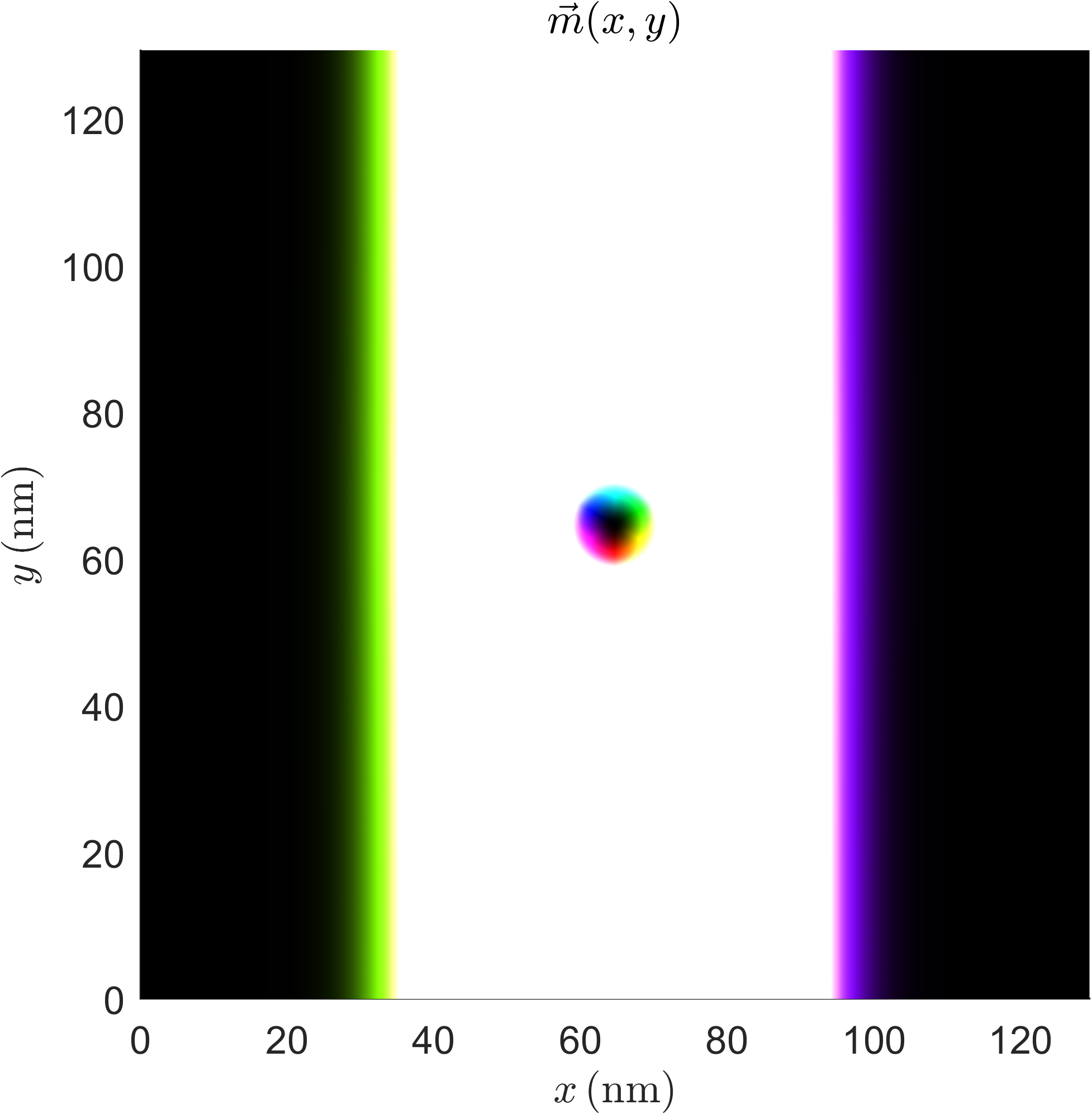} & $+1$ \\
        $(\pi/3,3\pi/2)$ & $+1$ & \includegraphics[width=25mm]{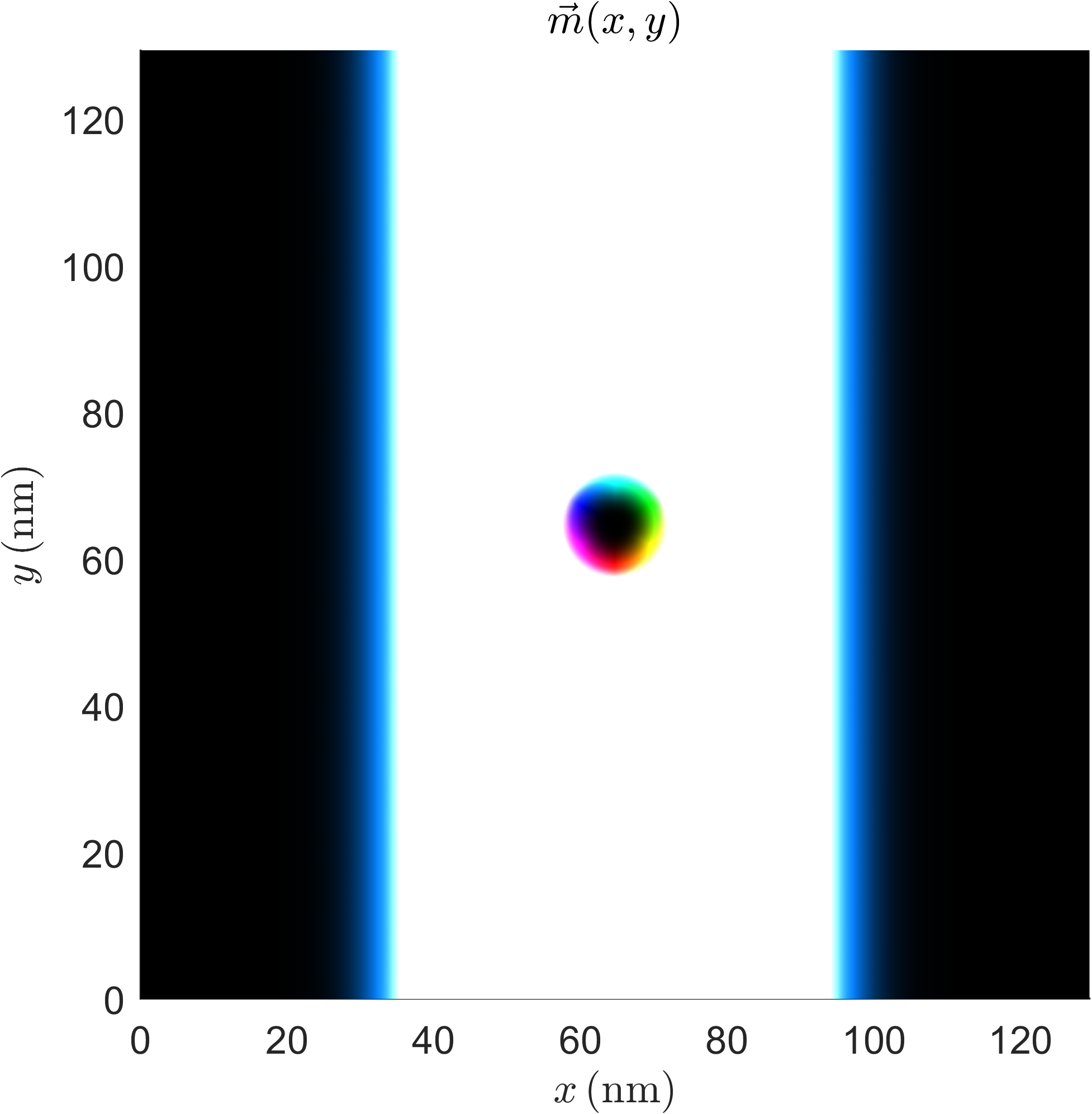} & \includegraphics[width=25mm]{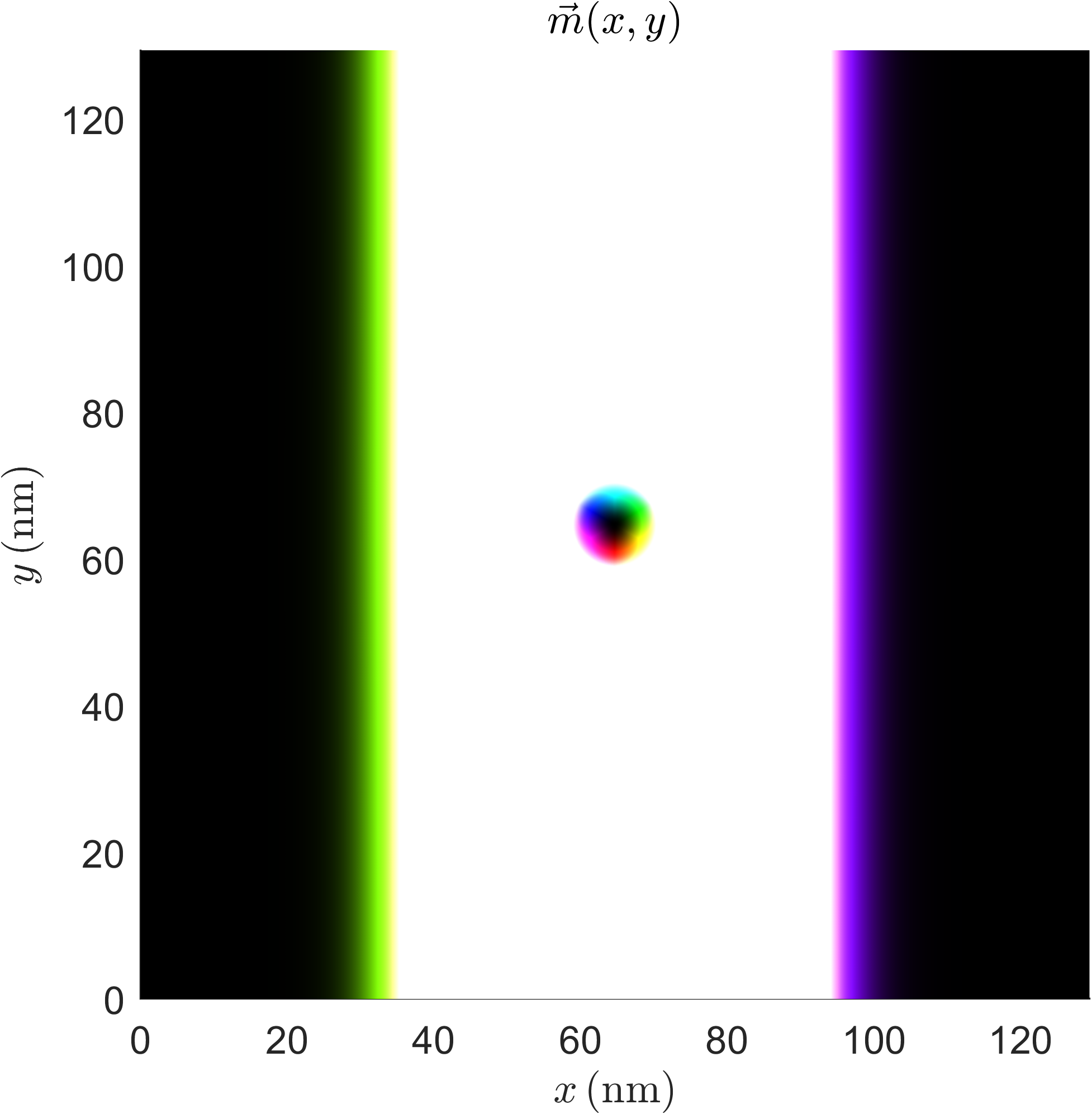} & $+1$ \\
        $(\pi/3,5\pi/3)$ & $+1$ & \includegraphics[width=25mm]{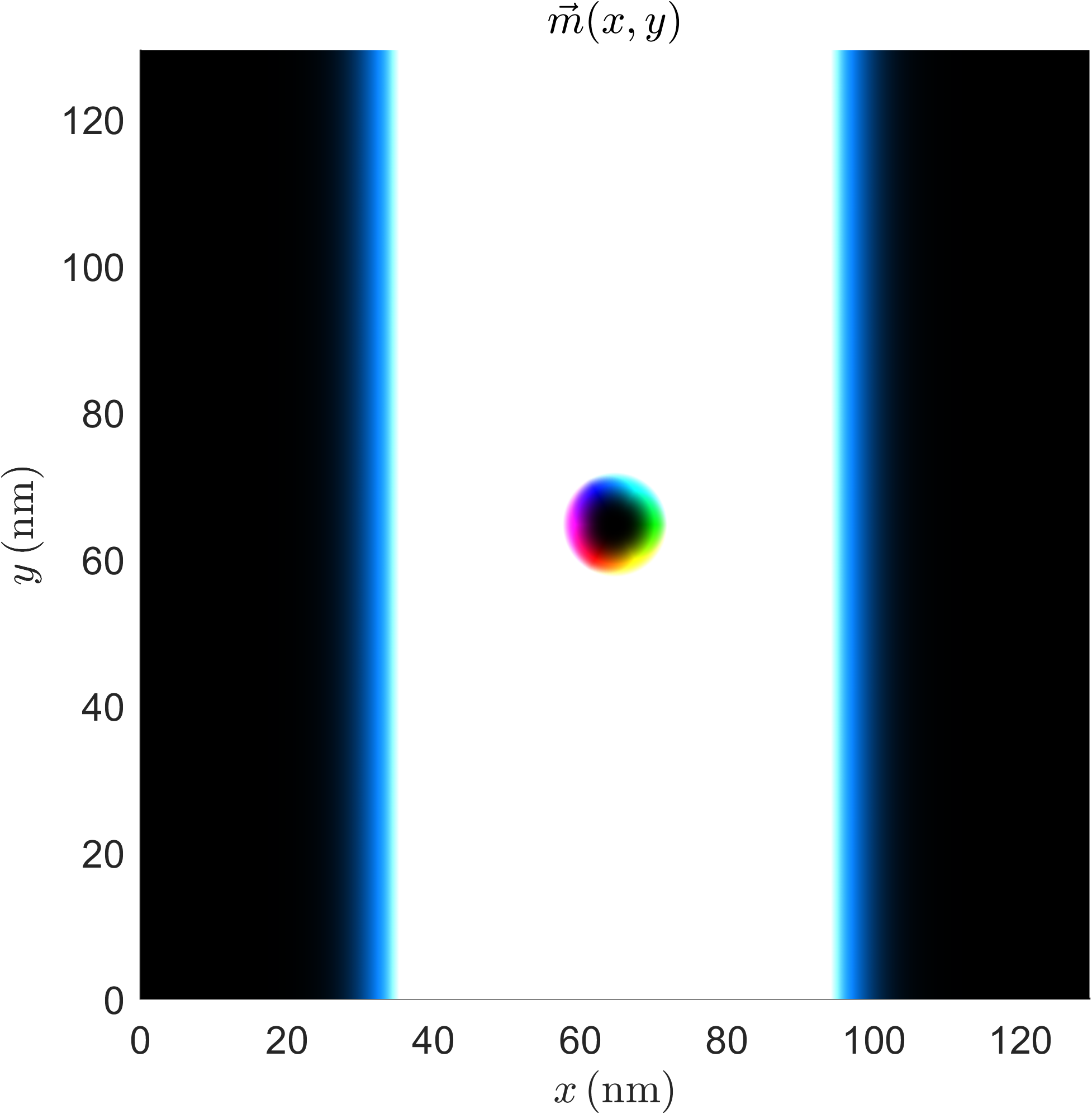} & \includegraphics[width=25mm]{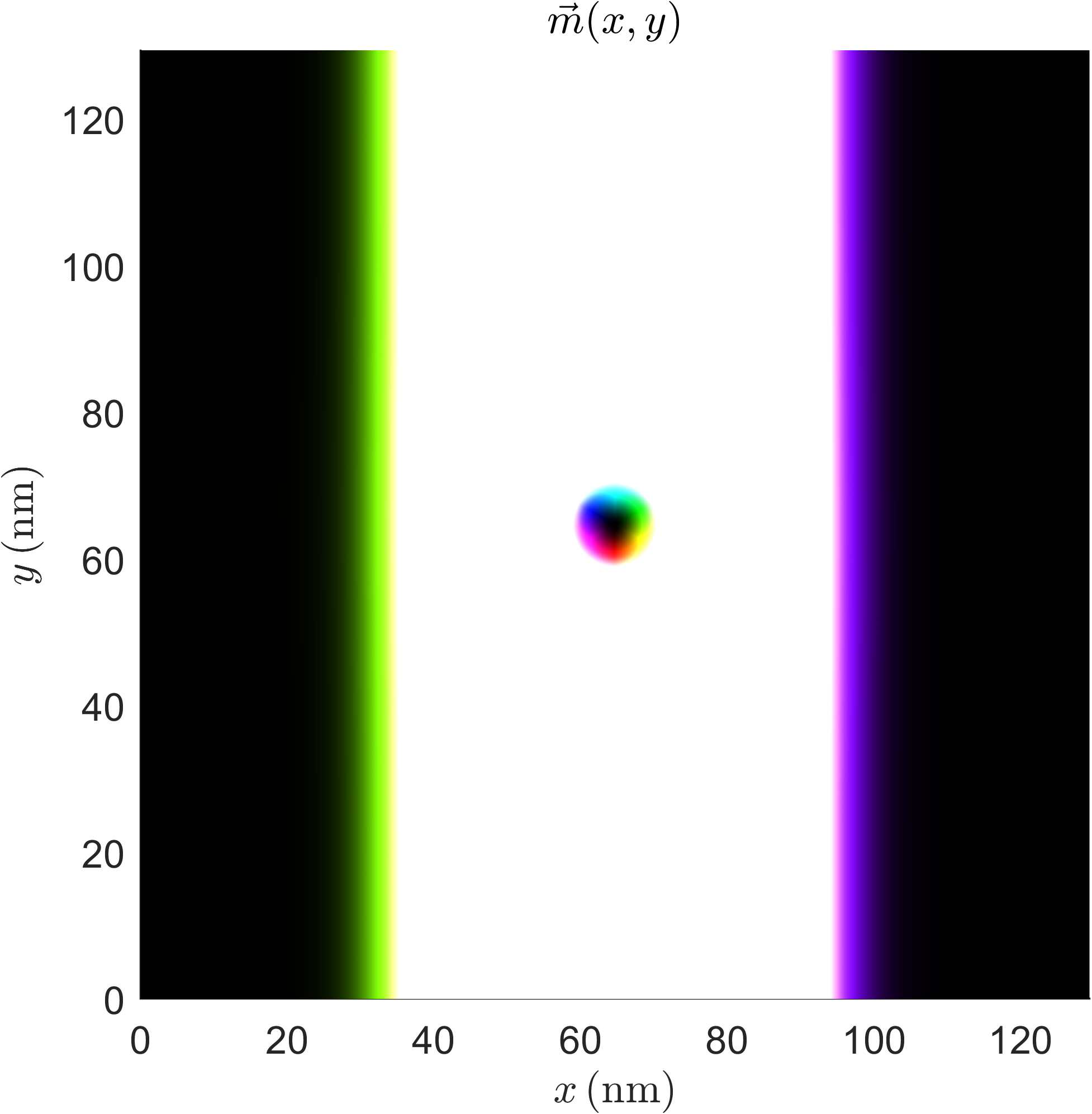} & $+1$ \\
        \bottomrule
    \end{tabular}
    \caption{Initial and final states for the domain wall phase $\chi=\pi/3$ as the skyrmion is rotated by $\pi/3$ from $\phi=0$ until $\phi=5\pi/3$.}
    \label{tbl: chi = pi/3}
\end{table}

\begin{table}
    \centering
    \begin{tabular}{ccM{40mm}M{40mm}c}
        \toprule
        $(\chi,\phi)$ & $Q_{\textup{i}}$ & Initial Configuration & Final Configuration & $Q_{\textup{f}}$ \\
        \midrule
        $(\pi/2,0)$ & $+1$ & \includegraphics[width=25mm]{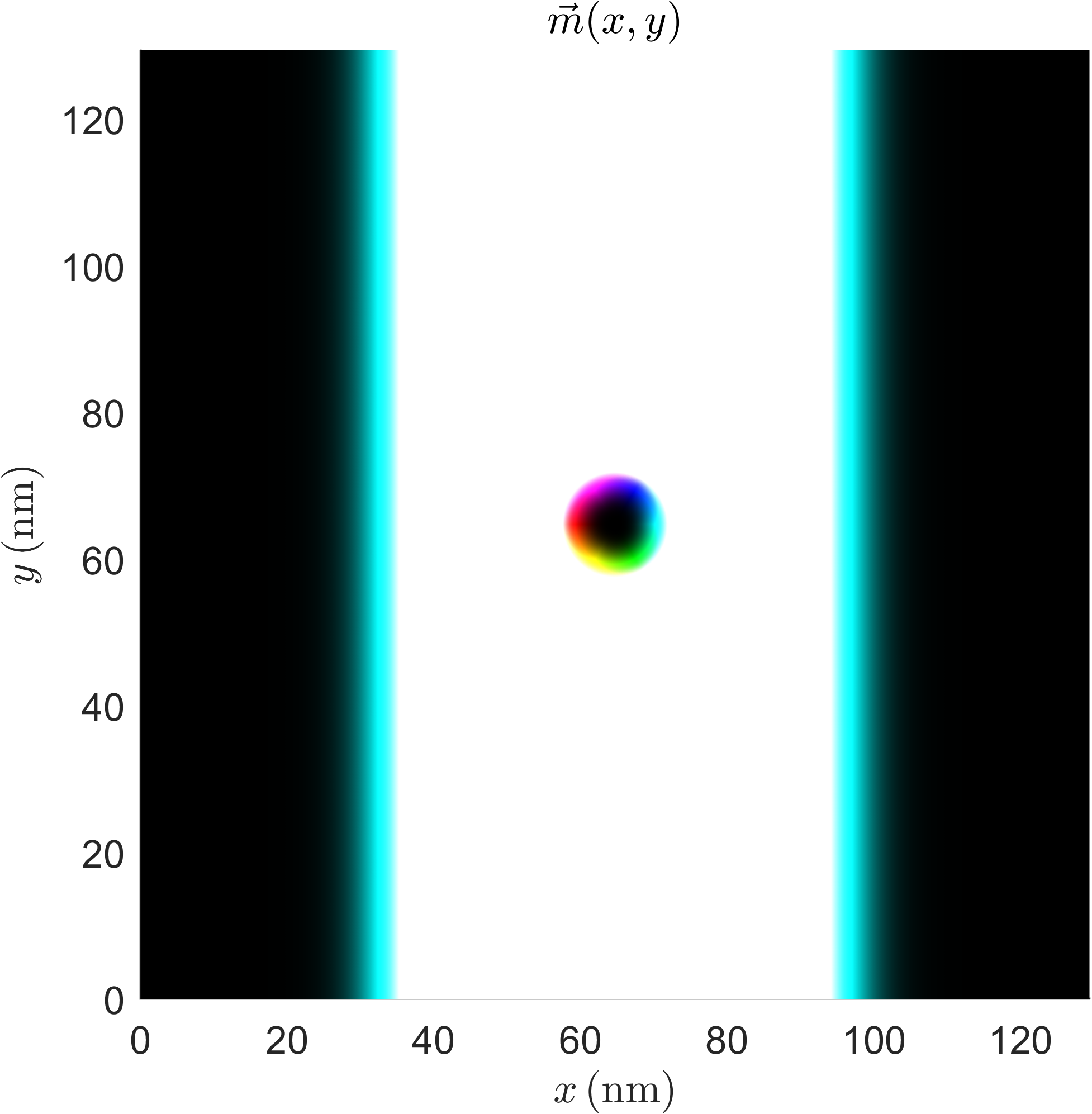} & \includegraphics[width=25mm]{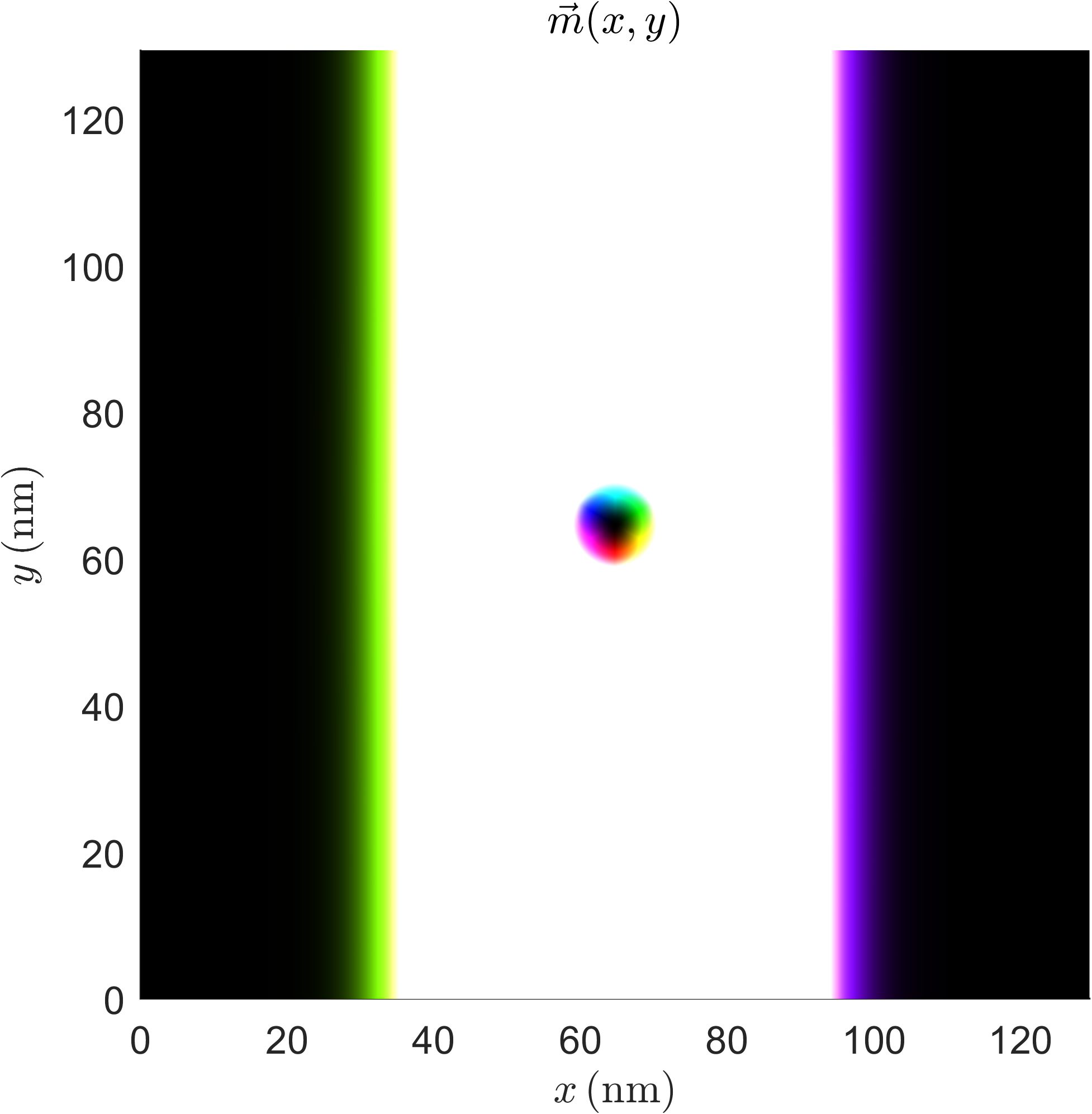} & $+1$ \\
        $(\pi/2,\pi/3)$ & $+1$ & \includegraphics[width=25mm]{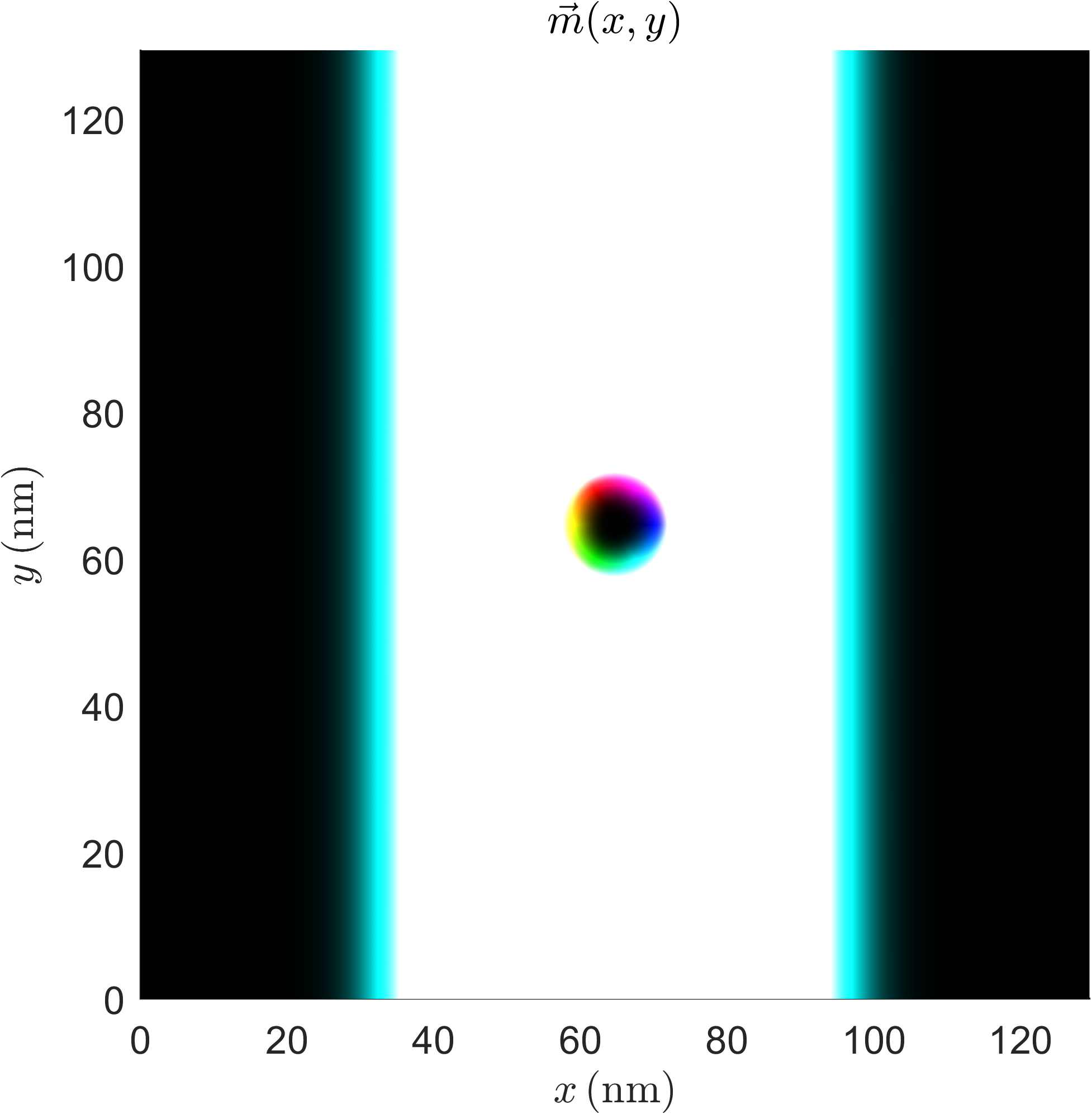} & \includegraphics[width=25mm]{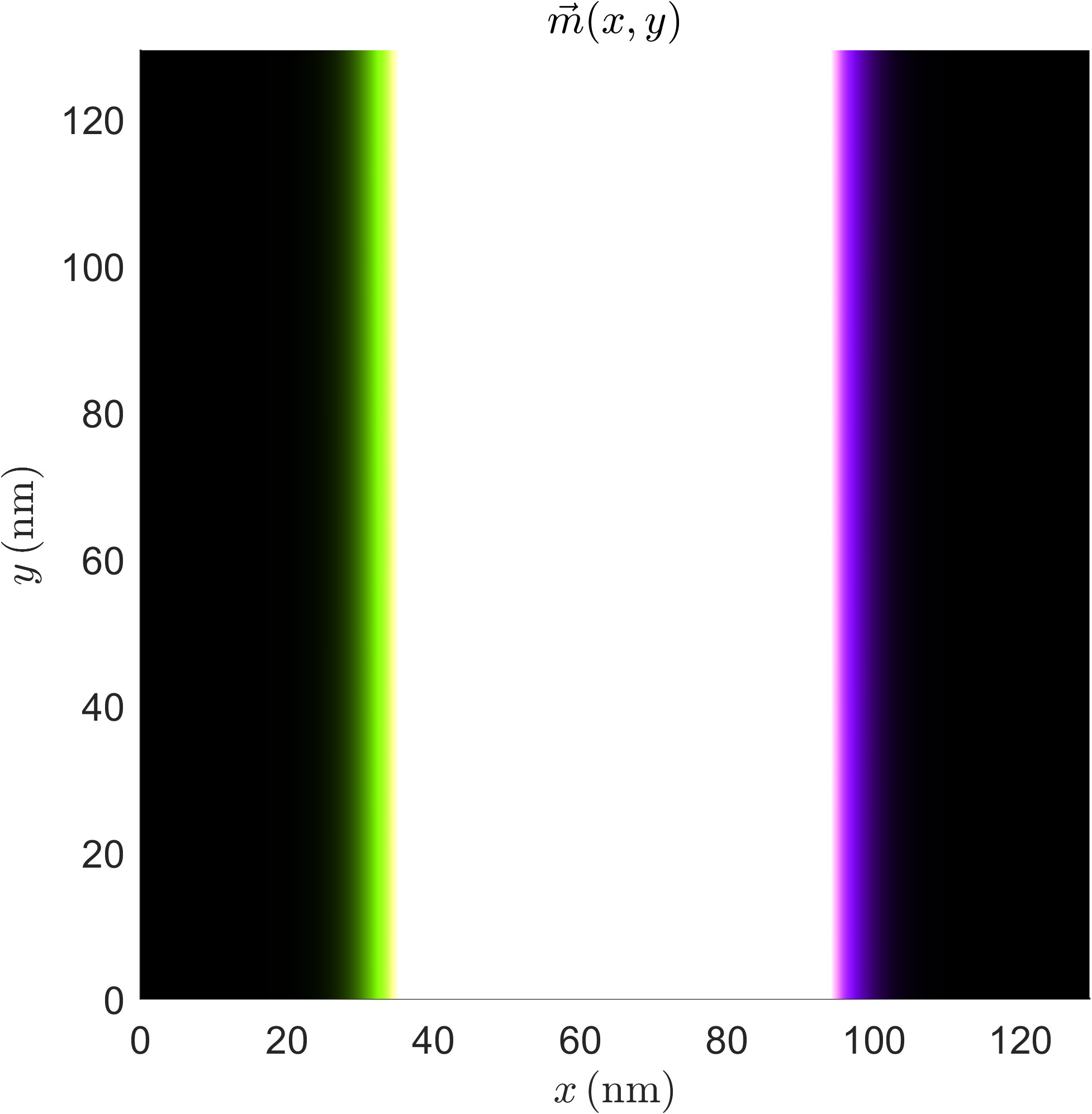} & $0$ \\
        $(\pi/2,\pi/2)$ & $+1$ & \includegraphics[width=25mm]{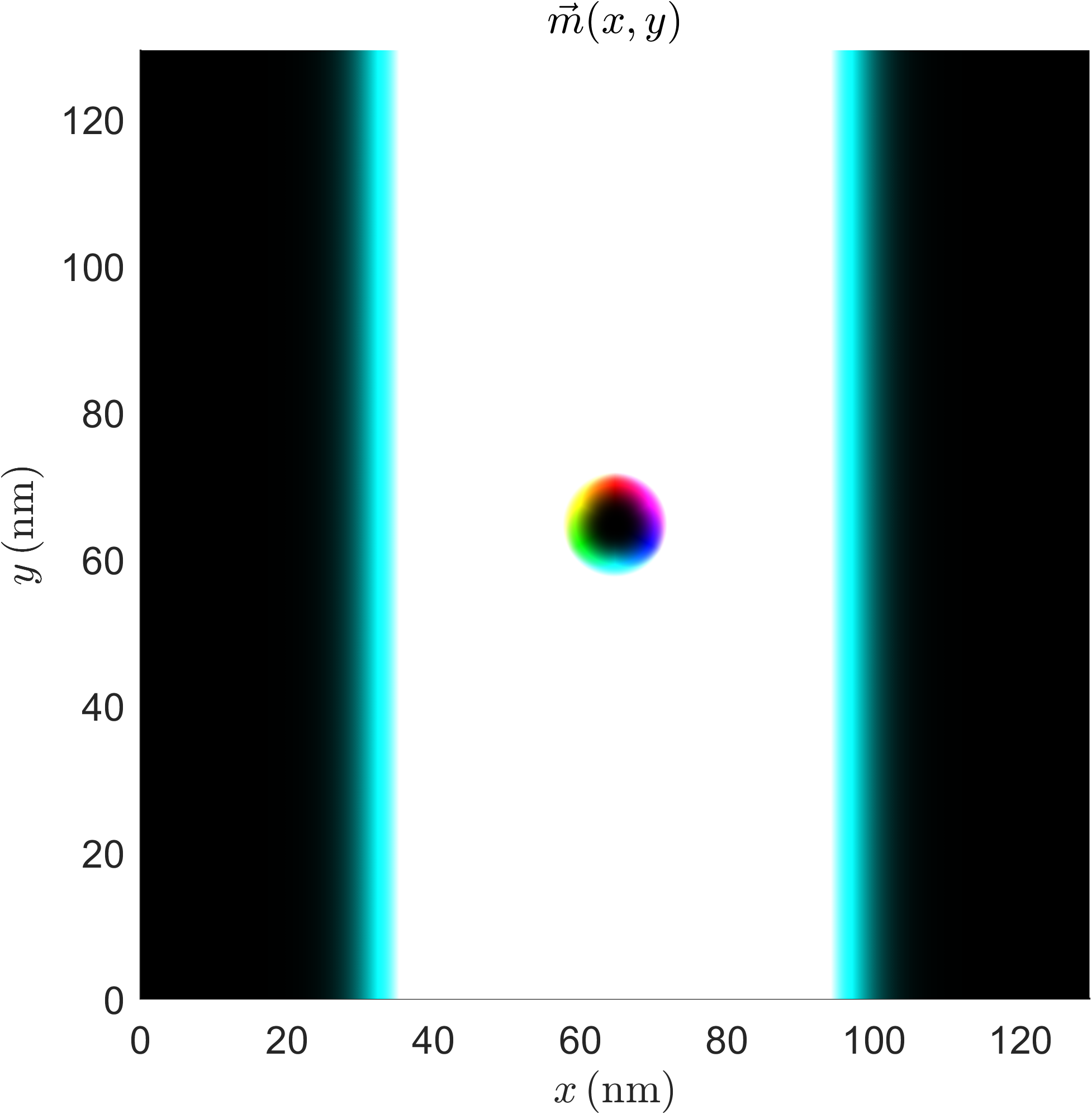} & \includegraphics[width=25mm]{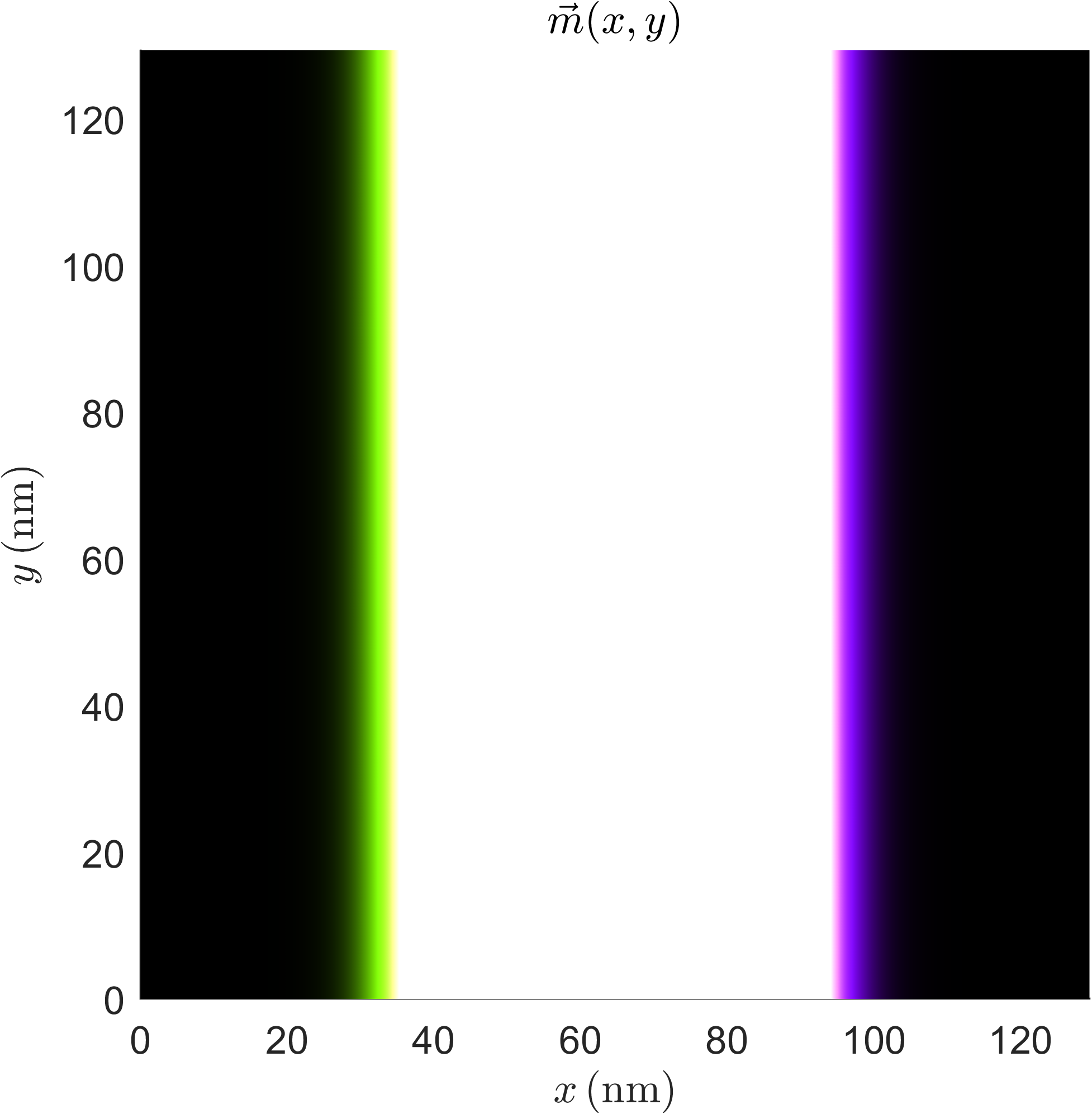} & $0$ \\
        $(\pi/2,2\pi/3)$ & $+1$ & \includegraphics[width=25mm]{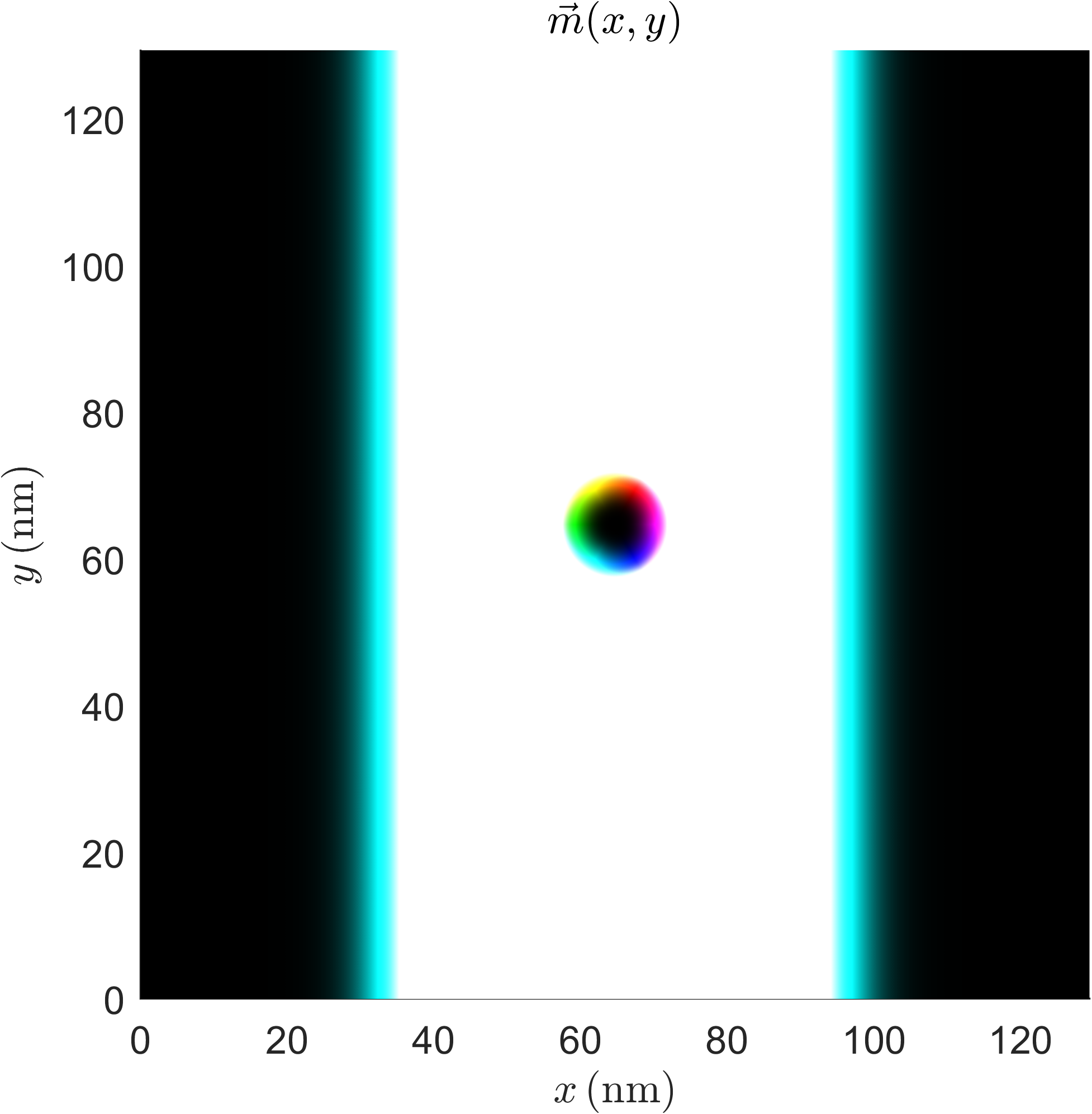} & \includegraphics[width=25mm]{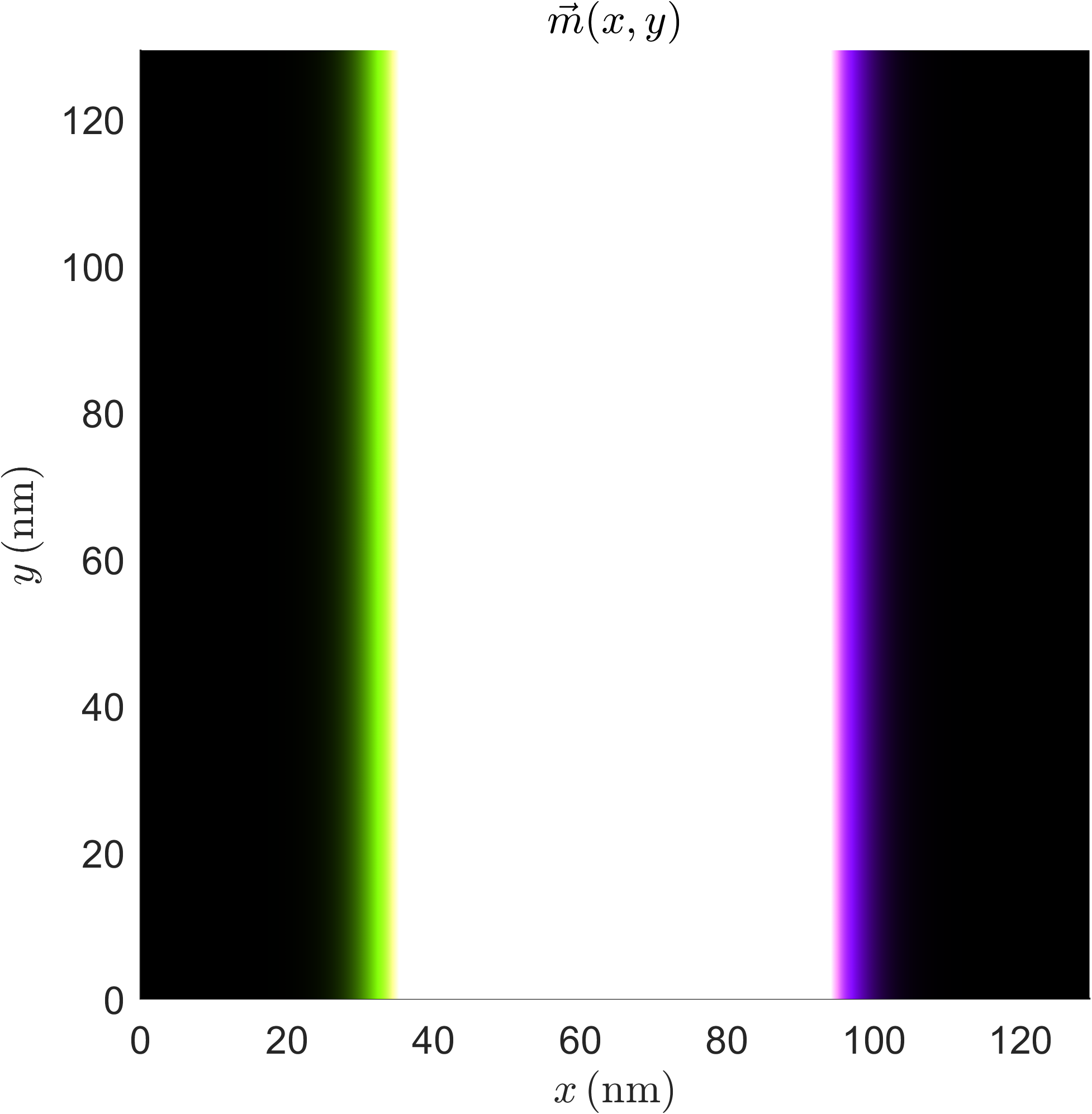} & $0$ \\
        $(\pi/2,\pi)$ & $+1$ & \includegraphics[width=25mm]{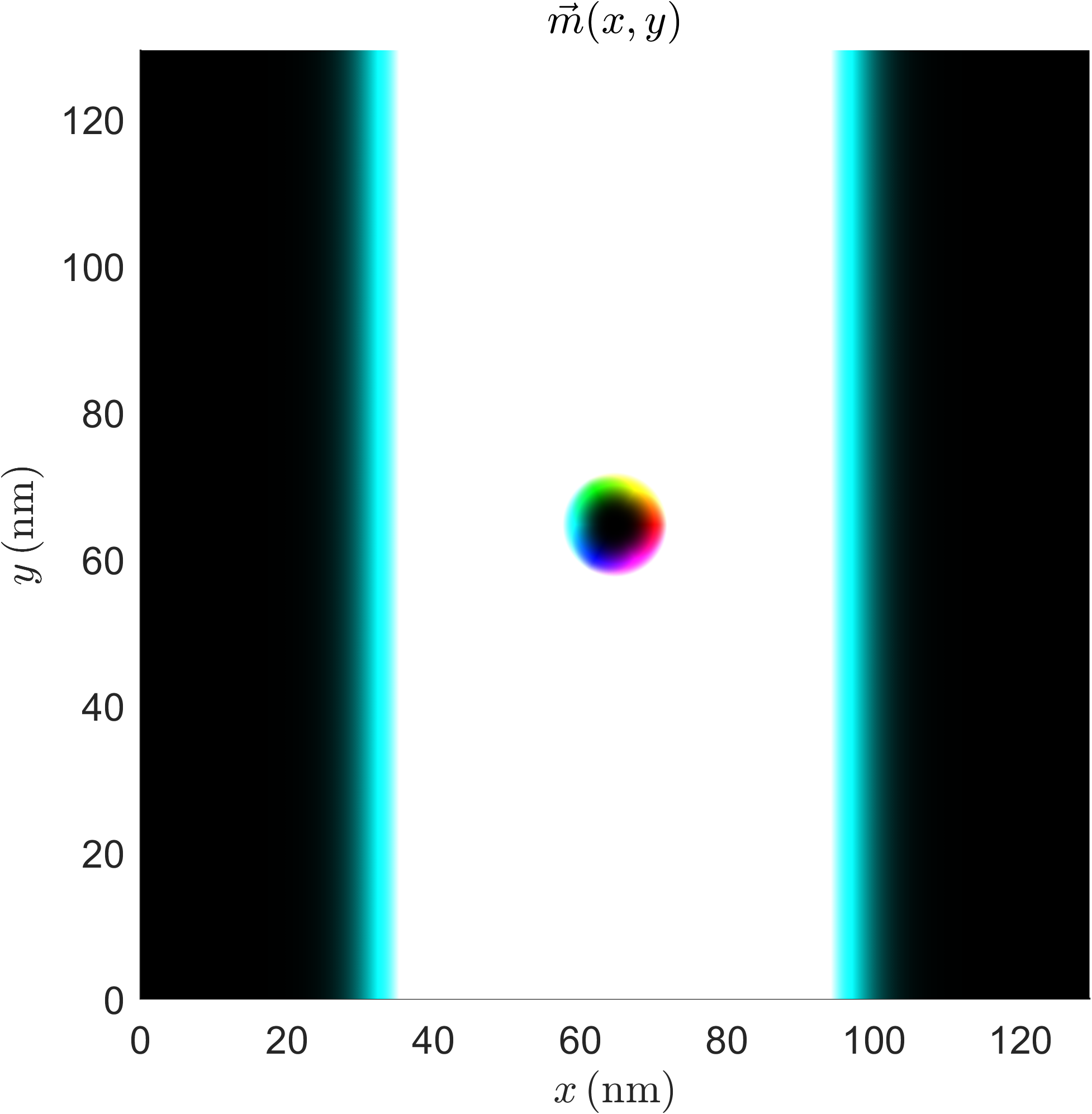} & \includegraphics[width=25mm]{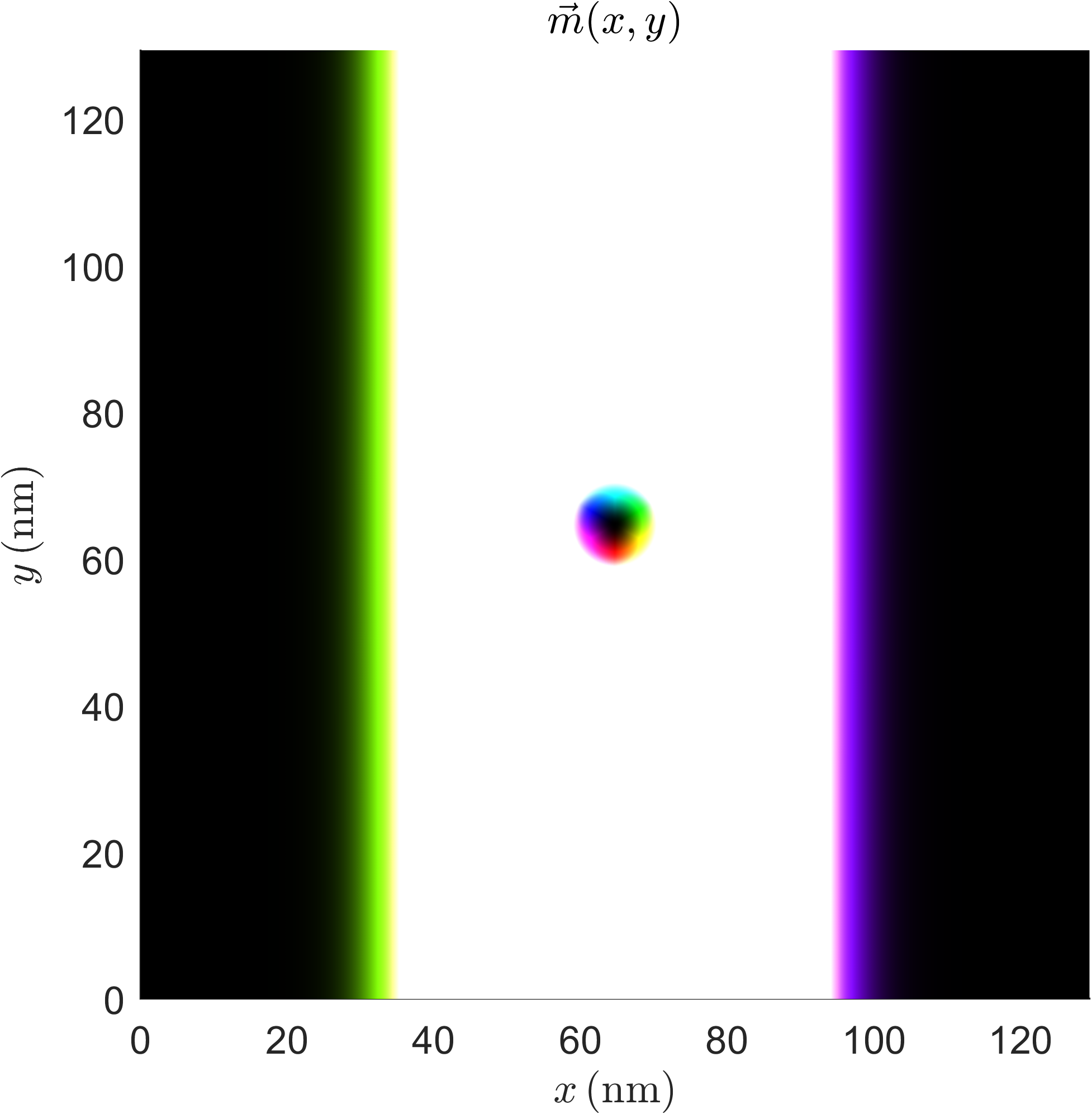} & $+1$ \\
        $(\pi/2,4\pi/3)$ & $+1$ & \includegraphics[width=25mm]{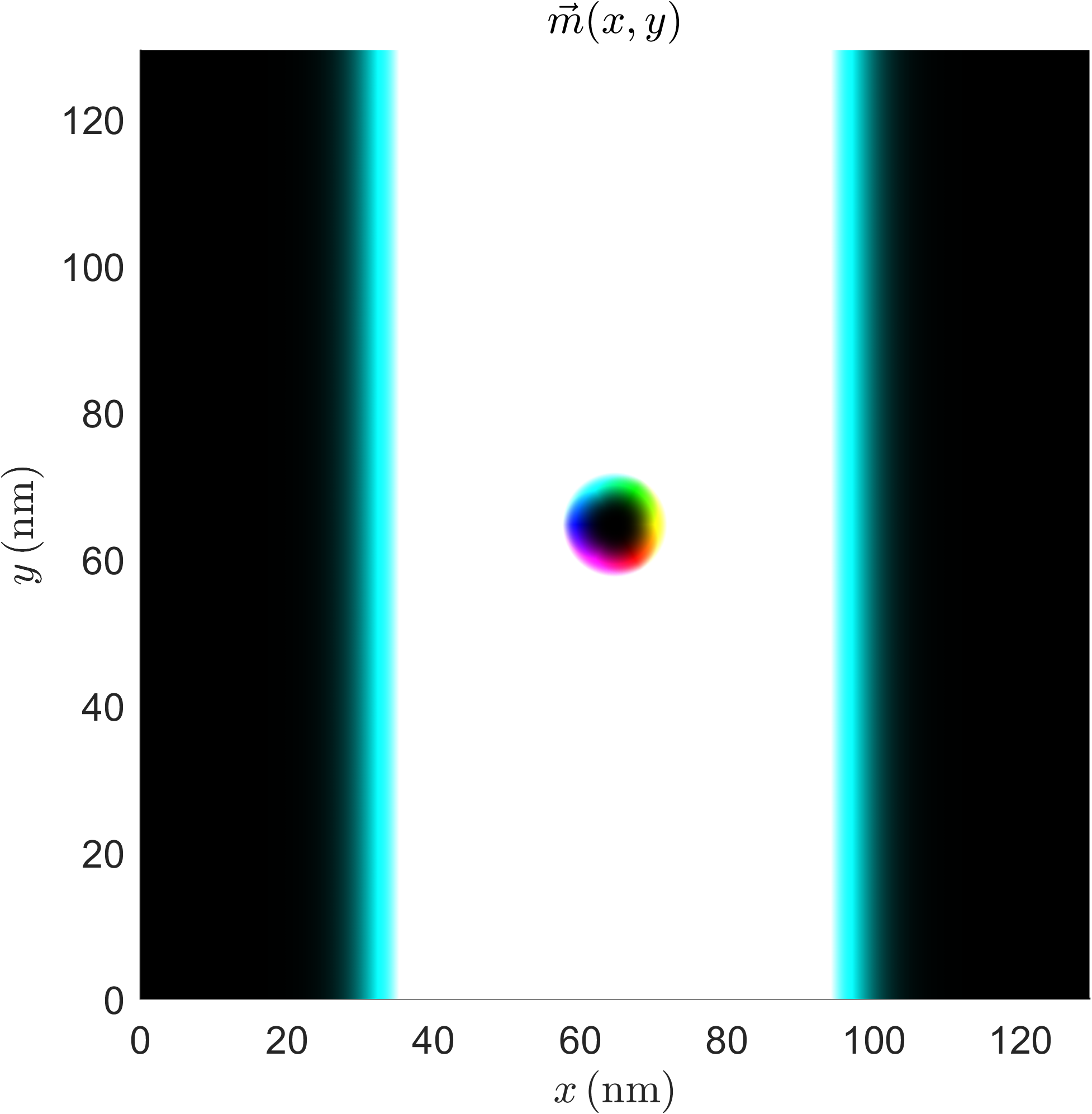} & \includegraphics[width=25mm]{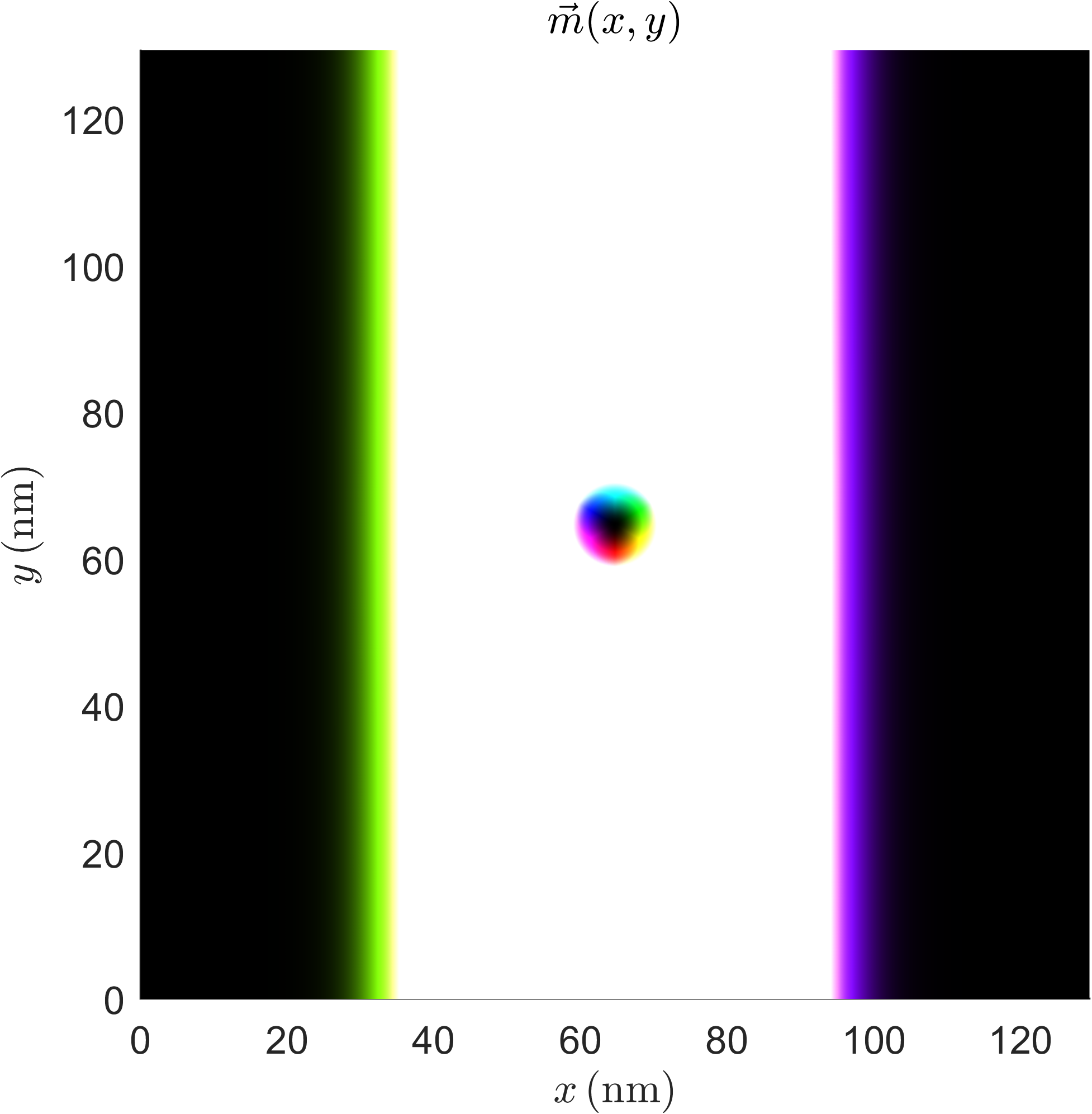} & $+1$ \\
        $(\pi/2,3\pi/2)$ & $+1$ & \includegraphics[width=25mm]{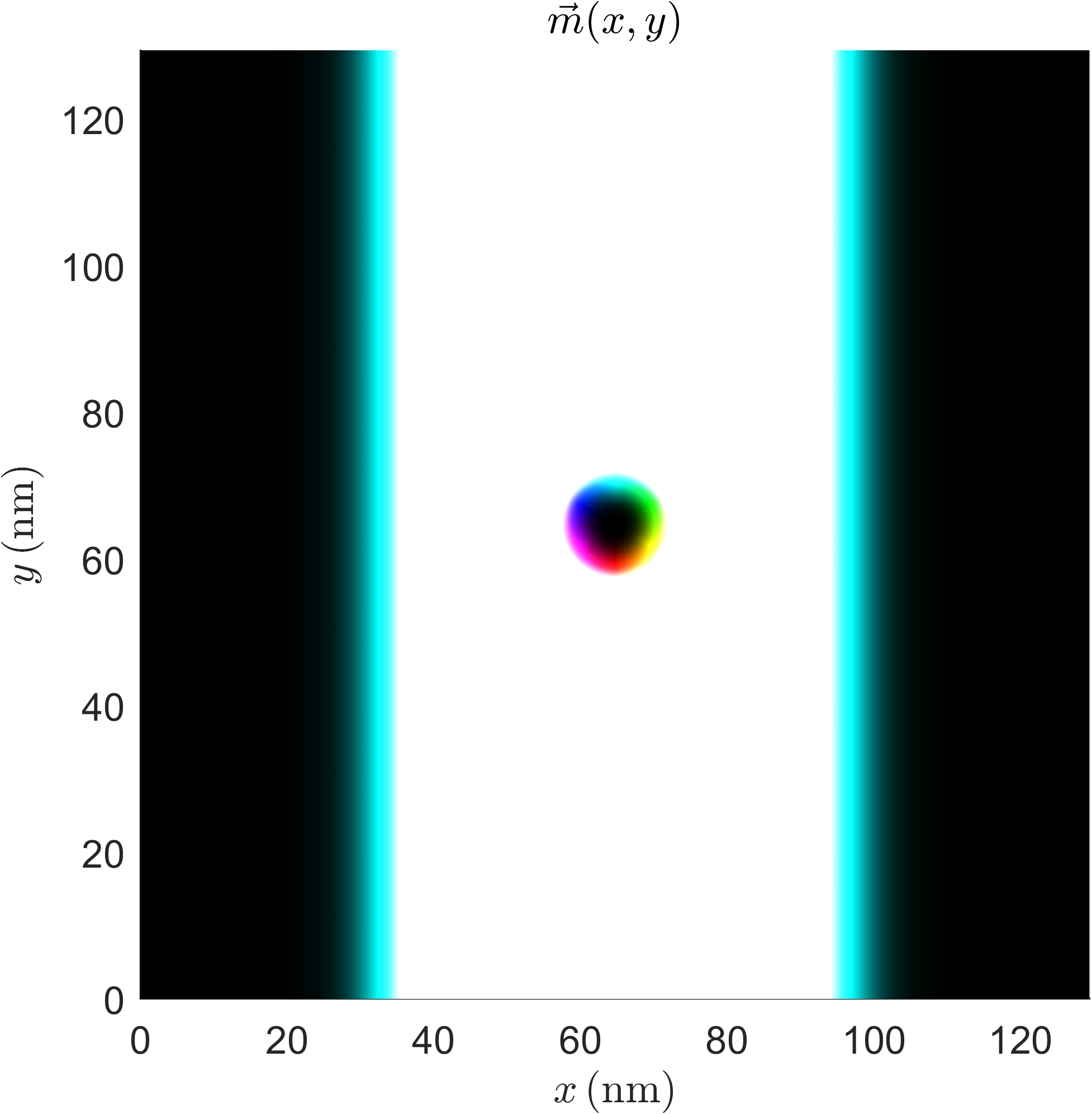} & \includegraphics[width=25mm]{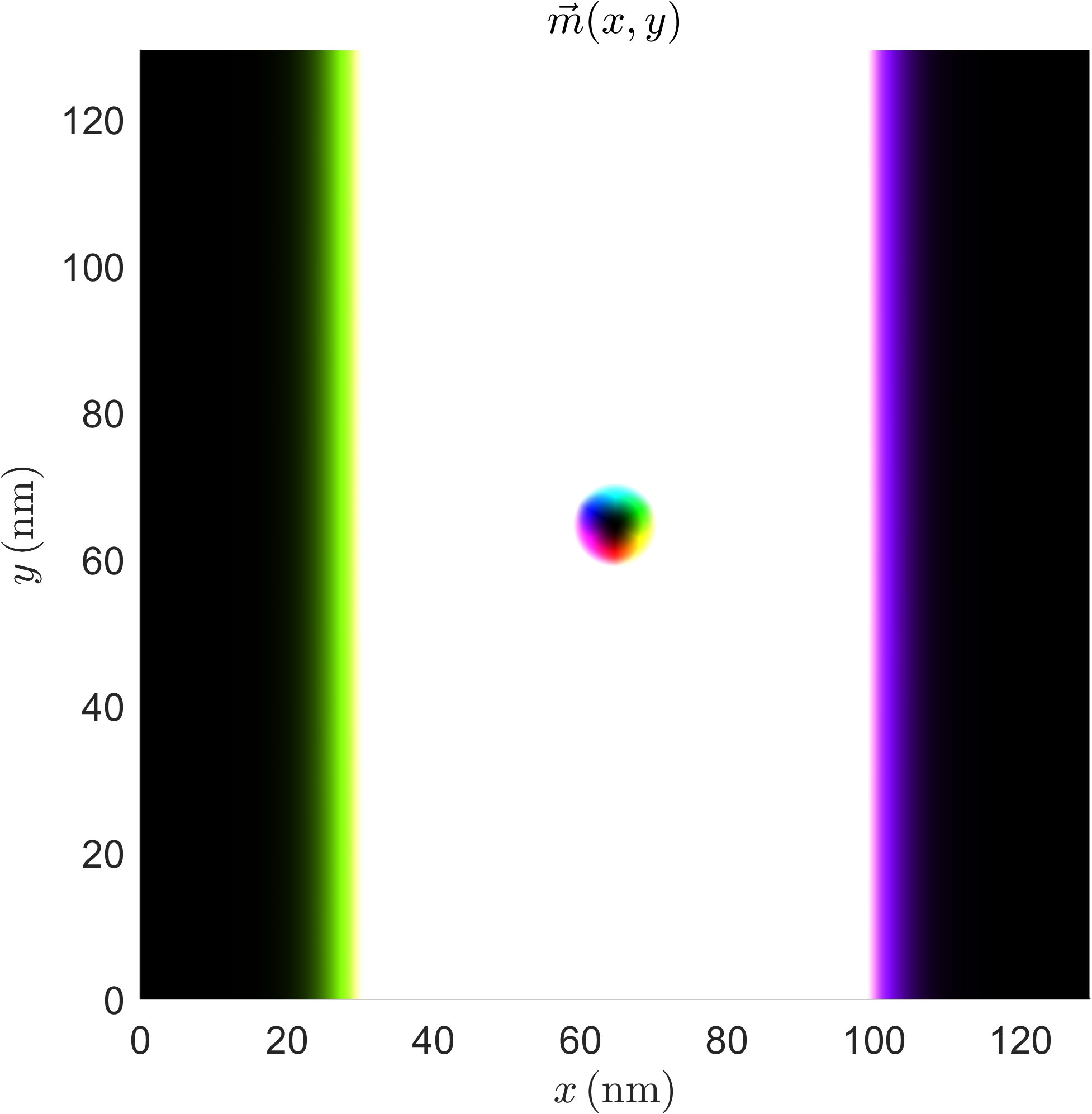} & $+1$ \\
        $(\pi/2,5\pi/3)$ & $+1$ & \includegraphics[width=25mm]{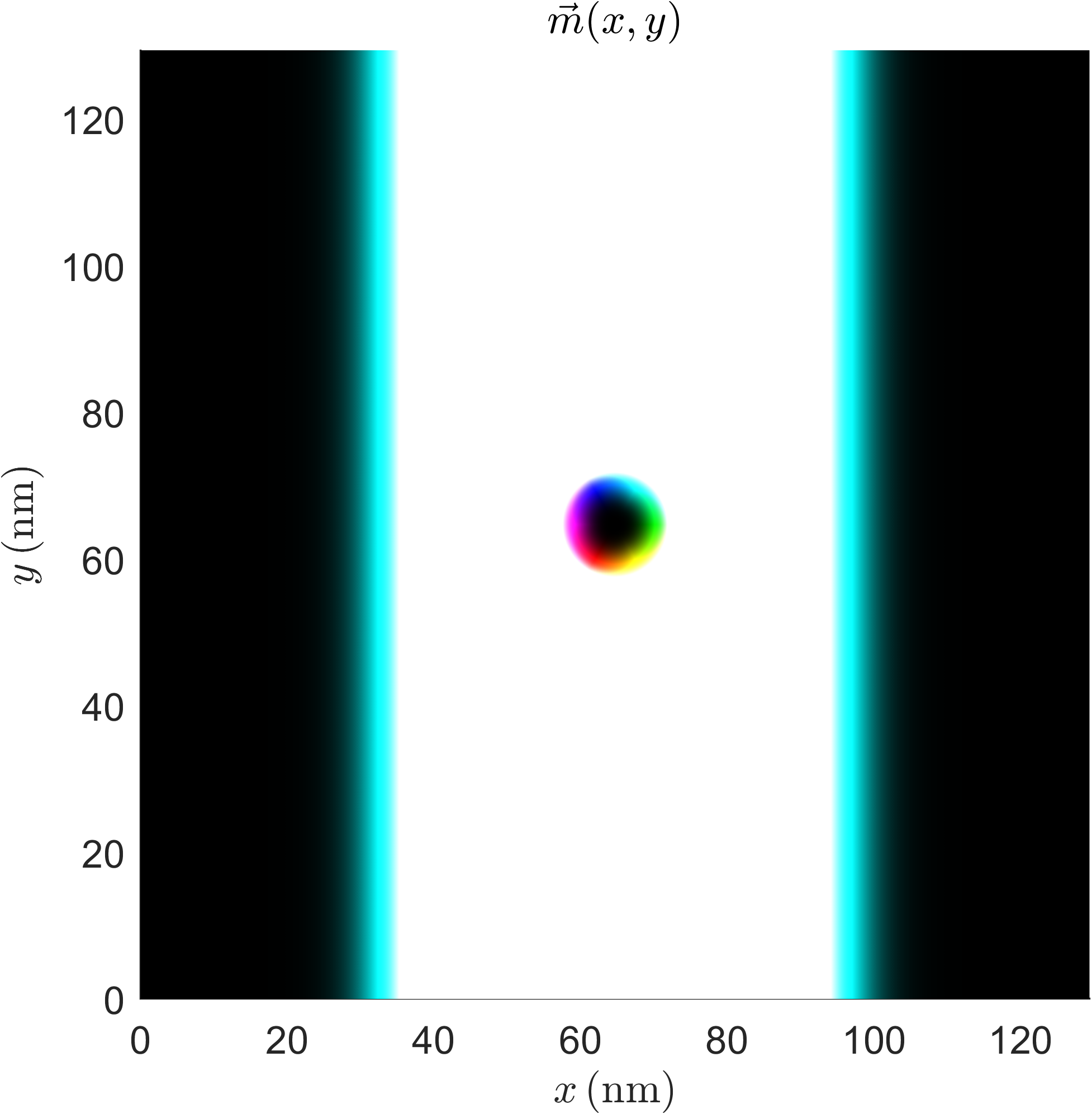} & \includegraphics[width=25mm]{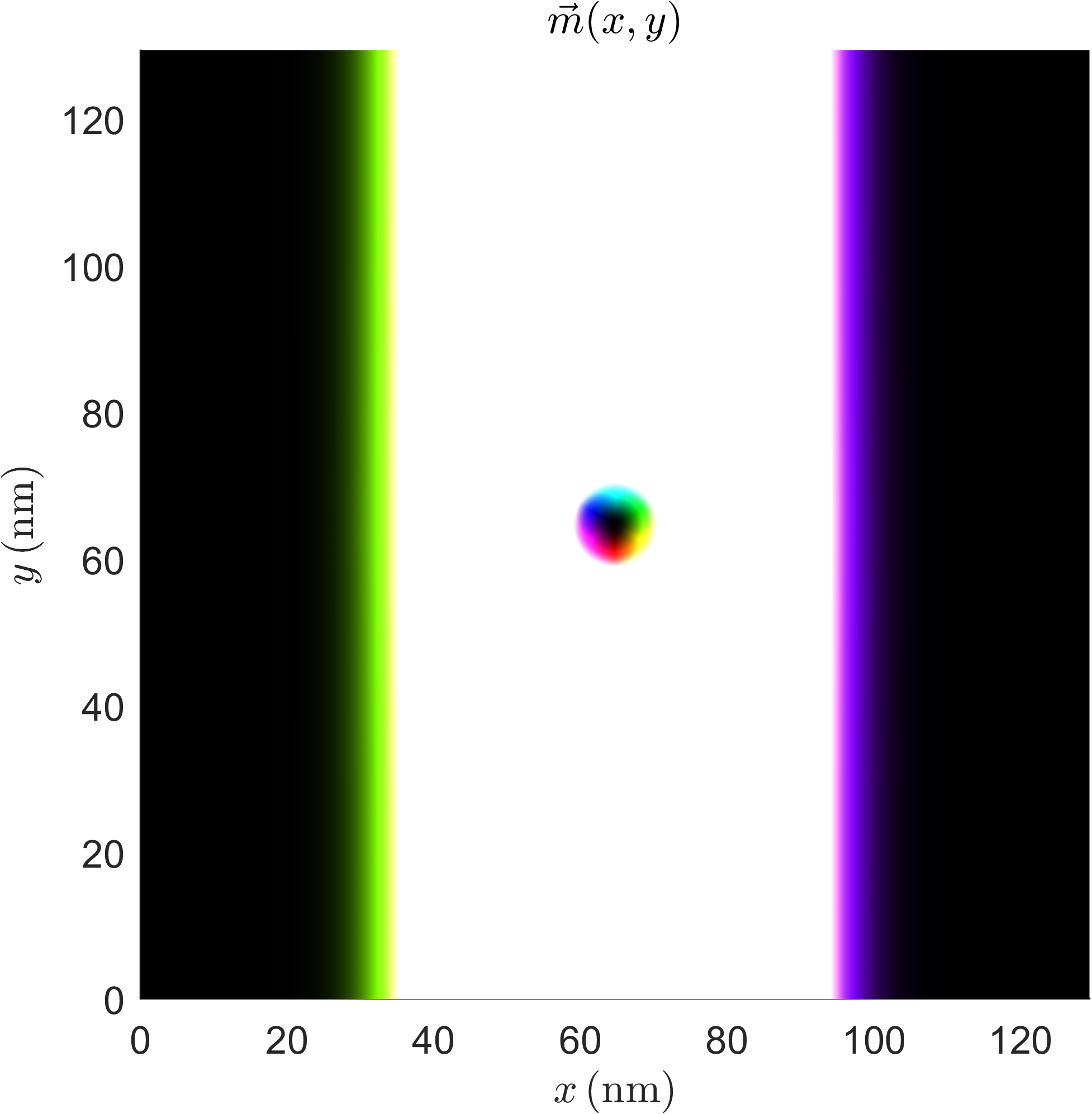} & $+1$ \\
        \bottomrule
    \end{tabular}
    \caption{Initial and final states for the domain wall phase $\chi=\pi/2$ as the skyrmion is rotated by $\pi/3$ from $\phi=0$ until $\phi=5\pi/3$.}
    \label{tbl: chi = pi/2}
\end{table}

\begin{table}
    \centering
    \begin{tabular}{ccM{40mm}M{40mm}c}
        \toprule
        $(\chi,\phi)$ & $Q_{\textup{i}}$ & Initial Configuration & Final Configuration & $Q_{\textup{f}}$ \\
        \midrule
        $(2\pi/3,0)$ & $+1$ & \includegraphics[width=25mm]{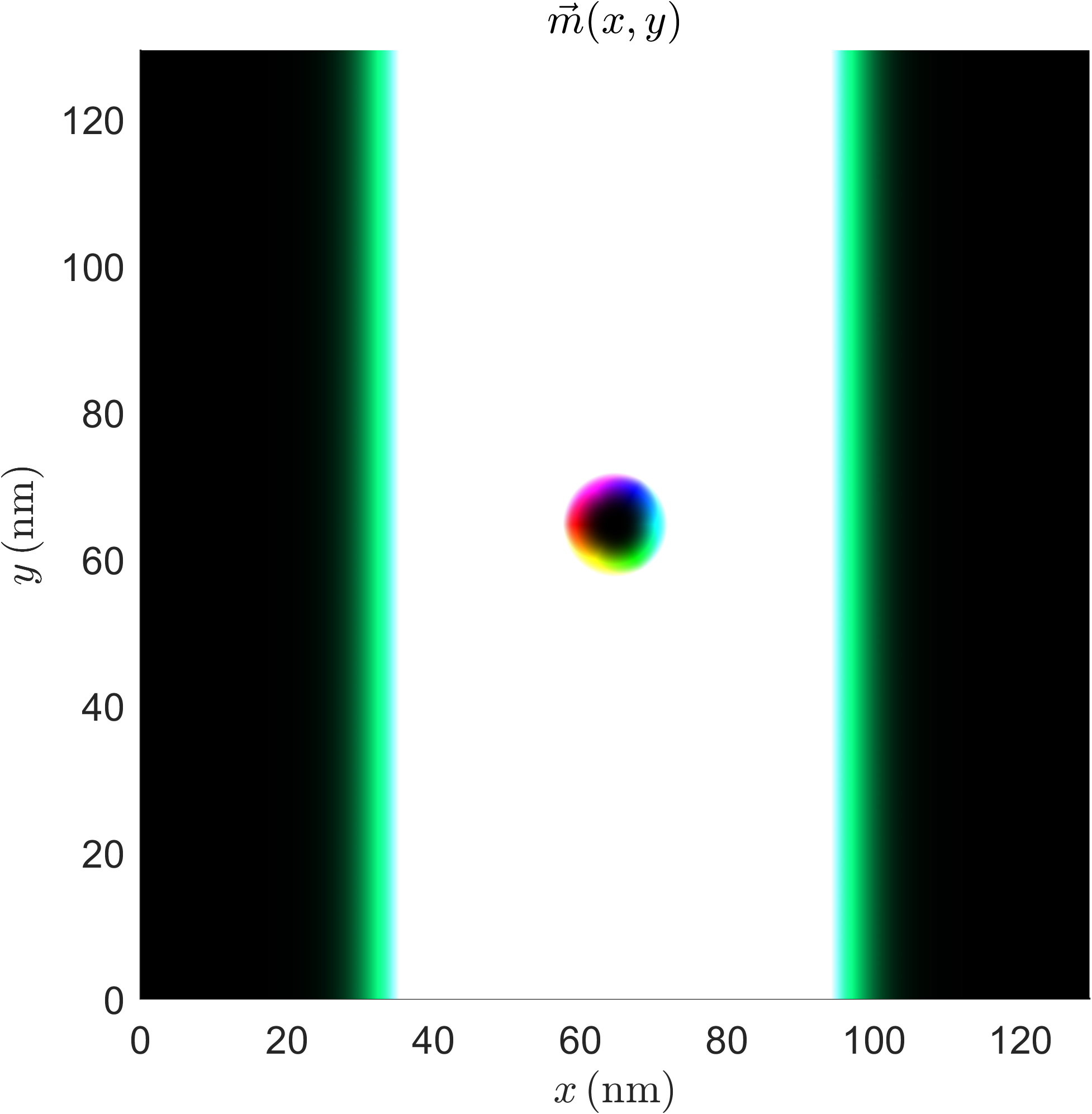} & \includegraphics[width=25mm]{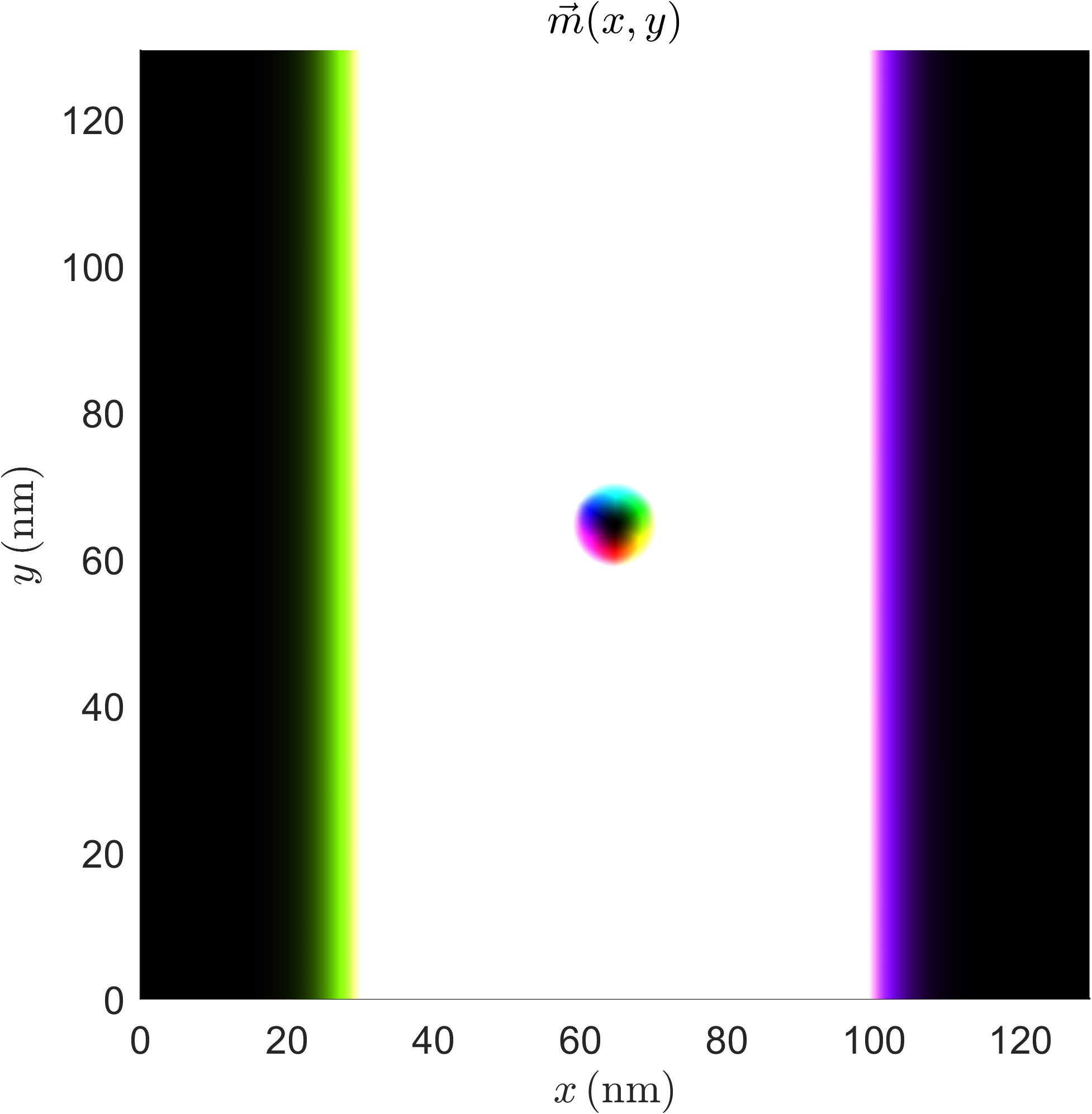} & $+1$ \\
        $(2\pi/3,\pi/3)$ & $+1$ & \includegraphics[width=25mm]{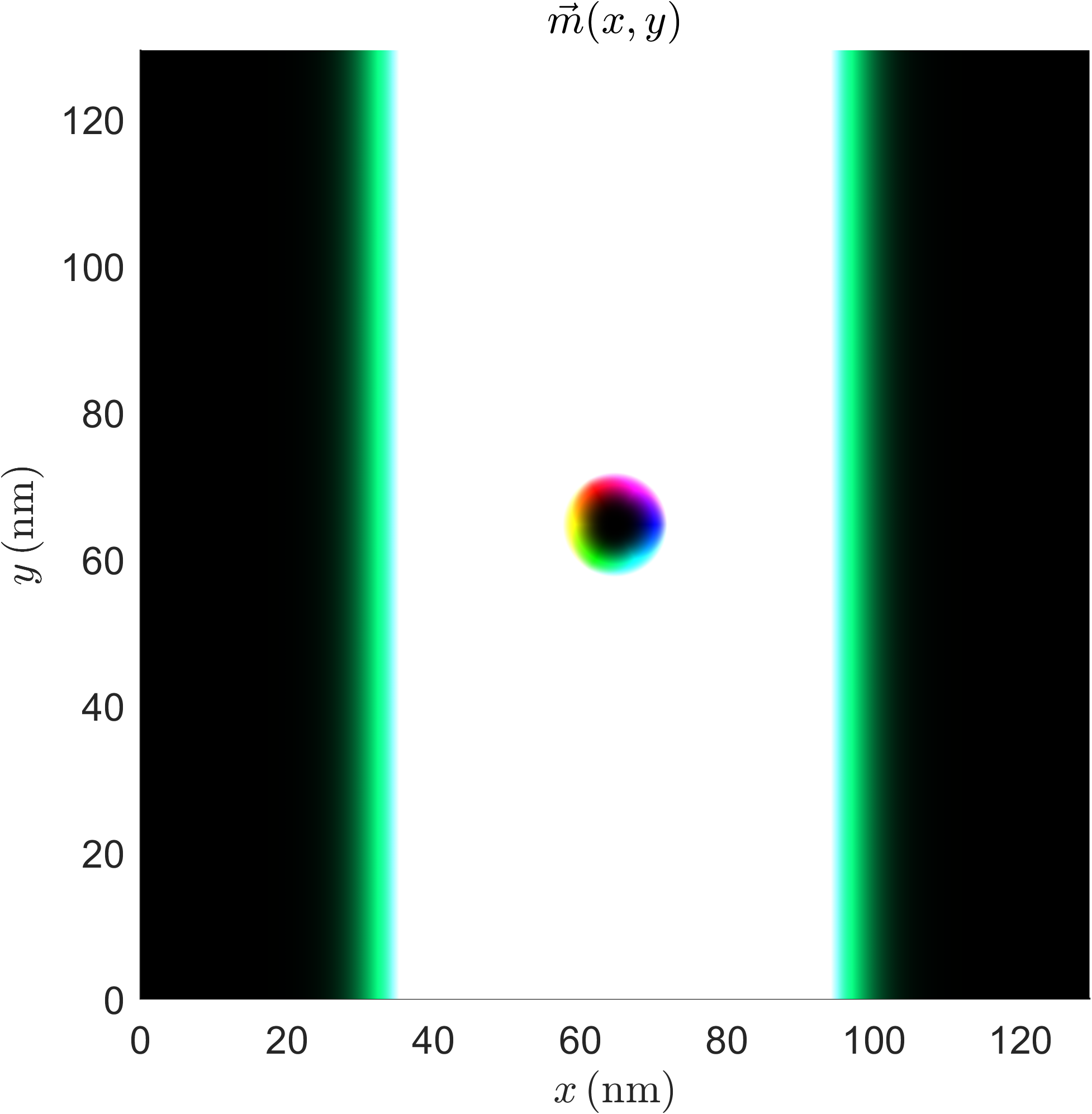} & \includegraphics[width=25mm]{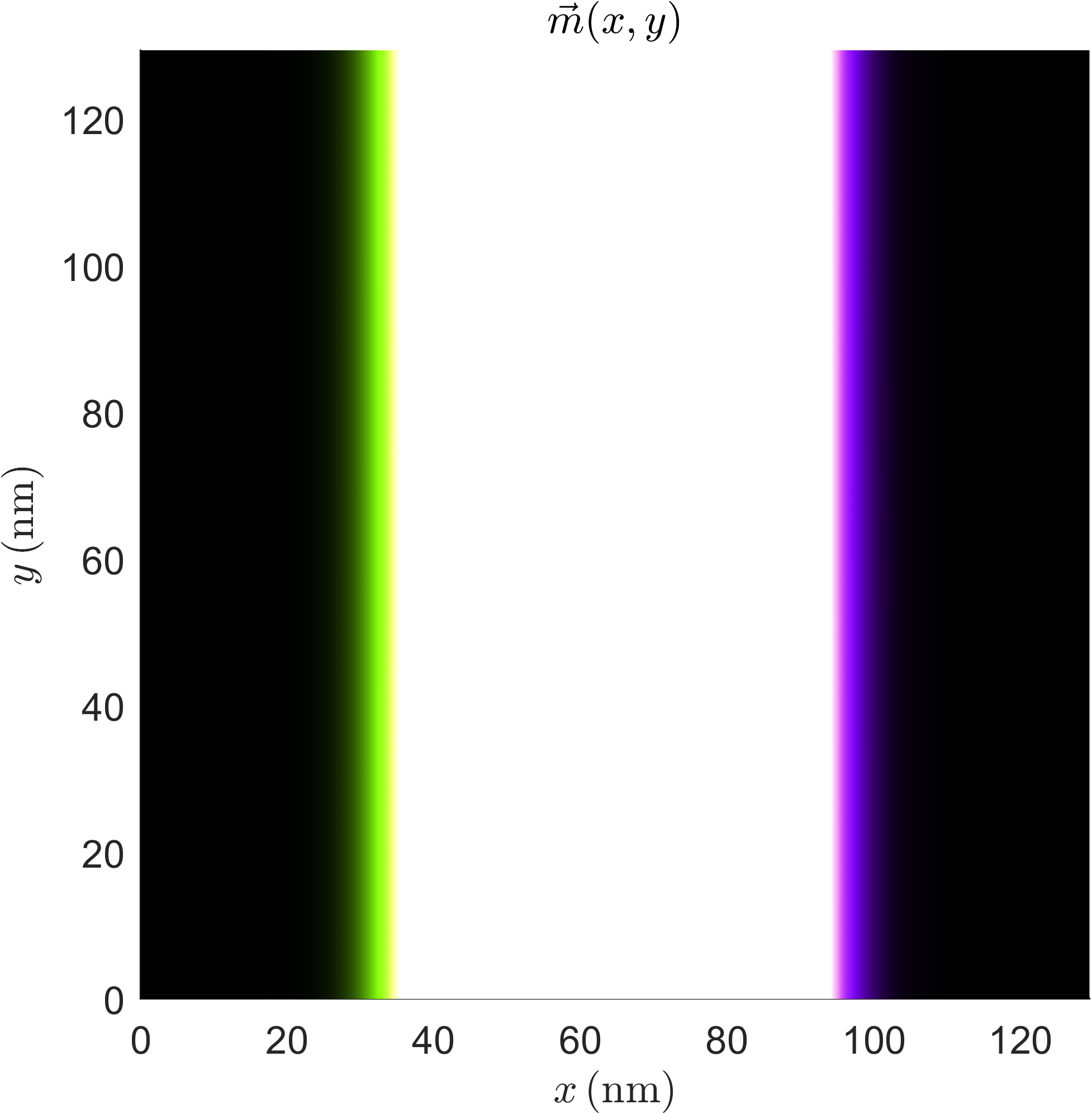} & $0$ \\
        $(2\pi/3,\pi/2)$ & $+1$ & \includegraphics[width=25mm]{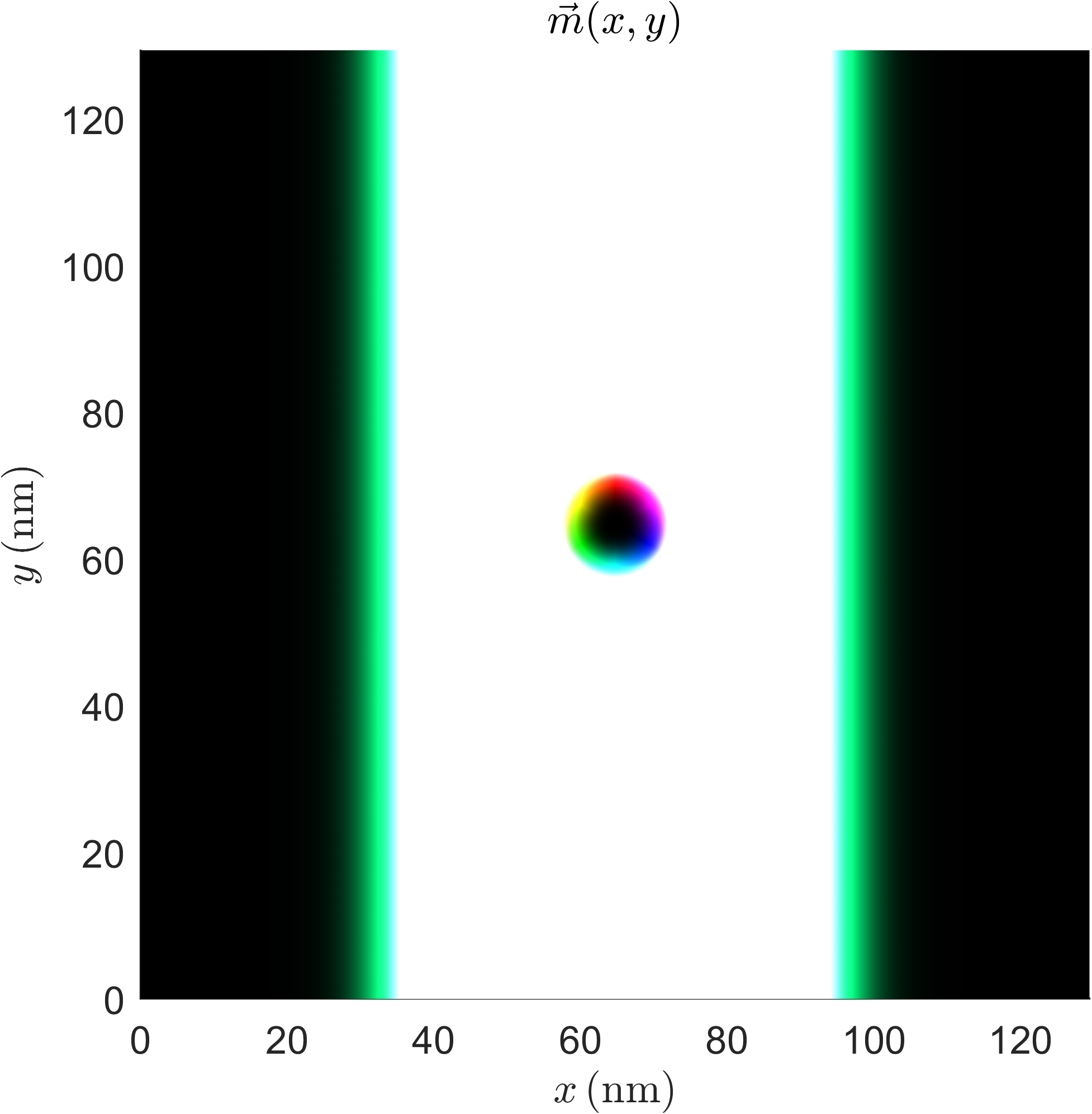} & \includegraphics[width=25mm]{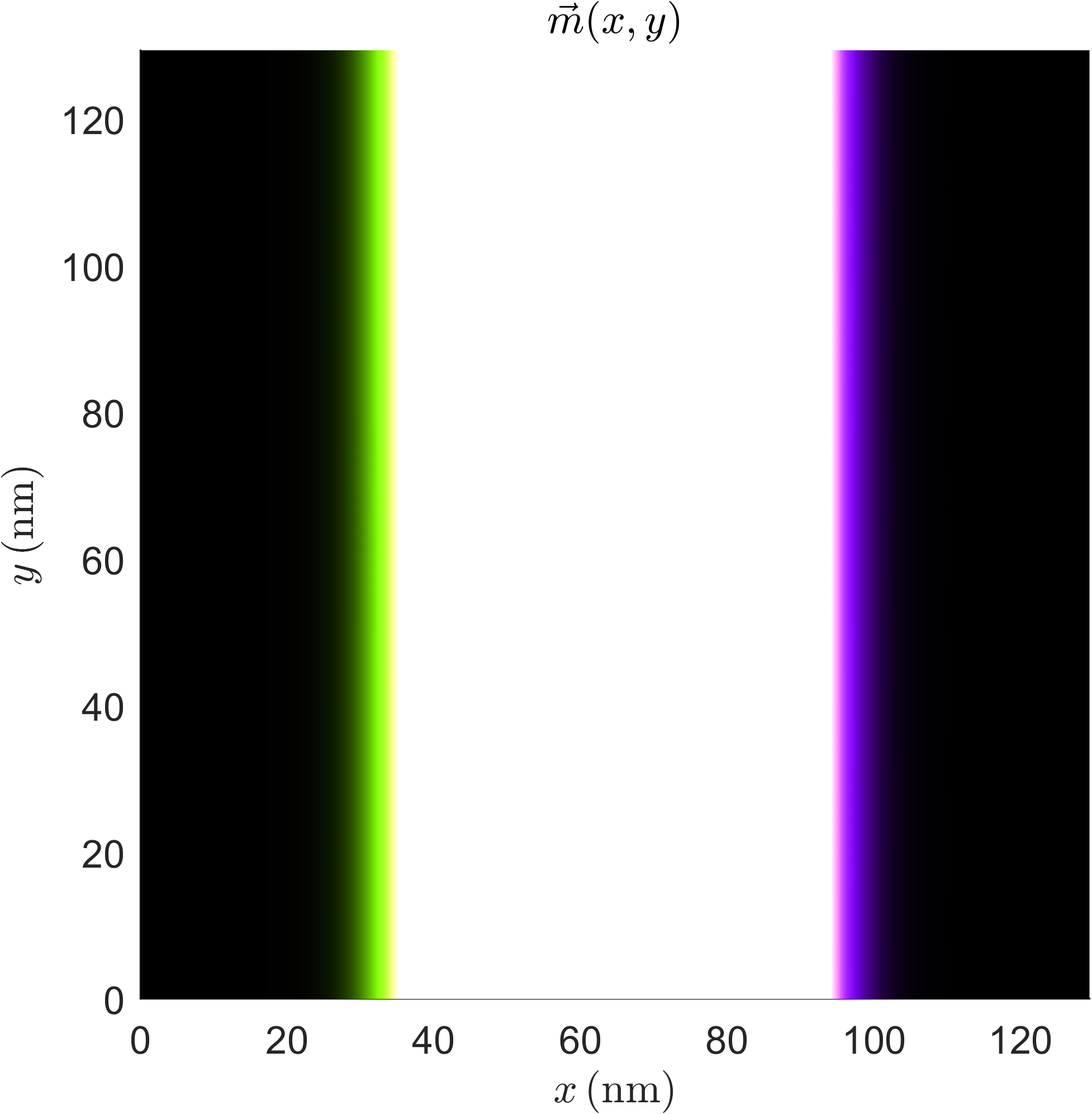} & $0$ \\
        $(2\pi/3,2\pi/3)$ & $+1$ & \includegraphics[width=25mm]{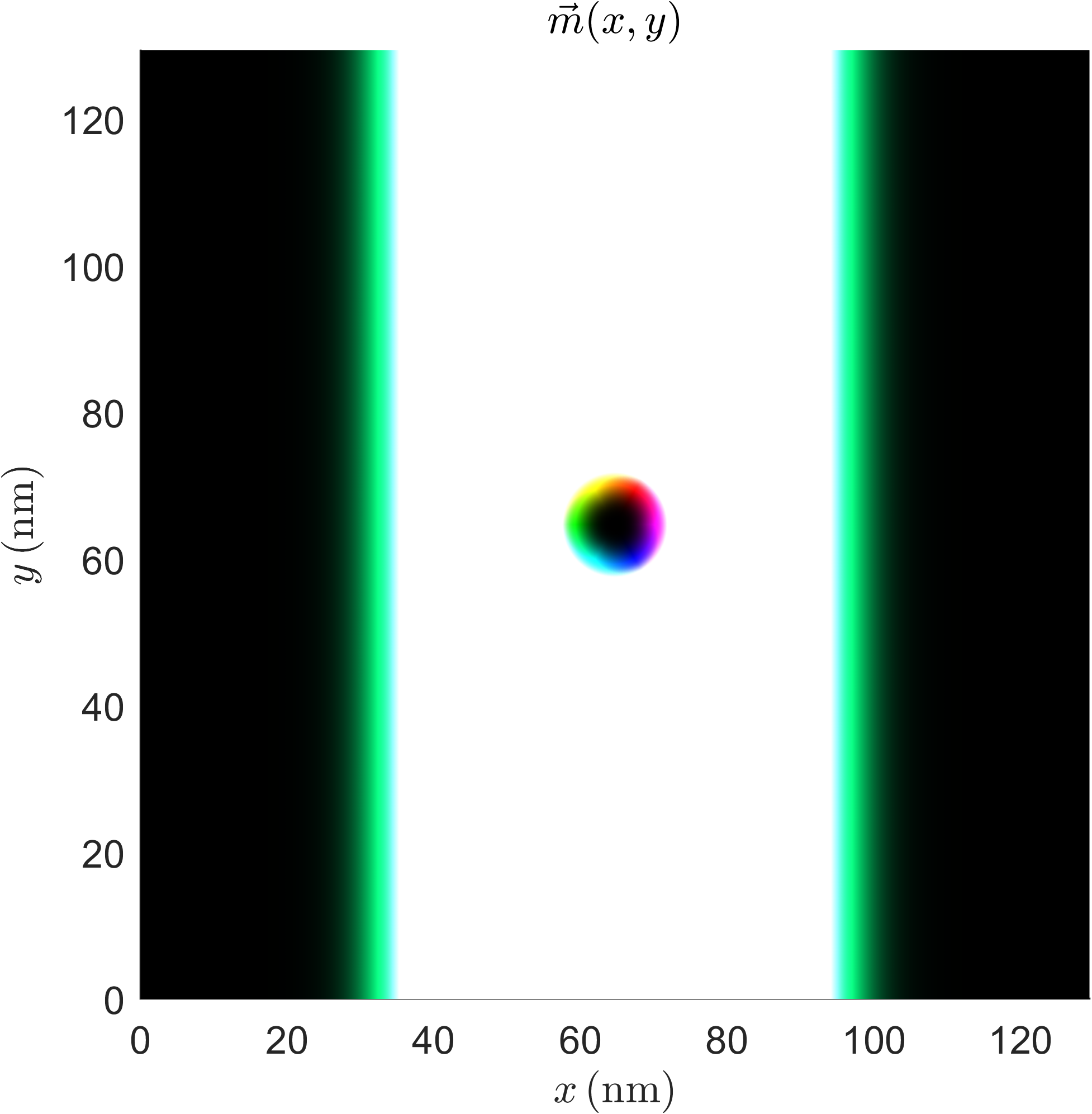} & \includegraphics[width=25mm]{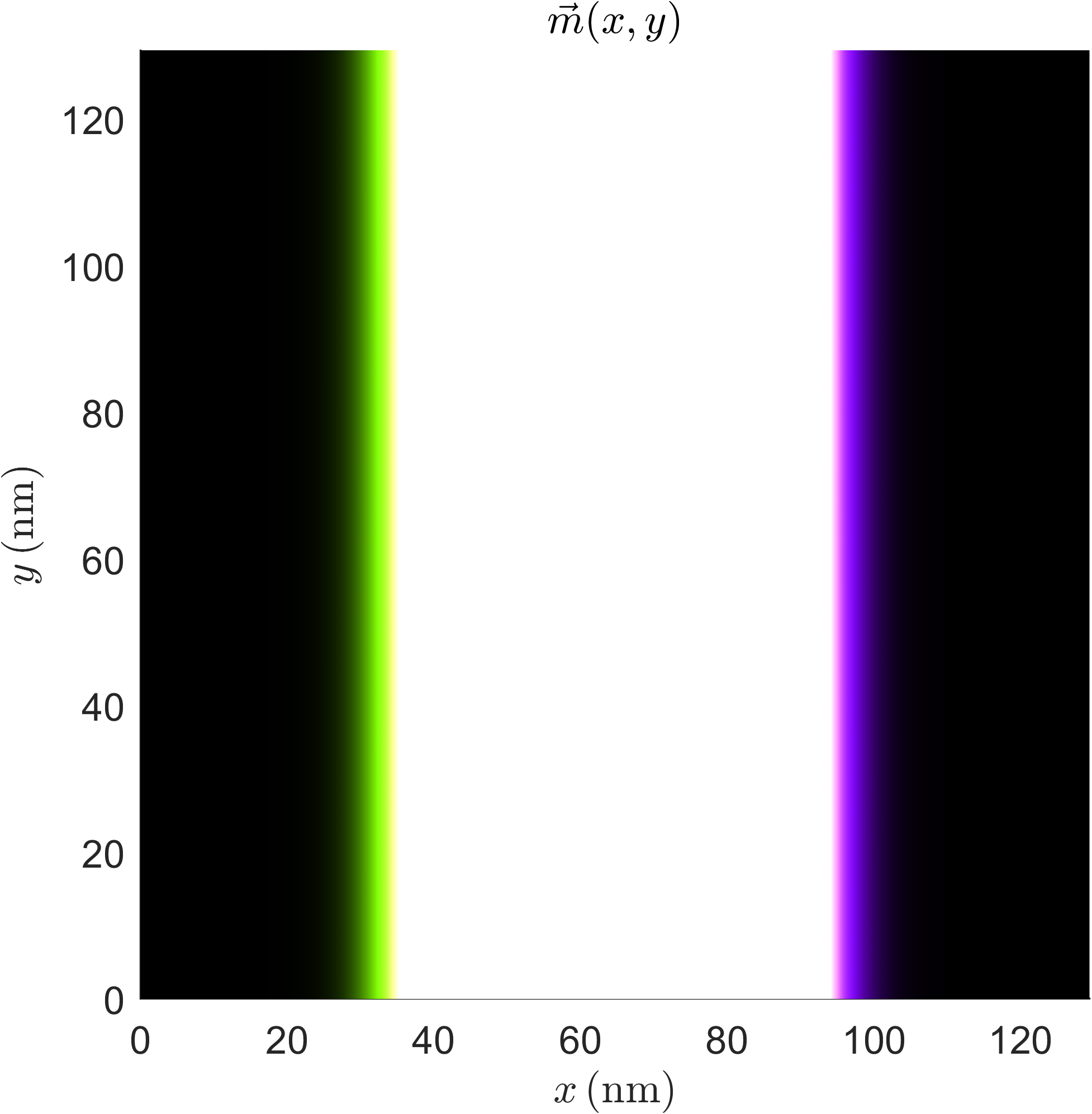} & $0$ \\
        $(2\pi/3,\pi)$ & $+1$ & \includegraphics[width=25mm]{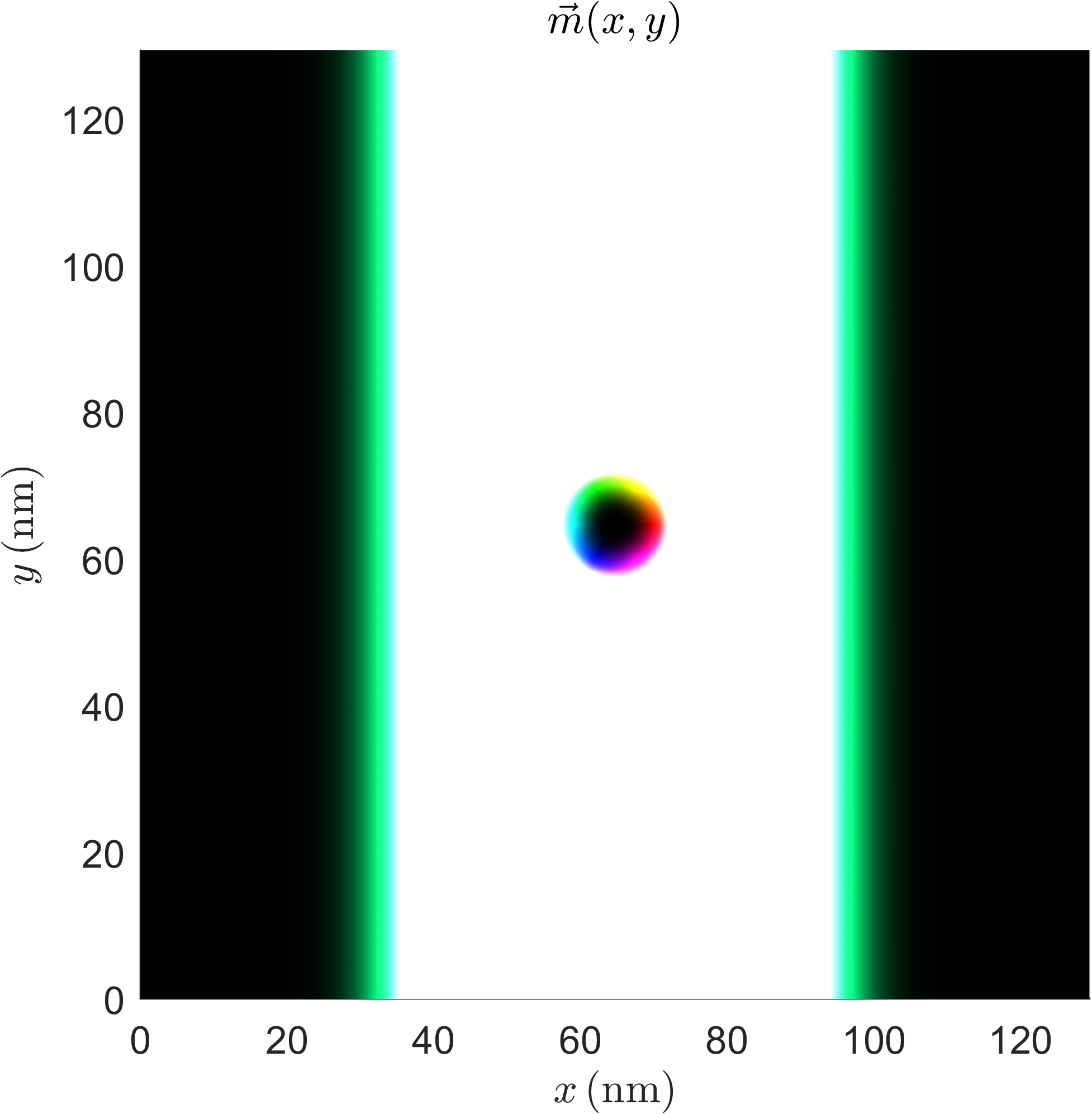} & \includegraphics[width=25mm]{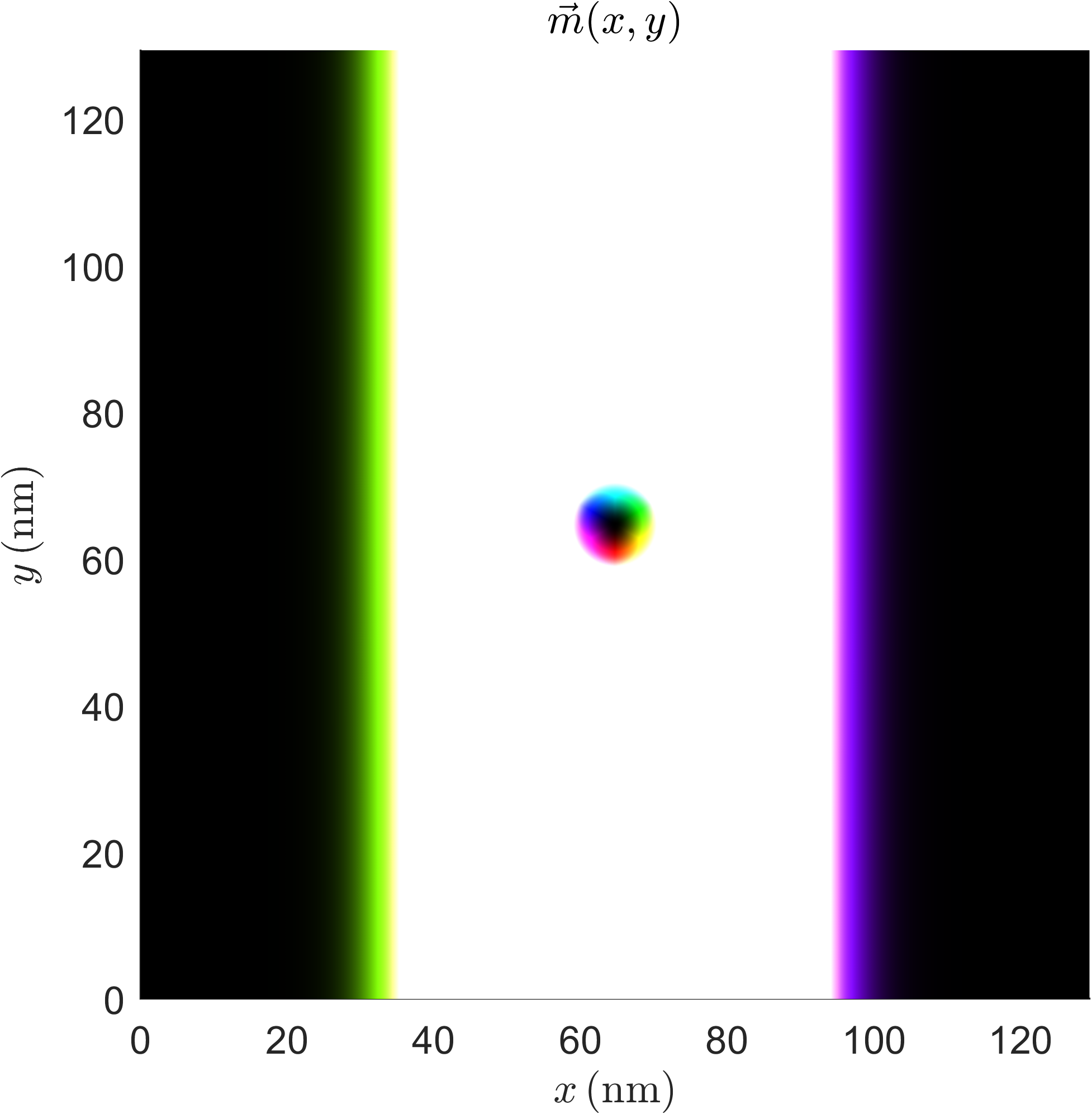} & $+1$ \\
        $(2\pi/3,4\pi/3)$ & $+1$ & \includegraphics[width=25mm]{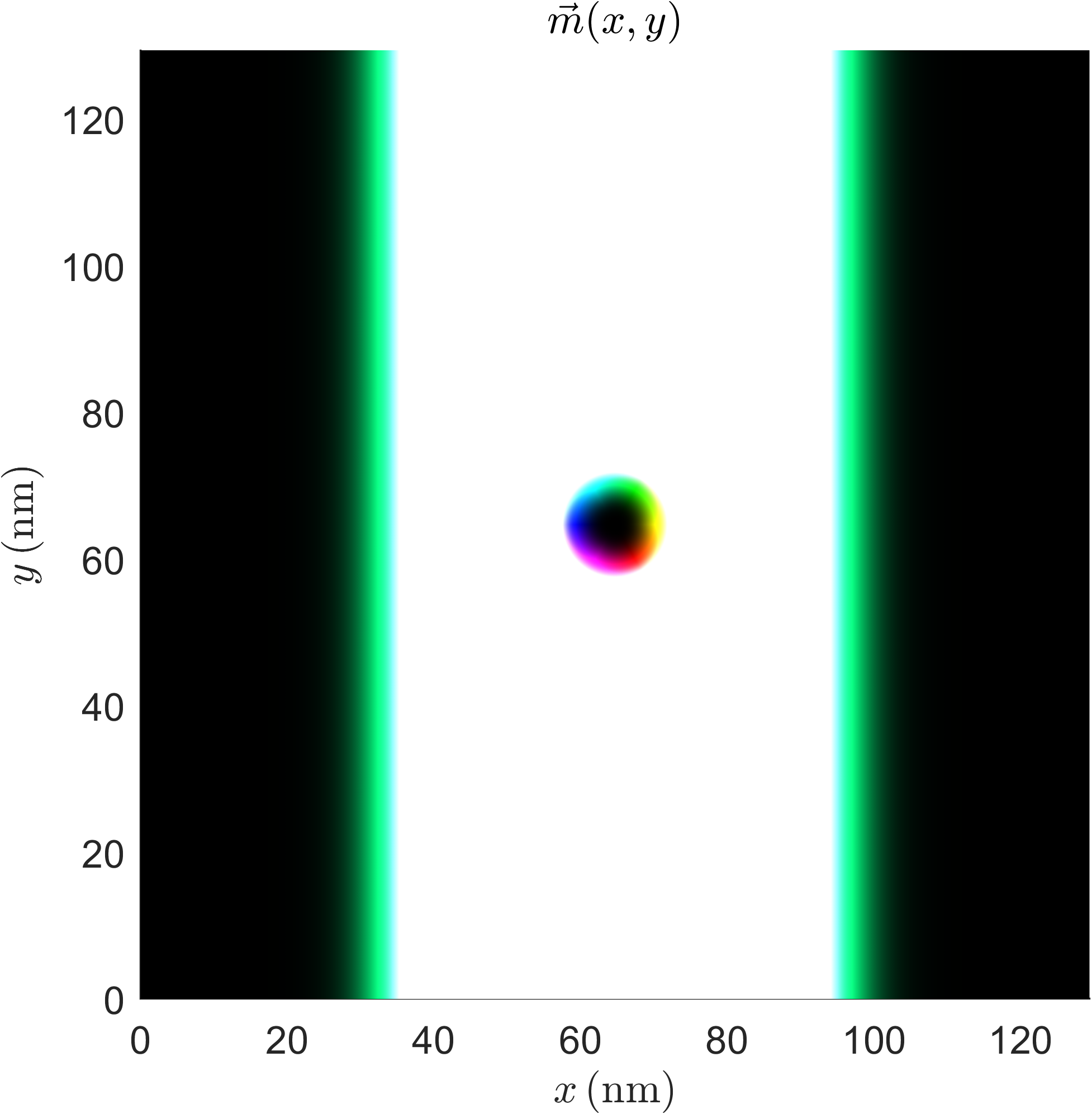} & \includegraphics[width=25mm]{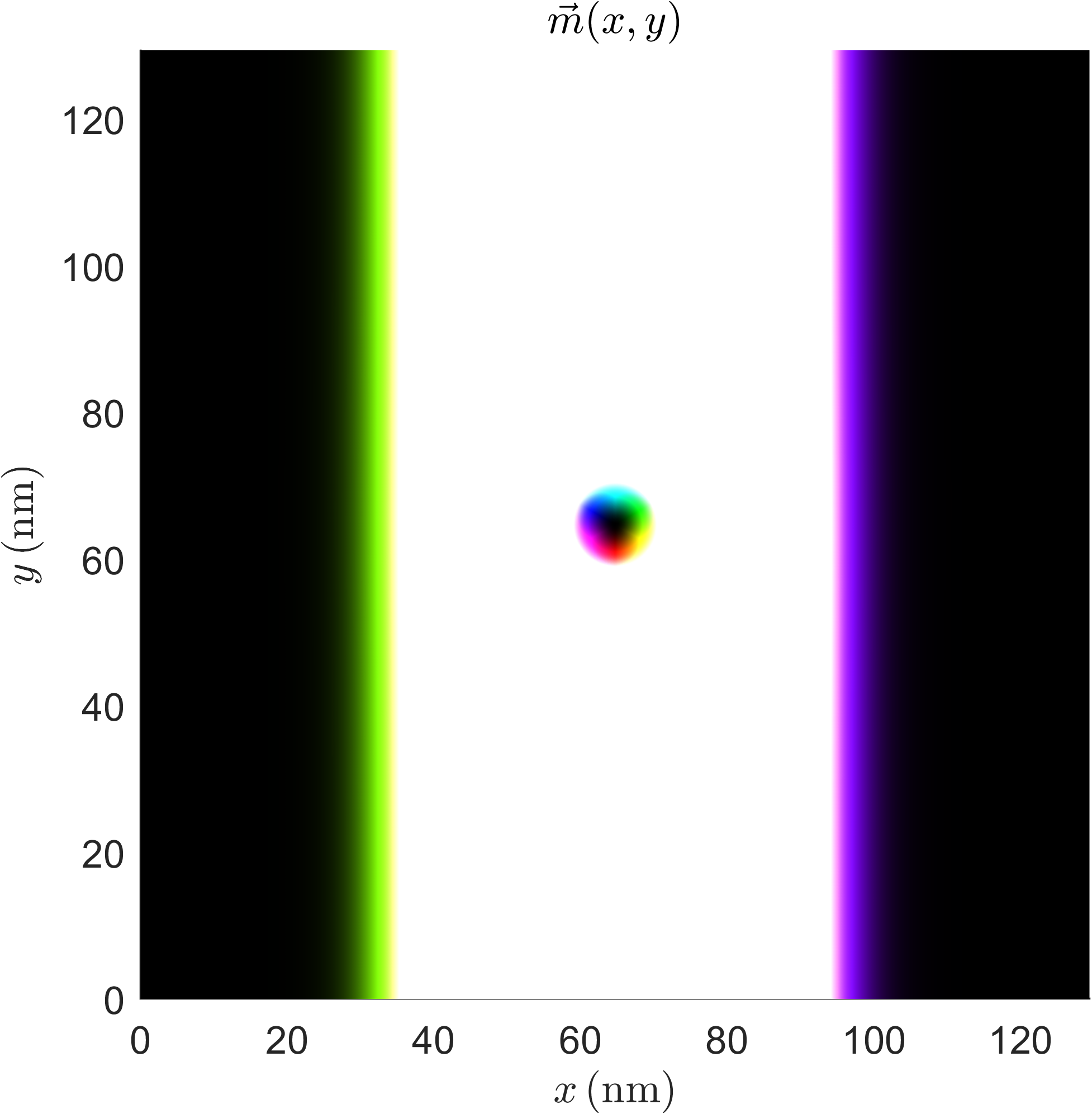} & $+1$ \\
        $(2\pi/3,3\pi/2)$ & $+1$ & \includegraphics[width=25mm]{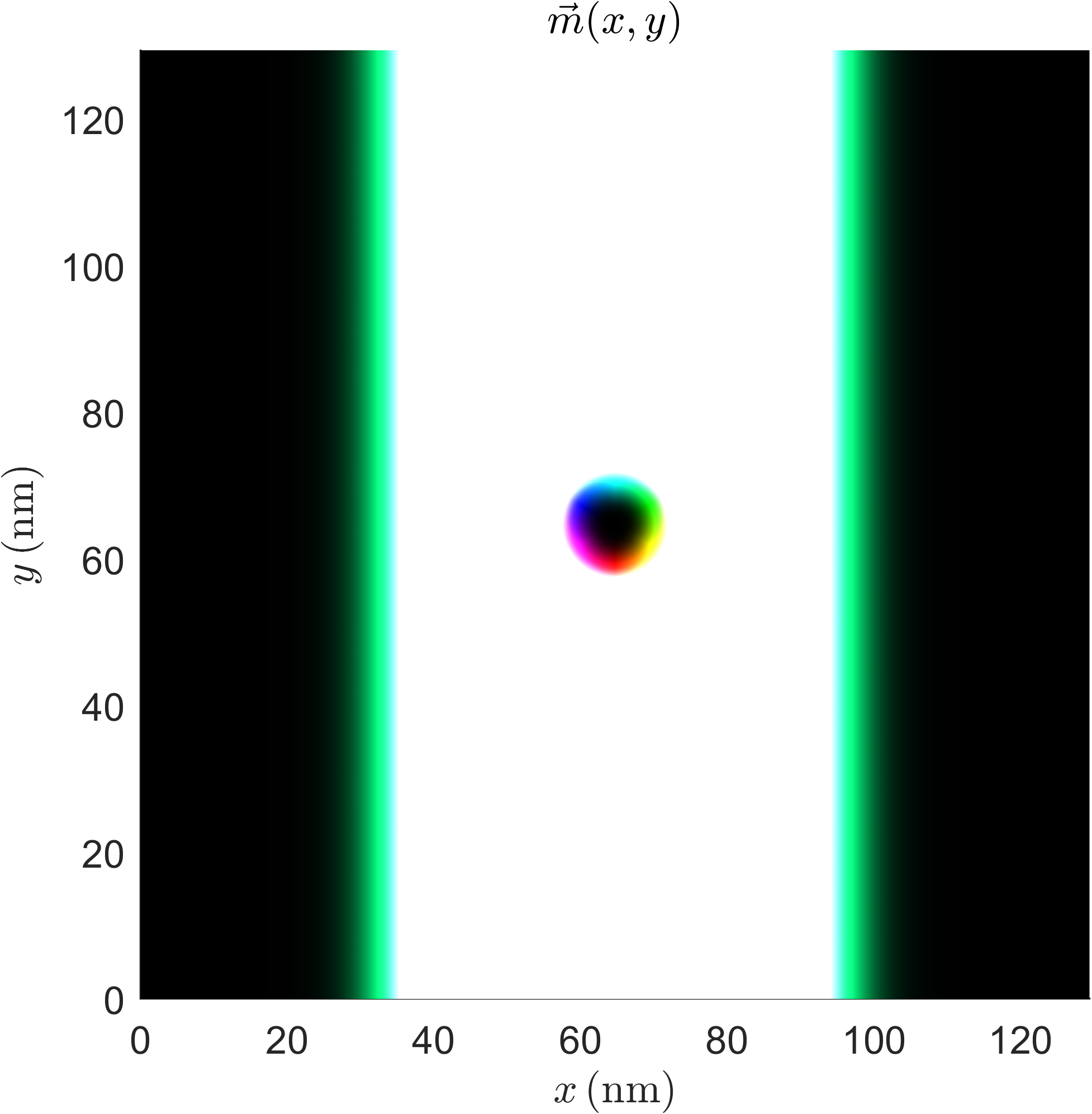} & \includegraphics[width=25mm]{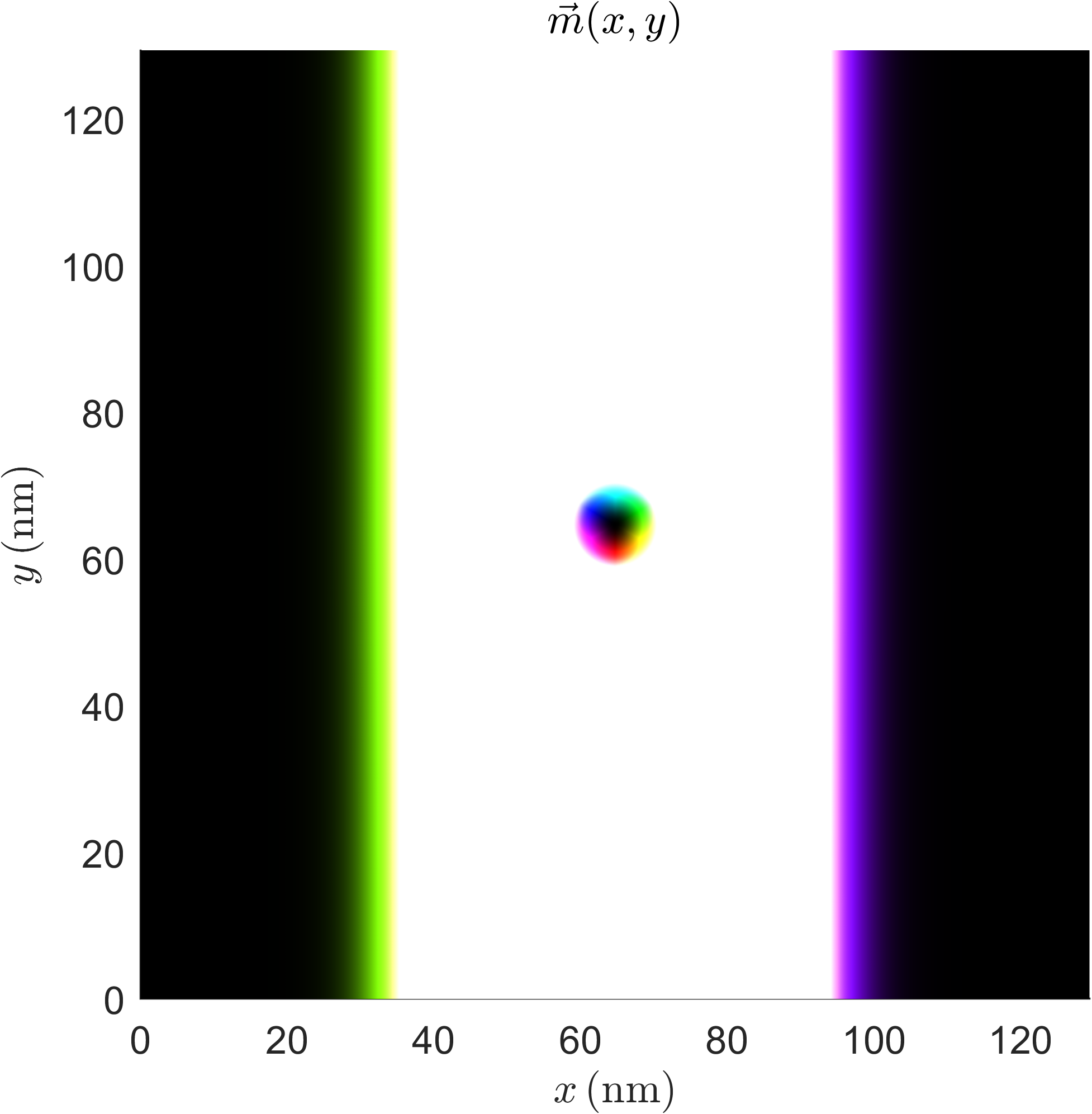} & $+1$ \\
        $(2\pi/3,5\pi/3)$ & $+1$ & \includegraphics[width=25mm]{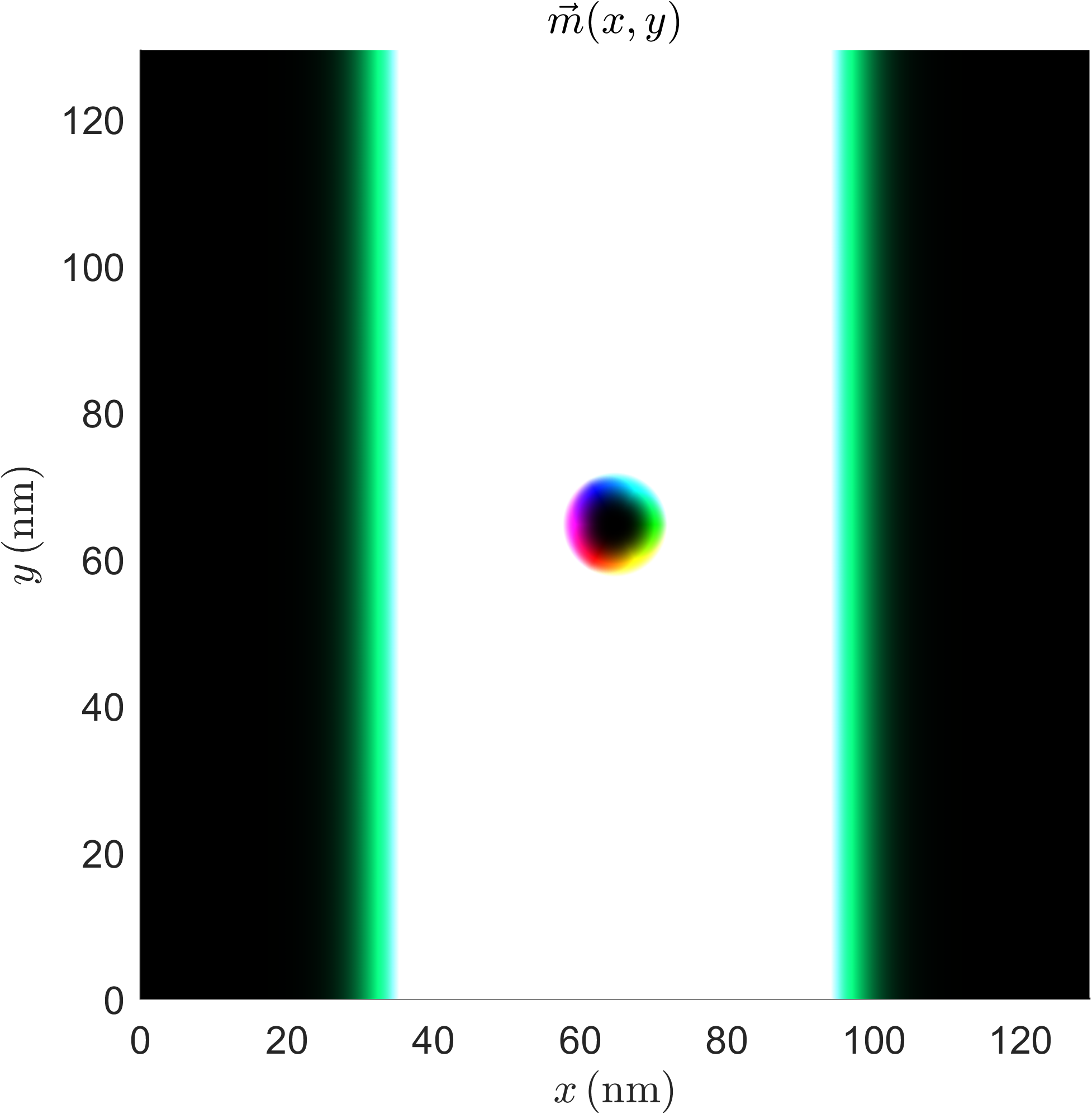} & \includegraphics[width=25mm]{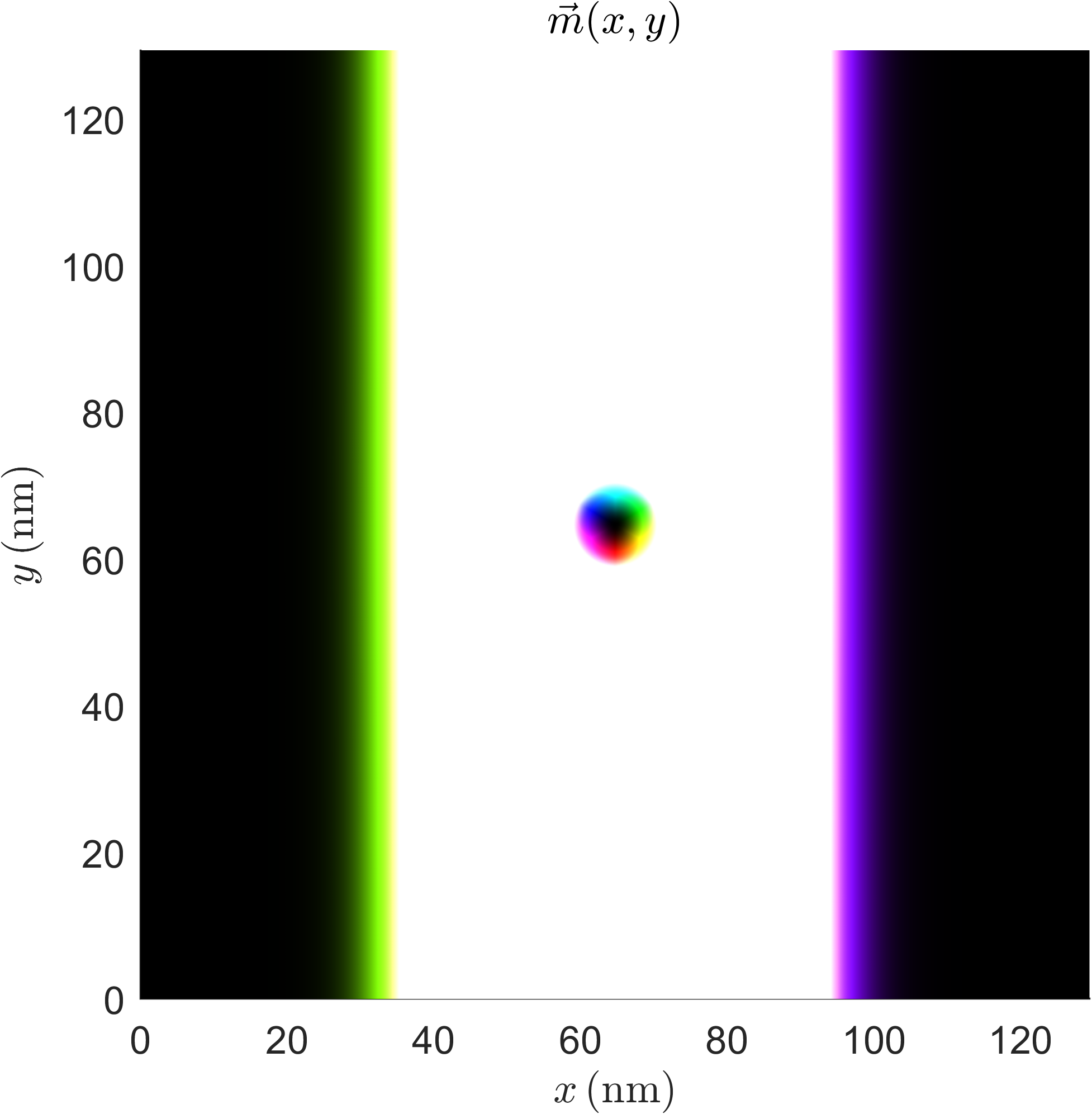} & $+1$ \\
        \bottomrule
    \end{tabular}
    \caption{Initial and final states for the domain wall phase $\chi=2\pi/3$ as the skyrmion is rotated by $\pi/3$ from $\phi=0$ until $\phi=5\pi/3$.}
    \label{tbl: chi = 2pi/3}
\end{table}

\begin{table}
    \centering
    \begin{tabular}{ccM{40mm}M{40mm}c}
        \toprule
        $(\chi,\phi)$ & $Q_{\textup{i}}$ & Initial Configuration & Final Configuration & $Q_{\textup{f}}$ \\
        \midrule
        $(\pi,0)$ & $+1$ & \includegraphics[width=25mm]{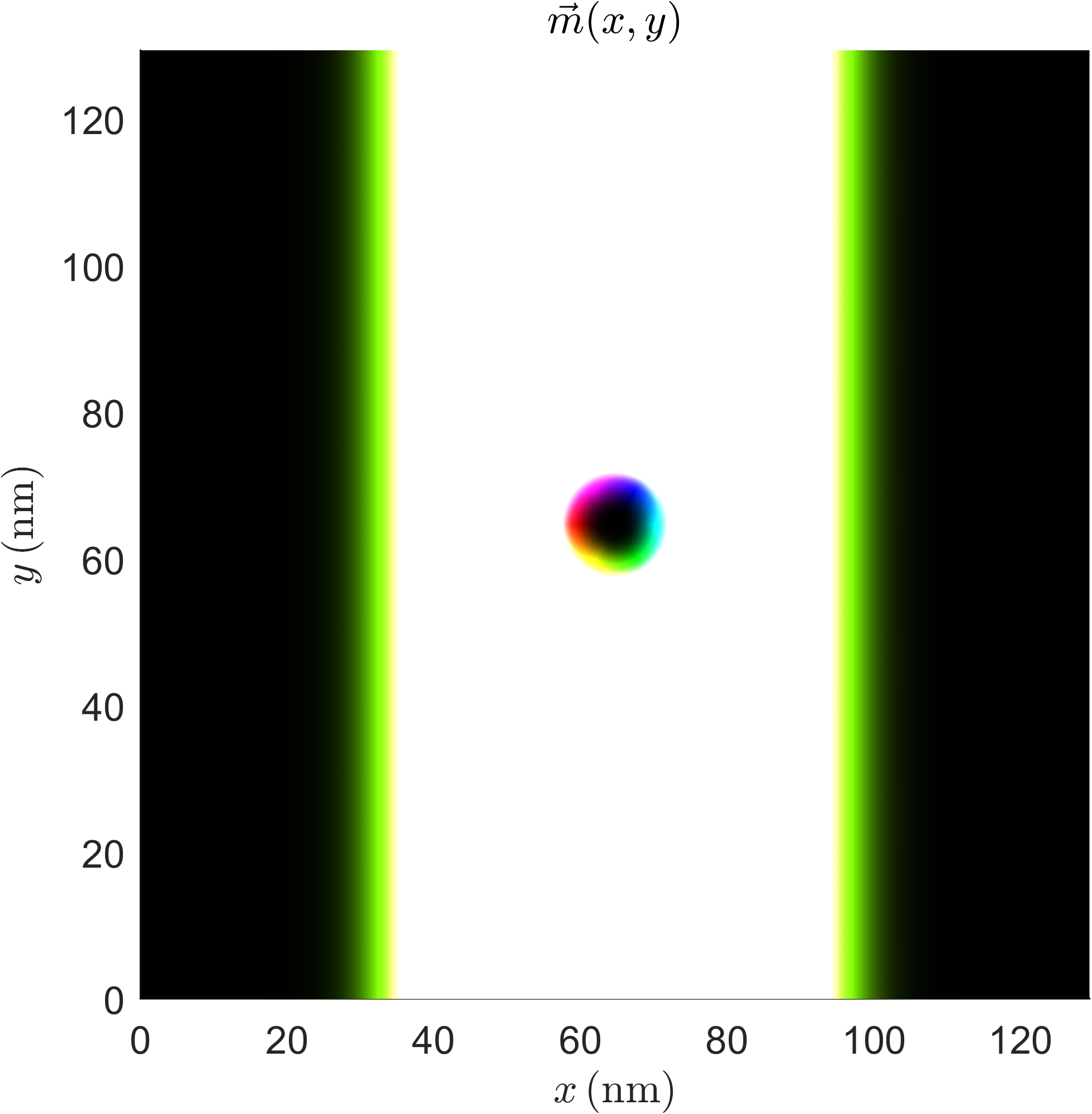} & \includegraphics[width=25mm]{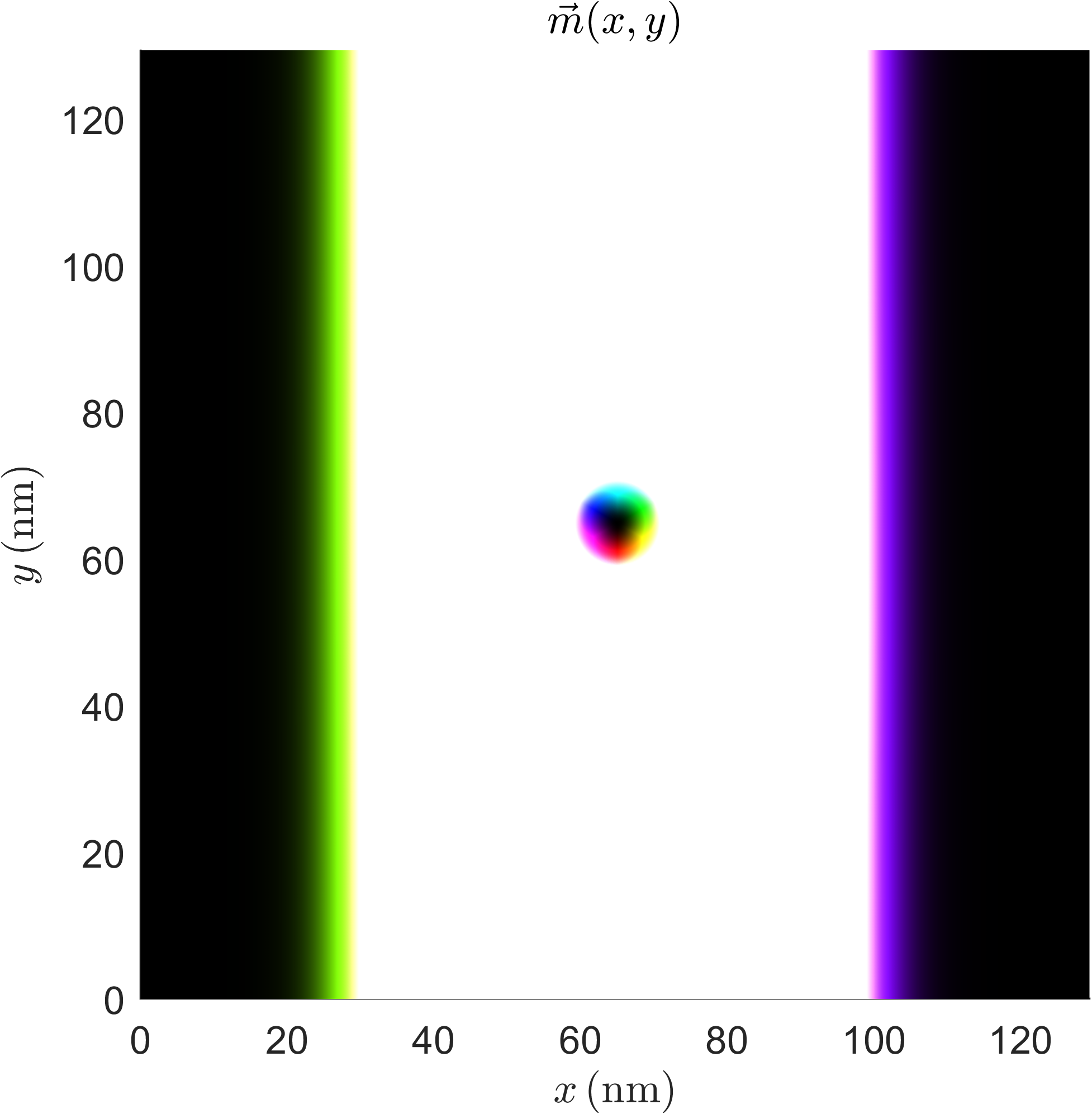} & $+1$ \\
        $(\pi,\pi/3)$ & $+1$ & \includegraphics[width=25mm]{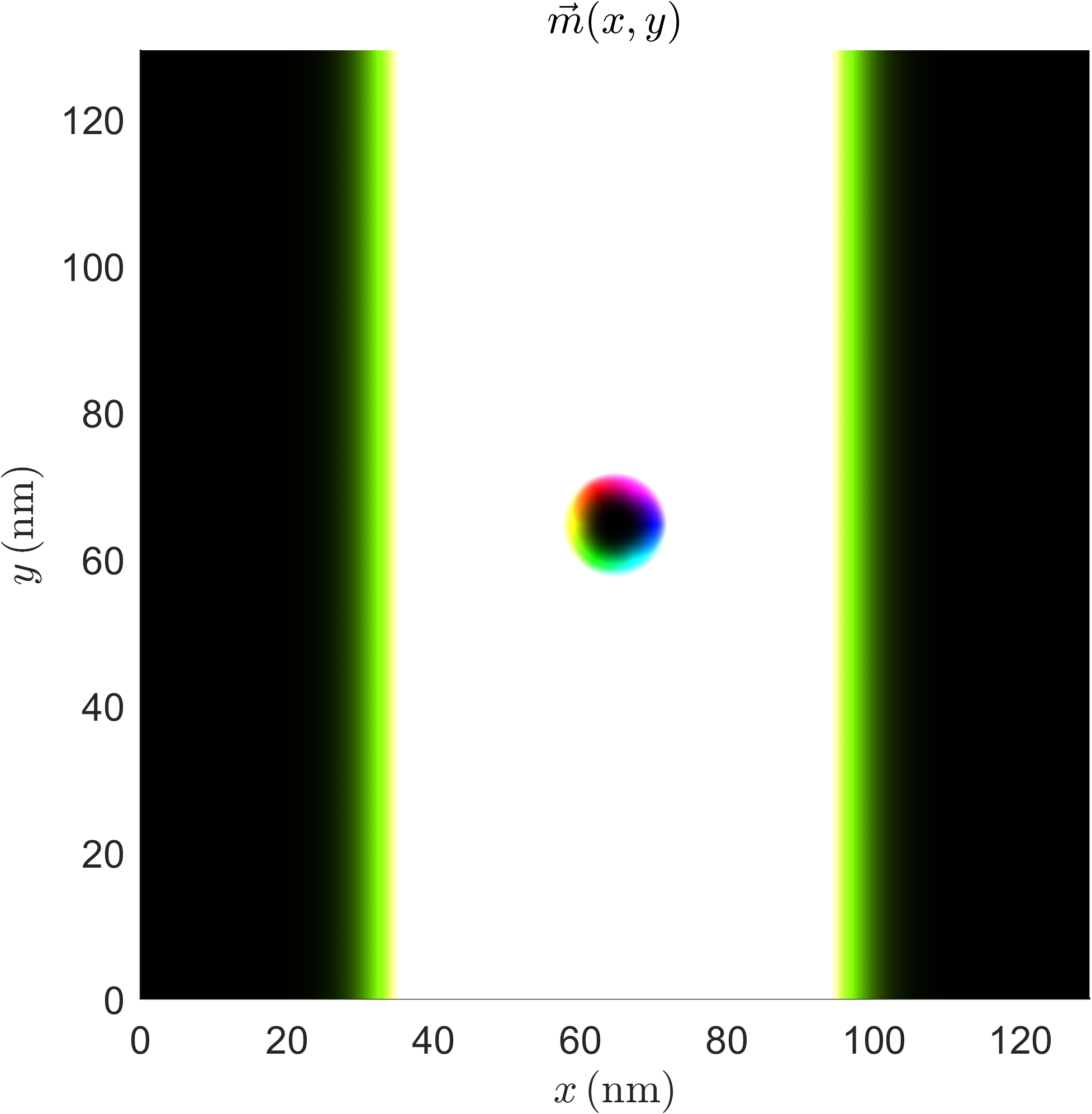} & \includegraphics[width=25mm]{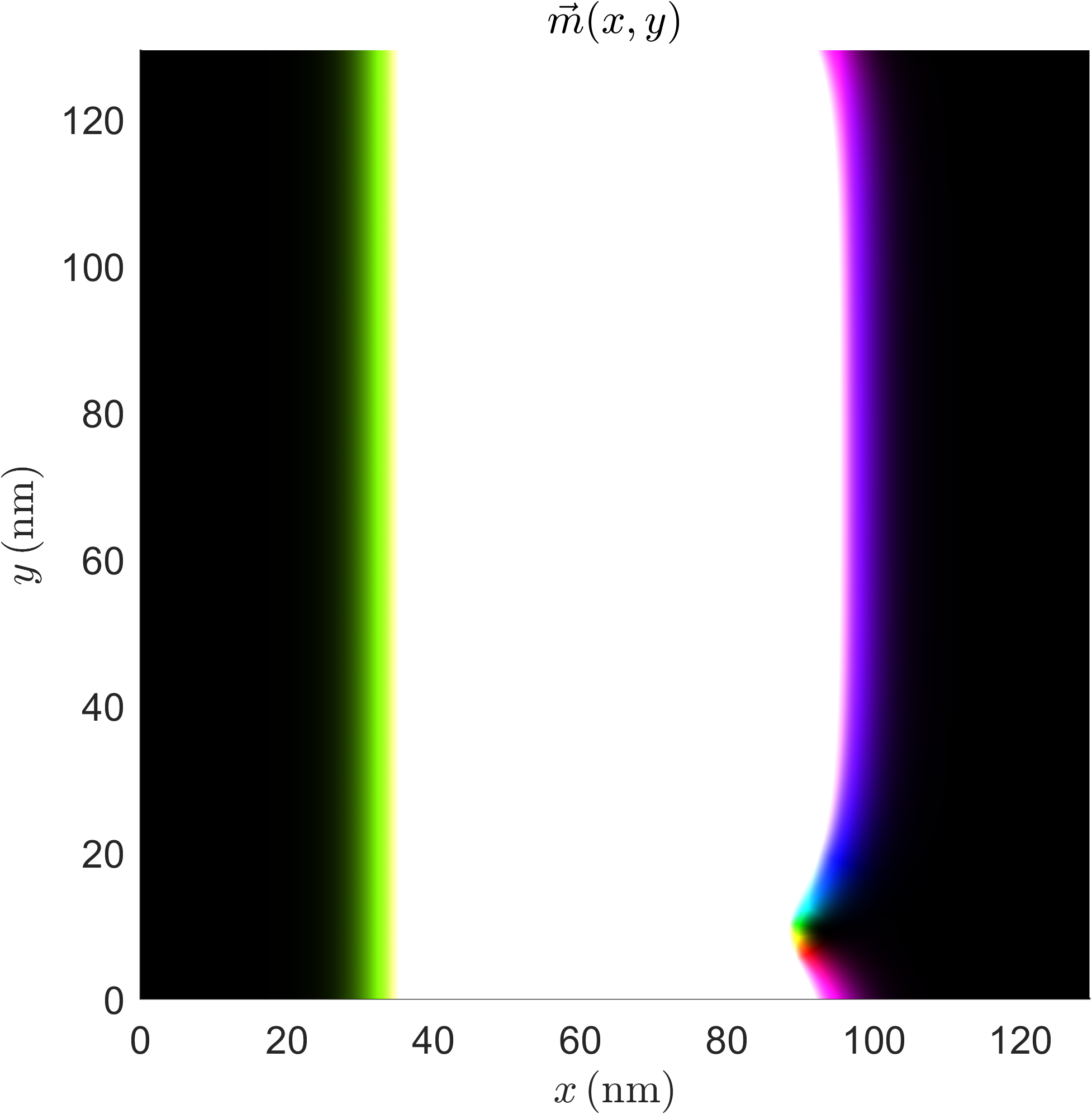} & $-1$ \\
        $(\pi,\pi/2)$ & $+1$ & \includegraphics[width=25mm]{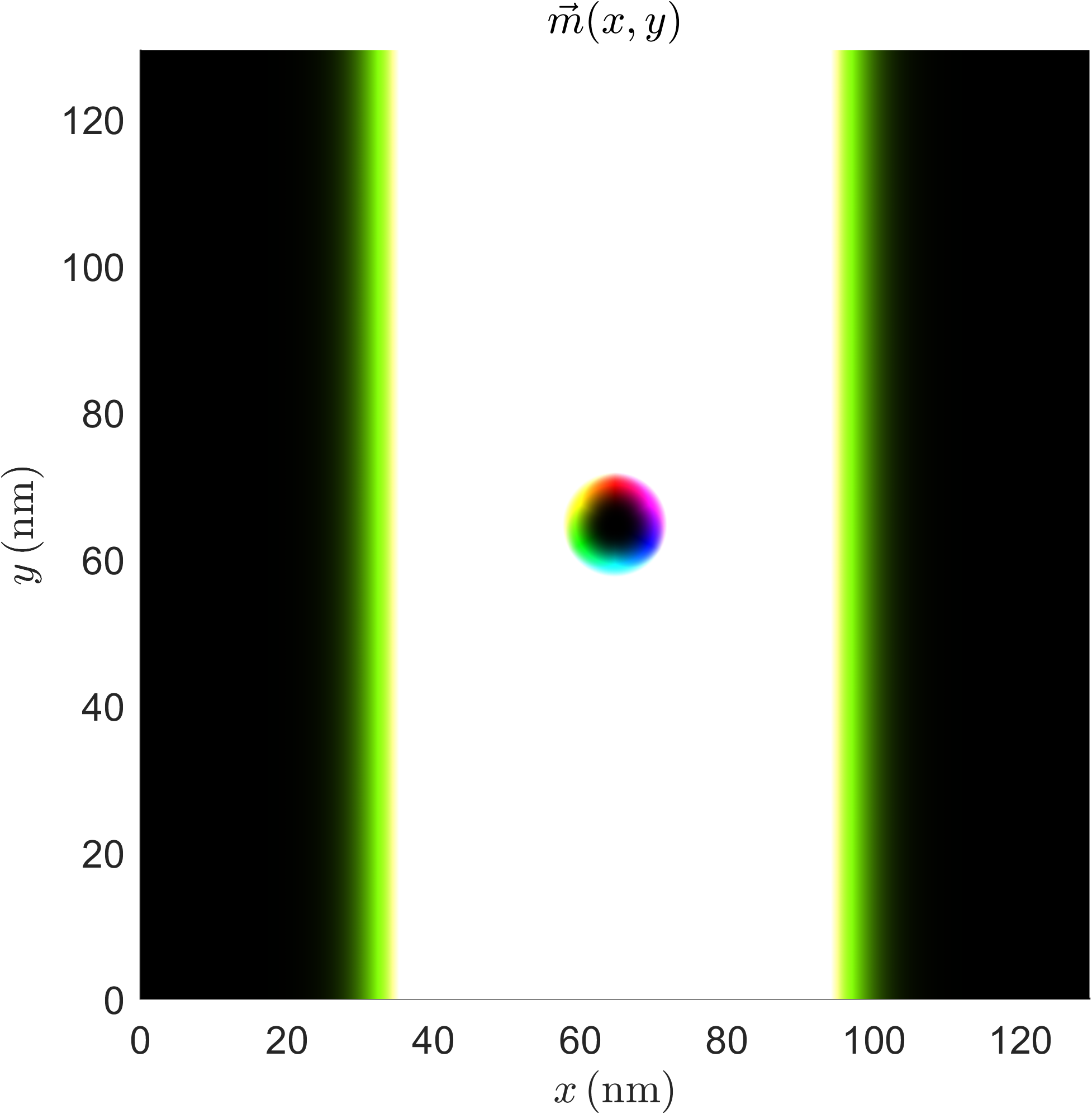} & \includegraphics[width=25mm]{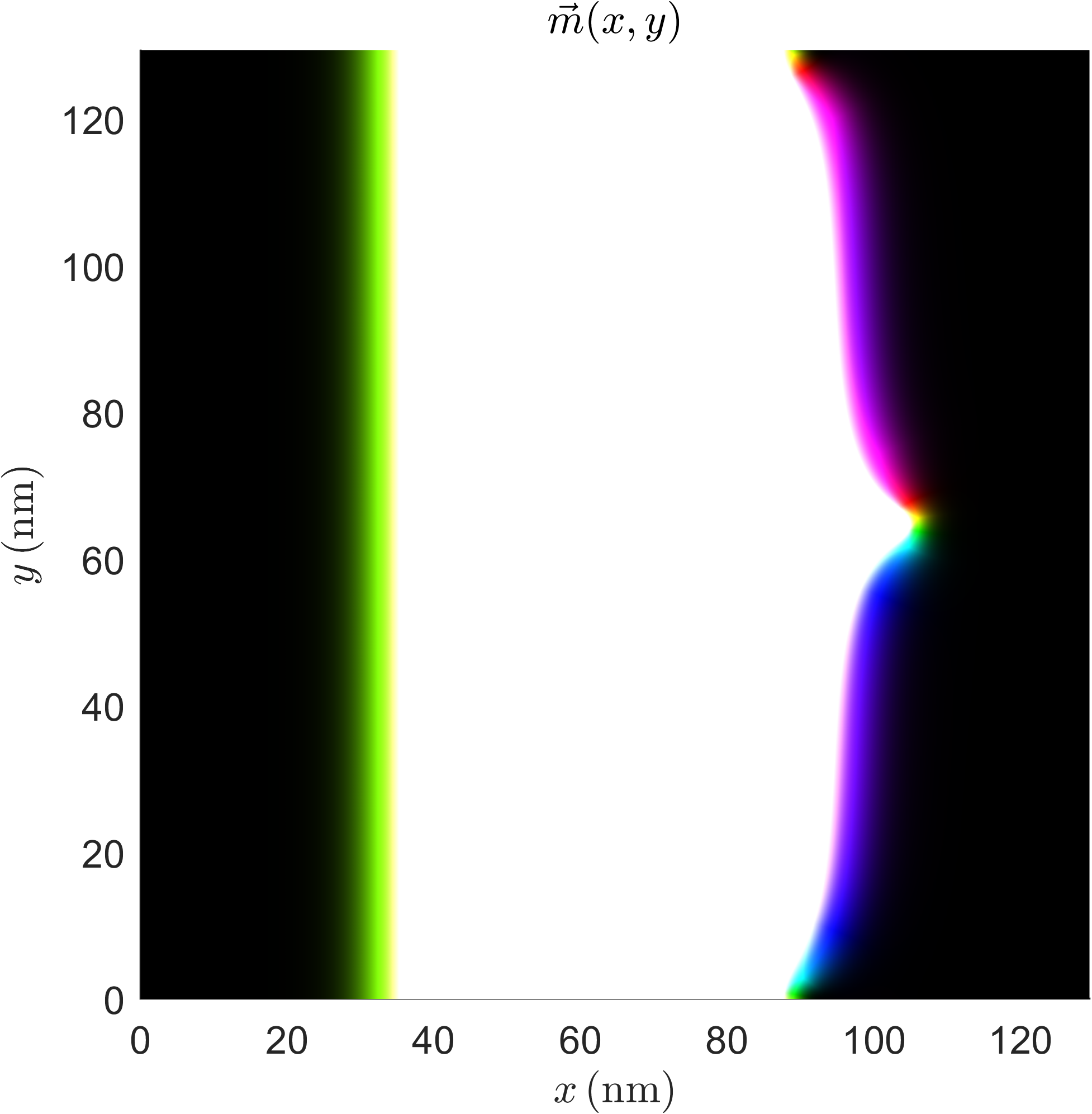} & $0$ \\
        $(\pi,2\pi/3)$ & $+1$ & \includegraphics[width=25mm]{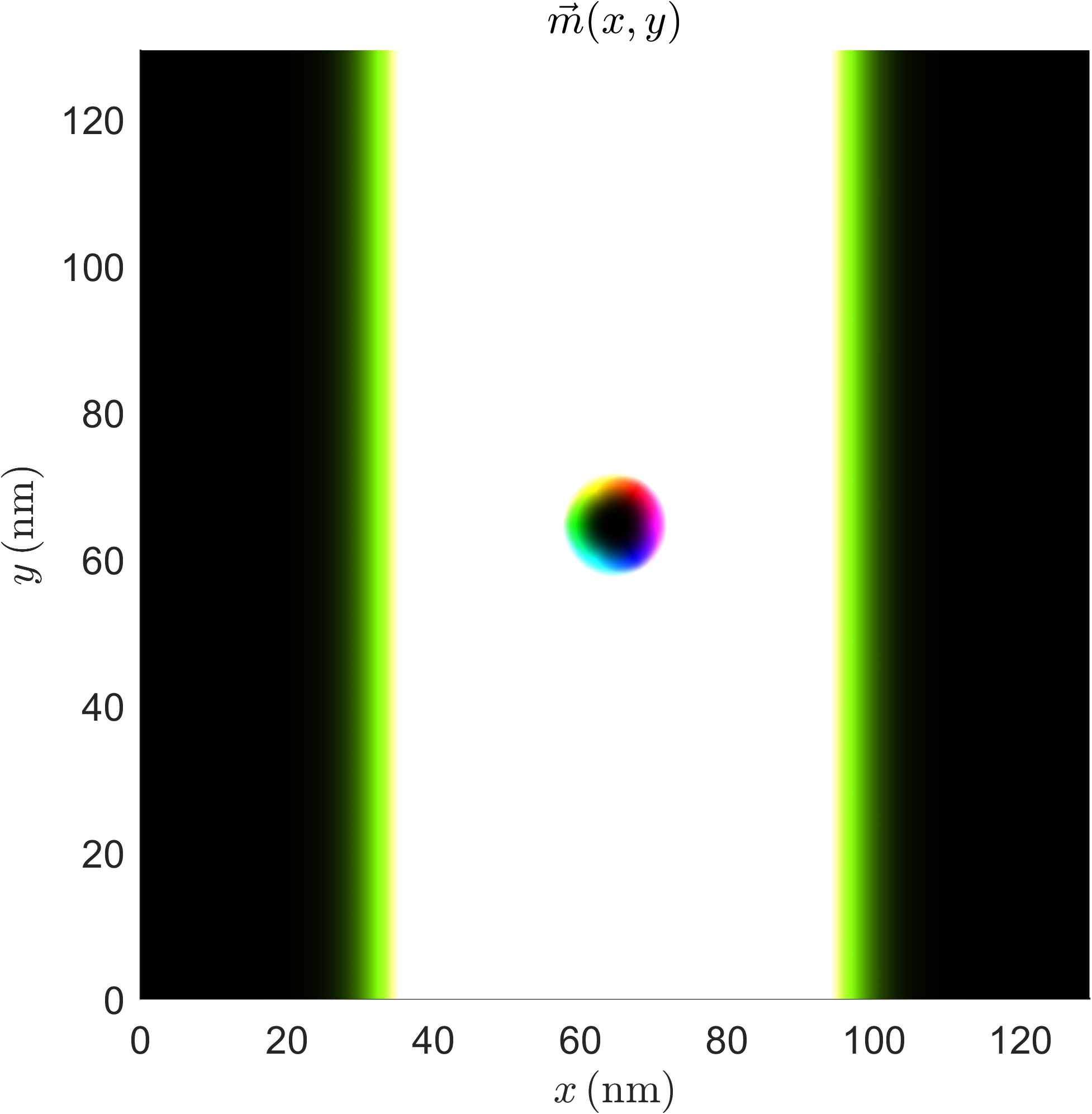} & \includegraphics[width=25mm]{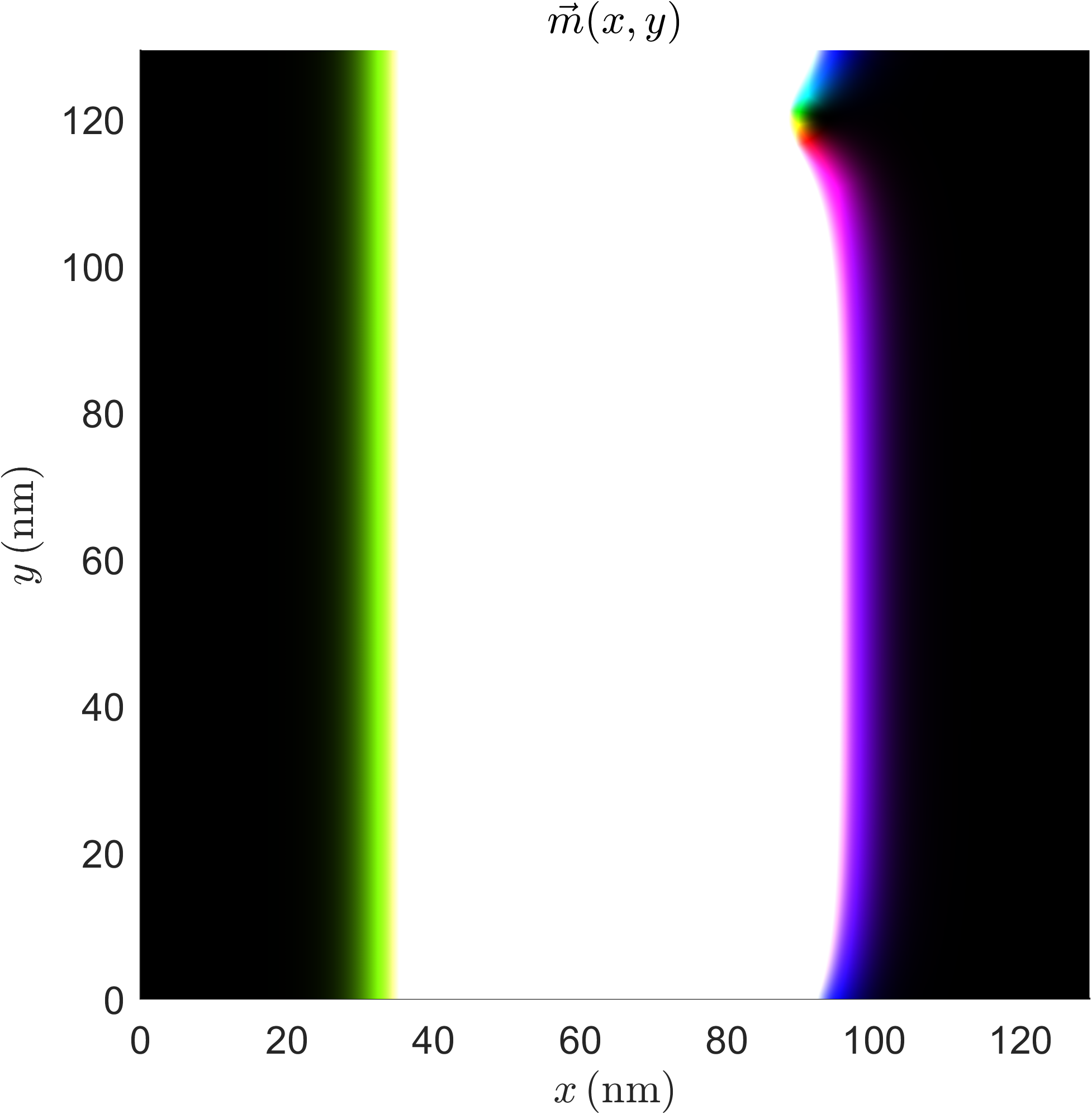} & $-1$ \\
        $(\pi,\pi)$ & $+1$ & \includegraphics[width=25mm]{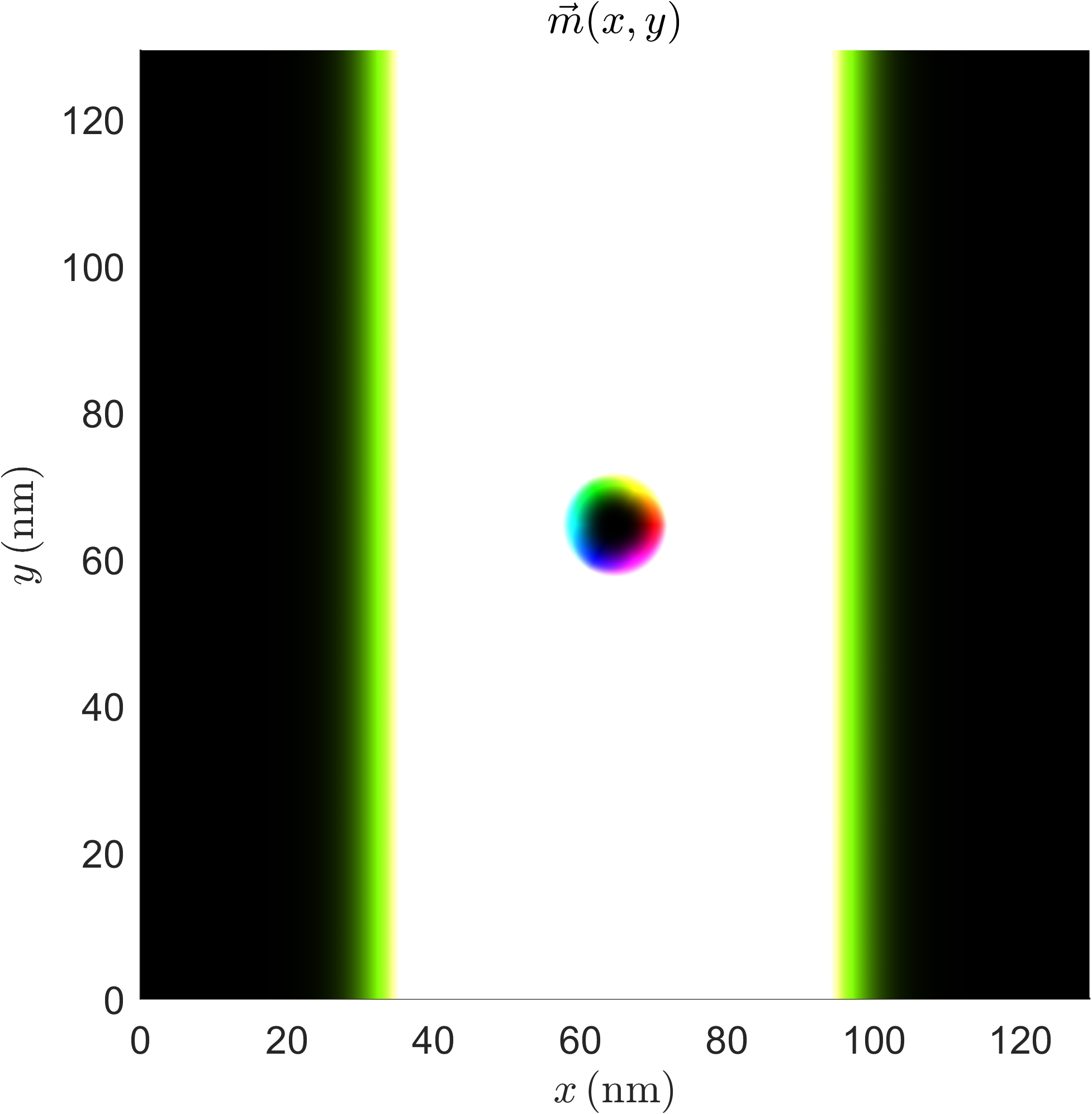} & \includegraphics[width=25mm]{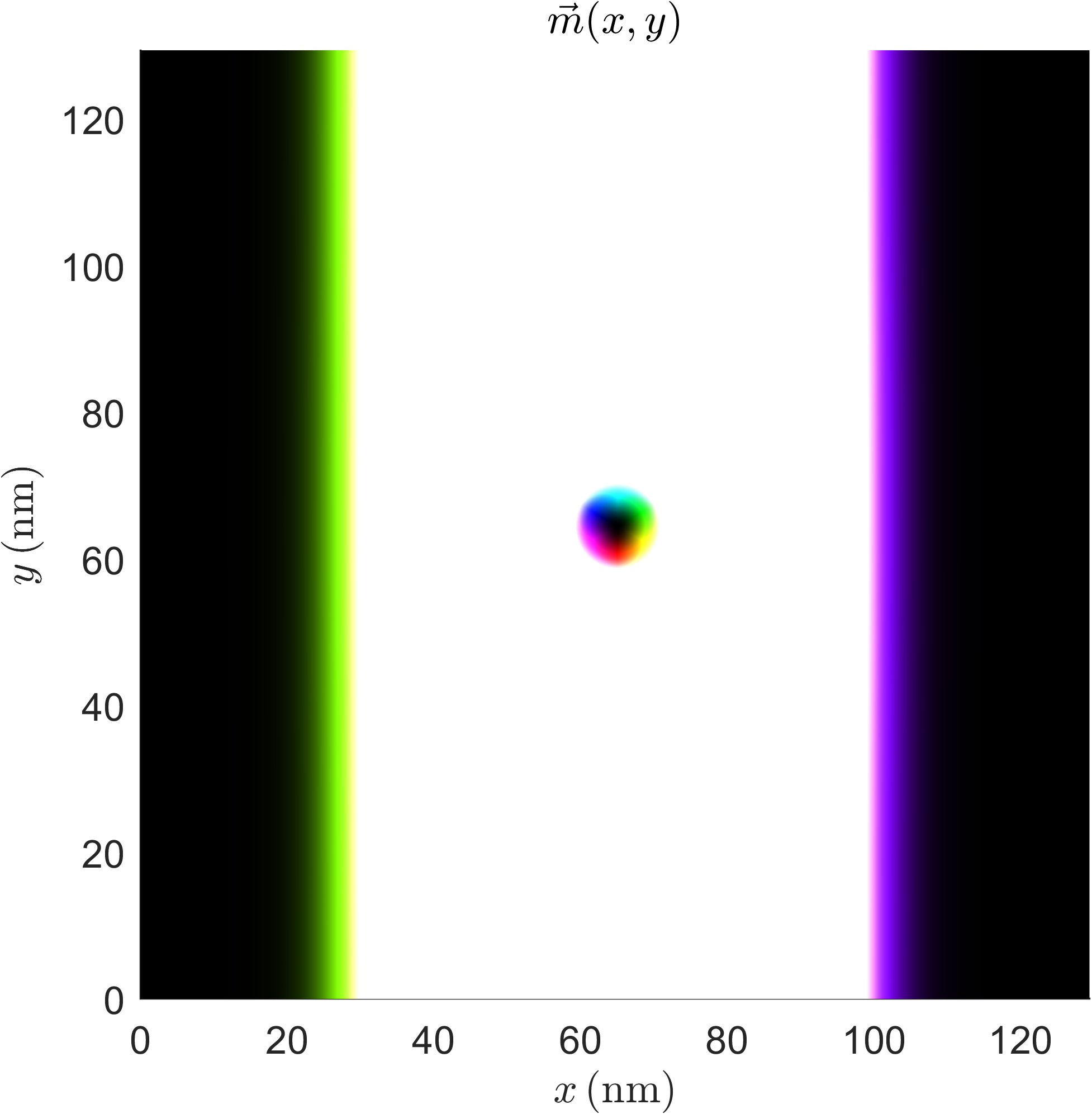} & $+1$ \\
        $(\pi,4\pi/3)$ & $+1$ & \includegraphics[width=25mm]{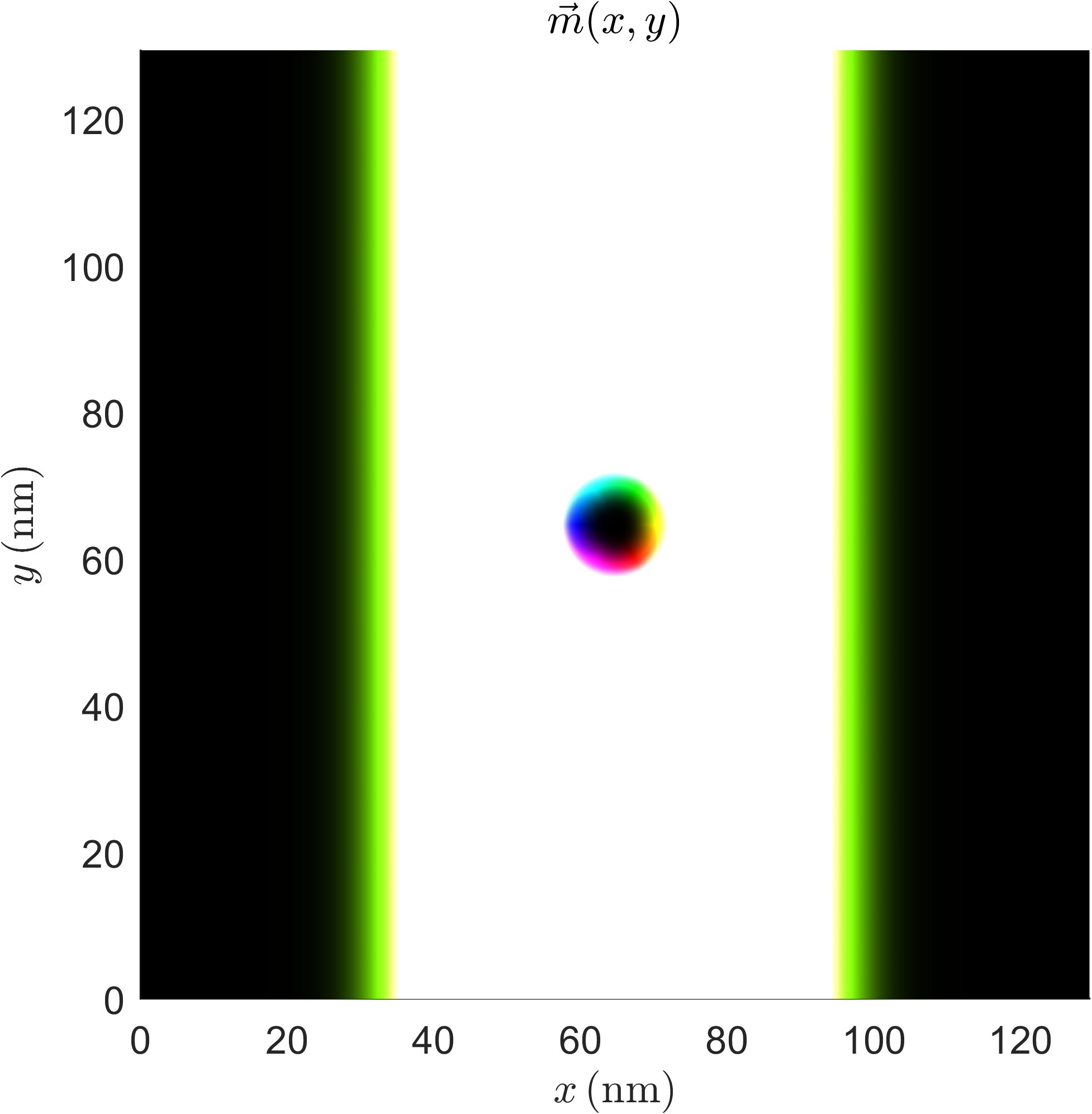} & \includegraphics[width=25mm]{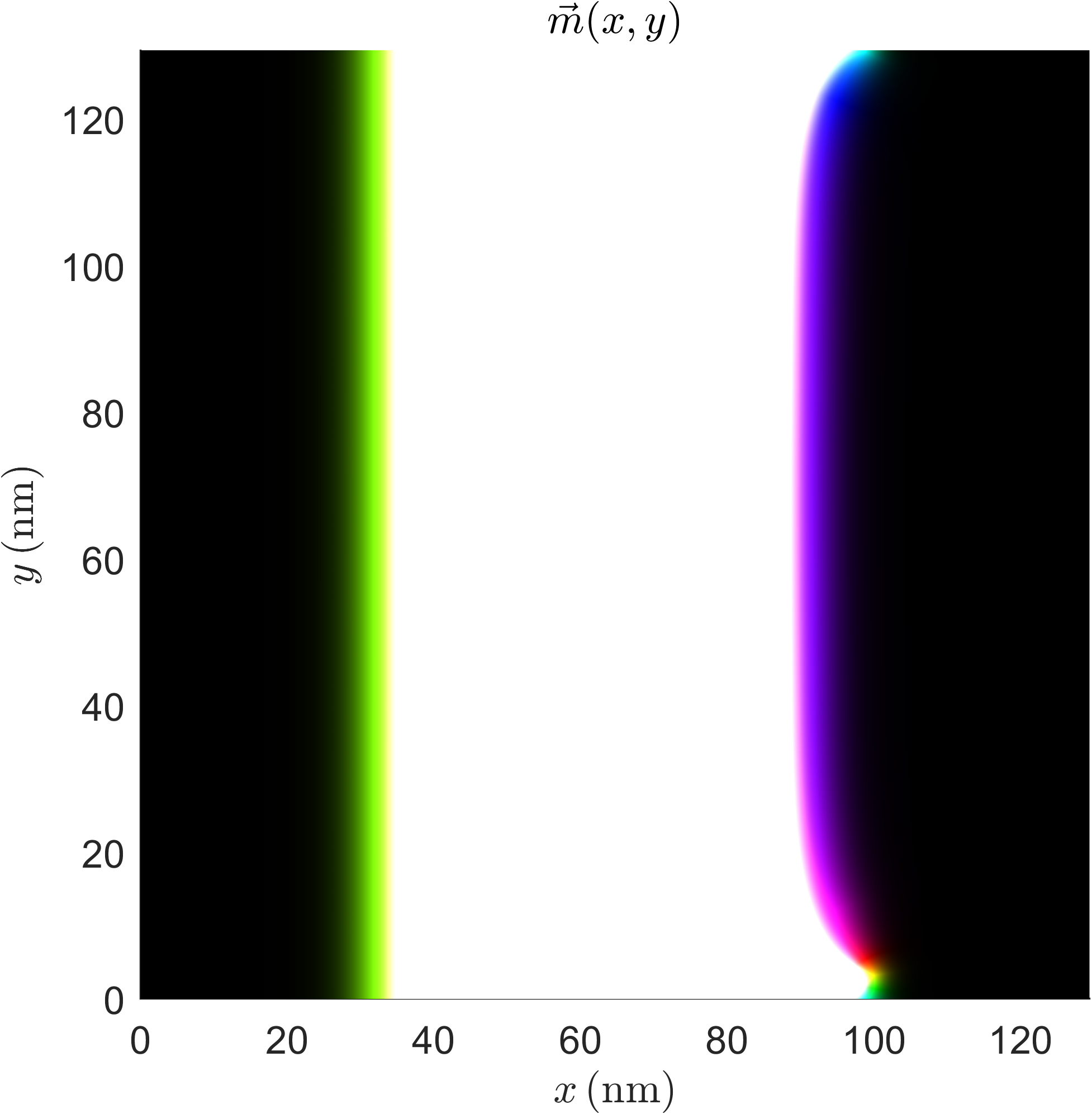} & $+1$ \\
        $(\pi,3\pi/2)$ & $+1$ & \includegraphics[width=25mm]{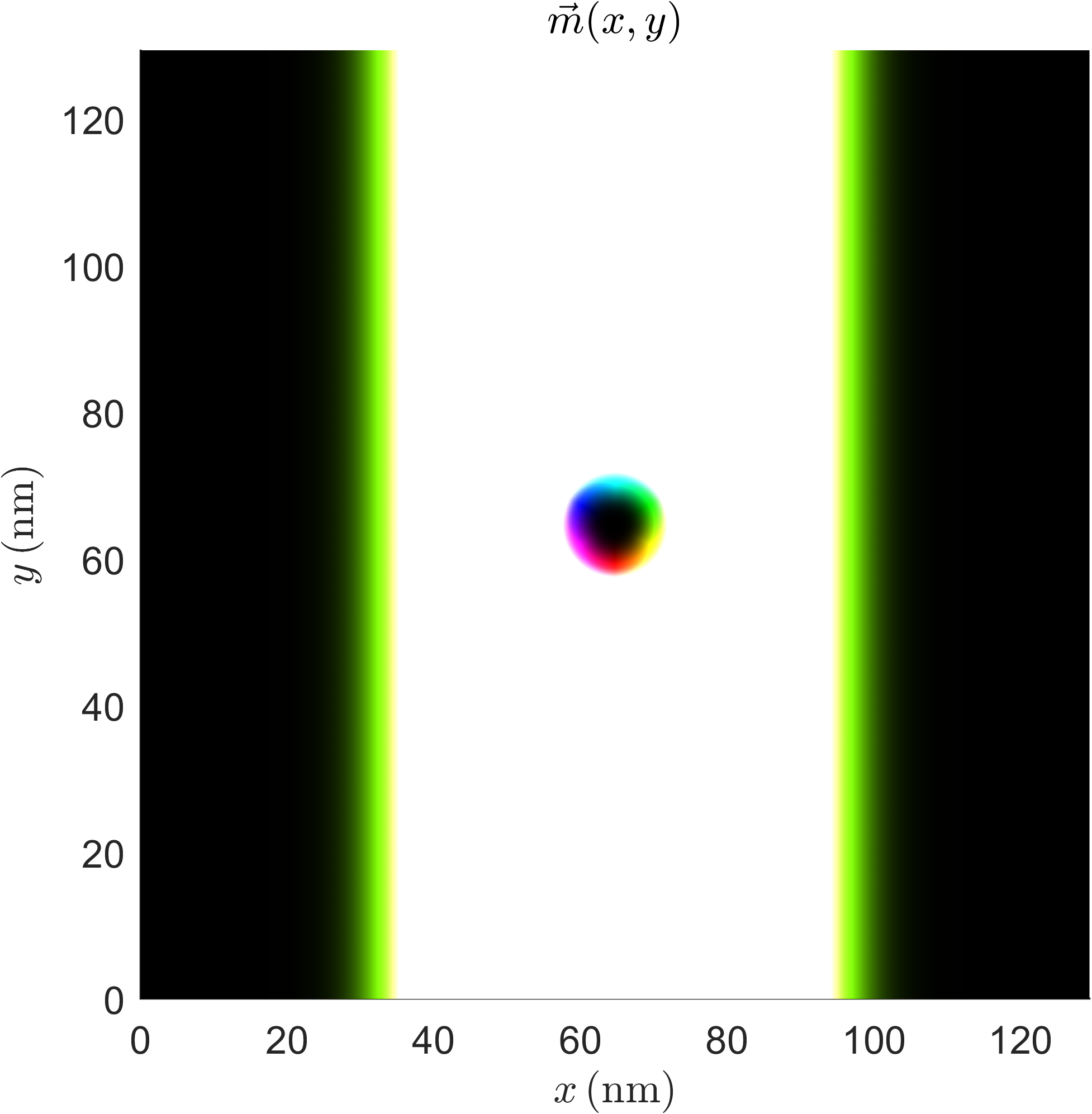} & \includegraphics[width=25mm]{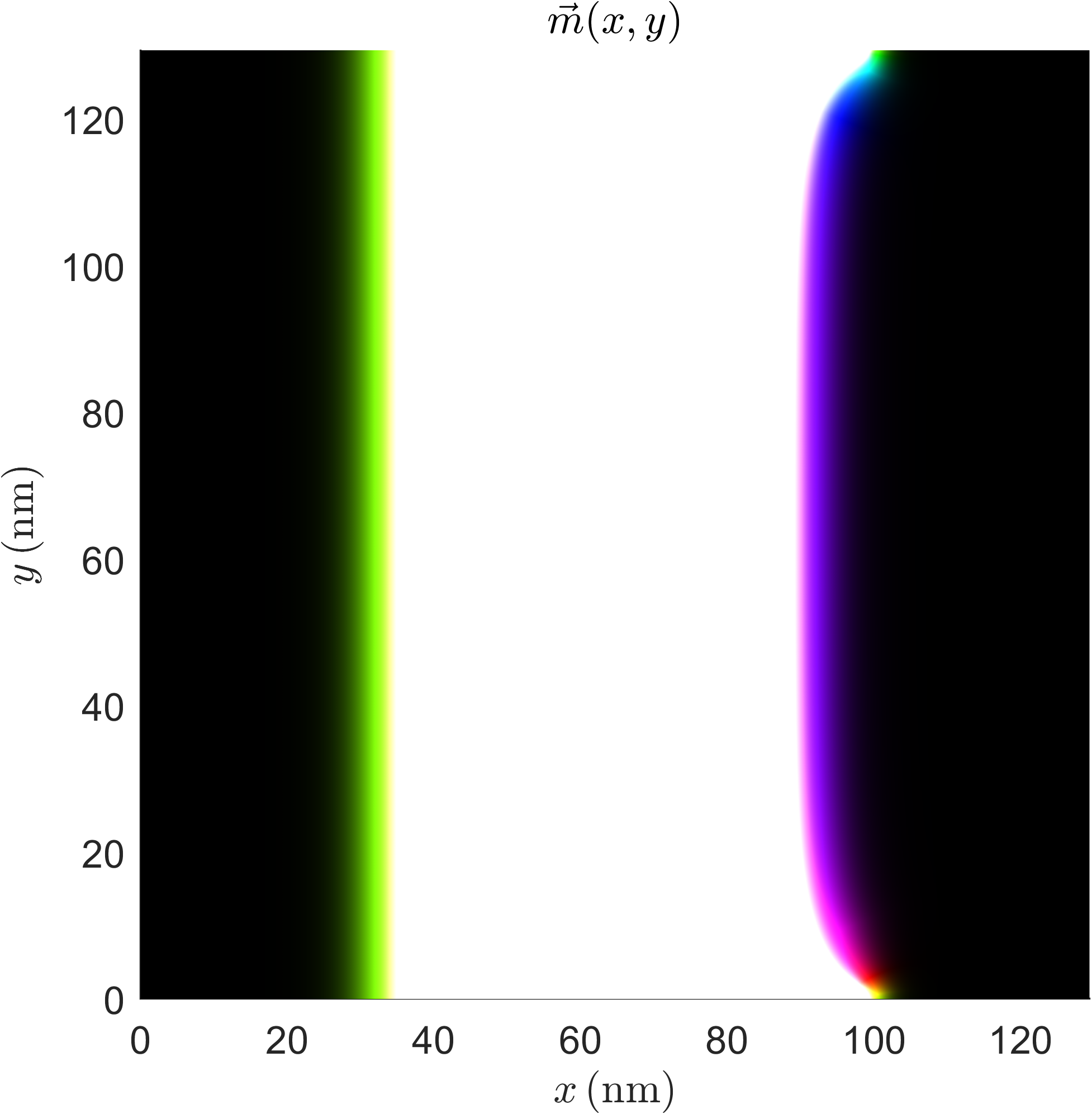} & $+1$ \\
        $(\pi/3,5\pi/3)$ & $+1$ & \includegraphics[width=25mm]{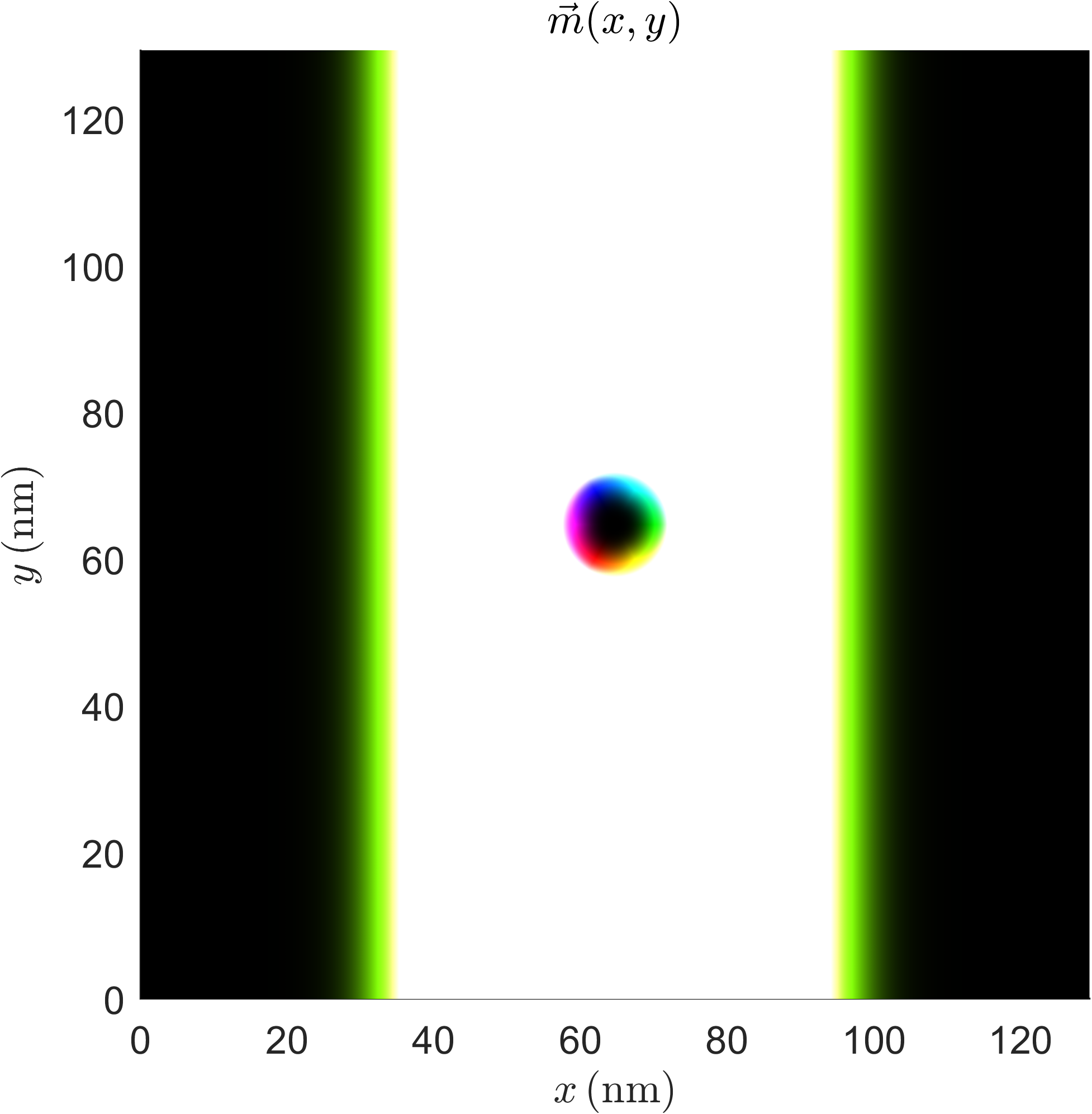} & \includegraphics[width=25mm]{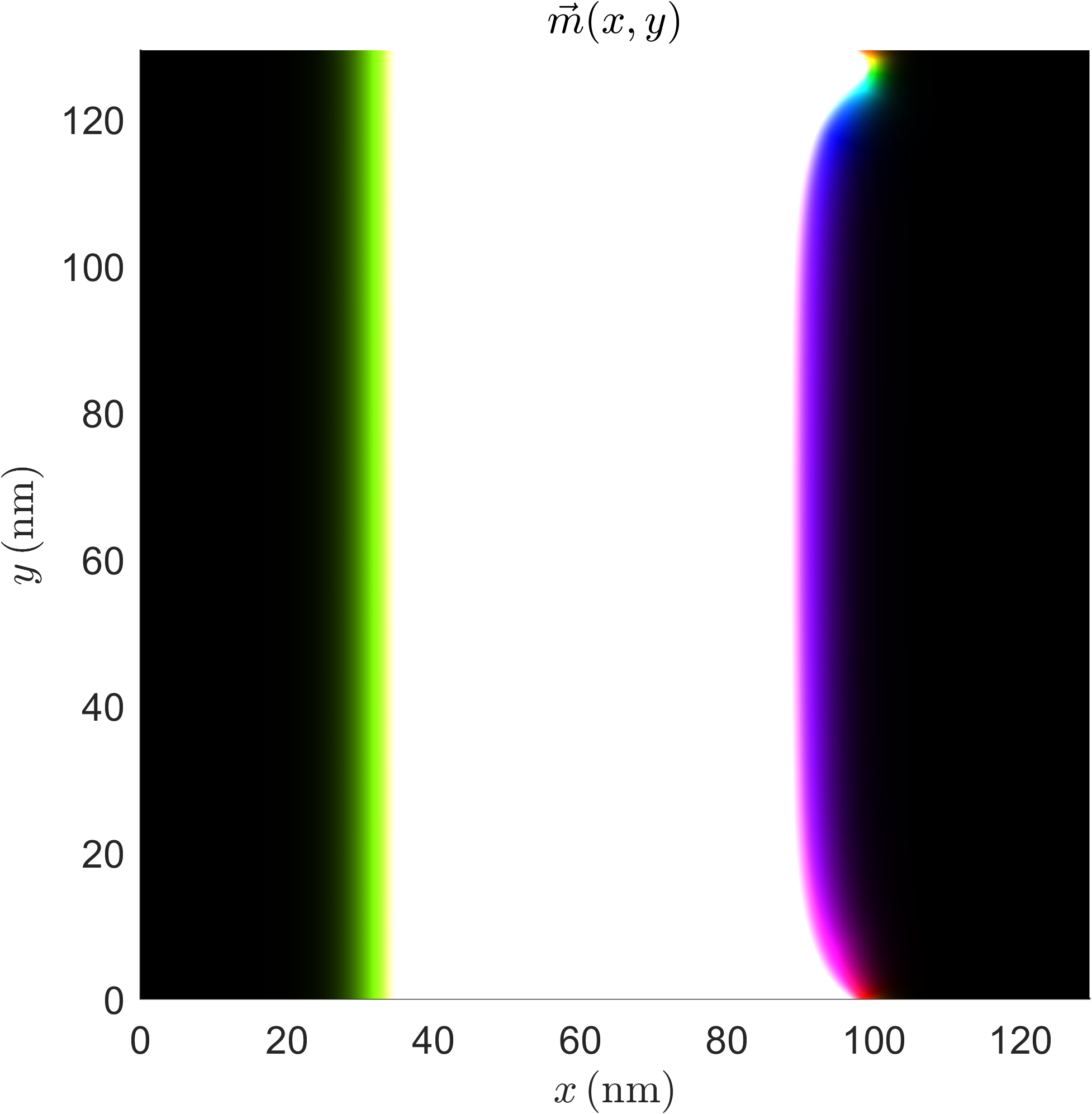} & $+1$ \\
        \bottomrule
    \end{tabular}
    \caption{Initial and final states for the domain wall phase $\chi=\pi$ as the skyrmion is rotated by $\pi/3$ from $\phi=0$ until $\phi=5\pi/3$.}
    \label{tbl: chi = pi}
\end{table}

\begin{table}
    \centering
    \begin{tabular}{ccM{40mm}M{40mm}c}
        \toprule
        $(\chi,\phi)$ & $Q_{\textup{i}}$ & Initial Configuration & Final Configuration & $Q_{\textup{f}}$ \\
        \midrule
        $(4\pi/3,0)$ & $+1$ & \includegraphics[width=25mm]{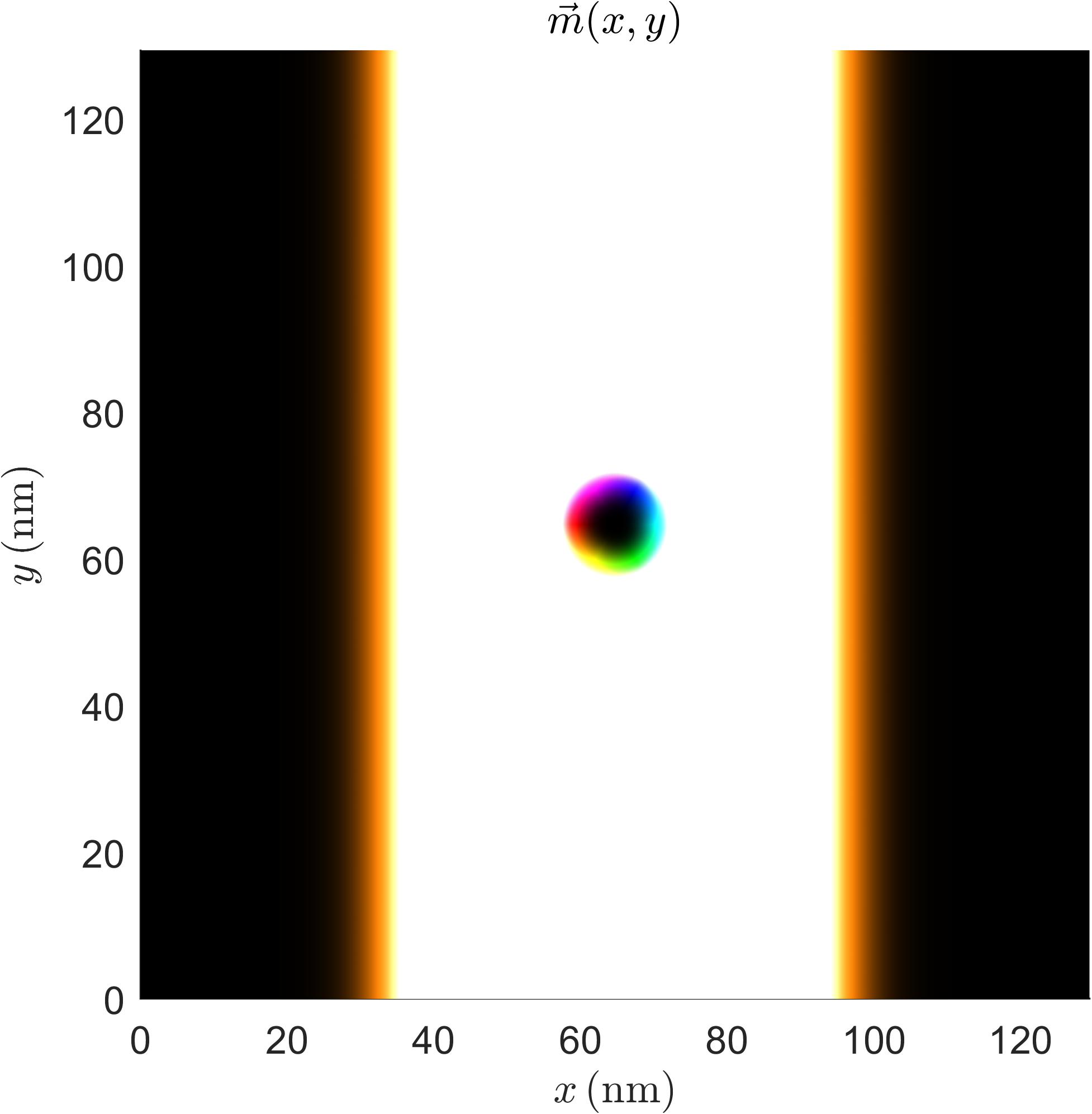} & \includegraphics[width=25mm]{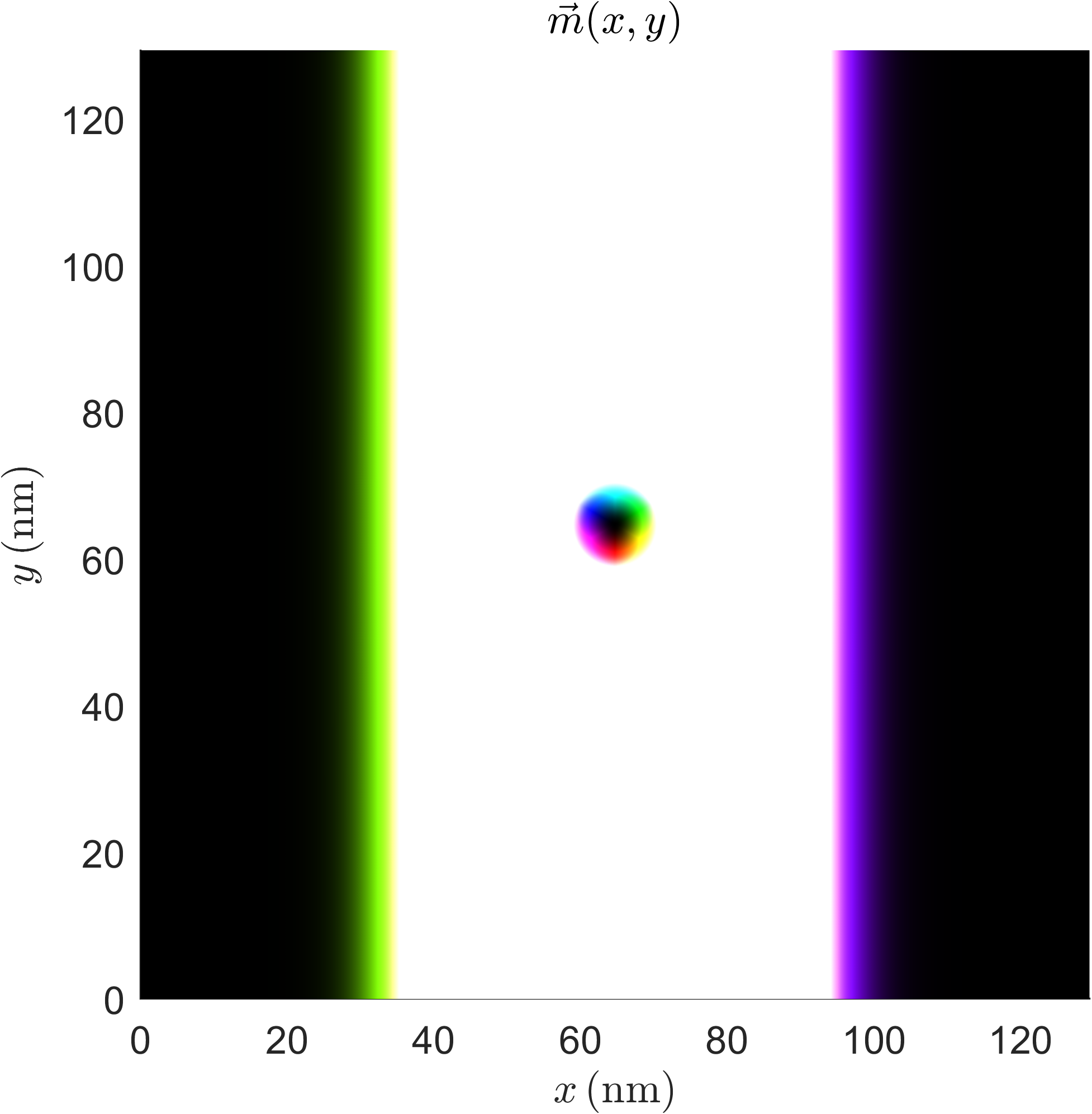} & $+1$ \\
        $(4\pi/3,\pi/3)$ & $+1$ & \includegraphics[width=25mm]{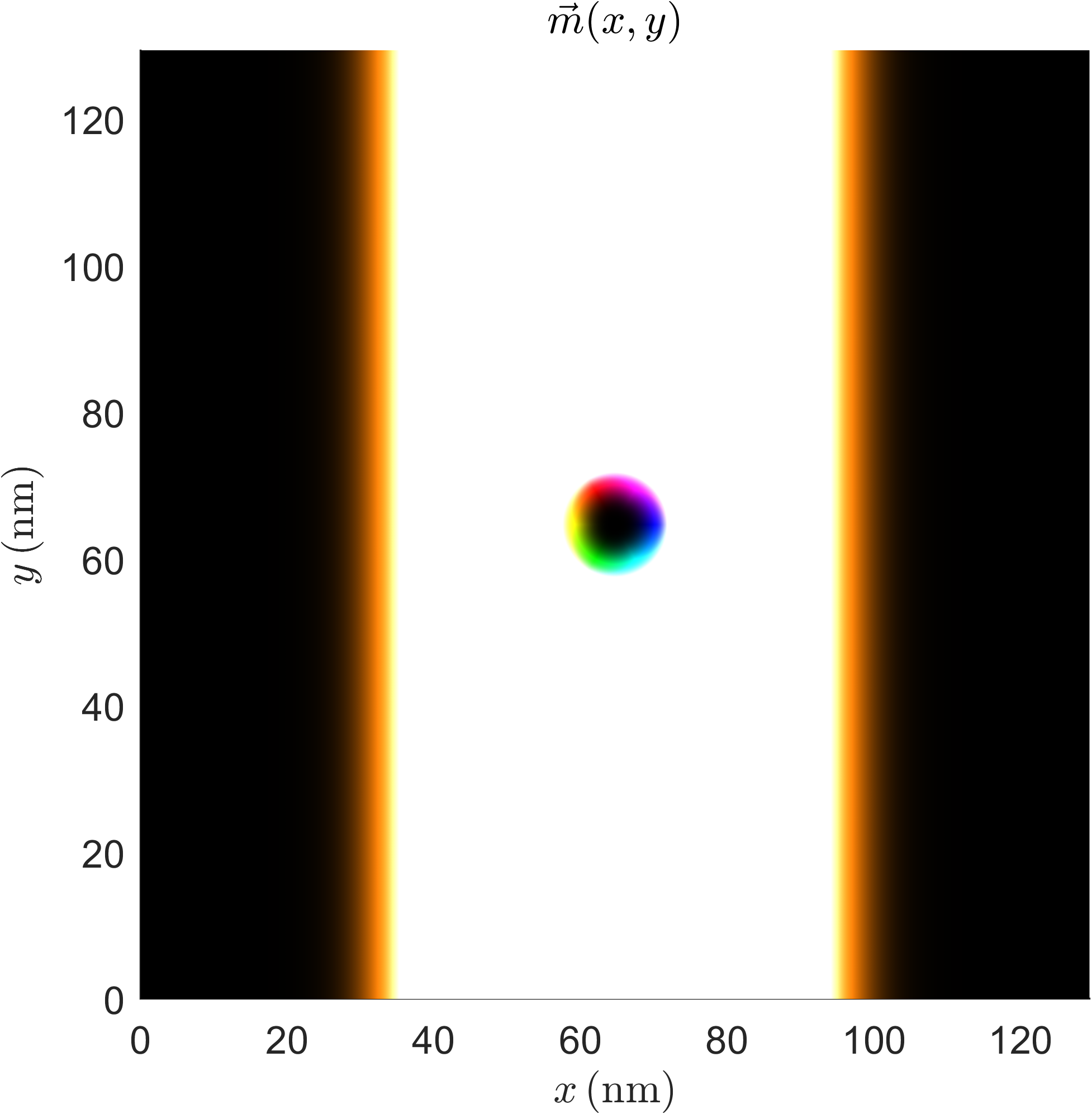} & \includegraphics[width=25mm]{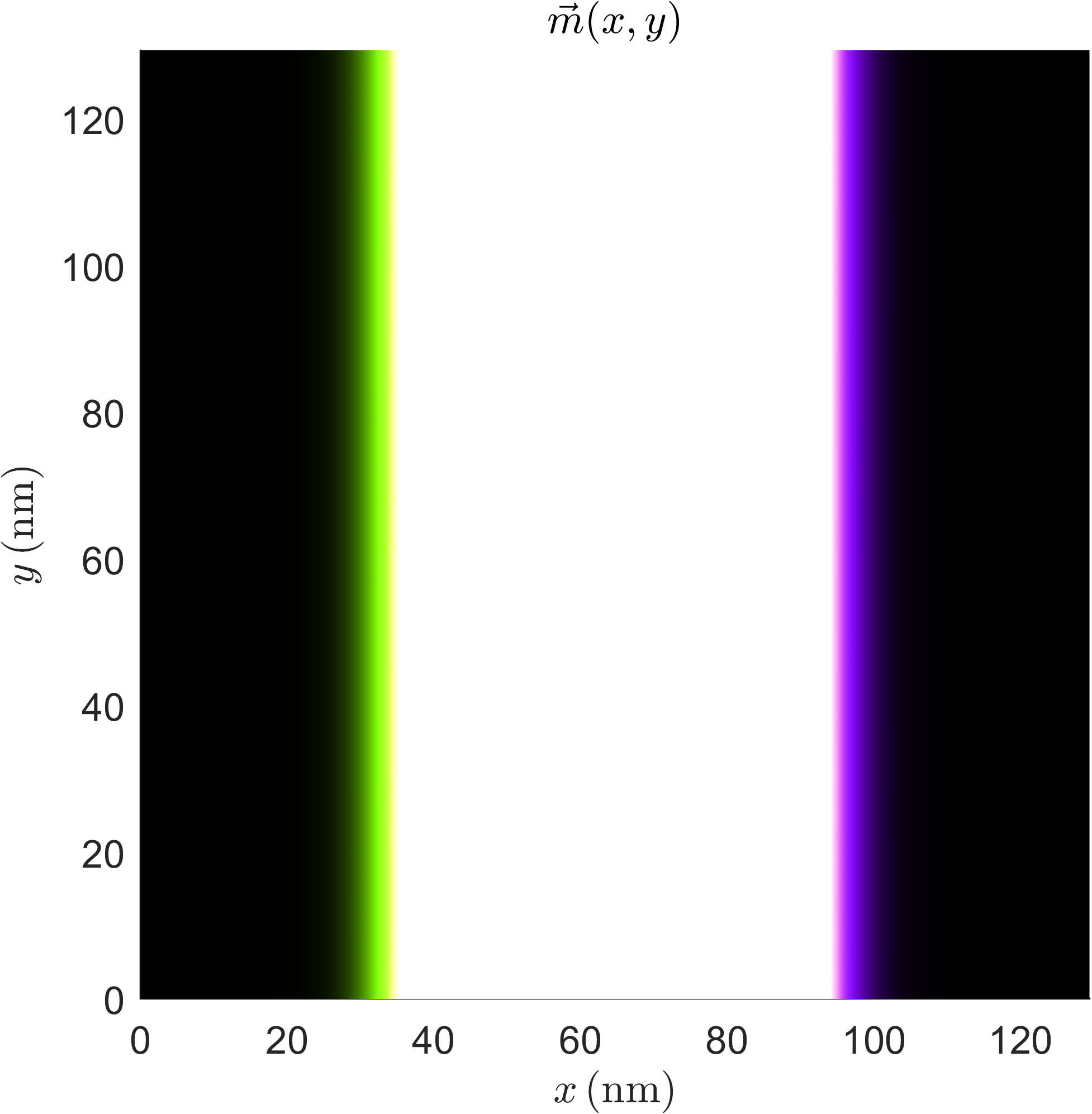} & $0$ \\
        $(4\pi/3,\pi/2)$ & $+1$ & \includegraphics[width=25mm]{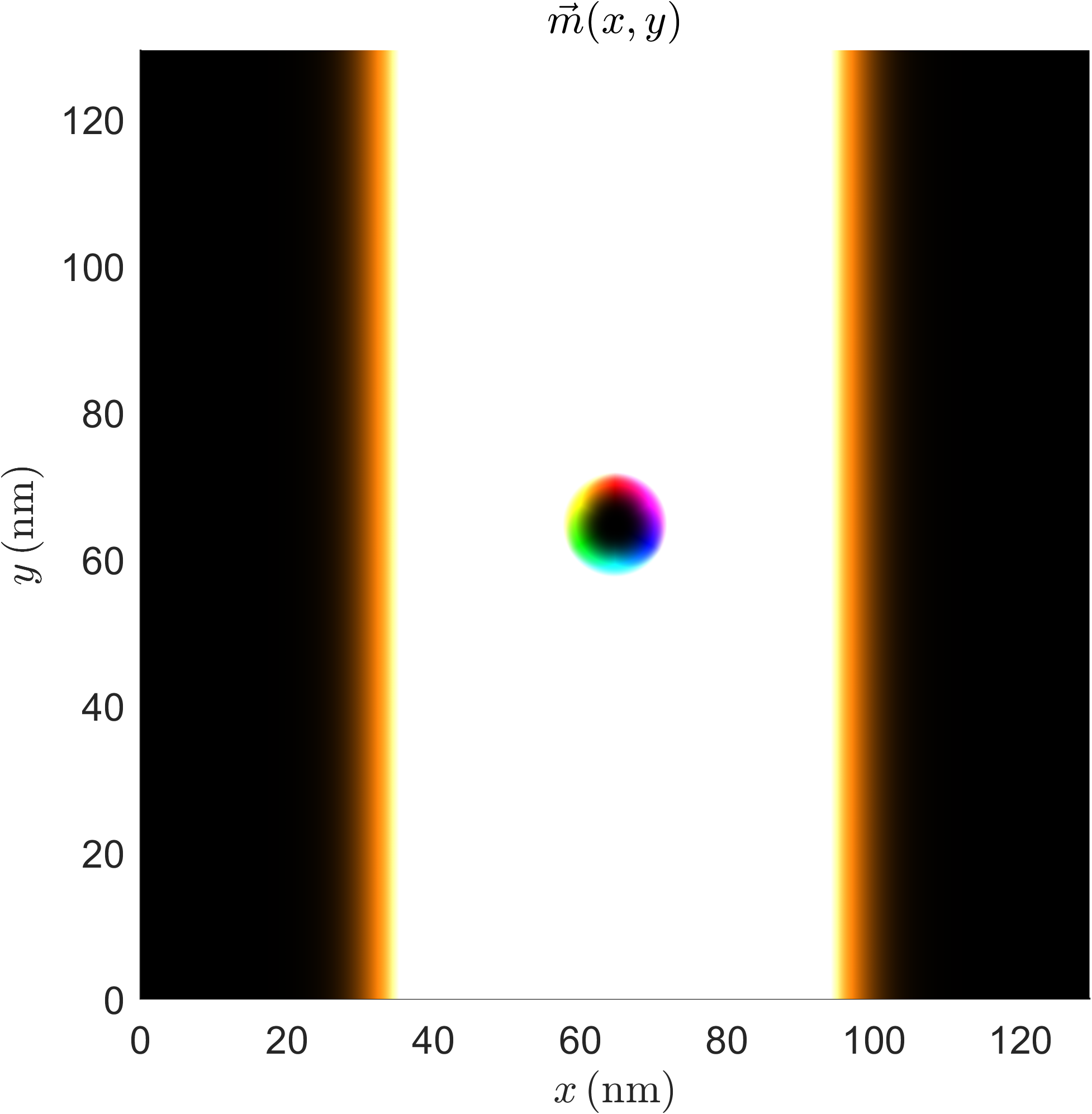} & \includegraphics[width=25mm]{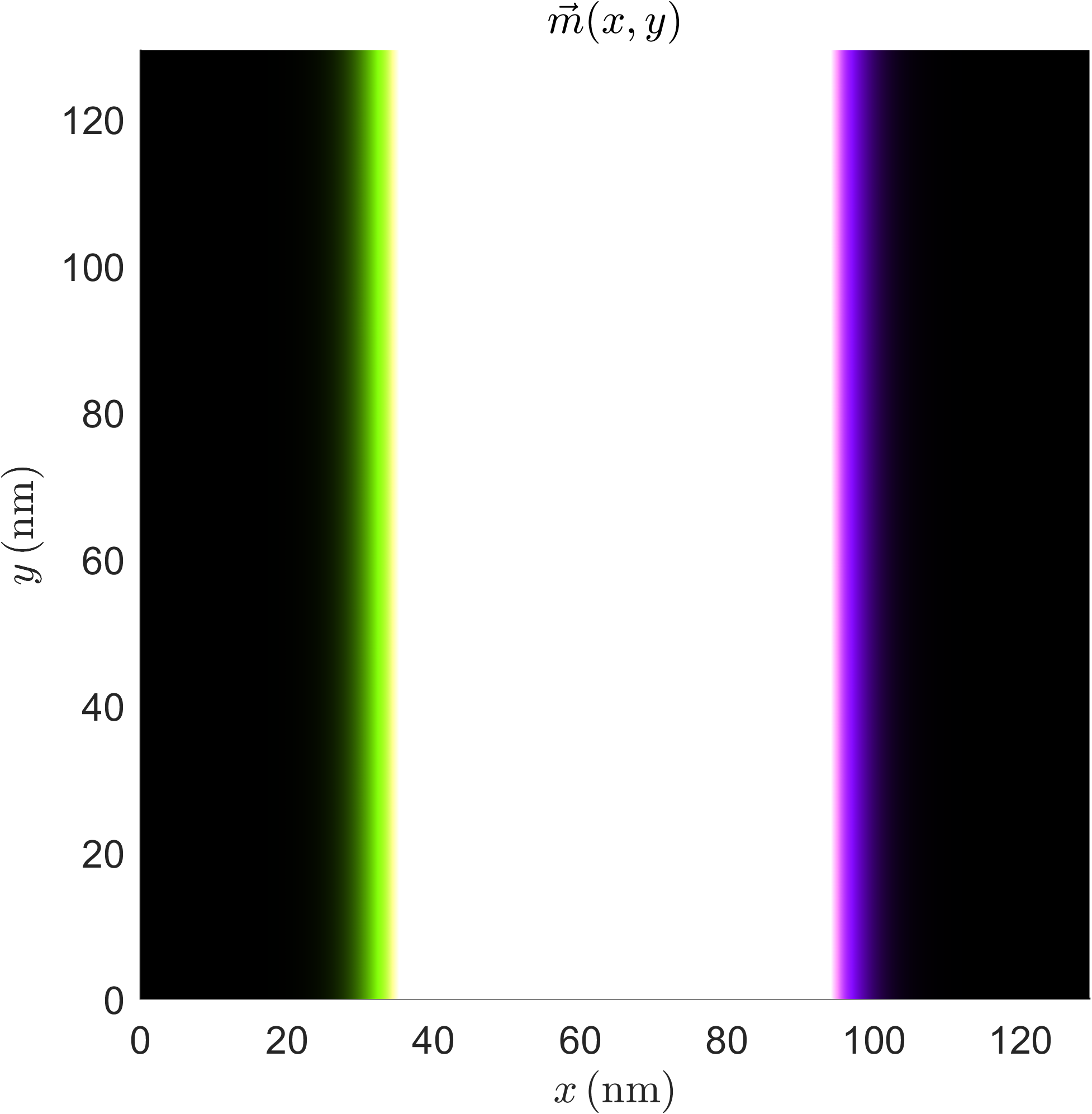} & $0$ \\
        $(4\pi/3,2\pi/3)$ & $+1$ & \includegraphics[width=25mm]{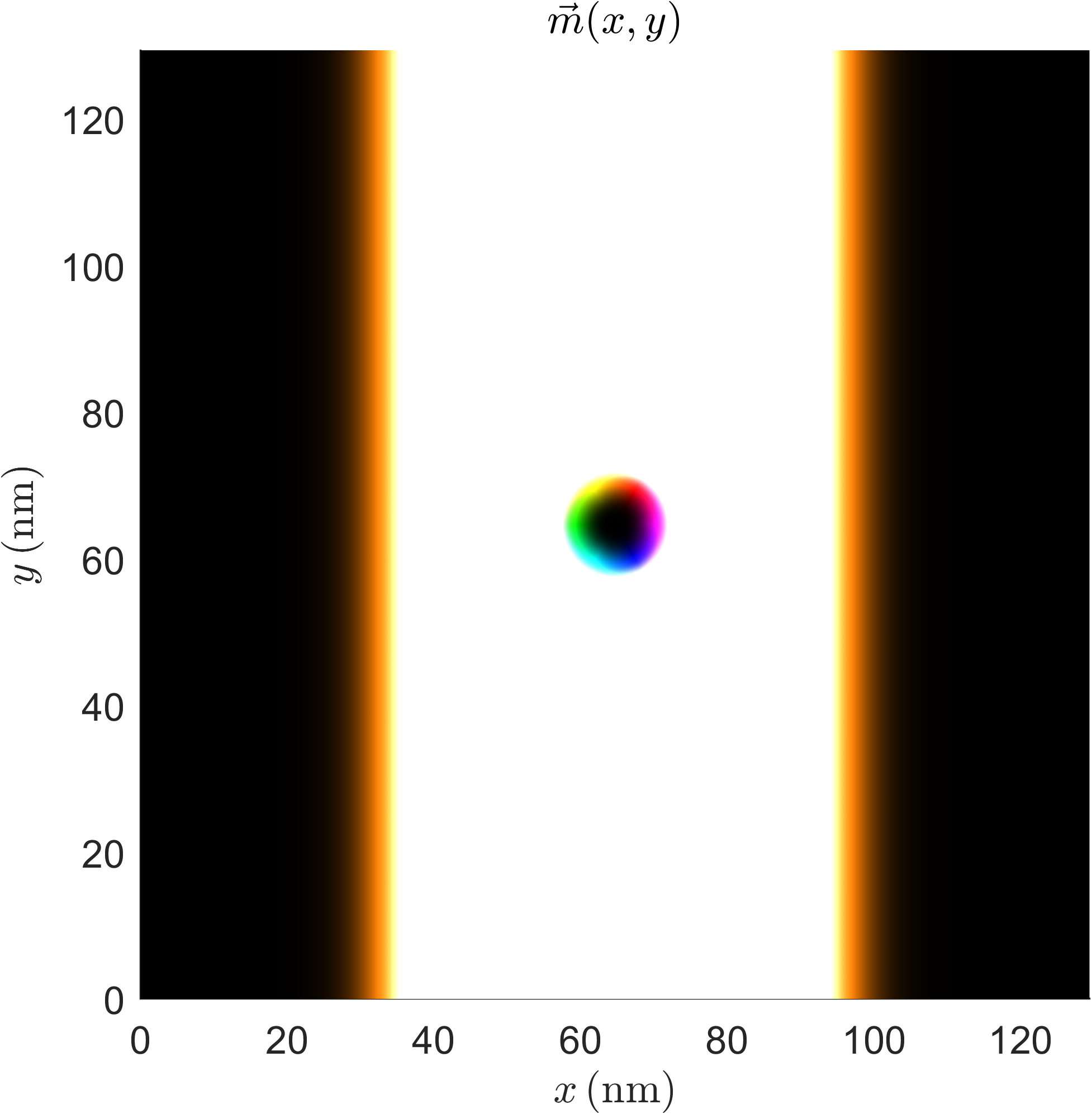} & \includegraphics[width=25mm]{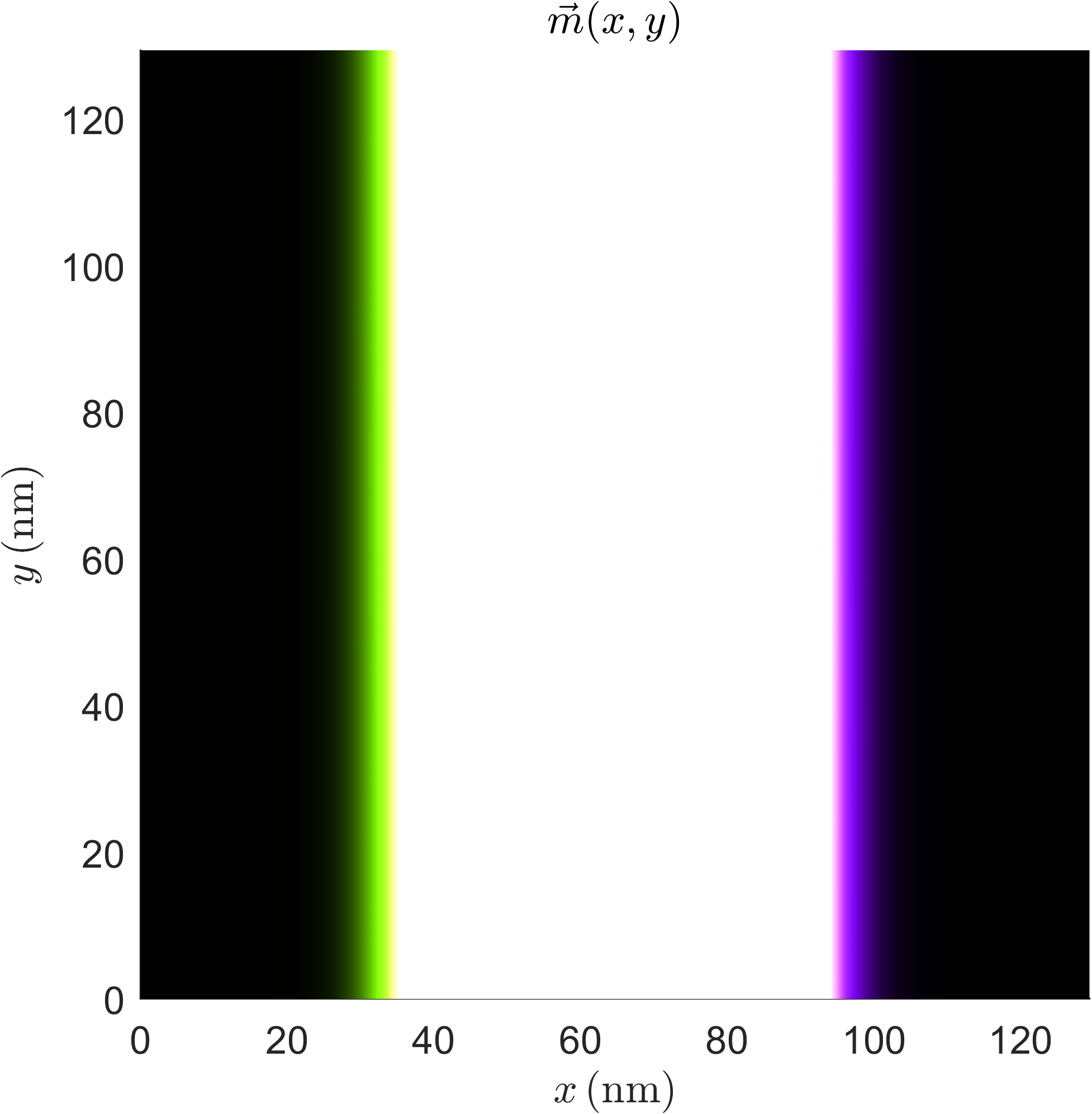} & $0$ \\
        $(4\pi/3,\pi)$ & $+1$ & \includegraphics[width=25mm]{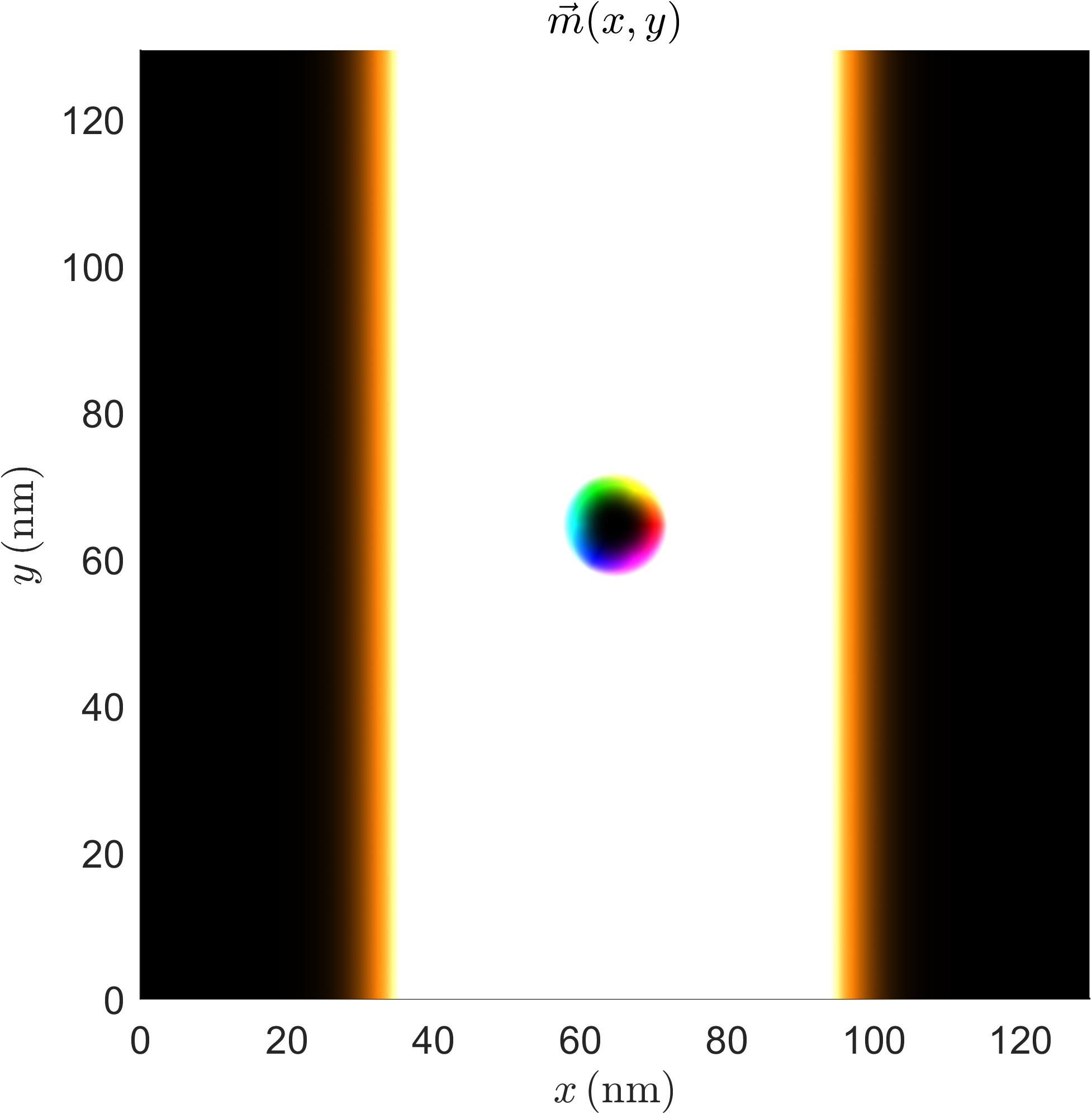} & \includegraphics[width=25mm]{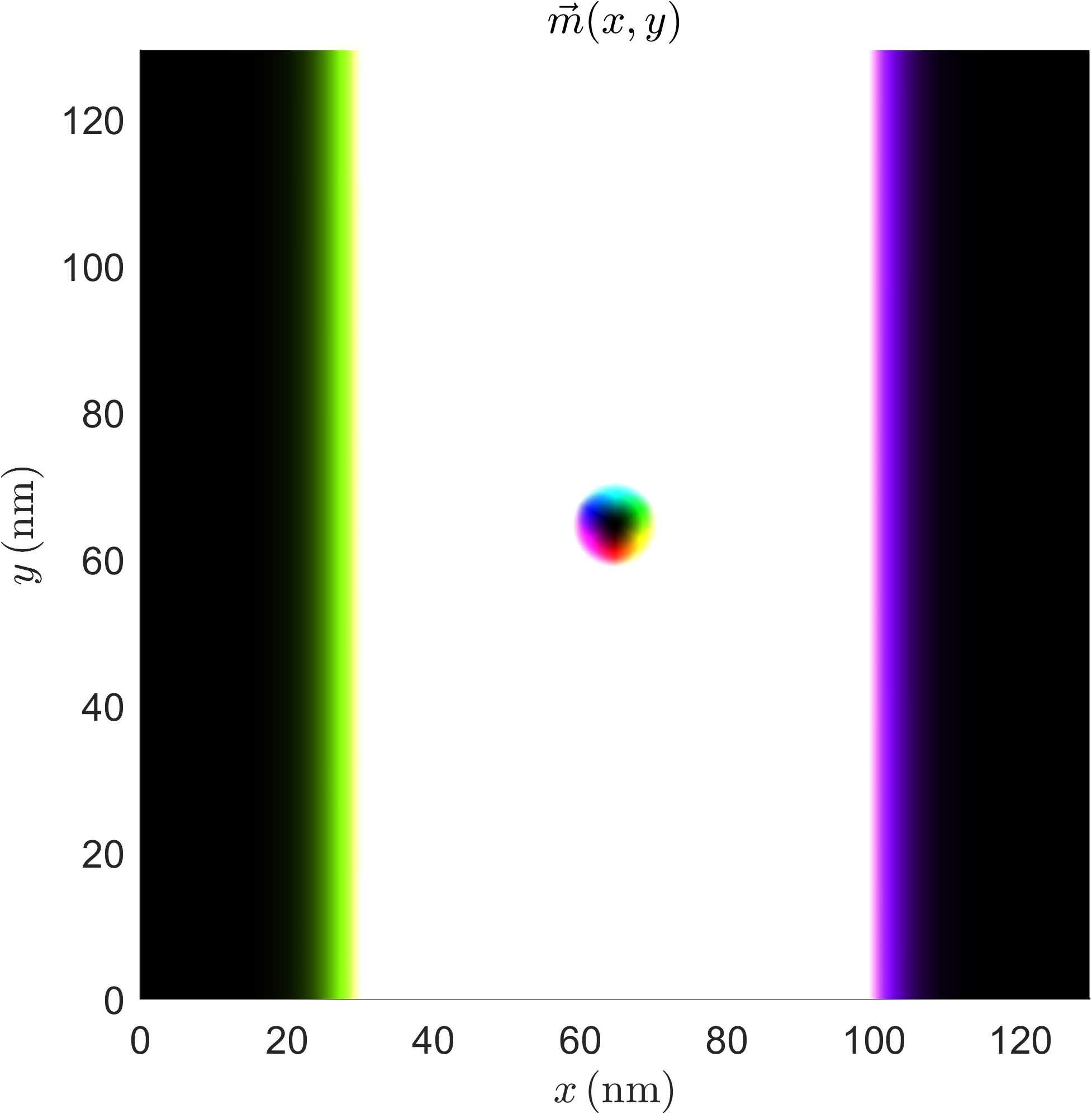} & $+1$ \\
        $(4\pi/3,4\pi/3)$ & $+1$ & \includegraphics[width=25mm]{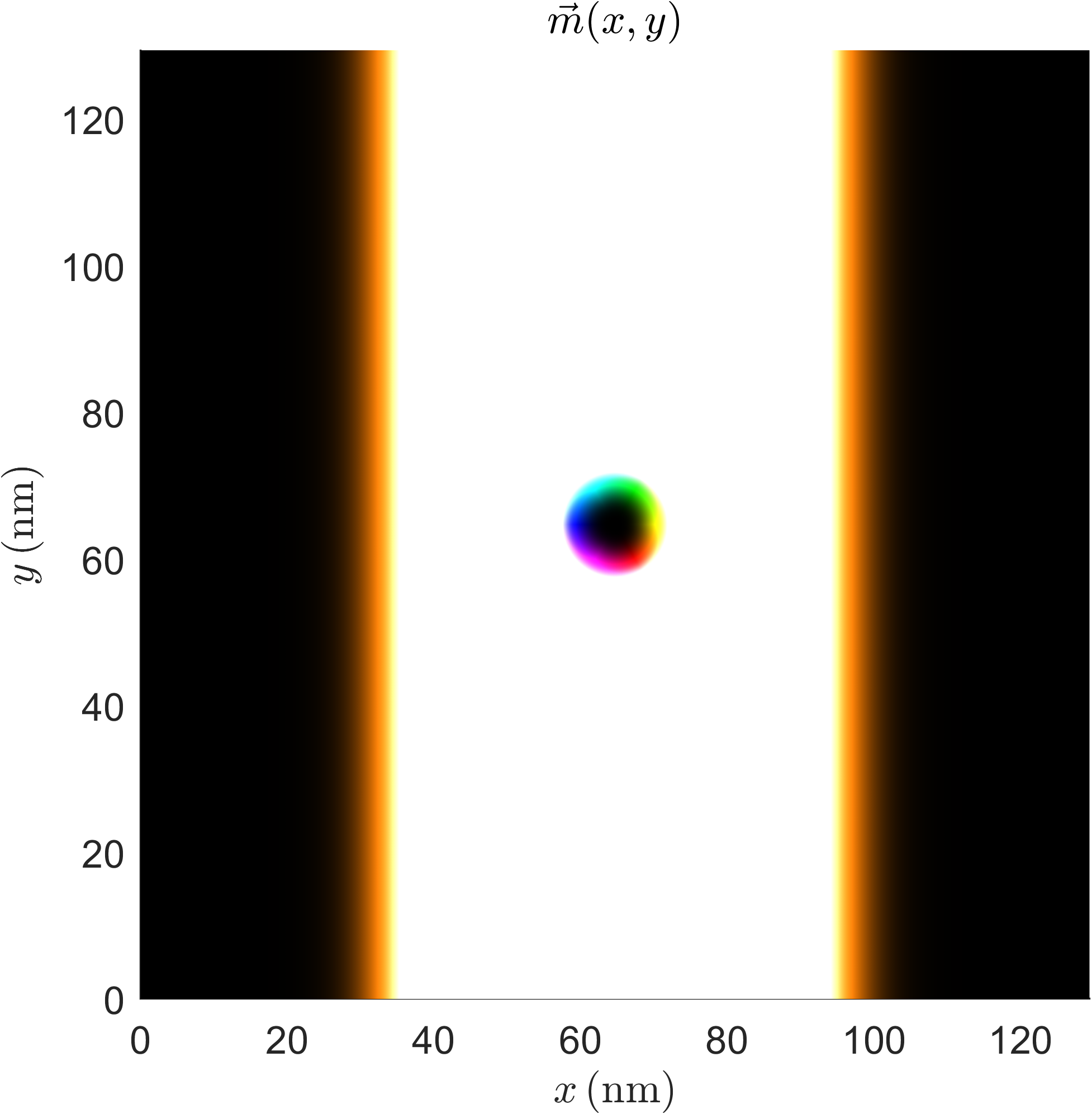} & \includegraphics[width=25mm]{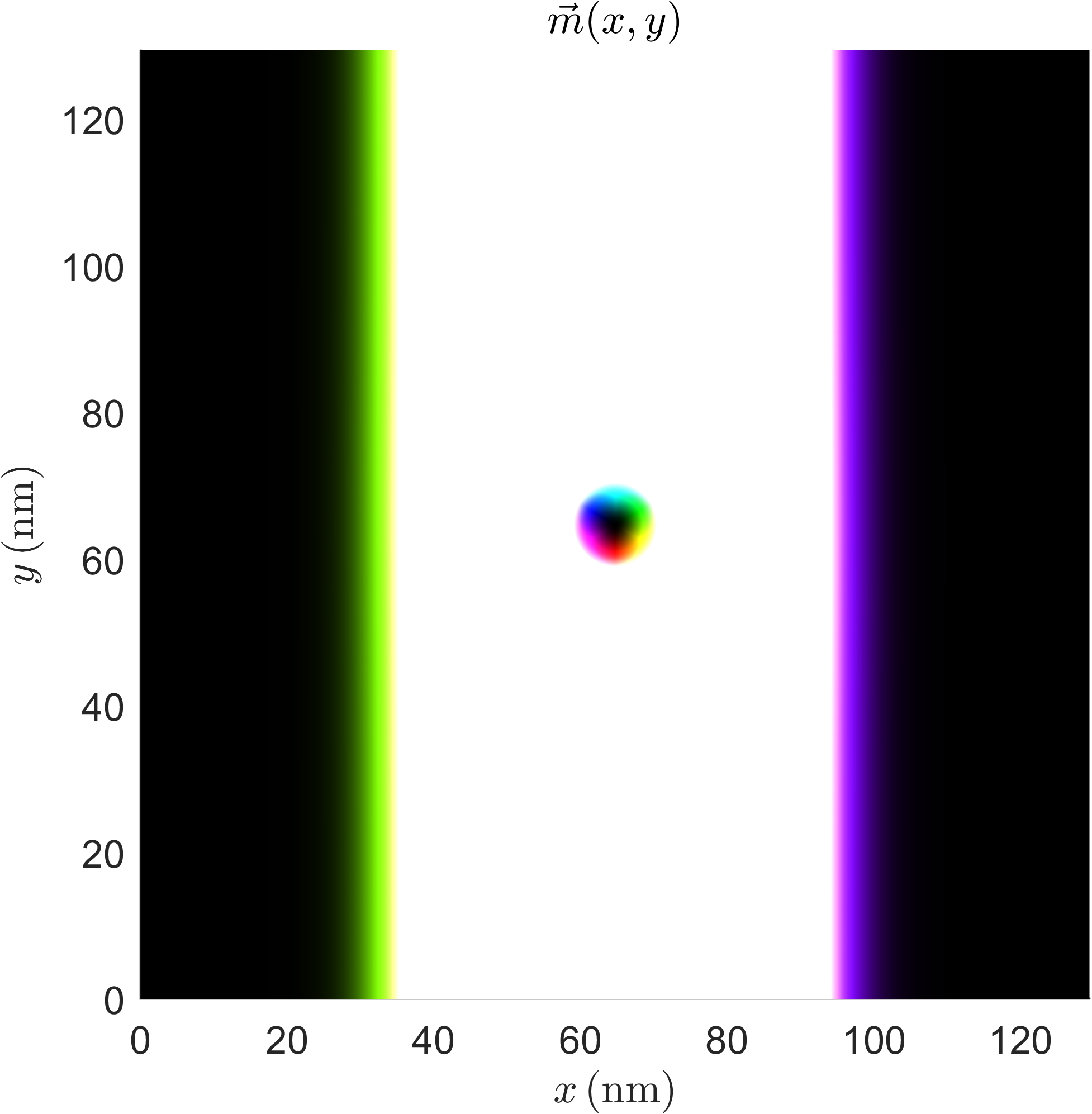} & $+1$ \\
        $(4\pi/3,3\pi/2)$ & $+1$ & \includegraphics[width=25mm]{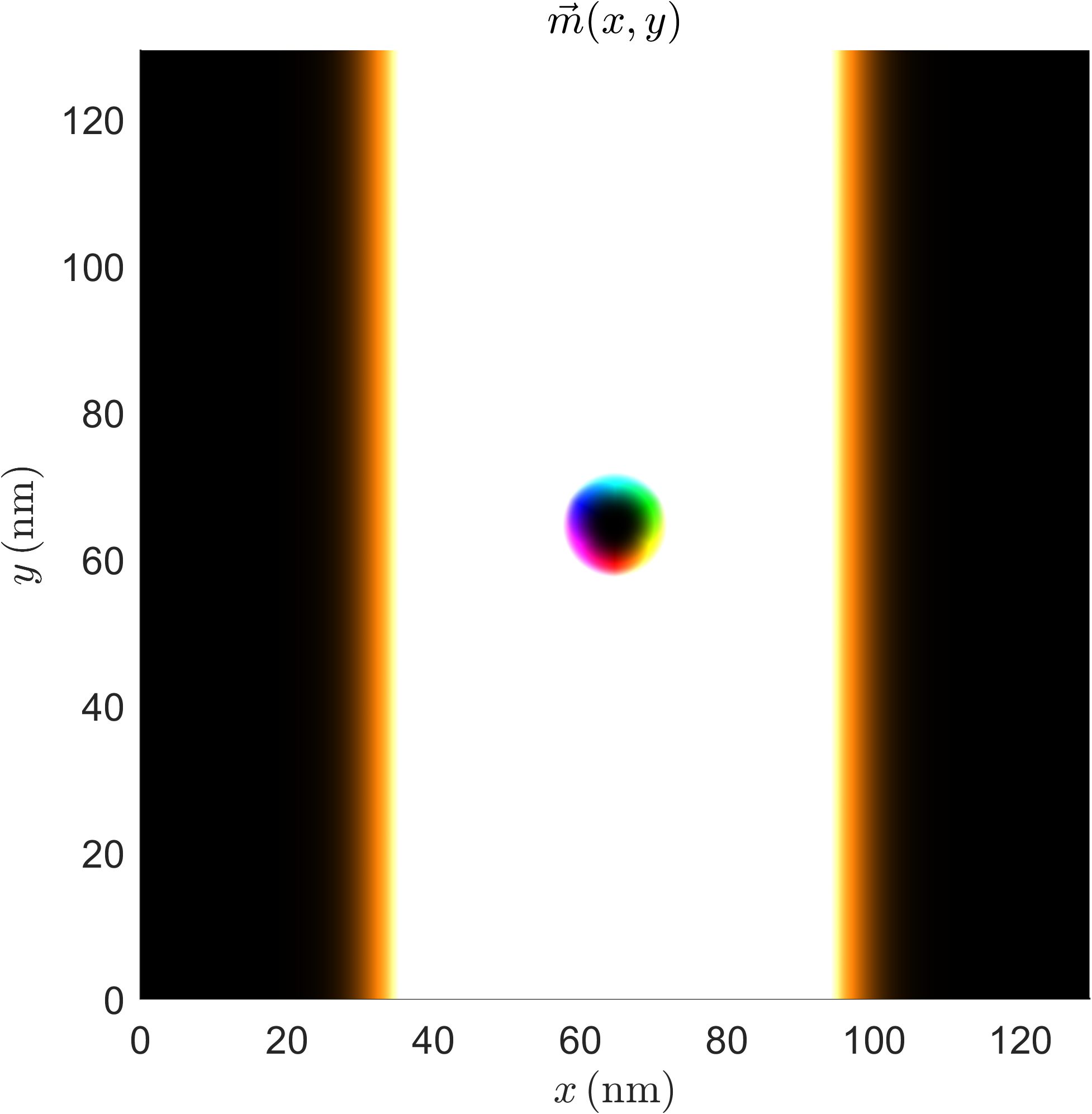} & \includegraphics[width=25mm]{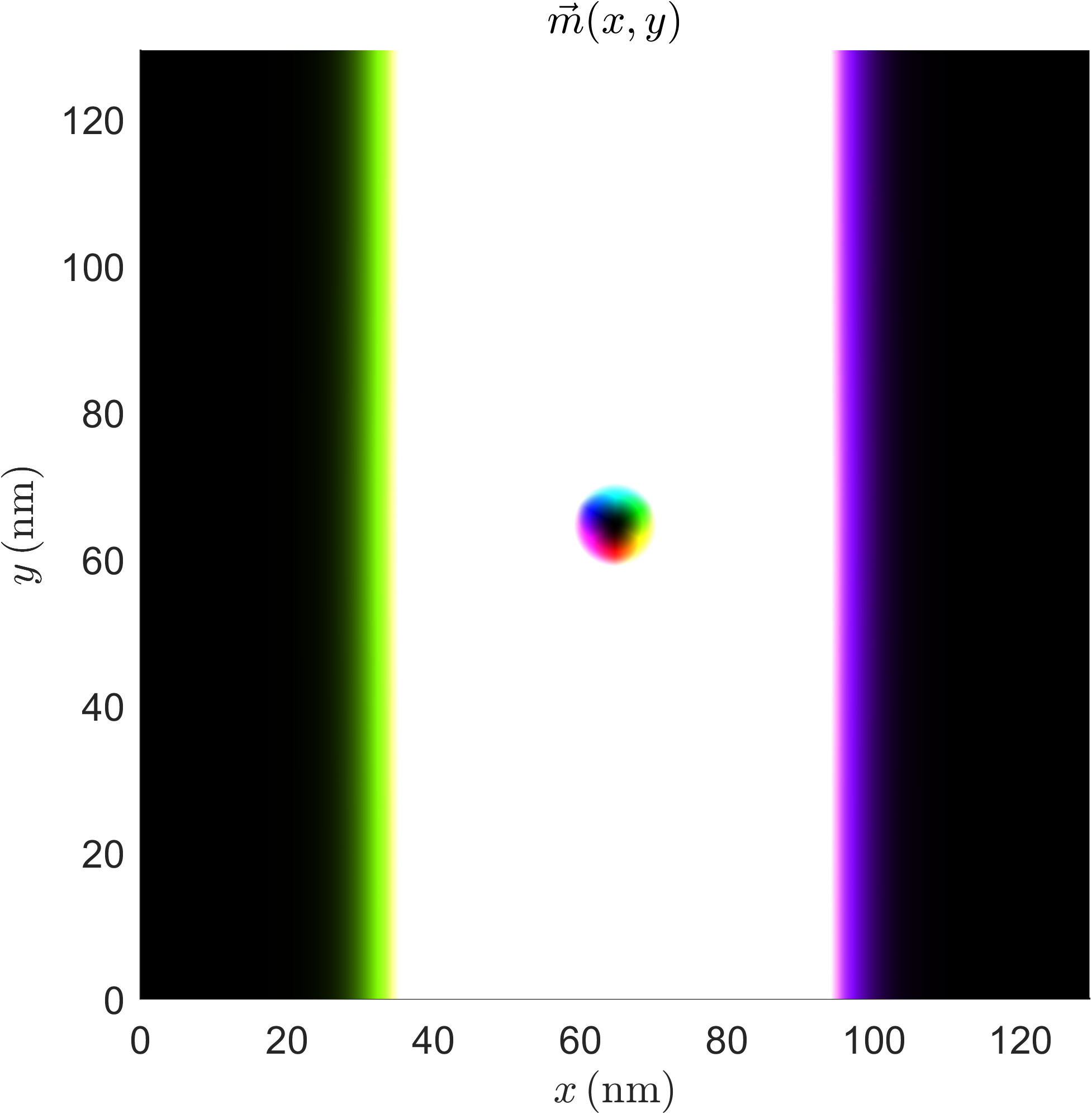} & $+1$ \\
        $(4\pi/3,5\pi/3)$ & $+1$ & \includegraphics[width=25mm]{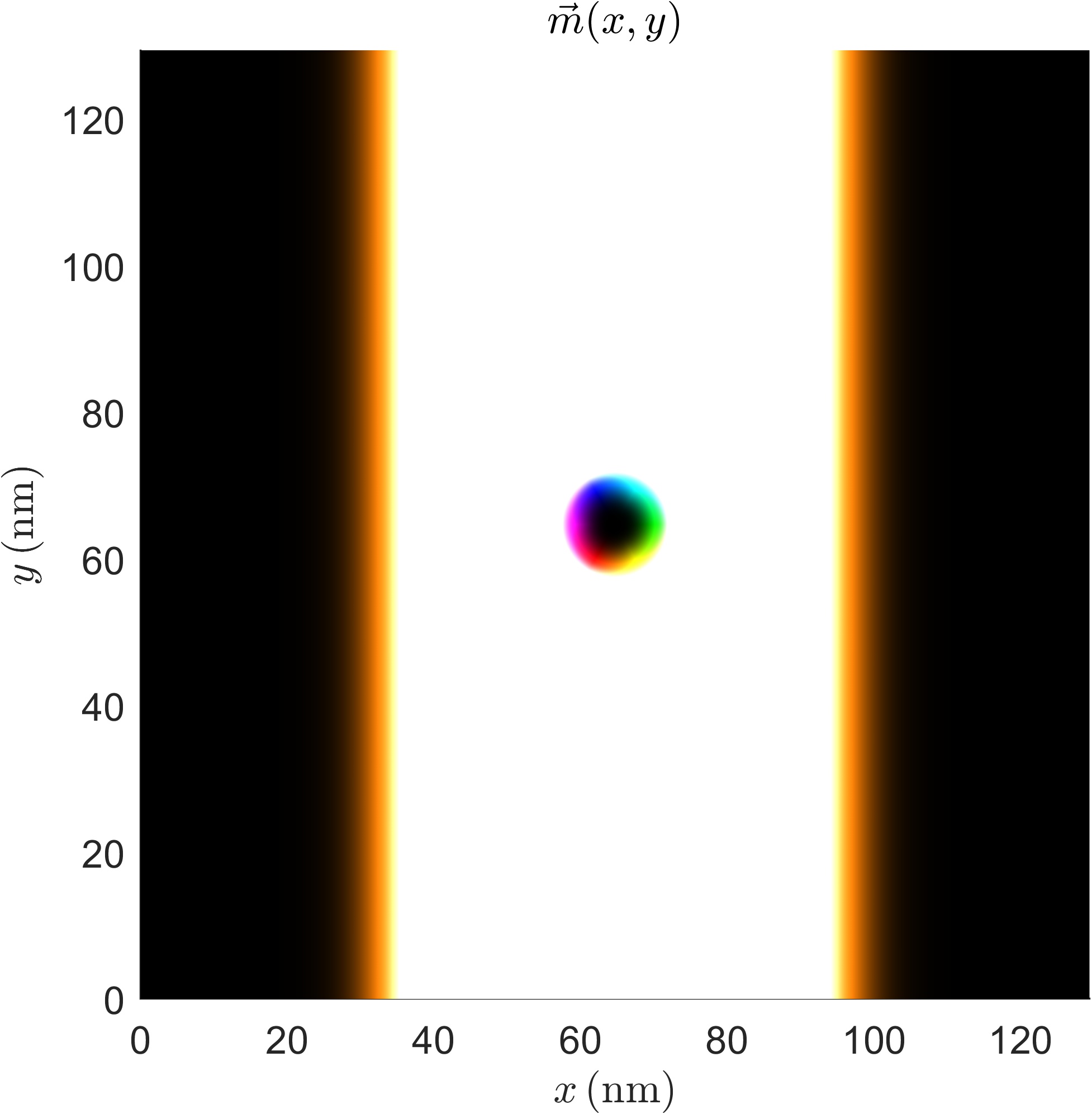} & \includegraphics[width=25mm]{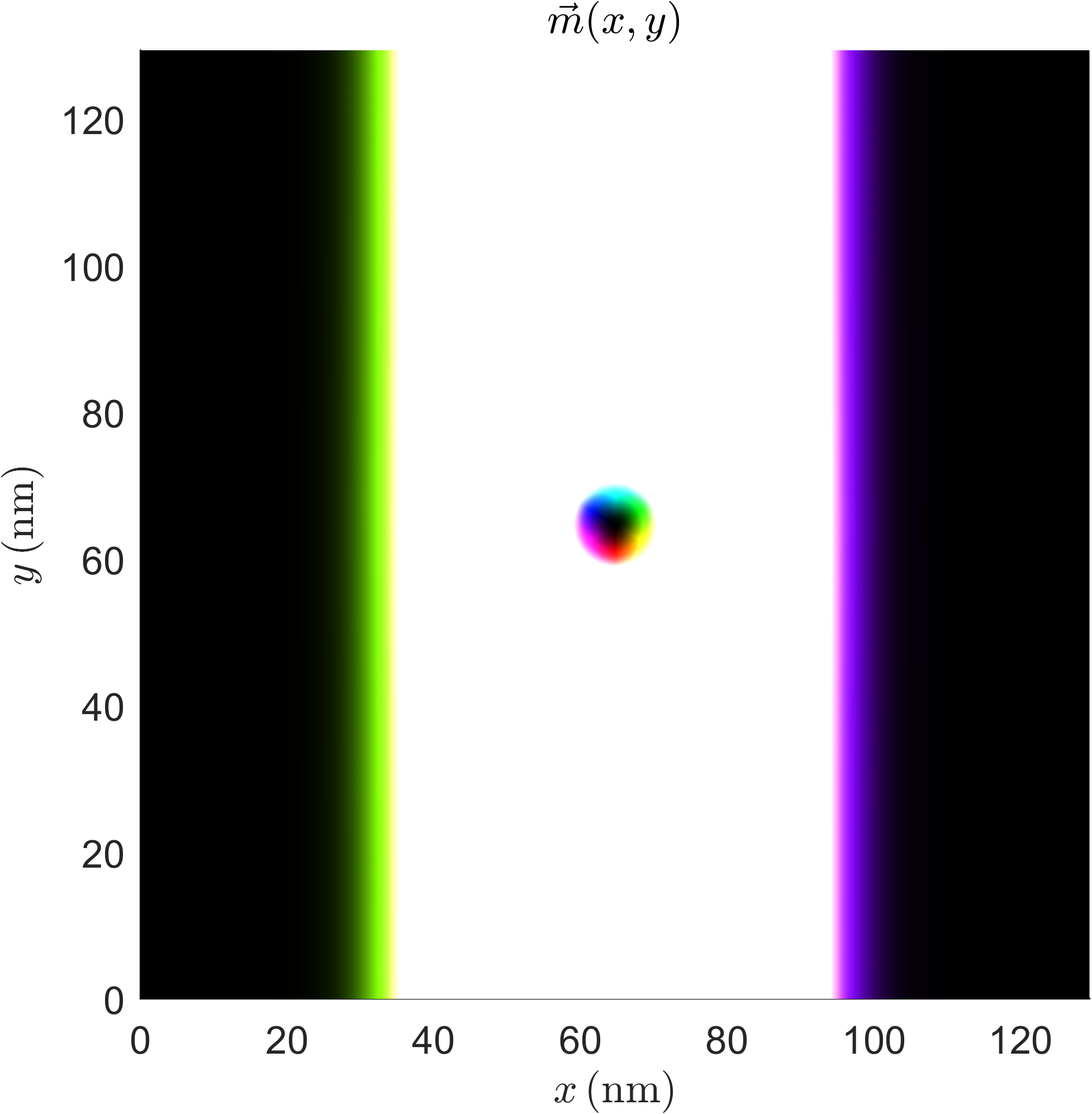} & $+1$ \\
        \bottomrule
    \end{tabular}
    \caption{Initial and final states for the domain wall phase $\chi=4\pi/3$ as the skyrmion is rotated by $\pi/3$ from $\phi=0$ until $\phi=5\pi/3$.}
    \label{tbl: chi = 4pi/3}
\end{table}

\begin{table}
    \centering
    \begin{tabular}{ccM{40mm}M{40mm}c}
        \toprule
        $(\chi,\phi)$ & $Q_{\textup{i}}$ & Initial Configuration & Final Configuration & $Q_{\textup{f}}$ \\
        \midrule
        $(3\pi/2,0)$ & $+1$ & \includegraphics[width=25mm]{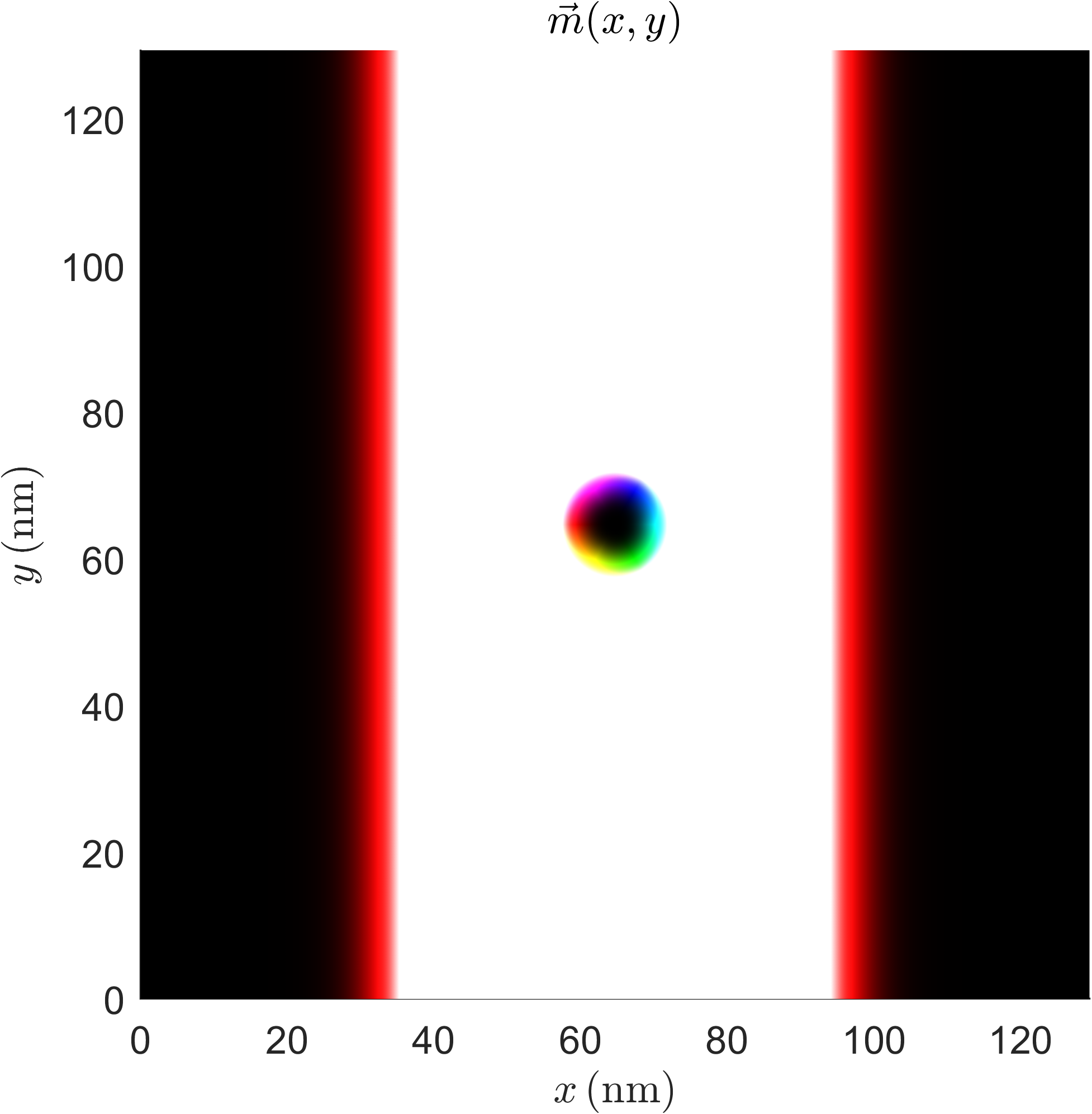} & \includegraphics[width=25mm]{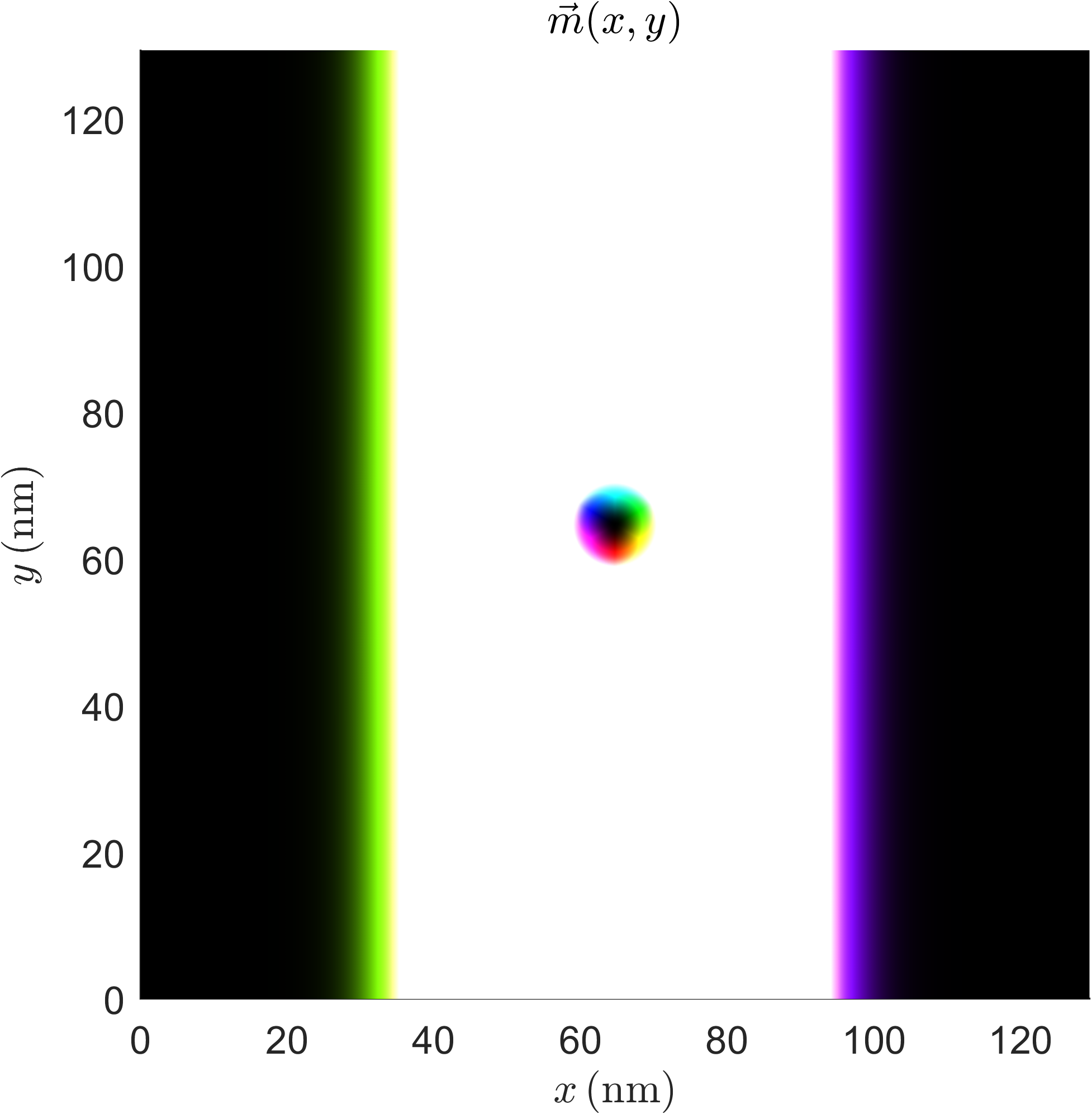} & $+1$ \\
        $(3\pi/2,\pi/3)$ & $+1$ & \includegraphics[width=25mm]{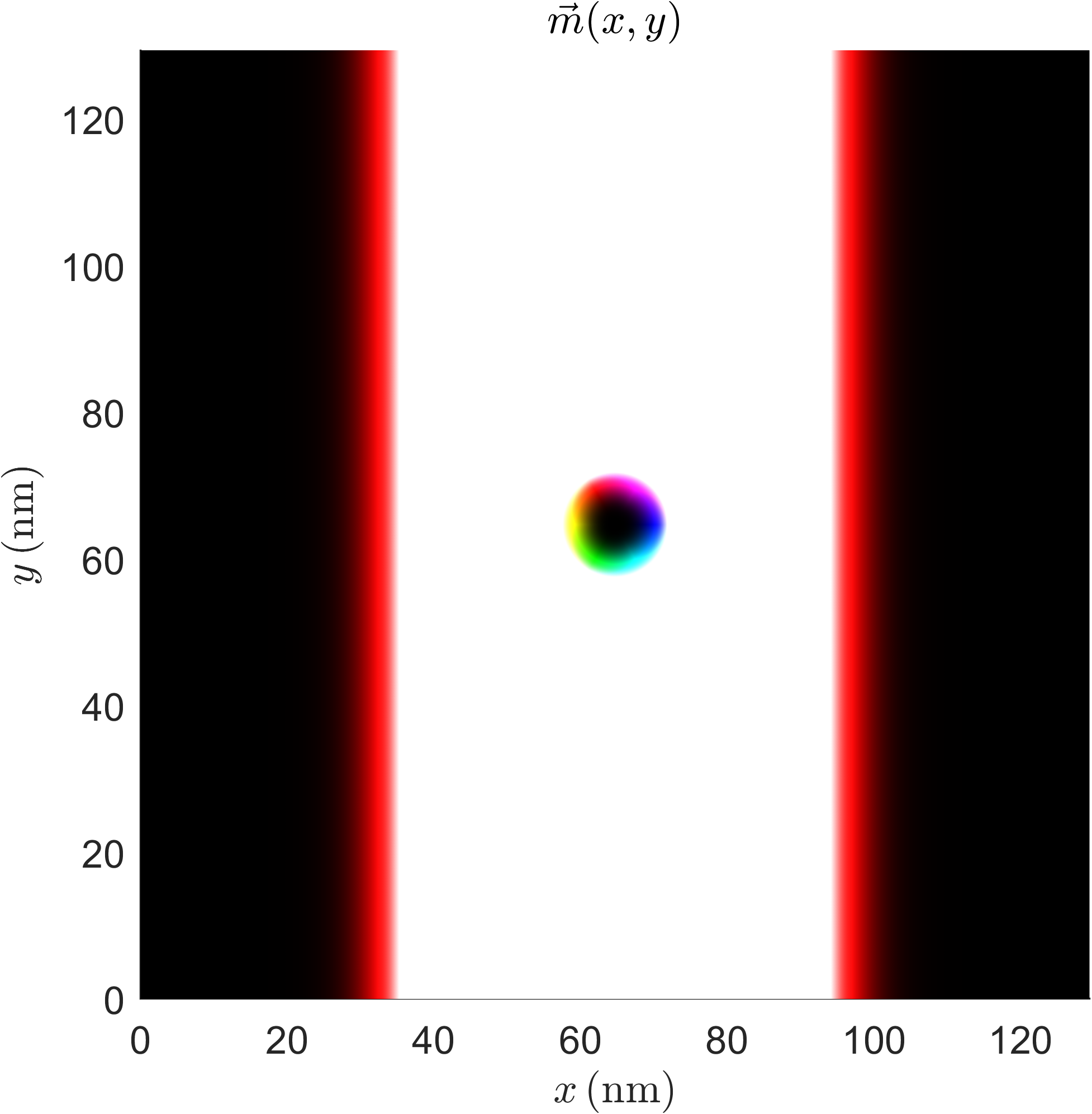} & \includegraphics[width=25mm]{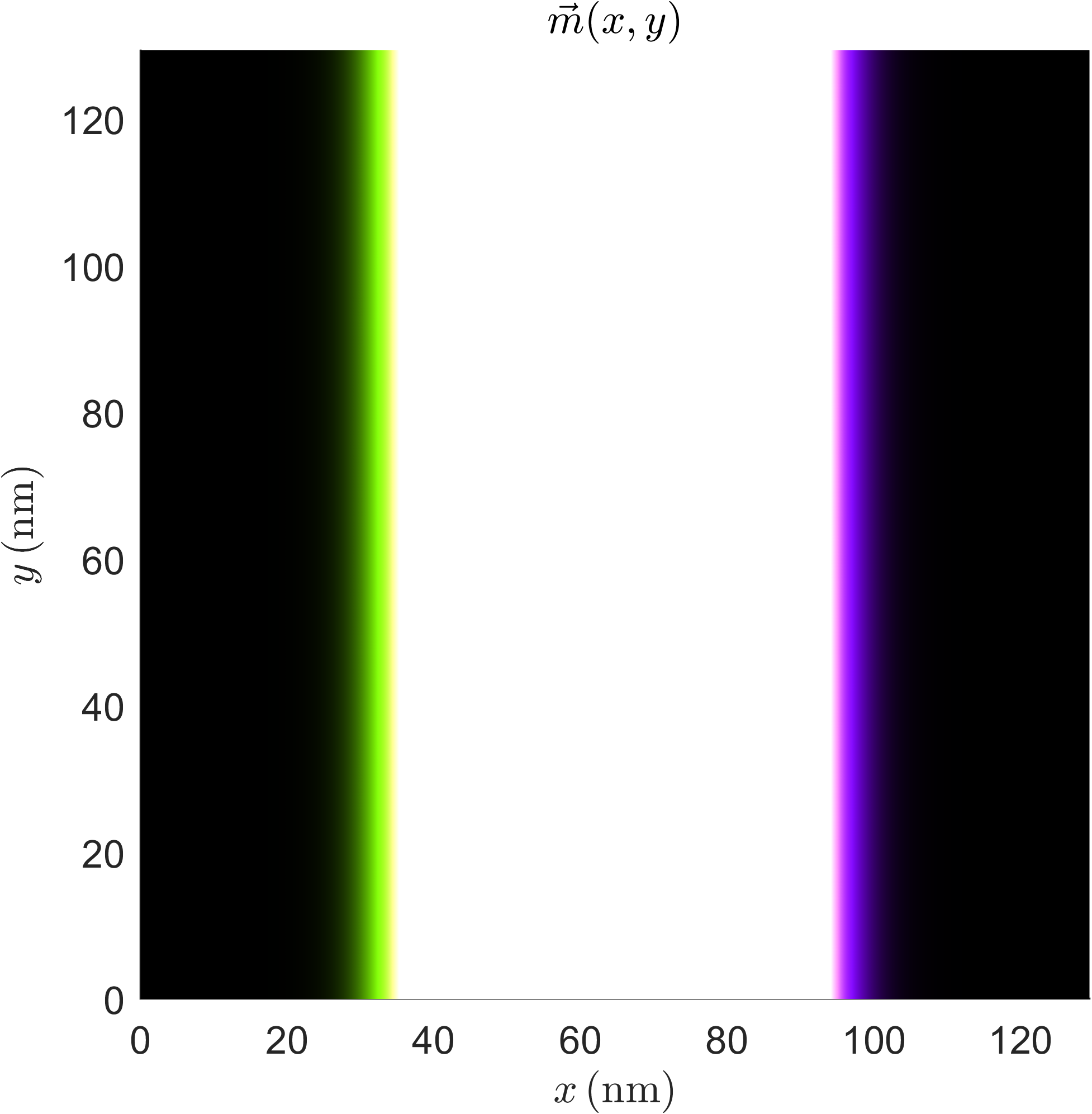} & $0$ \\
        $(3\pi/2,\pi/2)$ & $+1$ & \includegraphics[width=25mm]{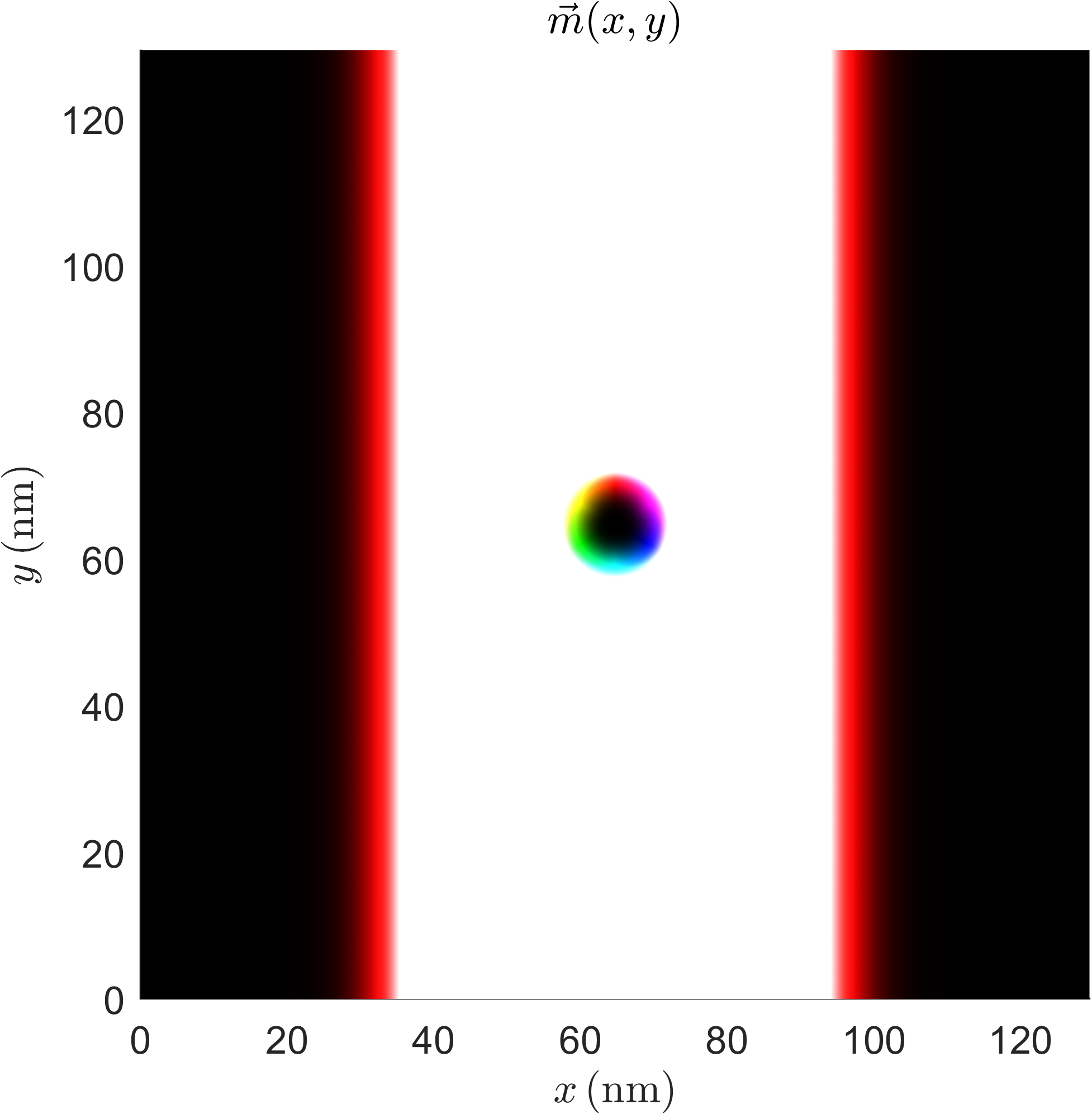} & \includegraphics[width=25mm]{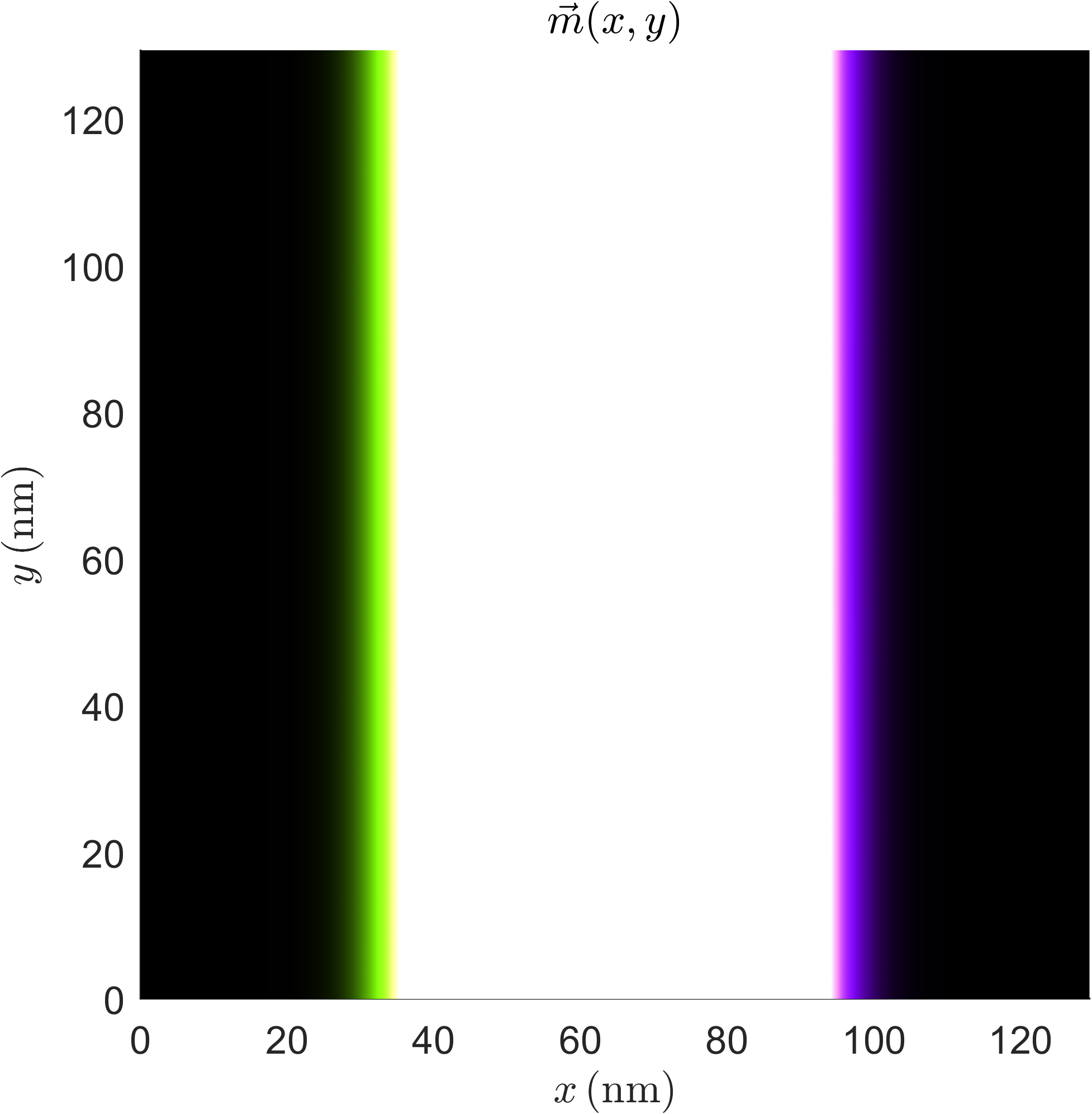} & $0$ \\
        $(3\pi/2,2\pi/3)$ & $+1$ & \includegraphics[width=25mm]{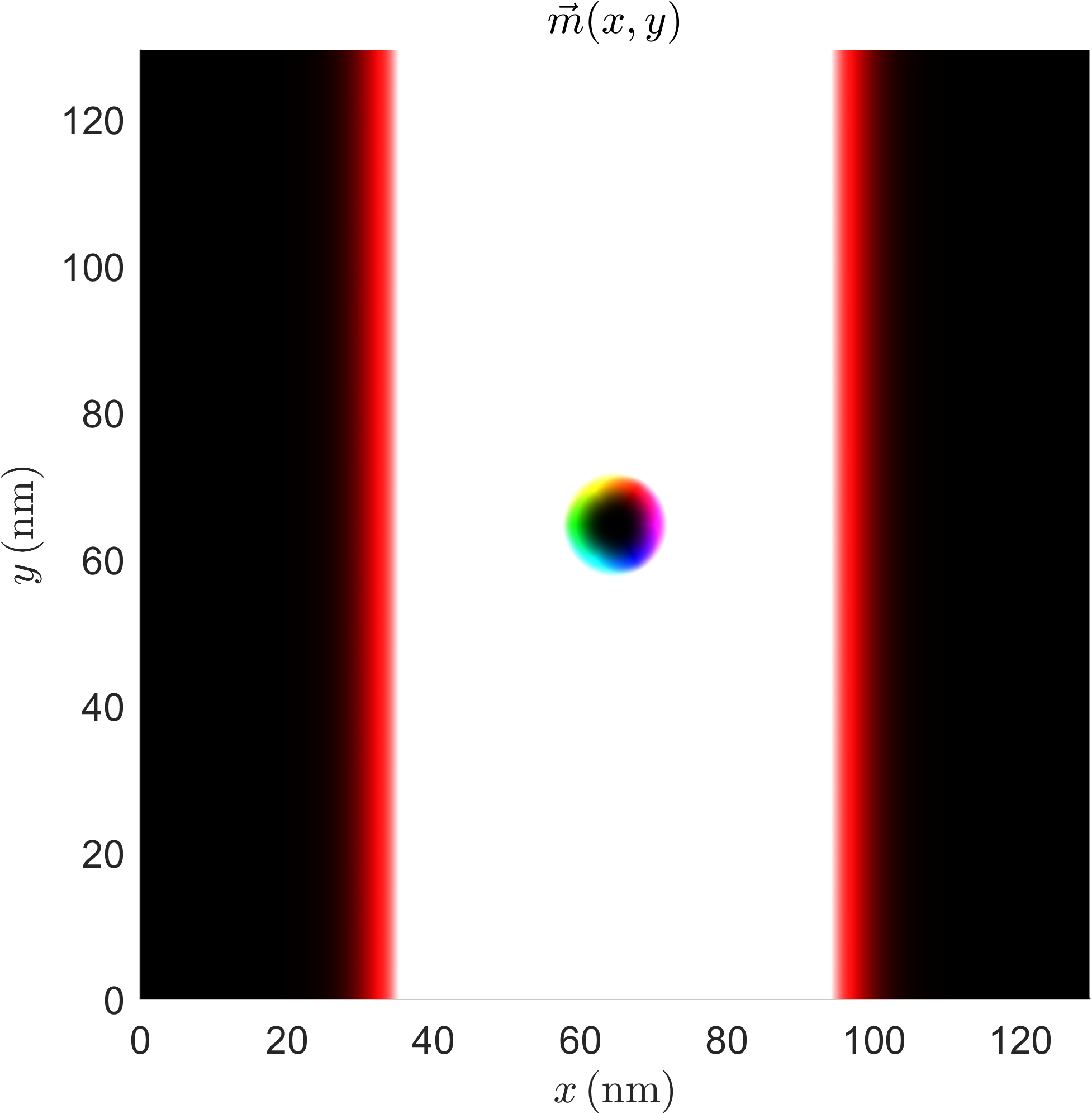} & \includegraphics[width=25mm]{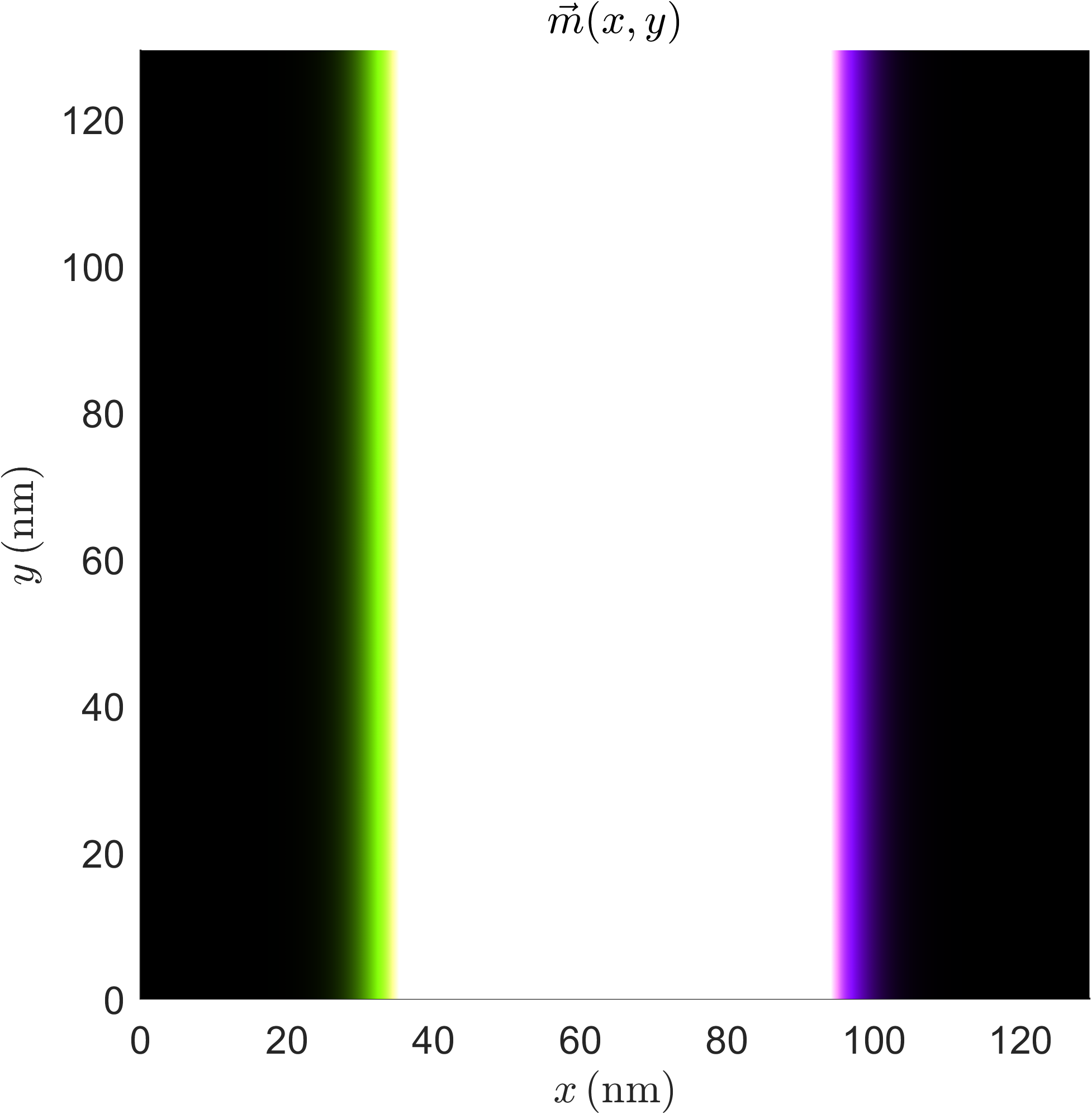} & $0$ \\
        $(3\pi/2,\pi)$ & $+1$ & \includegraphics[width=25mm]{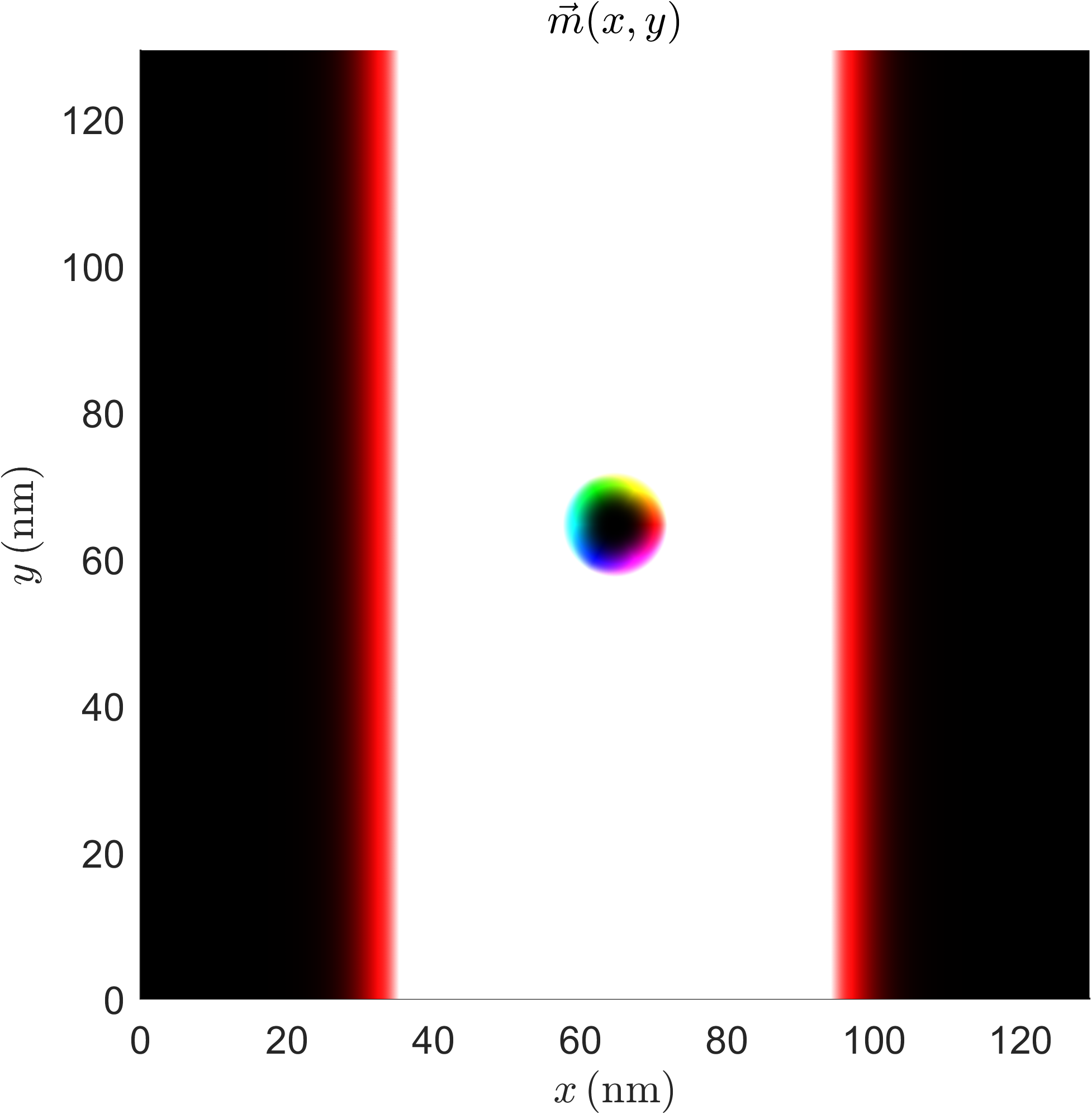} & \includegraphics[width=25mm]{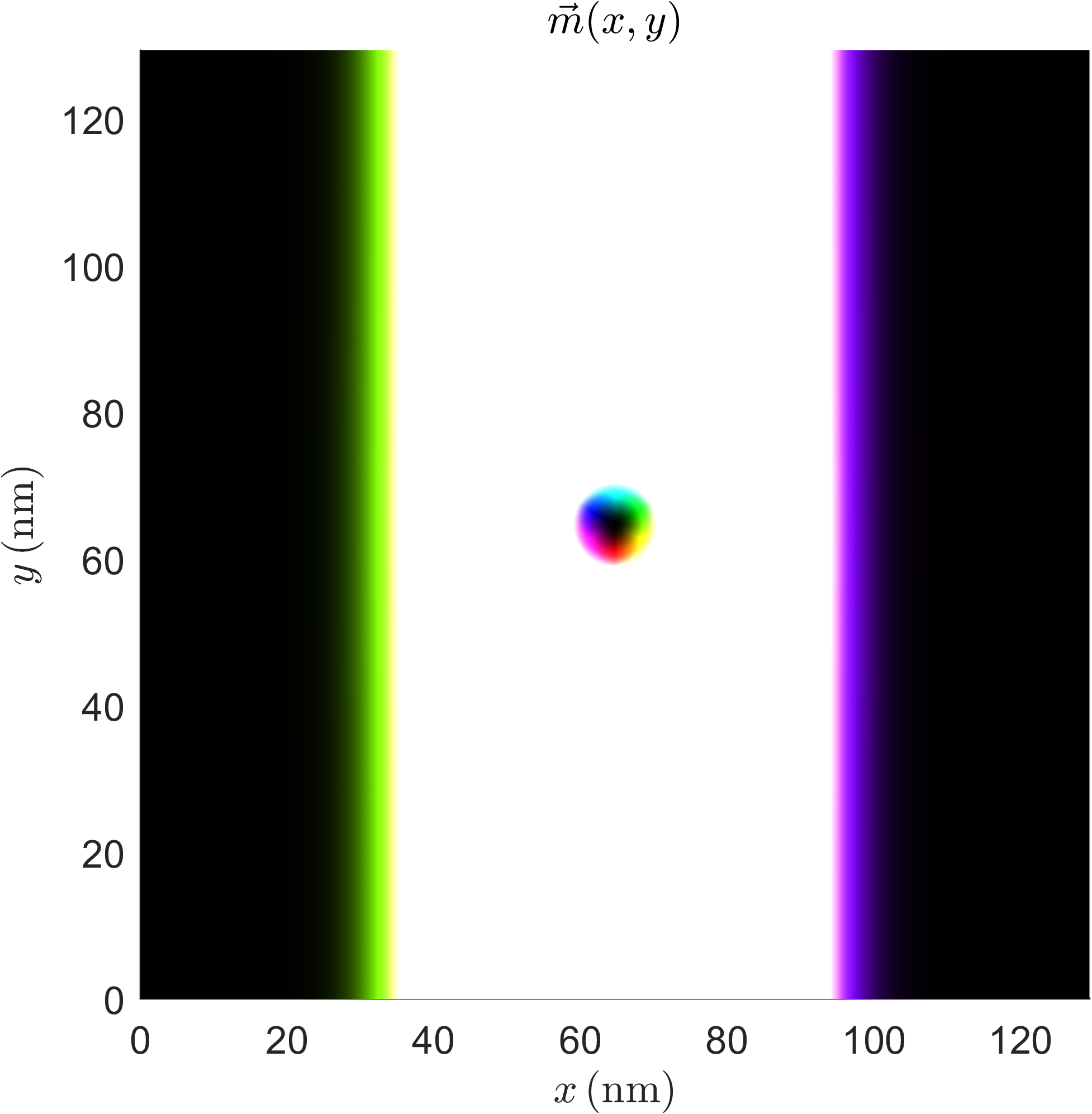} & $+1$ \\
        $(3\pi/2,4\pi/3)$ & $+1$ & \includegraphics[width=25mm]{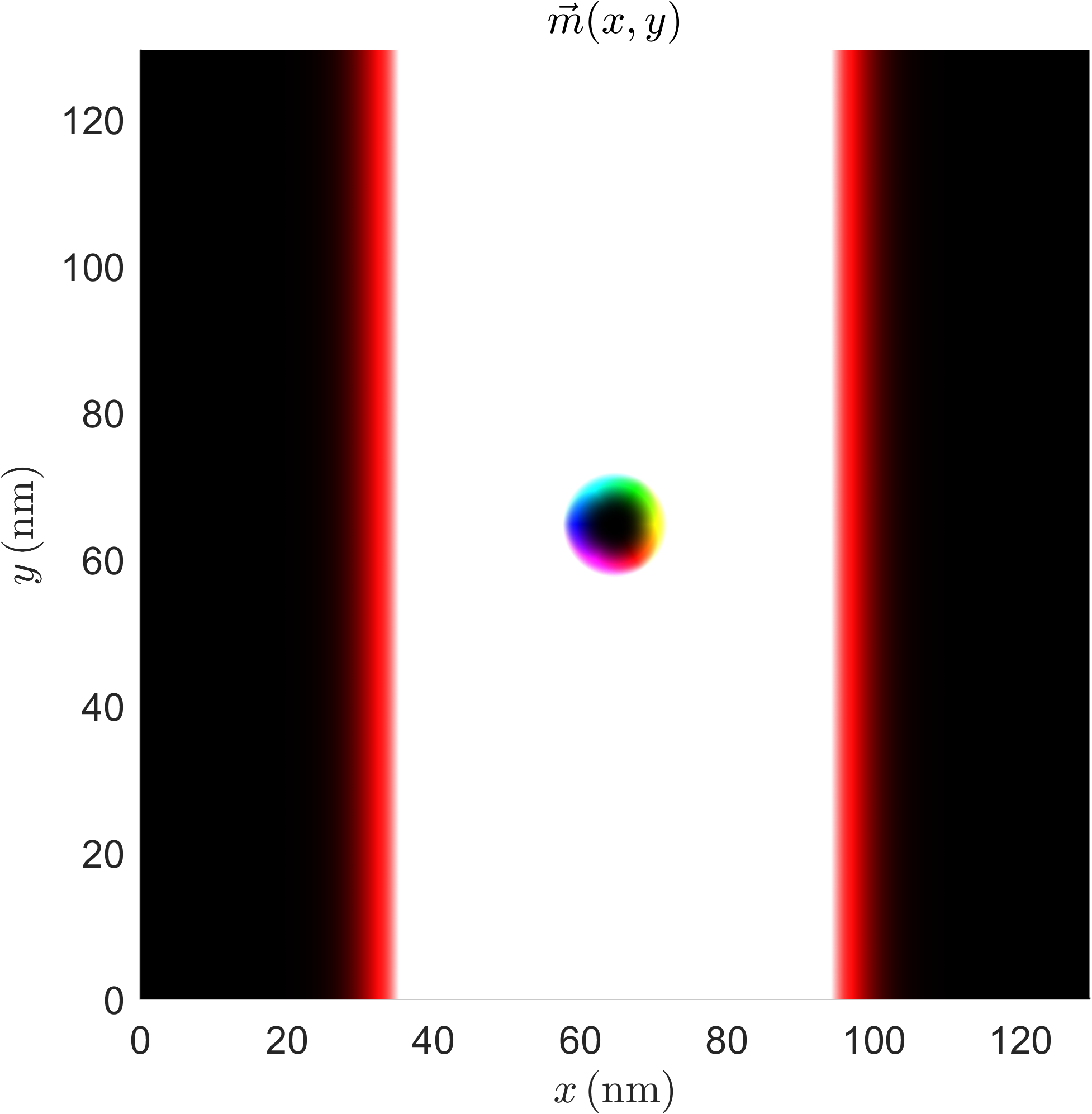} & \includegraphics[width=25mm]{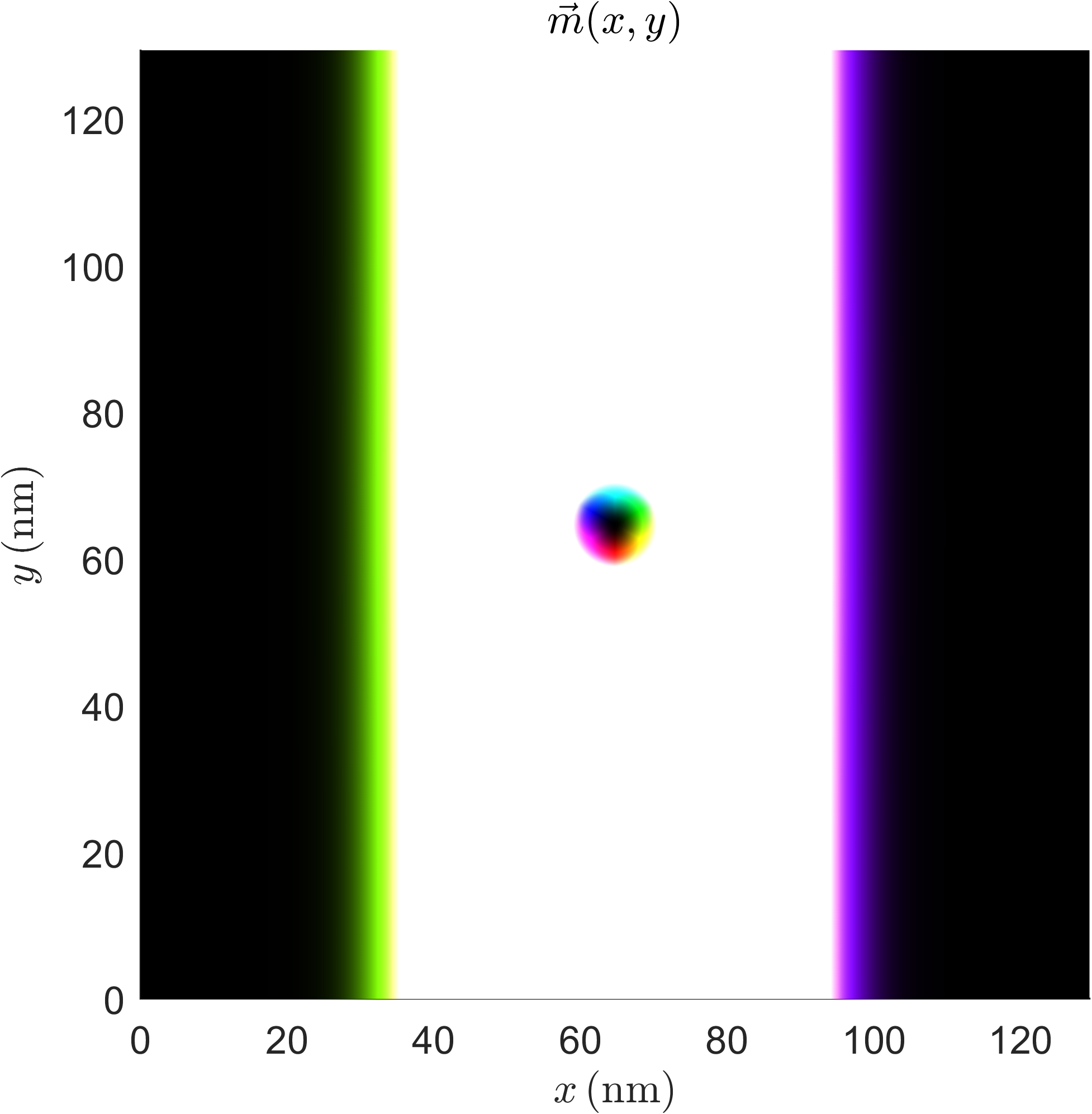} & $+1$ \\
        $(3\pi/2,3\pi/2)$ & $+1$ & \includegraphics[width=25mm]{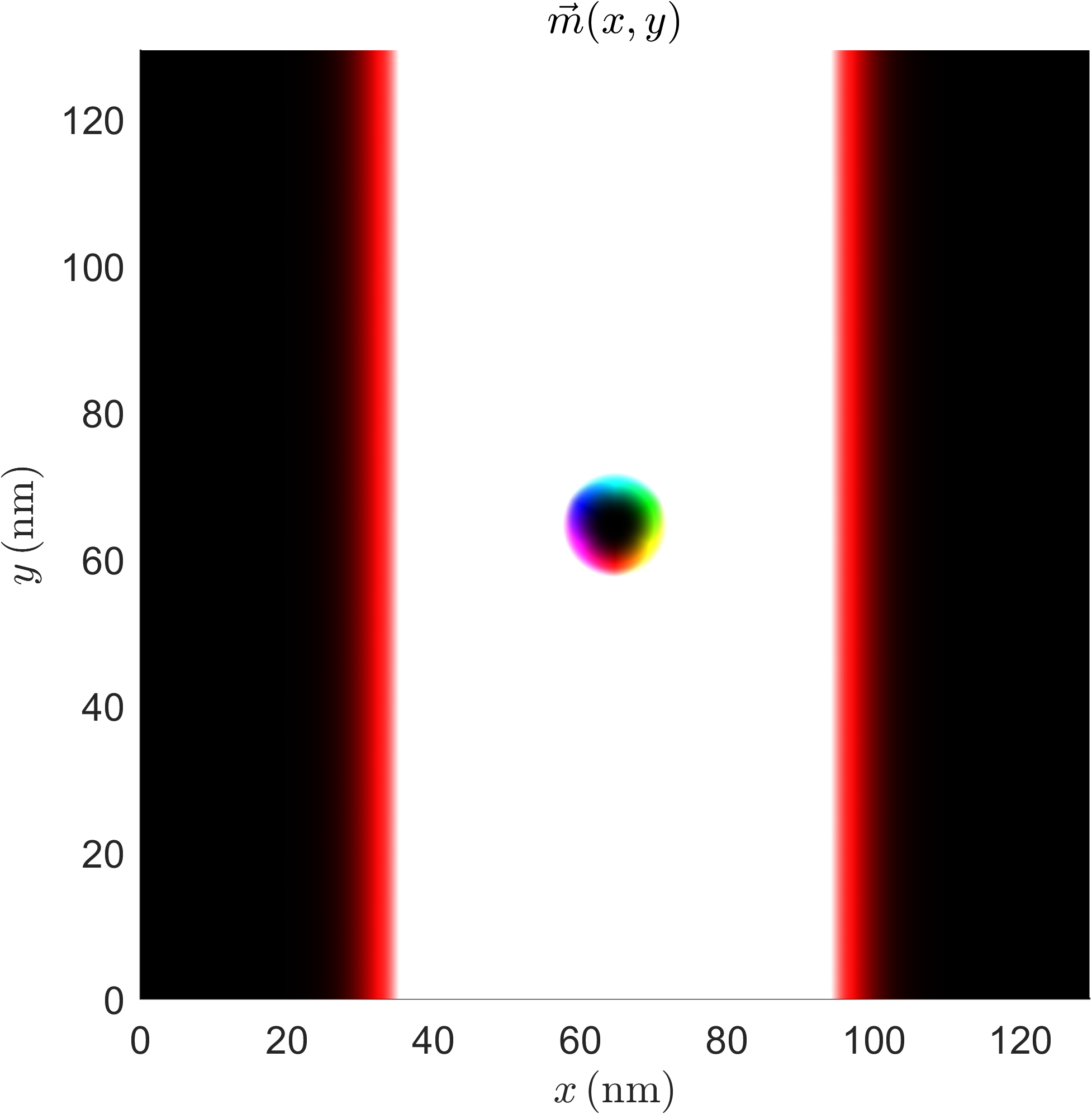} & \includegraphics[width=25mm]{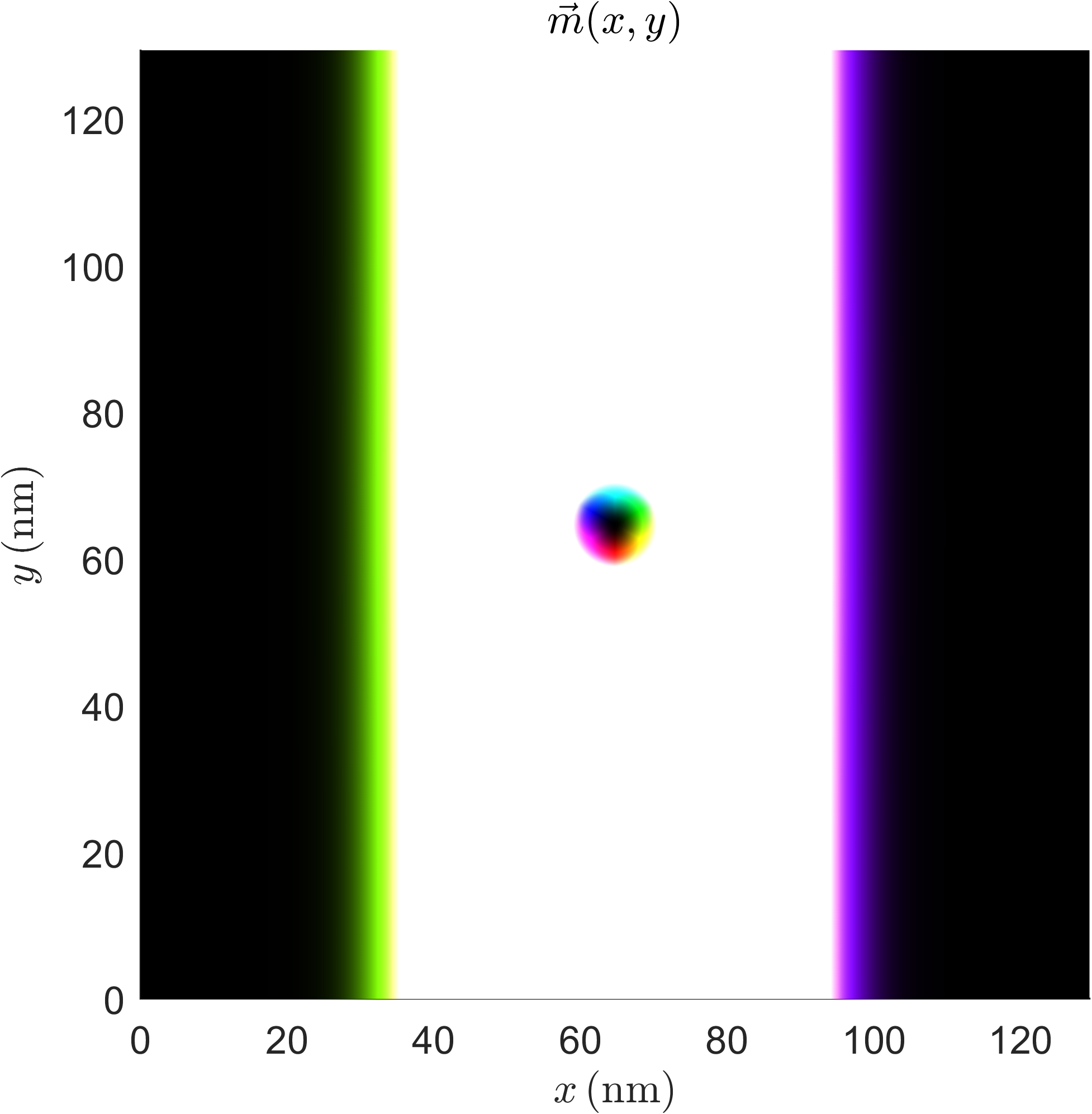} & $+1$ \\
        $(3\pi/2,5\pi/3)$ & $+1$ & \includegraphics[width=25mm]{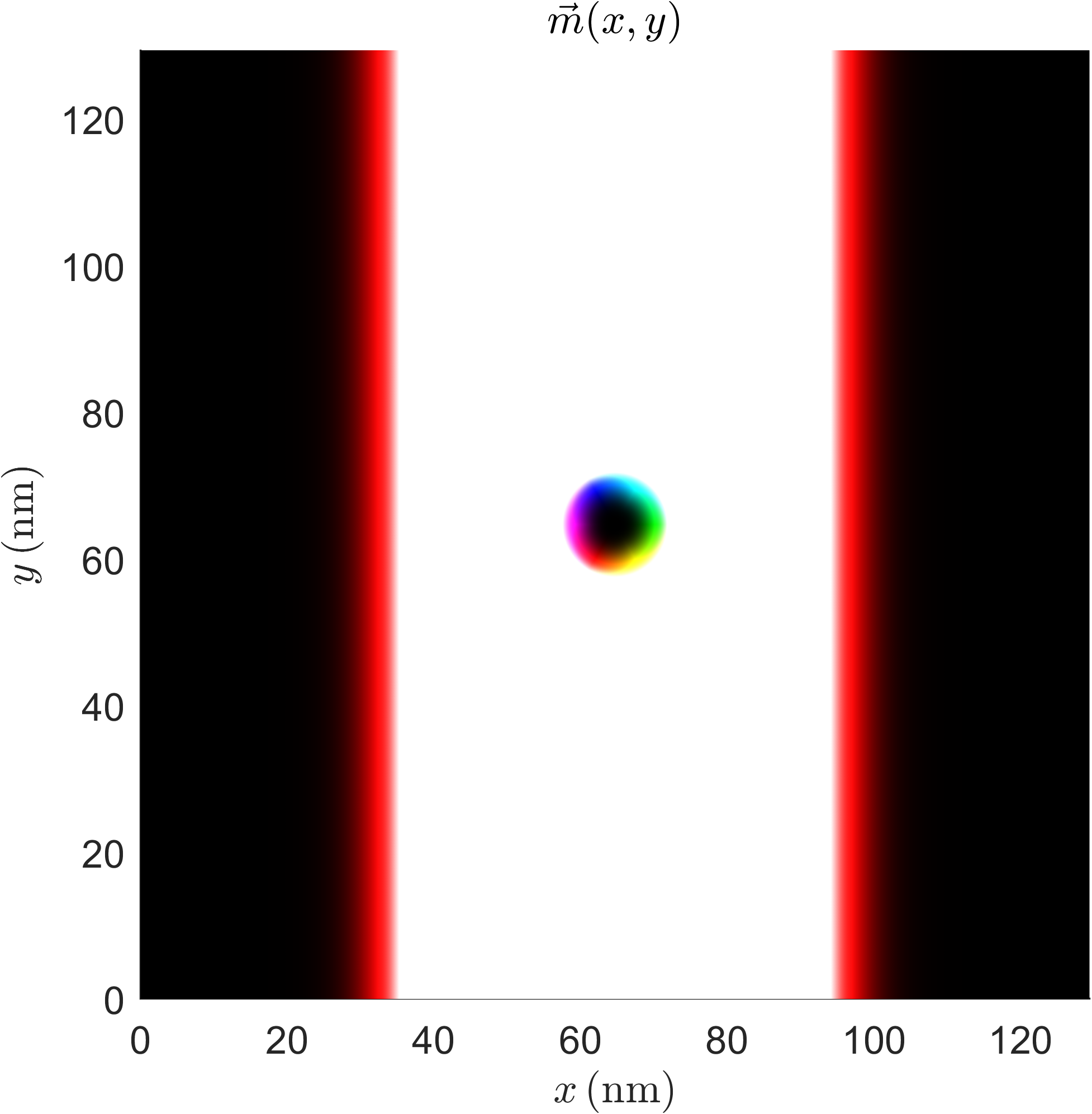} & \includegraphics[width=25mm]{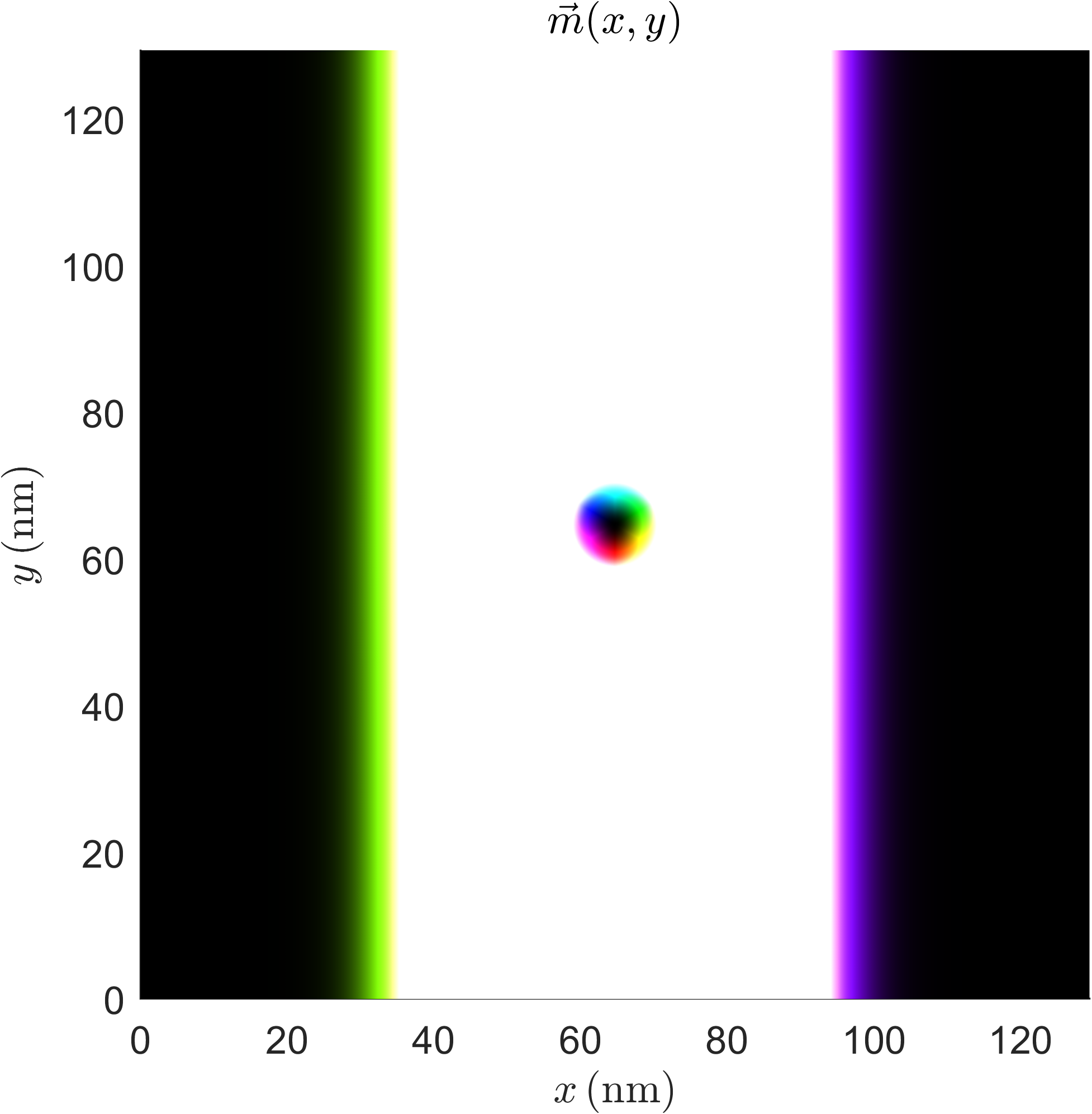} & $+1$ \\
        \bottomrule
    \end{tabular}
    \caption{Initial and final states for the domain wall phase $\chi=3\pi/2$ as the skyrmion is rotated by $\pi/3$ from $\phi=0$ until $\phi=5\pi/3$.}
    \label{tbl: chi = 3pi/2}
\end{table}

\begin{table}
    \centering
    \begin{tabular}{ccM{40mm}M{40mm}c}
        \toprule
        $(\chi,\phi)$ & $Q_{\textup{i}}$ & Initial Configuration & Final Configuration & $Q_{\textup{f}}$ \\
        \midrule
        $(5\pi/3,0)$ & $+1$ & \includegraphics[width=25mm]{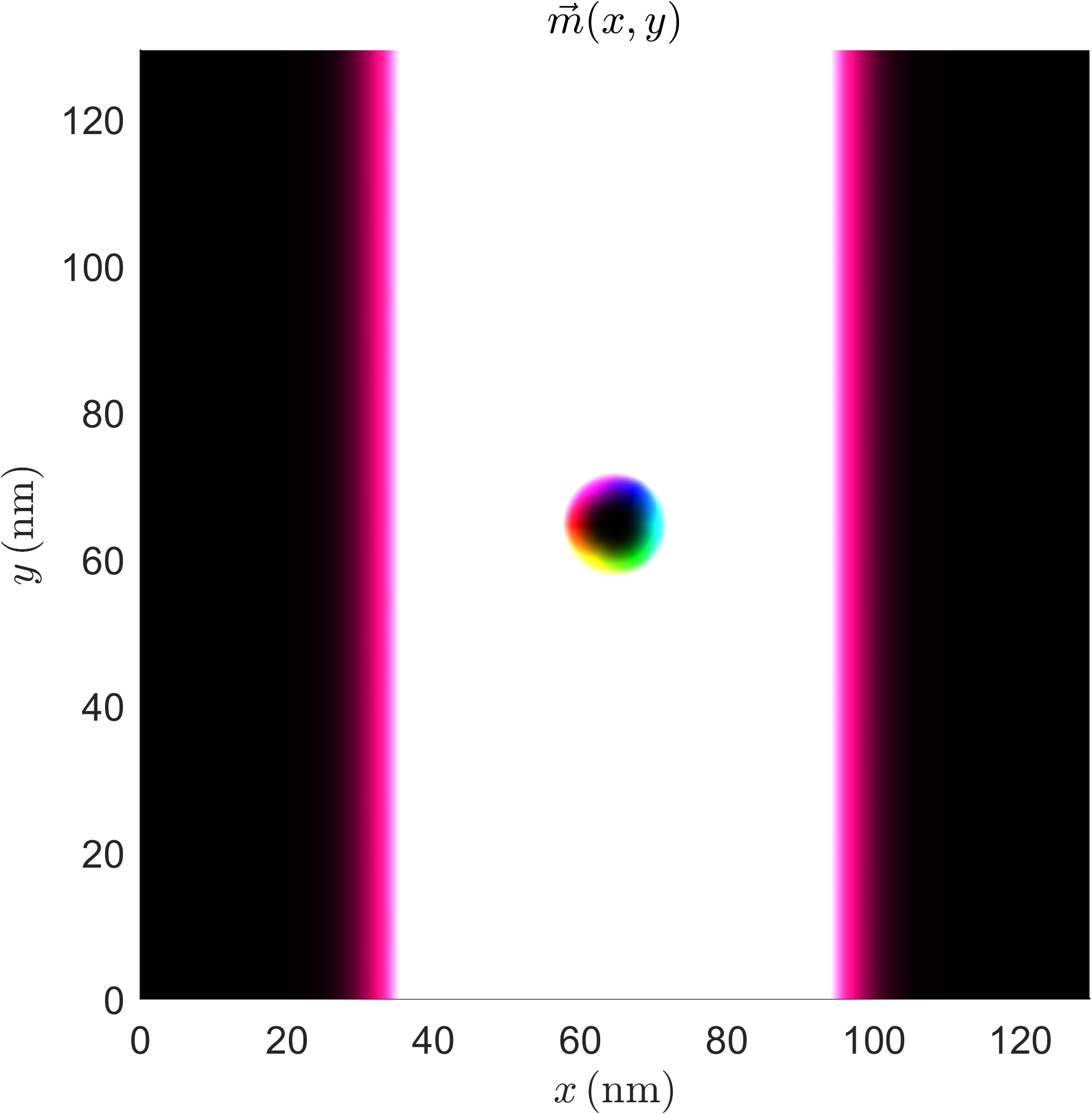} & \includegraphics[width=25mm]{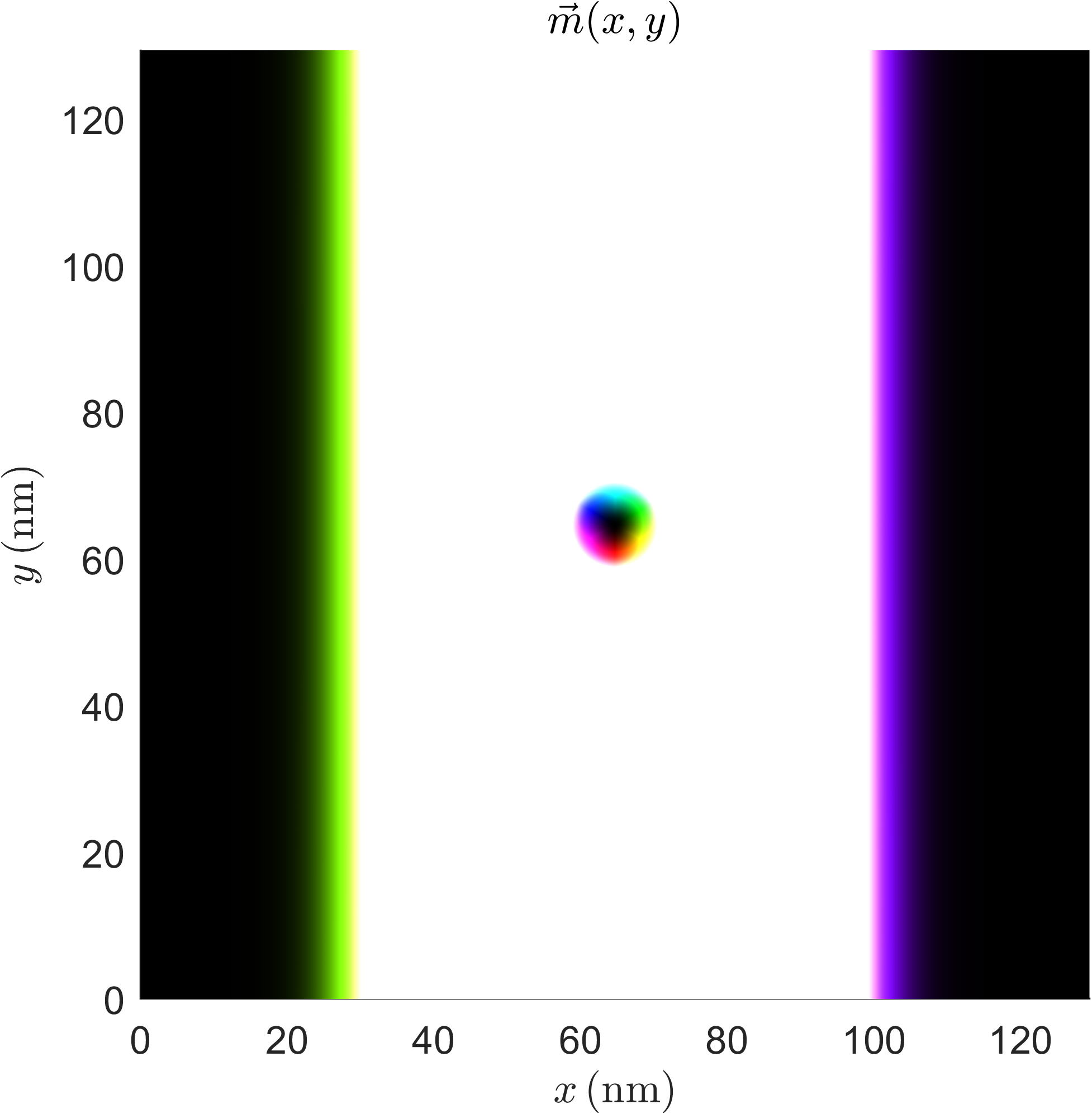} & $+1$ \\
        $(5\pi/3,\pi/3)$ & $+1$ & \includegraphics[width=25mm]{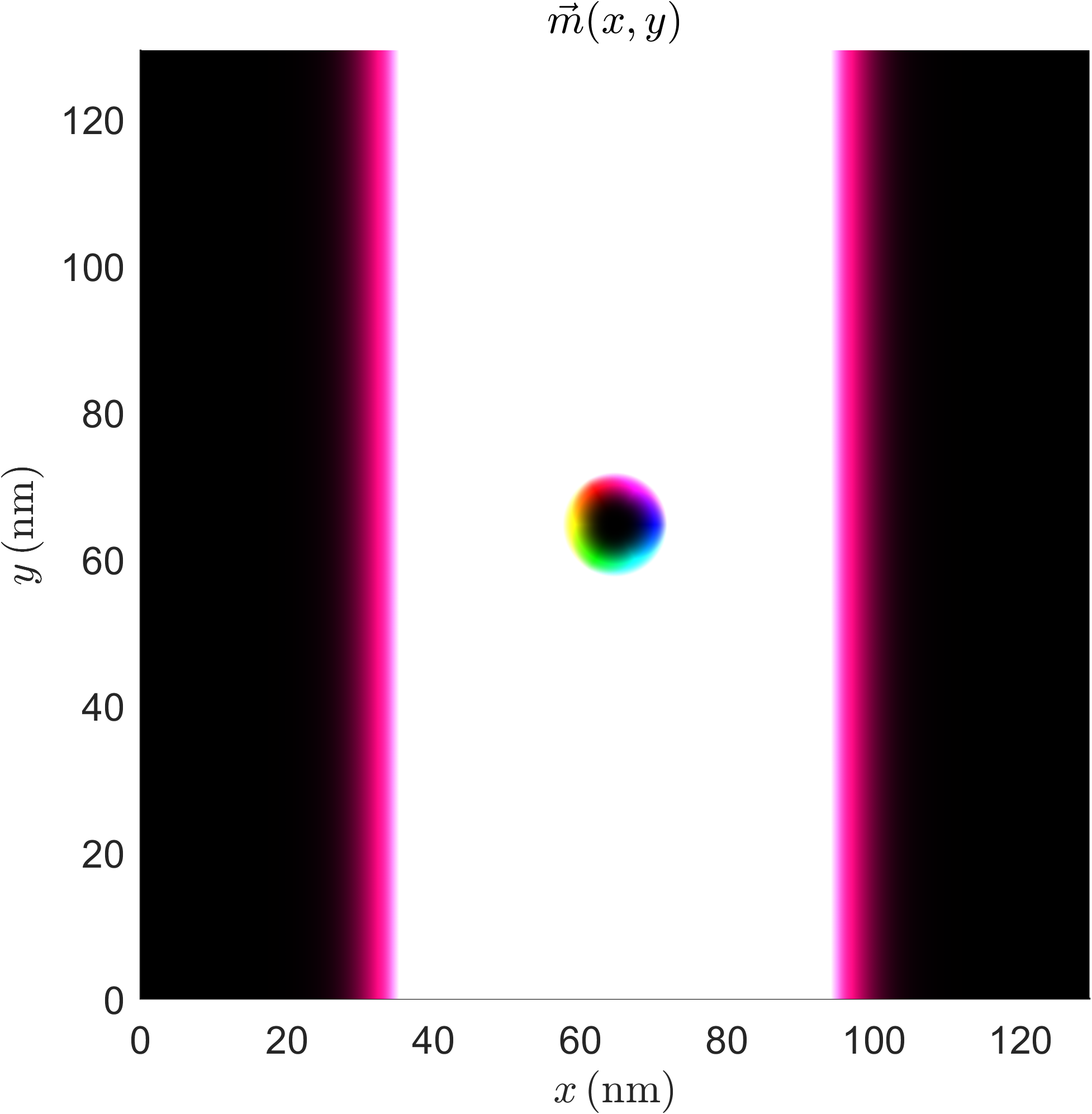} & \includegraphics[width=25mm]{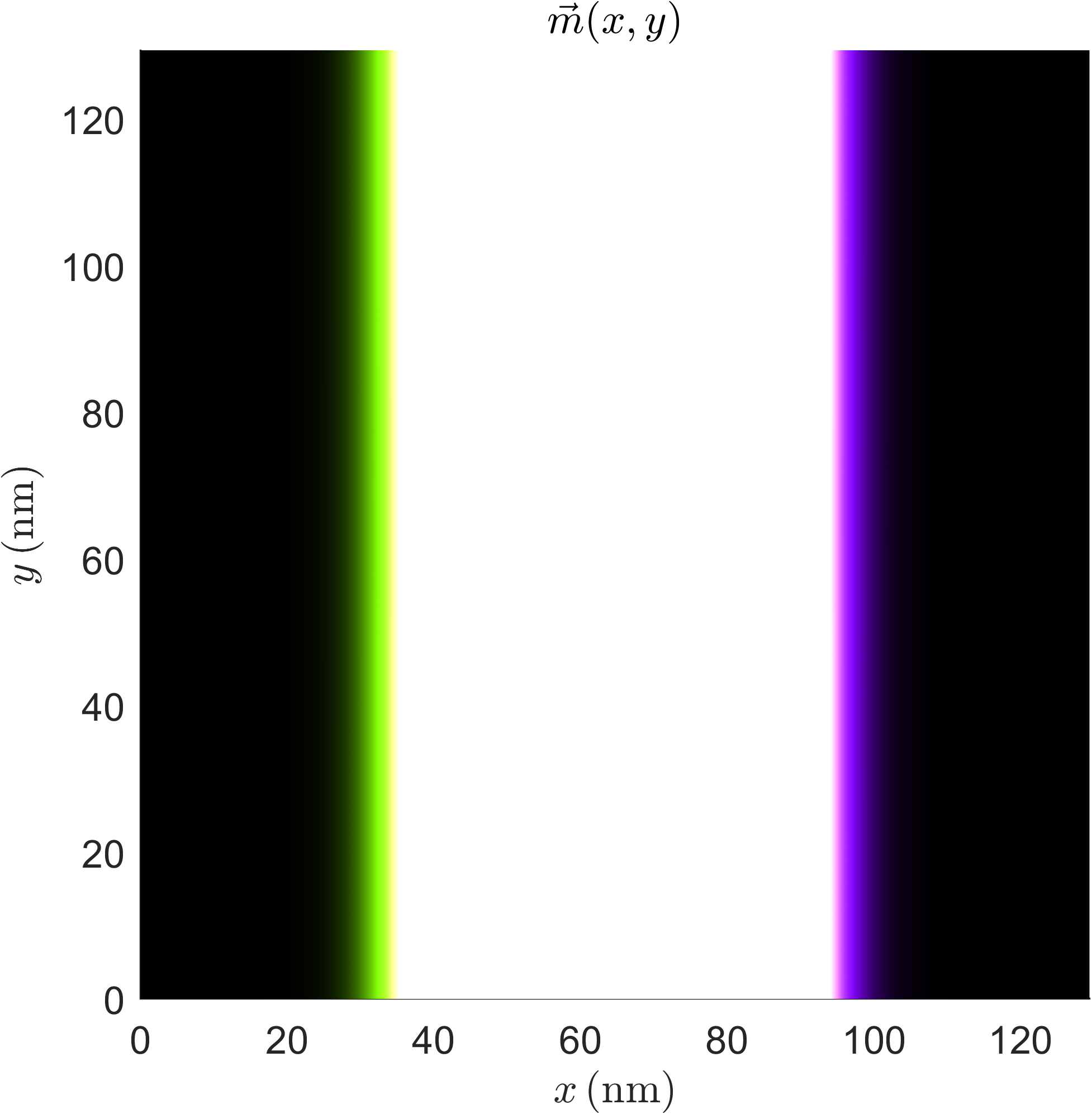} & $0$ \\
        $(5\pi/3,\pi/2)$ & $+1$ & \includegraphics[width=25mm]{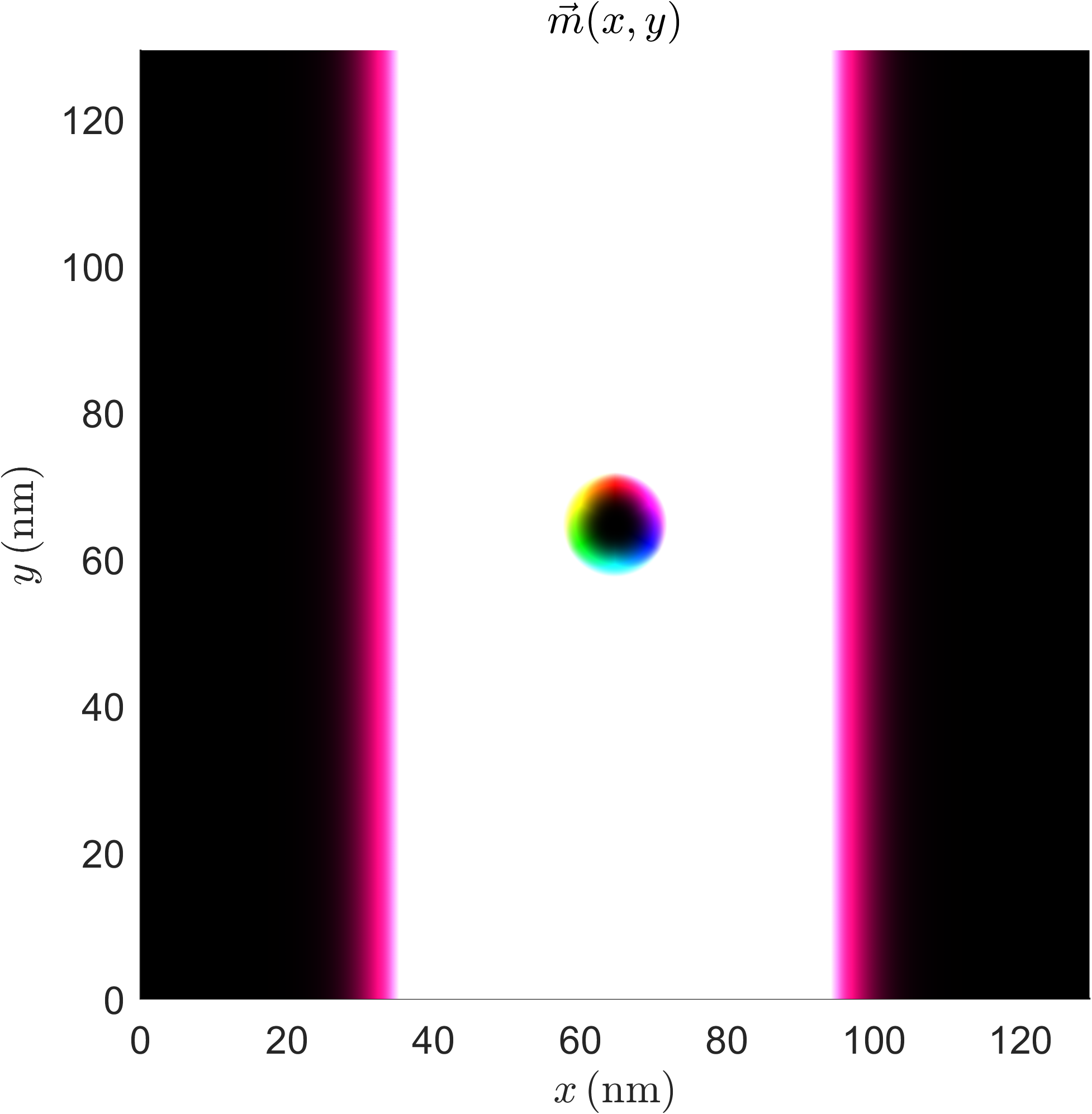} & \includegraphics[width=25mm]{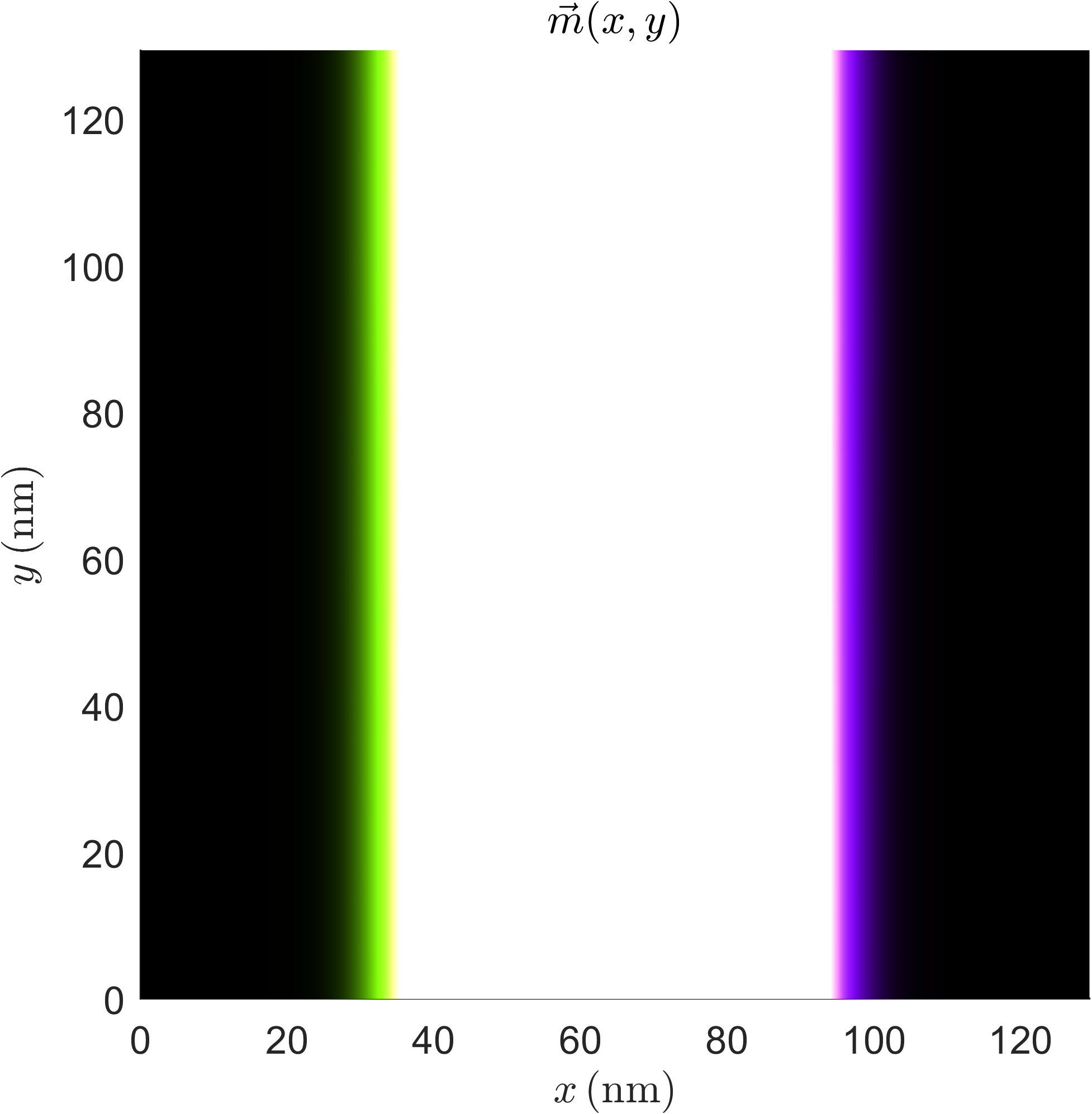} & $0$ \\
        $(5\pi/3,2\pi/3)$ & $+1$ & \includegraphics[width=25mm]{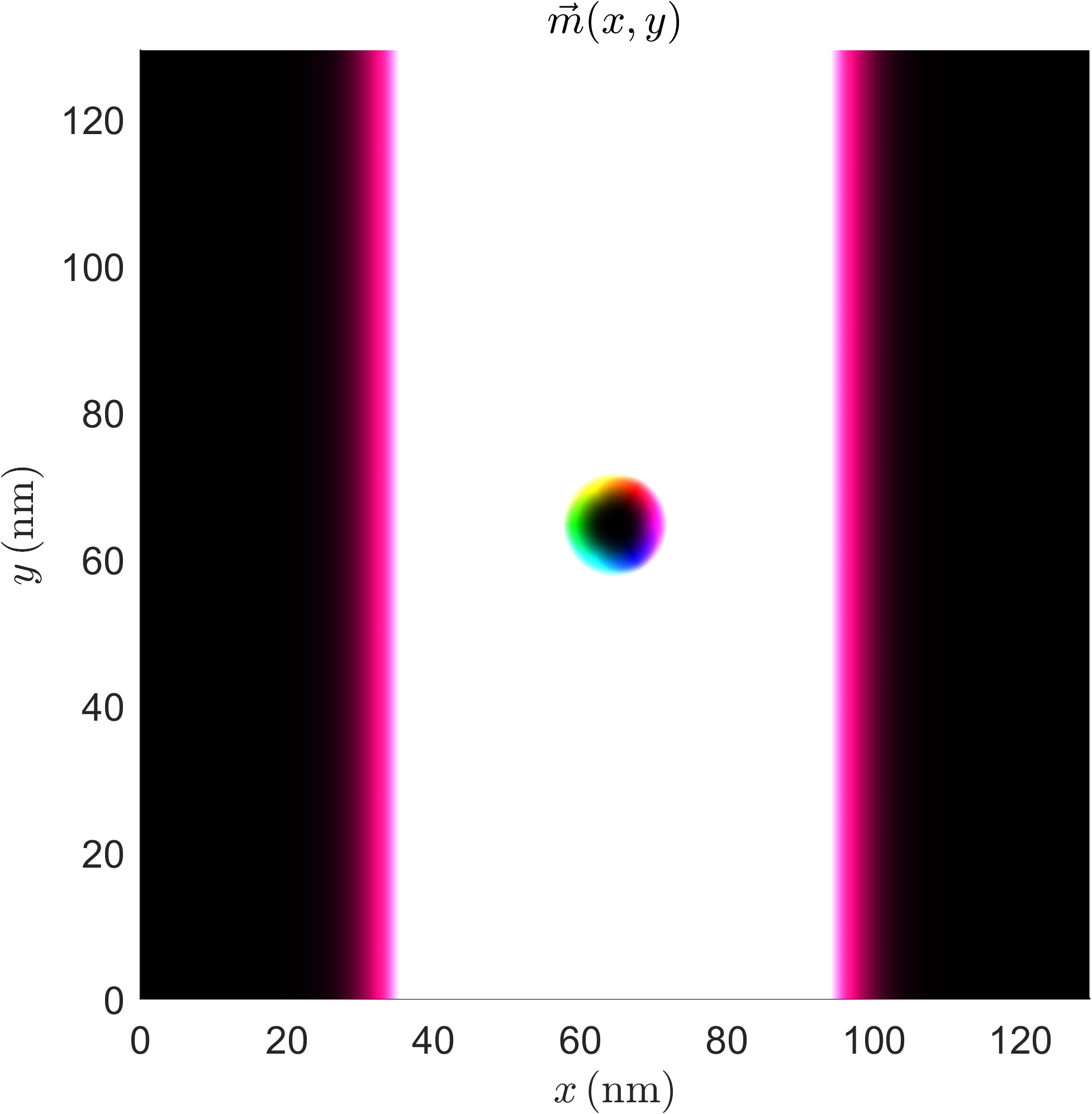} & \includegraphics[width=25mm]{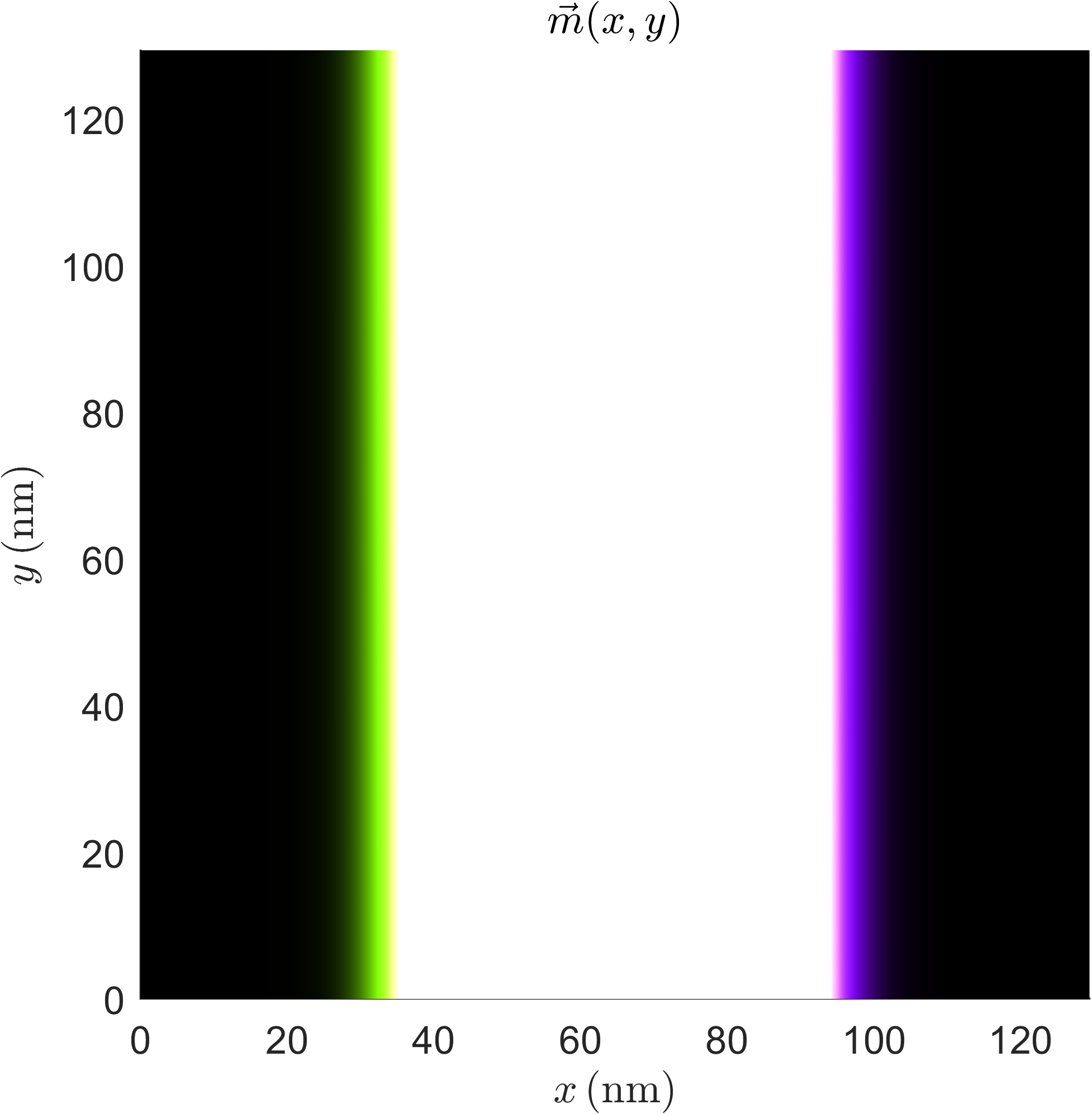} & $0$ \\
        $(5\pi/3,\pi)$ & $+1$ & \includegraphics[width=25mm]{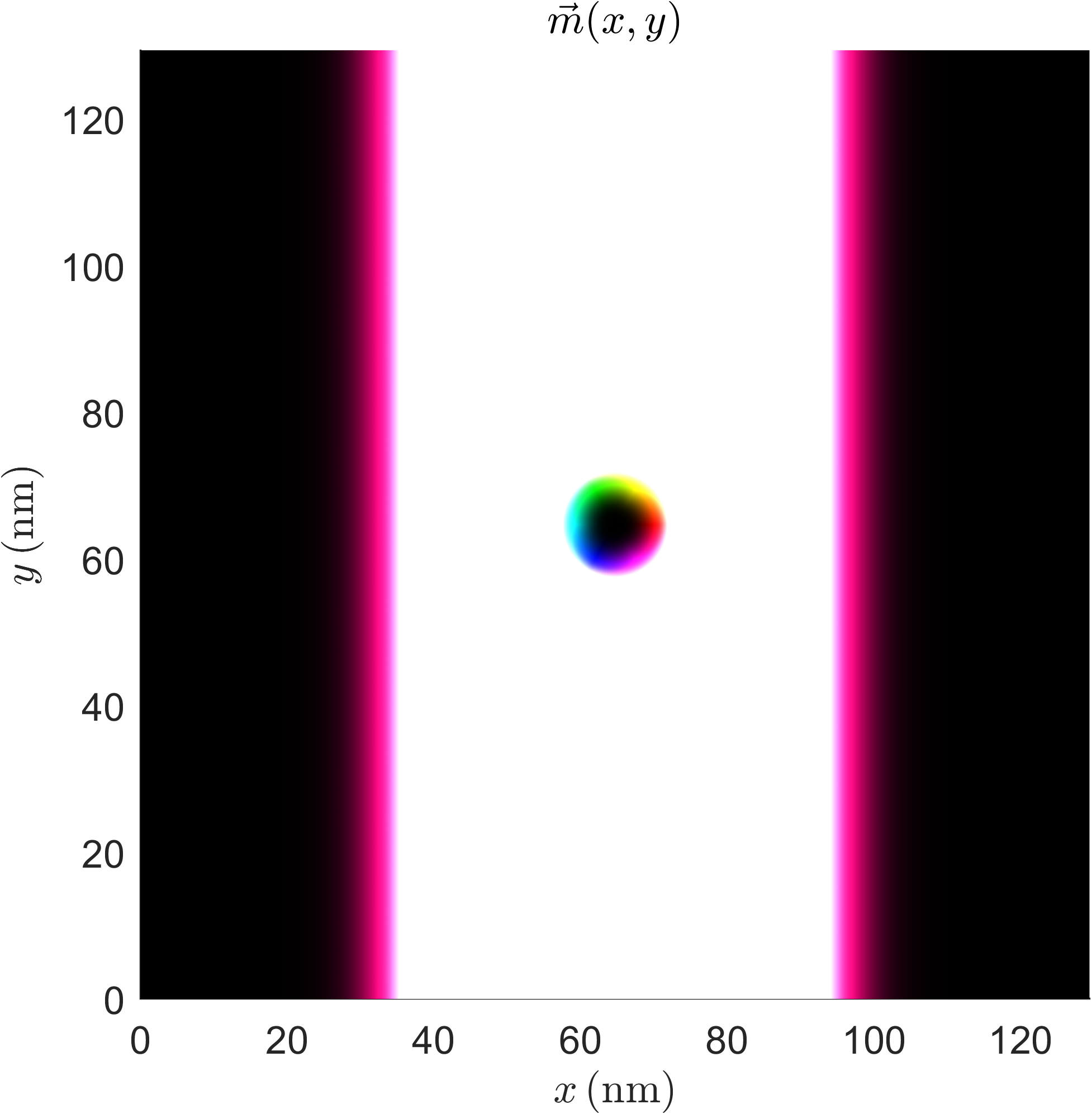} & \includegraphics[width=25mm]{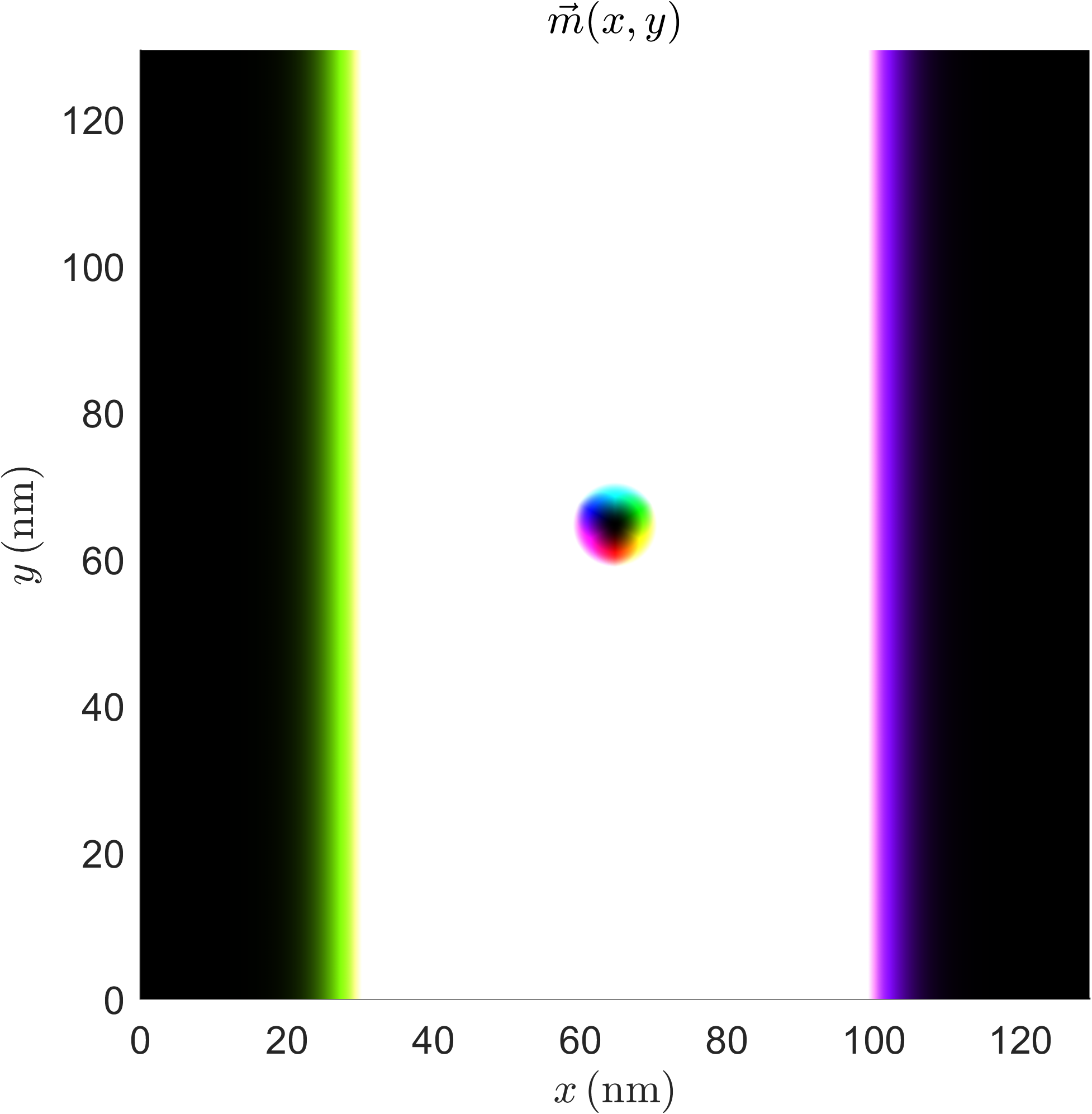} & $+1$ \\
        $(5\pi/3,4\pi/3)$ & $+1$ & \includegraphics[width=25mm]{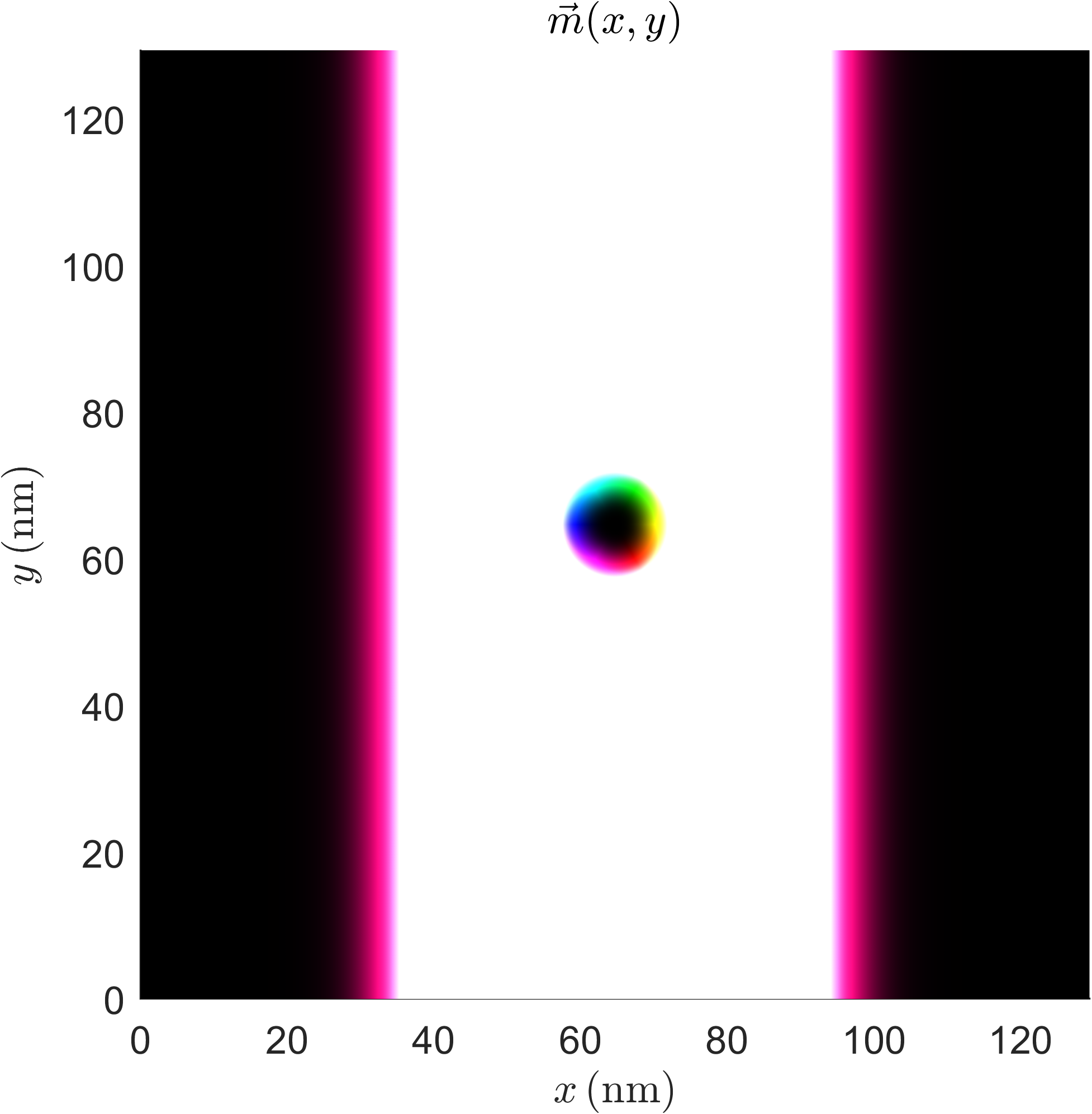} & \includegraphics[width=25mm]{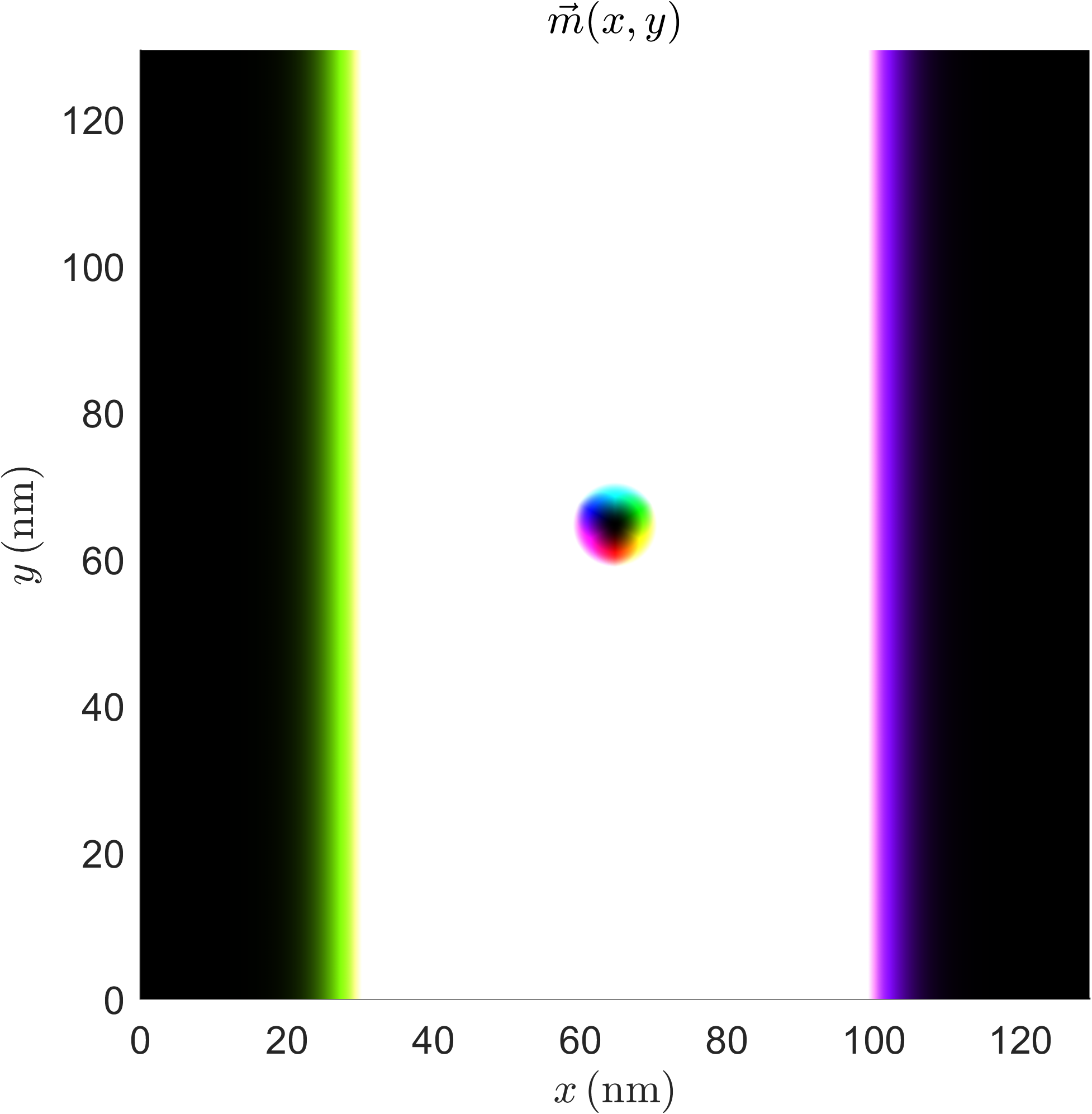} & $+1$ \\
        $(5\pi/3,3\pi/2)$ & $+1$ & \includegraphics[width=25mm]{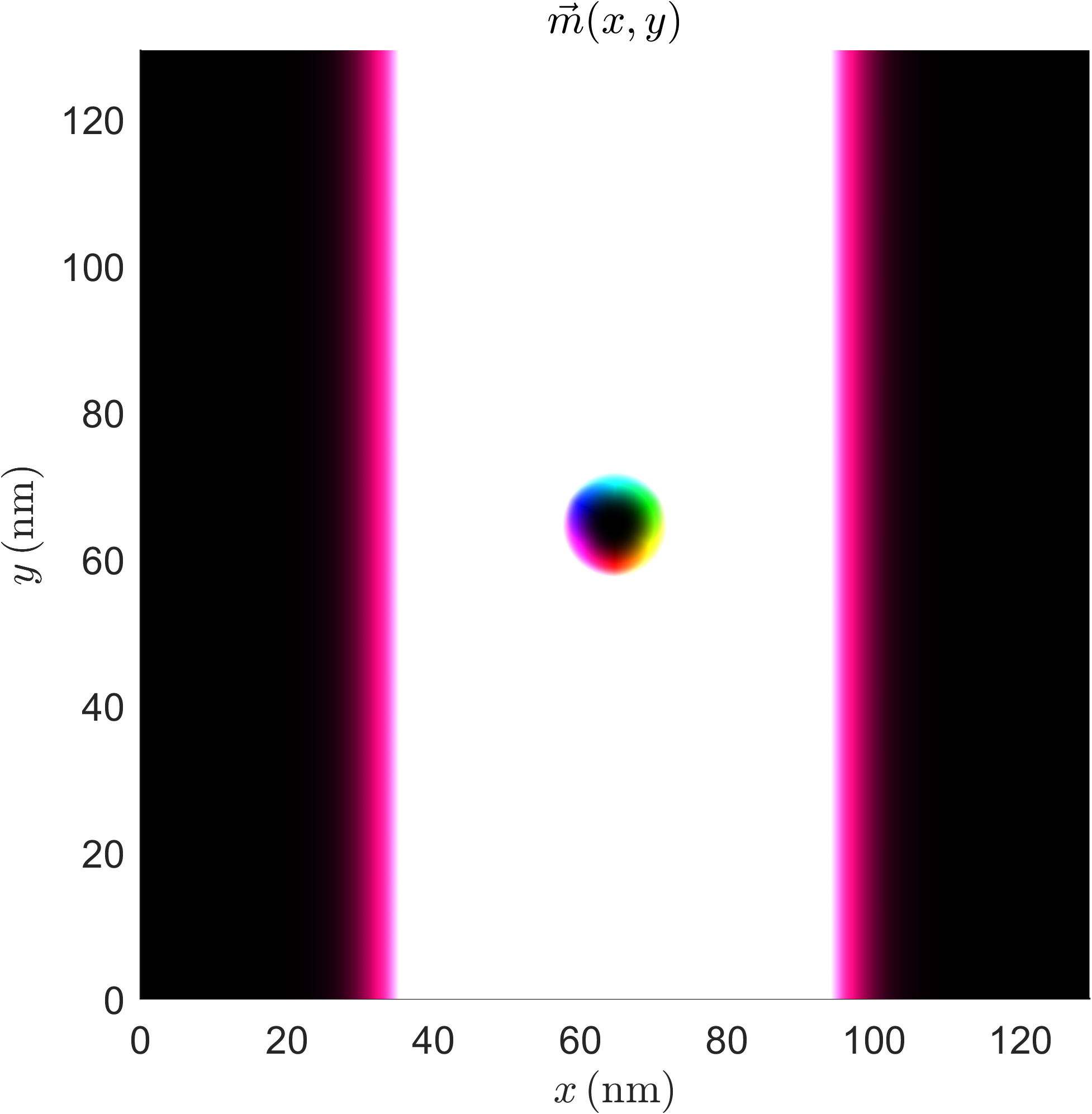} & \includegraphics[width=25mm]{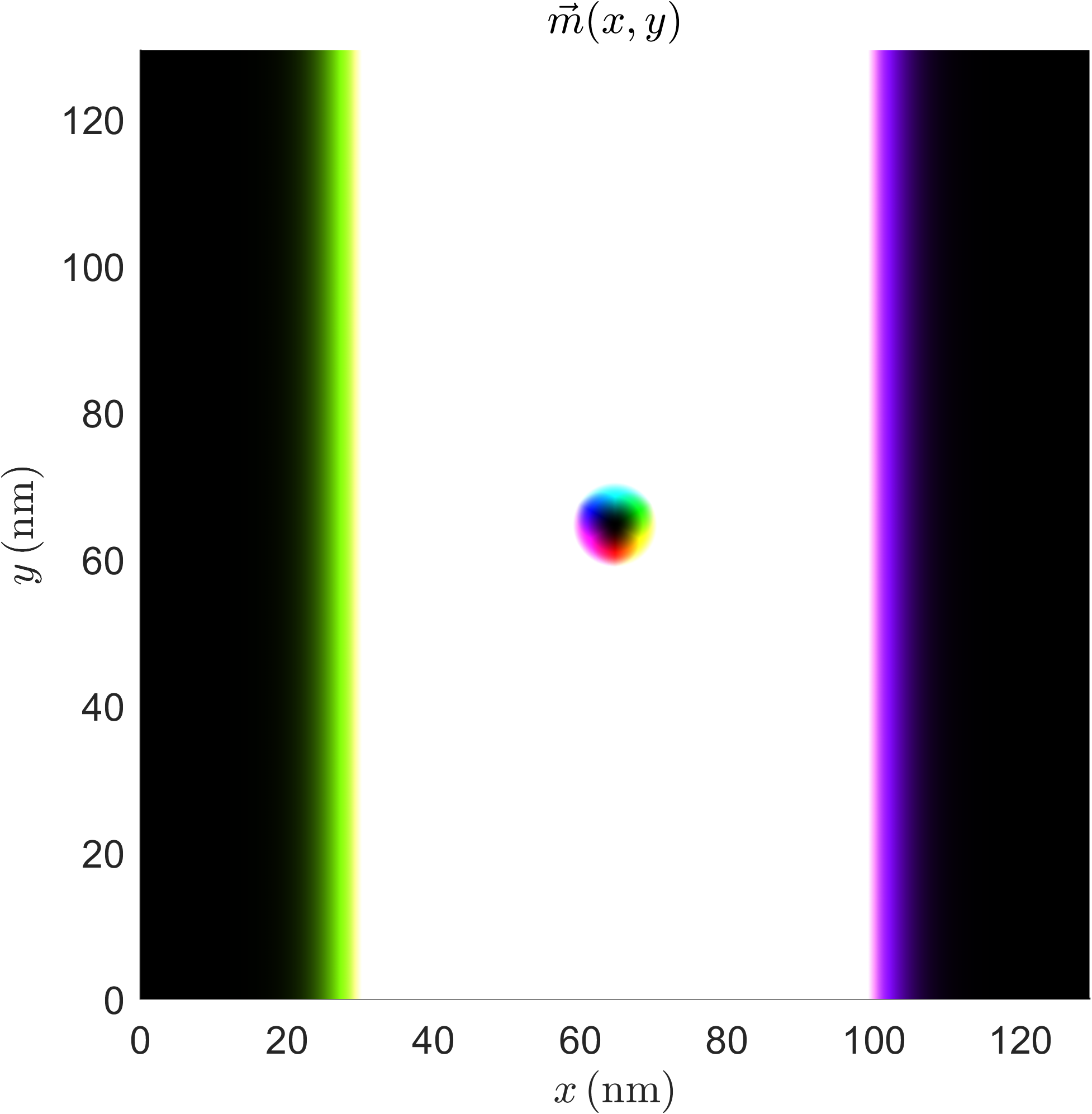} & $+1$ \\
        $(5\pi/3,5\pi/3)$ & $+1$ & \includegraphics[width=25mm]{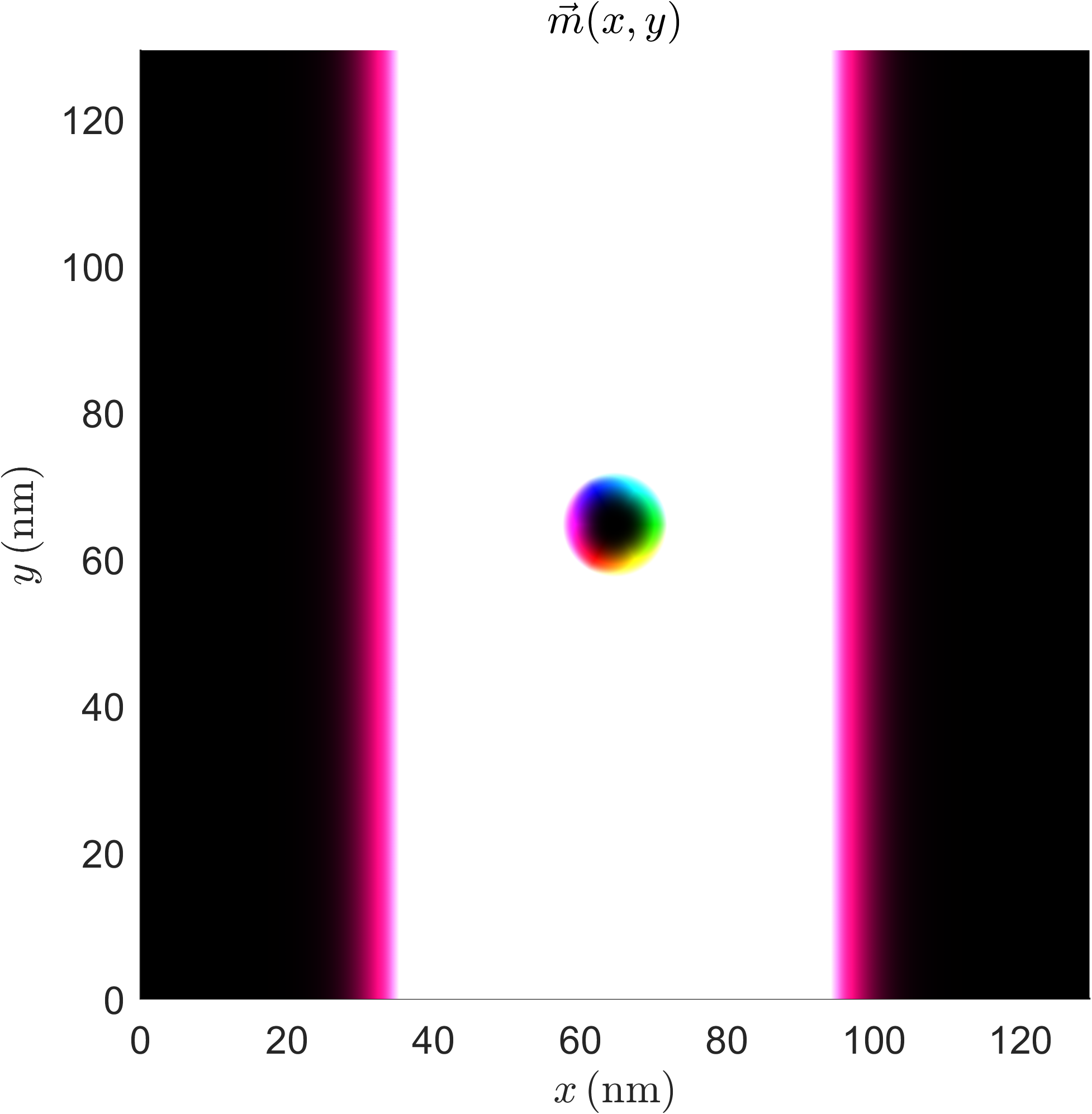} & \includegraphics[width=25mm]{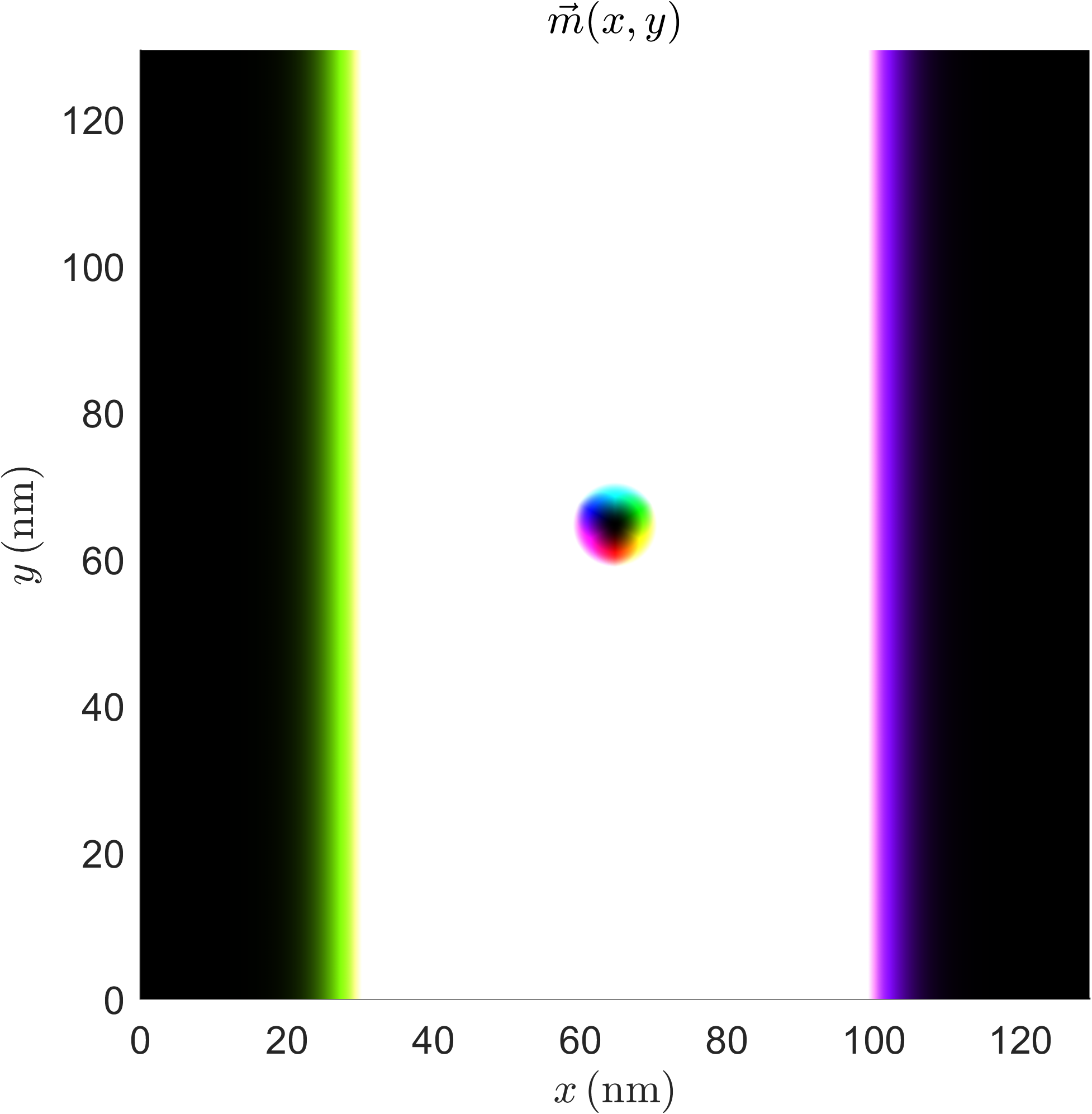} & $+1$ \\
        \bottomrule
    \end{tabular}
    \caption{Initial and final states for the domain wall phase $\chi=5\pi/3$ as the skyrmion is rotated by $\pi/3$ from $\phi=0$ until $\phi=5\pi/3$.}
    \label{tbl: chi = 5pi/3}
\end{table}


\bibliographystyle{JHEP.bst}
\bibliography{bib.bib}

\end{document}